\documentclass[12pt]{article}
\usepackage{a4wide}
\usepackage{amssymb,amsmath}
\usepackage{graphicx}
\begin{document}
{\renewcommand{\thefootnote}{\fnsymbol{footnote}}
\begin{center}
{\LARGE  Emergent modified gravity coupled to scalar matter }\\
\vspace{1.5em}
Martin Bojowald\footnote{e-mail address: {\tt bojowald@psu.edu}}
and Erick I.\ Duque   \footnote{e-mail address: {\tt eqd5272@psu.edu}} 
\\
\vspace{0.5em}
Institute for Gravitation and the Cosmos,\\
The Pennsylvania State
University,\\
104 Davey Lab, University Park, PA 16802, USA\\
\vspace{1.5em}
\end{center}
}

\setcounter{footnote}{0}

\begin{abstract}
  Emergent modified gravity presents a new set of generally covariant
  gravitational theories in which the space-time metric is not directly given
  by one of the fundamental fields. A metric compatible with the modified
  dynamics of gravity is instead derived from covariance conditions for
  space-time in canonical form. By staying within the canonical setting
  throughout all the required steps, several assumptions about space-time made
  implicitly in modified action principles can be relaxed. This paper presents
  a significant extension of existing vacuum models to the case of a scalar
  field coupled to emergent modified gravity in a spherically symmetric setting. Unlike
  in previous attempts for instance in models of loop quantum gravity, it is
  possible to maintain general covariance in the presence of modified
  gravity-scalar couplings. In general, however, the emergent space-time
  metric depends not only on the phase-space degrees of freedom of the
  gravitational part of the coupled theory, but also on the scalar
  field. Matter therefore directly and profoundly affects the geometry of
  space-time, not only through the well-known dynamical coupling of
  stress-energy to curvature as in Einstein's equation, but even on a
  kinematical level before equations of motion are imposed. In addition to the
  covariance condition, this paper introduces further physical requirements
  that may be imposed in order to reduce modified gravity-scalar theories to
  more specific classes. In some cases, coupling emergent modified gravity to
  a scalar field eliminates some of the modifications that would be possible
  in a vacuum situation. Moreover, certain results about the removal of
  classical black-hole singularities in vacuum emergent modified gravity are
  found to be unstable under the inclusion of matter fields. However,
  alternative modifications exist in which singularities are removed even in
  the presence of matter. Emergent modified gravity is seen to provide a large
  class of new scalar-tensor theories with second-order field equations.
\end{abstract}

\section{Introduction}

The search for modified theories of gravity is motivated by both observational
considerations as well as deep theoretical developments. Examples of the
former are the desire to compare the increasing number of strong-field
measurements with a sufficiently large class of consistent parameterized
theoretical descriptions of black holes, or to explain puzzling cosmological
features such as dark matter and dark energy.The latter are relevant in
particular in the context of quantum gravity, or in the quest to find
singularity-free models of black holes and the big bang. General covariance is
a crucial property that makes it possible to introduce a space-time
description for these phenomena and define the horizon of a black hole or the
expanding geometry of the universe. However, general covariance applied to an
action principle for the gravitational field appears to be a strong and very
restrictive consistency condition that does not seem to allow sufficiently many
interesting and viable alternatives to general relativity
\cite{MultiMess1,MultiMess2,MultiMess3,MultiMess4}. 

Emergent modified gravity \cite{HigherCov,SphSymmMinCoup} presents a new
version that can include modifications of general relativity even without
going beyond second derivative order and without including extra fields, so
far at least in a spherically symmetric setting. Its key observation is that
the metric used to describe space-time geometrically need not be one of the
fundamental fields, which allows us to weaken some of the usual assumptions
that lead for instance to higher-curvature effective actions as the main
source of generally covariant modifications of vacuum general
relativity. Moreover, emergent modified gravity works on a canonical level and
therefore does not require assumptions about space-time integrations and
4-volume measures. It provides a new source of modified gravity for
phenomenological studies, and it helps to analyze questions such as whether
proposed quantum effects, for instance in models of loop quantum gravity
\cite{ReviewEff}, have a chance of being consistent with space-time
covariance. Emergent modified gravity is well suited to the latter
applications because it is inherently canonical, and it is able to test the
covariance for instance of holonomy modifications suggested by the eponymous
loop integrations in loop quantum gravity \cite{LoopRep}.

While consistency conditions for a space-time formulation within canonical
quantum gravity, such as having first-class constraints that may generate
hypersurface deformations, have been imposed to varying degrees in vacuum
spherically symmetric models, attempts to include scalar matter
\cite{SphSymmCov} (or, more generally, local physical degrees of freedom
\cite{GowdyCov}) had for some time led only to no-go results. This outcome
presents a major challenge to models of canonical or loop quantum gravity, not
only conceptually because including matter with local degrees of freedom is
important to test whether proposed modifications have a chance of being
sufficiently general for physical applications, but also for important
practical questions of how to study matter collapse or Hawking radiation in
the presence of such modifications. For instance, if quantum effects may avoid
the classical black-hole singularity only in vacuum models, they would be of
little use when it comes to the physical question of stellar
collapse. Recently, the constructions in \cite{SphSymmMatter,SphSymmMatter2}
showed that the first-class nature of spherically symmetric constraints can be
maintained in the presence of matter if a specific coupling term to spatial
derivatives of the gravitational momenta is included. Such terms are not
directly suggested by loop quantum gravity and therefore had been omitted in
earlier attempts.

The canonical analysis underlying emergent modified gravity includes the
conditions that the gravitational constraints remain first class, but
\cite{HigherCov}, building on \cite{EffLine}, has also shown that this
property is not sufficient for the theory to be consistent with a geometrical
space-time interpretation of its solutions. First-class modifications of the
constraints in general imply modifications not only of the constraint
functionals themselves but also of the structure function in their Poisson
brackets. Classically, and in all standard higher-curvature effective actions,
this structure function equals the inverse spatial part of a space-time metric
compatible with its solutions, reflecting a general geometrical property of
deformations of embedded hypersurfaces. If the structure function is modified,
the compatible space-time geometry in which hypersurfaces can be embedded must
therefore be adapted to the new theory; it must be derived from the modified
constraints through the structure function. Classically, the structure
function is closely related to the configuration variables among the
gravitational phase-space degrees of freedom, given by metric or triad
components, but this need not be the case in a modified theory in which the
structure function could also depend on the gravitational momenta. Even if the
modified constraints are first class, it is not guaranteed that the modified
structure function can be part of a consistent space-time metric. There are
therefore first-class modifications of the classical gravitational constraints
that are not compatible with a covariant space-time interpretation.

Imposing the condition that modified canonical solutions can be used to
describe space-time geometrically therefore goes beyond the algebraic
condition that the constraints remain first class. In spherically symmetric
models, including scalar matter, the first-class property has been analyzed in
\cite{SphSymmMatter,SphSymmMatter2}, but the condition that the structure
function be compatible with a space-time interpretation remains to be
analyzed. The recent \cite{SphSymmMinCoup} proposed a minimal coupling for
scalar matter to modified gravity in canonical form, but it is not
sufficiently general to encompass all possibilities of modified structure
functions of interest in emergent modified gravity. Moreover, the possibility
of minimal coupling for momentum-dependent structure functions is non-trivial
and requires a proof of existence, which we provide in this paper as a
corollary of our general theory.

We present the required analysis for covariant scalar-field couplings in
spherically symmetric emergent modified gravity in the main part of this
paper, with several surprising outcomes. In particular, even in the presence
of matter with local degrees of freedom it is still possible to find new
versions of emergent modified gravity that are not of higher-curvature
form. The emergent space-time geometry is determined by a line element whose
components, expressed as functions of the original phase-space degrees of
freedom, depend on the gravitational as well as matter fields. It is therefore
impossible to separate the geometrical roles of gravitational and matter
degrees of freedom on phase space, as initially defined by their appearance in
different contributions to the constraints. Instead, for a given theory of
modified gravity, the covariance condition predicts a unique combination of
these fields that can serve as the spatial part of a space-time line element.

In Section~\ref{sec:Covariance in modified gravity}, we review the vacuum
covariance conditions for a consistent space-time geometry in emergent
modified gravity, following \cite{HigherCov}, and formulate a new covariance
condition for the scalar field. The same section contains our proof that
minimal coupling of a scalar field is consistent in emergent modified
gravity. We formulate and discuss several additional requirements of physical
interest in Section~\ref{sec:Conditions on the modified theory}.  These
covariance and other conditions, specialized to spherical symmetry, are
evaluated in various combinations in Section~\ref{sec:Spherically symmetric
  sector}. In order to manage the large space of possible theories, we will
take a viewpoint of effective field theory in which generic constraints are
formulated by including terms up to a certain order in spatial derivatives,
and then subjected to several consistency conditions.
Section~\ref{sec:Classes} contains the derivation of three classes of modified
theories with different physically desirable properties, and
Section~\ref{sec:Homogeneous spacetime} discusses some of their equations of
motion and some solutions with the additional assumption of spatial
homogeneity, drawing conclusions about the potential to resolve classical
singularities.  After giving an outlook on new possibilities for the
phenomenology of scalar-tensor theories in Section~\ref{sec:Outlook}, our main
results, several characteristic properties, and possible future applications
are discussed in Section~\ref{sec:Discussion}.

\section{Covariance in canonical gravity}
\label{sec:Covariance in modified gravity}

New theories of canonical gravity can be formulated by modifying the classical
Hamiltonian constraint $H$, derived from general relativity, to a new
Hamiltonian constraint $\tilde{H}$ such that the classical expression is
obtained in a specific limit of suitable parameters. One could also try to
modify the diffoemorphism constraint $\vec{H}$, but this is not necessary if
one is interested in new space-time structures that retain the well-understood
classical structure of space on spacelike hypersurfaces. A canonical
formulation also requires a phase space, providing the variables on which the
constraints depend.  In a minimal modification, one may assume that the phase
space remains unchanged, with configuration variables $q_{ab}$ with momenta
$p^{ab}$ for gravity that, in the classical limit, equal the spatial metric
and a $q_{qb}$-dependent linear combination of extrinsic-curvature
components. We will maintain this assumption and use it to identify $q_{ab}$
and $p^{ab}$ as the gravitational variables distinct from matter degrees of
freedom. Therefore, we will not allow for higher-derivative theories that
would require an extended phase space in canonical form. However, unlike other
approaches to modified gravity (where such a condition may appear only
implicitly if they are not formulated canonically), in emergent modified
gravity we impose the relationship between the gravitational phase-space
variables and the spatial metric and extrinsic curvature of spacelike
hypersurfaces only in the classical limit. In a modified theory of this form,
there is therefore no a-priori relationship between $q_{ab}$ and a spatial
metric and between momenta $p^{ab}$ and extrinsic curvature. Such
relationships and geometrical interpretations rather have to be derived (and
therefore emerge) from covariance conditions imposed on the modified constraints.

\subsection{General theory}

A modified Hamiltonian constraint in general has a non-classical Poisson
bracket with the diffeomorphism constraint, such that covariance will be
completely removed if the modification is not chosen with sufficient care. It
is therefore necessary to restrict modified constraints to a form that
preserves the classical brackets as much as possible, implementing the
classical gauge symmetry of hypersurface deformations. The constraints must
remain first class in order to ensure that the number of independent gauge
transformations is not reduced and still equals the required number of
independent infinitesimal space-time diffeomorphisms. Moreover, the brackets
should resemble the classical brackets of hypersurface deformations in order
for a space-time interpretation to remain possible.  These condition lead to
the requirement that the modified Hamiltonian constraint $\tilde{H}$ together
with the classical diffeomoprhism constraint $\vec{H}$ obey
\begin{subequations}
\begin{align}
    \{ \Vec{H} [ \Vec{N}] , \Vec{H} [ \Vec{M} ] \} &= \vec{H} [\mathcal{L}_{\Vec{N}} \Vec{M}]
    \ , \\
    \{ \tilde{H} [ N ] , \Vec{H} [ \Vec{N}]\} &= - \tilde{H} [ N^b \partial_b N ]
    \ , \\
    \{ \tilde{H} [ N ] , \tilde{H} [ M ] \} &= - \vec{H} [ \tilde{q}^{a b} ( N \partial_b M - M \partial_b N )]
\end{align}
\label{eq:Hypersurface deformation algebra - modified}
\end{subequations}
with a structure function $\tilde{q}^{a b}$ that approaches the inverse
$q^{ab}$ of the configuration variables $q_{ab}$ in the classical limit, but
not necessarily for all parameter choices in the modified Hamiltonian
constraint. If the first-class condition is satisfied and the general
structure of the hypersurface deformation brackets is maintained, the new structure
function $\tilde{q}^{ab}$ is uniquely determined by the modified $\tilde{H}$. 

The constraints generate gauge transformations of the phase-space variables in
the usual way, given by Poisson brackets
$\delta_{\epsilon}f(q_{ab},p^{ab})=\{f,H[\epsilon^0]+\vec{H}[\vec{\epsilon}]\}$. Since
evolution is a gauge transformation in a generally covariant theory, evolution
is generated by the same constraints, but with specific gauge functions $N$
and $\vec{N}$ for a given choice of a time-evolution vector field:
$\dot{f}=\{f,H[N]+\vec{H}[\vec{N}]\}$ for a phase-space function $f$.
Standard results in canonical gravity \cite{LapseGauge,CUP} show that the
evolutionary gauge functions $N$ and $\vec{N}$ in $\tilde{H}[N]$ and $\vec{H}[\vec{N}]$,
respectively, are subject to gauge transformations that follow from the
requirement that Hamiltonian evolution generated by
$\tilde{H}[N]+\vec{H}[\vec{N}]$ must be compatible with gauge transformations
generated by the same functionals $\tilde{H}$ and $\vec{H}$ but with different
gauge functions, $\tilde{H}[\epsilon^0]+\vec{H}[\vec{\epsilon}]$. The
evolution and gauge generators are the same because the theory is completely
constrained, but in a space-time interpretation (if it exists), the
multipliers $N$ and $\vec{N}$ for evolution play a different role than the gauge
functions $\epsilon^0$ and $\vec{\epsilon}$: The former appear as time
components of the space-time line element compatible with the constraints,
identified as the lapse function and the shift vector, while the latter
parameterize generic gauge changes or transformation between different
slicings in the resulting space-time. The compatibility condition that the
evolution of a gauge-transformed configuration be the gauge-transformation of
the evolved original configuration then implies that the lapse function and
shift vector transform as
\begin{subequations}
\begin{eqnarray}
    \delta_\epsilon N &=& \dot{\epsilon}^0 + \epsilon^a \partial_a N - N^a \partial_a \epsilon^0
    \ ,
    \label{eq:Off-shell gauge transformations for lapse - modified}
    \\
    \delta_\epsilon N^a &=& \dot{\epsilon}^a + \epsilon^b \partial_b N^a - N^b \partial_b \epsilon^a + \tilde{q}^{a b} \left(\epsilon^0 \partial_b N - N \partial_b \epsilon^0 \right)
    \label{eq:Off-shell gauge transformations for shift - modified}
\end{eqnarray}
\label{eq:Off-shell gauge transformations for lapse and shift - modified}
\end{subequations}
Since the condition involves the commutator of gauge and evolution equations,
it is sensitive to the structure function $\tilde{q}^{ab}$ in the constraint
brackets, which implies the only modification in these equations.

The geometrical structure of hypersurface deformations, algebraically
expressed by constraint brackets of a specific form, suggests that the modified theory is
compatible with a space-time interpretation of its solutions given in terms of
the emergent line element
\begin{align}
    {\rm d} s^2 =& - N^2 {\rm d} t^2 + \tilde{q}_{a b} ( {\rm d} x^a + N^a
                   {\rm d} t ) ( {\rm d} x^b + N^b {\rm d} t )
    \ ,
    \label{eq:ADM line element - Modified}
\end{align}
where the inverse $\tilde{q}_{a b}$ of the modified structure function appears as the
spatial metric. (If the modified structure function is not invertible, the
emergent line element may have to be split into separate expressions with
varying signatures depending on the sign of $\det(\tilde{q}^{ab})$, see \cite{HigherCov}.)
However, gauge transformations generated by the modified constraints, applied
to $\tilde{q}_{ab}$, $N$ and $N^z$, are not
guaranteed to be compatible with coordinate transformations applied to ${\rm
  d}t$ and ${\rm d}x^a$. If this is not the case, the expression (\ref{eq:ADM
  line element - Modified}) is not invariant and therefore meaningless as a
line element or distance measure. The condition that the emergent line element
be invariant imposes additional conditions on the modified Hamiltonian
constraint through conditions on the modified structure function implied by it.

We say that the modified canonical theory is generally covariant if the
emergent space-time line element is coordinate invariant. Coordinate changes
applied to ${\rm d}x^{\mu}$ must therefore be dual to canonical gauge
transformations applied to the components of (\ref{eq:ADM line element -
  Modified}). The case of the time components with coefficients given by $N$
and $N^a$ has been considered in \cite{EffLine}, but not the spatial part. As
a complete equation, this condition implies that gauge transformations in the
modified canonical theory have a strict correspondence with infinitesimal
space-time diffeomorphisms or space-time Lie derivatives, at least on-shell
when the constraints and equations of motion are satisfied (indicated by a
subscript $O.S.$):
\begin{align}
    \delta_\epsilon \tilde{g}_{\mu \nu} \big|_{\text{O.S.}} =&
    \mathcal{L}_{\xi} \tilde{g}_{\mu \nu}
    \,.
    \label{eq:Covariance condition of spacetime - modified}
\end{align}
(There is an analogous relationship between gauge transformations and
infinitesimal diffeomorphisms acting on extrinsic curvature of spacelike
hypersurfaces in the emergent space-time, but, as shown in \cite{HigherCov},
it does not imply a new covariance condition in addition to the equation for
$\tilde{q}_{ab}$.)

The canonical gauge transformations with gauge functions
$(\epsilon^0, \epsilon^a)$, taken on-shell, then reproduce space-time
diffeomorphisms with a space-time vector $\xi$ related to the gauge functions
by
\begin{subequations}
\begin{align}
    \xi^\mu =& \epsilon^0 n^\mu + \epsilon^a s^\mu_a
    = \xi^t t^\mu + \xi^a s^\mu_a
    \ ,
    \\
    \xi^t =& \frac{\epsilon^0}{N}
    \quad ,
    \quad
    \xi^a = \epsilon^a - \frac{\epsilon^0}{N} N^a
\end{align}
\label{eq:Diffeomorphism generator projection}
\end{subequations}
because the former has components referring to the time direction in
space-time, while the latter refer to the normal direction of embedded spacelike hypersurfaces.
Following \cite{EffLine}, the timelike components of the covariance condition are
satisfied by virtue of the hypersurface-deformation brackets,
\eqref{eq:Hypersurface deformation algebra - modified}, if we use
\eqref{eq:Off-shell gauge transformations for lapse and shift - modified} and
assume that the spatial metric is covariant,
$\delta_\epsilon q_{a b} |_{\text{O.S.}} = \mathcal{L}_{\xi} q_{a b}$.
This latter equation is not true for any first-class modification of the  constraints,
but only if \cite{HigherCov}
\begin{align}
    \frac{\partial (\delta_{\epsilon^0} \tilde{q}^{a b})}{\partial (\partial_c
  \epsilon^0)} \bigg|_{\text{O.S.}} 
    = \frac{\partial (\delta_{\epsilon^0} \tilde{q}^{a b})}{\partial
  (\partial_c \partial_d \epsilon^0)} \bigg|_{\text{O.S.}} 
    = \cdots
    = 0
    \ ,
    \label{eq:Covariance condition of 3-metric - modified - reduced}
\end{align}
where $\delta_{\epsilon^0} \tilde{q}^{a b} = \{ \tilde{q}^{a b} ,
H[\epsilon^0]\}$ without a spatial shift.

\subsection{Scalar fields}

As a new result, we now extend the covariance condition to scalar fields. We
begin with the case of a single-component scalar 
$\phi$ with momentum $P_{\phi}$, introduced as an additional phase-space
degree of freedom that couples to the gravitational degrees of freedom through
a matter Hamiltonian added to $\tilde{H}$, and then consider additional
structures available in the case of scalar multiplets.

\subsubsection{Single scalar field}
\label{sec:Single}

For a canonical theory with hypersurface-deformation brackets
\eqref{eq:Hypersurface deformation algebra - modified} for the combined
constraints of gravitational and matter variables,
$\tilde{H}_{\rm grav}[N]+\tilde{H}_{\rm matter}[N]$ and
$\vec{H}_{\rm grav}[\vec{N}]+\vec{H}_{\rm matter}[\vec{N}]$, we say that a
scalar field $\phi$ is covariant if
\begin{align}
    \delta_\epsilon \phi \big|_{\text{O.S.}} =&
    \mathcal{L}_{\xi} \phi
    \ .
    \label{eq:Covariance condition of phi - modified1}
\end{align}
Just as the gravitational configuration variable $q_{ab}$, the canonical
scalar field $\phi$ is initially defined only on a spatial slice. However,
on-shell we can use equations of motion to relate the momentum of $\phi$ to
its time derivative, defined by
$\dot{\phi}= \{\phi,\tilde{H}[N]+\vec{H}[\vec{N}]\}$. This time derivative,
available on-shell, can then be used in a comparison with the time component
of the space-time Lie derivative.

Written in the basis adjusted to the foliation into spacelike hypersurfaces,
the scalar covariance condition takes the form
\begin{equation}
  \delta_\epsilon \phi |_{\text{O.S.}} = \frac{\epsilon^0}{N} \dot{\phi} +
  \left( \epsilon^a - \frac{\epsilon^0}{N} \epsilon^a \right) \partial_a \phi\,.
\end{equation}
Using of the canonical gauge transformation
$\delta_ \epsilon \phi = \{ \phi , \tilde{H}[\epsilon^0] + H_a [\epsilon^a]\}$
on the left-hand side and Hamilton's equation of motion
$\dot{\phi} = \{ \phi , \tilde{H}[N] + H_a [N^a]\}$ on the right-hand side,
the equation can be simplified to
\begin{equation}
  \frac{1}{\epsilon^0} \{ \phi , \tilde{H}[\epsilon^0]\} |_{\text{O.S.}} = \frac{1}{N} \{ \phi ,
  \tilde{H}[N]\}
\end{equation}
using the assumption that the diffeomorphism constraint is
unmodified, and the fact that a scalar field $\phi$ has spatial density weight zero.

The normal gauge transformation of the scalar field has the generic form
$\{\phi , \tilde{H} [\epsilon^0]\} = \Phi \epsilon^0 + \Phi^c \partial_c \epsilon^0 +
\Phi^{c d} \partial_c \partial_d \epsilon^0 + \cdots$, where the $\Phi$
tensors are phase-space functions.  Substituting this expansion into the
covariance condition, we obtain
\begin{align}
    \Phi^c \frac{\partial_c \epsilon^0}{\epsilon^0}
    + \Phi^{c d} \frac{\partial_c \partial_d \epsilon^0}{\epsilon^0}
    + \cdots \bigg|_{O.S.}
    =&
    \Phi^c \frac{\partial_c N}{N}
    + \Phi^{c d} \frac{\partial_c \partial_d N}{N}
    + \cdots \bigg|_{O.S.}
    \label{eq:Spatial covariance condition - first reduced form}
\end{align}
for independent $\epsilon^0$ and $N$.
We can now use
$\{\phi,H[\epsilon^0]\} = \delta \tilde{H} [\epsilon^0] / \delta P_\phi$ to write
the $\Phi$ tensors in \eqref{eq:Spatial covariance condition - first reduced form}
as
\begin{subequations}
\begin{align}
    \Phi^c =& - \frac{\partial \tilde{H}}{\partial ( \partial_c P_\phi)}
    + \partial_d \left( \frac{\partial \tilde{H}}{\partial ( \partial_c \partial_d P_\phi)} \right)
    - \cdots 
    \ , \\
    \Phi^{c d} =& \frac{\partial \tilde{H}}{\partial ( \partial_c \partial_d P_\phi)}
    - \partial_d \partial_e \left( \frac{\partial \tilde{H}}{\partial (
                  \partial_c \partial_d \partial_e P_\phi)} \right) 
    + \cdots
    \ ,
\end{align}
\end{subequations}
and so on. The space-time Lie derivative of a scalar field of density weight
zero does not contain terms with spatial derivatives of the lapse
function. Therefore, $\Phi^c$, $\Phi^{cd}$ and so on must vanish on-shell,
such that
\begin{align}
    \frac{\partial \tilde{H}}{\partial ( \partial_c P_\phi)}
    = \frac{\partial \tilde{H}}{\partial ( \partial_c \partial_d P_\phi)}
    = \cdots
    = 0
    \ .
    \label{eq:Spatial covariance condition - second reduced form}
\end{align}

These equations, as derived, are required to hold on-shell, but since partial
derivatives of the Hamiltonian constraint by spatial derivatives of the
momentum are neither constraints nor equations of motion, they must vanish
identically.  Therefore, no derivatives of the scalar momentum $P_\phi$ are
allowed in a modified Hamiltonian constraint.

\subsubsection{General scalar multiplets}

It is straightforward to generalize the covariance condition from a single
scalar field $\phi$ with momentum $P_{\phi}$ to a scalar multiplet $\phi^I$
with momenta $P_J$ suitable for instance for the Higgs field. While we will
consider only single-scalar models in our specific examples, there is an
additional non-trivial property of multiplets given by global symmetries that
can be used to formulate physical conditions on admissible modified
theories. For purposes such as quantum field theory on a curved background, it
is important to know the gauge current as a space-time vector, which is not
directly related to general covariance but implies additional conditions from
the Poisson brackets of the gauge generator with the hypersurface-deformation
generators. There is a remnant of this important property in models with a
single scalar field, which we will make use of in some of our explicit
constructions.

Consider a scalar field multiplet $\phi^I$ with internal indices
$I = 1 , 2 , \dots , n$.  The scalar field's indices denote its components as
a vector in the representation $\mathcal{R}$ of some Lie group
$\mathcal{G} = {\rm SU} (N)$ of dimension $n$.  Its Lie algebra $\mathfrak{g}$
then has ${\rm dim}(\mathfrak{g}) = N^2-1$ generators $\tau_i$ with
$i = 1 , \dots , {\rm dim}(\mathfrak{g})$ satisfying the algebra
$[\tau_i , \tau_j] = f_{i j k} \tau_k$, where $f_{i j k} = f_{[i j k]}$ are
the structure constants.  Given a Lie-algebra generator
$\tau_i \in \mathfrak{g}$, the associated Lie-group element
$g = \exp ( \alpha^i \tau_i) \in \mathcal{G}$, $\alpha^i \in \mathbb{R}$ acts
on the scalar field by
\begin{eqnarray}
    \phi^I \to (e^{ \alpha^i \tau_i})^I{}_J \phi^J
    \ .
\end{eqnarray}

The classical Higgs-type action in curved space-time with metric $g_{\mu\nu}$
and its canonical decomposition are given by
\begin{eqnarray}
    S_{\rm scalar} [\phi]
    &=&
    - \int {\rm d}^4 x\ \sqrt{- \det g}  \left( \frac{1}{2}\delta_{I J} g^{\mu
        \nu} (\nabla_\mu \phi^I) (\nabla_\nu \phi^J) + V ( \delta_{I J} \phi^I
        \phi^I) \right) 
    \nonumber\\
    &=&
    \int {\rm d}^4 x\ \Bigg[
    P_I \Dot{\phi}^I
    - N^a \left( P_I \partial_a \phi^I \right)
    \label{eq:Classical action for scalar field}\\
    &&
    -  N\left( \frac{1}{2}\frac{\delta^{I J} P_I P_J}{\sqrt{\det q}} +
       \frac{1}{2}\delta_{I J} \sqrt{\det q}\; q^{a b} (\partial_a \phi^I) (\partial_b
       \phi^J) + \sqrt{\det q}\; V ( \delta_{I J} \phi^I \phi^I) \right) 
    \Bigg]
    \nonumber
\end{eqnarray}
where $V (\delta_{IJ}\phi^I\phi^J)$ is the potential and the momenta are given by
\begin{eqnarray}
    P_I &=& \frac{\delta S_{\rm scalar} [\phi]}{\delta \Dot{\phi}^I}
    =
    \delta_{I J} \sqrt{\det q} \:n^\mu \partial_\mu \phi^J
    \ .
\end{eqnarray}
The scalar field therefore implies a contribution
\begin{equation}
  \vec{H}_{\rm scalar}[\vec{N }]= \int{\rm d}^3x N^aP_I\partial_a \phi^I
\end{equation}
to the diffeomorphism constraint $\vec{H}[\vec{N}]$, and a contribution
\begin{equation}
  H_{\rm scalar}[N]=  \int{\rm d}^3x N \left( \frac{1}{2}\frac{\delta^{I J}
      P_I P_J}{\sqrt{\det q}} + \frac{1}{2}\delta_{I J} \sqrt{\det q}\; q^{a b} (\partial_a
    \phi^I) (\partial_b \phi^J) + \sqrt{\det q}\; V ( \delta_{I J} \phi^I
    \phi^I) \right)
\end{equation}
to the Hamiltonian constraint $H[N]$.

Elements of the Lie group and Lie algebra act on the momentum and field values as
\begin{eqnarray}
    P_I &\to&
    P_J (e^{- \alpha^i \tau_i})^J{}_I
    \approx
    P_J \left( \delta^J{}_I - \alpha^i (\tau_i)^J{}_I \right)\nonumber
  \\
  \phi^I &\to& (e^{\alpha^i \tau_i})^I{}_J \phi_J\approx \left( \delta^I{}_J + \alpha^i (\tau_i)^I{}_J \right)
               \phi^J\label{eq:Internal transformation of scalar field}\\ 
    \phi_I = \delta_{I J} \phi^J &\to& \phi_J (e^{- \alpha^i \tau_i})^J{}_I\approx
    \phi_J \left( \delta^J{}_I - \alpha^i (\tau_i)^J{}_I \right)\nonumber
\end{eqnarray}
where we used $\tau_i^{\rm T} = \tau_{i}^{-1} = - \tau_i$.  Thus, the action
(\ref{eq:Classical action for scalar field}) is invariant under
transformations generated by the Lie group $\mathcal{G}$ and, thus, also under
infinitesimal transformations generated by the Lie algebra $\mathfrak{g}$,
leading to a Noether current. For later applications, we derive this conserved
current from the Hamiltonian perspective with due attention to applications of
non-trivial covariance conditions that are required for a meaningful
space-time current.

In canonical terms, the transformation (\ref{eq:Internal transformation of
  scalar field}) is generated by the phase-space function
\begin{eqnarray}
    G [\alpha] = \int {\rm d}^3 x\ \alpha^i P_I (\tau_i)^I{}_J \phi^J
    \ ,
    \label{eq:Symmetry generator of scalar field}
\end{eqnarray}
smeared with a $\mathfrak{g}$-valued constant $\alpha^i$. Thus, $G[\alpha]$
generates a global symmetry, which could be generalized to a local one by the
usual introduction of gauge fields but we leave this step for future treatments
as it would complicate our analysis. The global symmetry generator commutes, up to
possible boundary terms, with the Hamiltonian and diffeomorphism constraints,
\begin{eqnarray} \label{HHG}
    \{ H [N] , G[\alpha] \} = \{ \Vec{H} [\Vec{N}] , G[\alpha] \} = 0
    \ , \label{eq:G-invariance}
\end{eqnarray}
and it reproduces brackets of the Lie algebra it is based on,
\begin{eqnarray}
    \{G[\alpha_{1}] , G[\alpha_{2}]\} = \int {\rm d}^3 x\ \alpha_1^i
  \alpha_2^j f_{i j k} G_k = G \left[ [ \alpha_1 , \alpha_2] \right] 
\end{eqnarray}
with the Lie commutator $[\alpha_1,\alpha_2]$.
The brackets (\ref{eq:G-invariance}) imply $\mathcal{G}$-gauge
invariance of the theory. Therefore, the non-local
phase-space function $G[\alpha]$ is conserved during evolution, and the
smearing constant transforms in the adjoint representation
$\alpha_1^i \to \alpha_1^i + f_{jk}{}^i \alpha_2^j \alpha_1^k$.  The local
phase-space function $G_i = \tau_i P_I \phi^I$ then evolves according to an
equation of the form $\Dot{G}_i = - \partial_a J_i^a$ where $J^a$ are obtained
from possible boundary terms in (\ref{HHG}).  In a covariant theory, the
spatial vector $J_i^a$ must be part of a space-time current $J_i^\mu$ with
density weight one, satisfying the covariant conservation equation
$\partial_\mu J_i^\mu = \nabla_\mu J_i^\mu = 0$. The completion of $J_i^a$ to a
space-time vector allows us to identify the charge density $J_i^t$ (which
turns out to equal $G_i$) as a function of the canonical fields.

An explicit computation with the classical constraints yields
\begin{eqnarray}
    \{ G_i , H [N] + H_a [N^a]\}
    &=& - (\tau_i)^I{}_J \partial_a \left( N\sqrt{\det q} \left( \delta_{I K} q^{a b} \phi^J \partial_b \phi^K - \phi^J \frac{N^a}{N} \frac{P_I}{\sqrt{\det q}} \right) \right)
    \nonumber\\
    &=& - \partial_a J_i^a \label{currentboundary}
\end{eqnarray}
with
\begin{equation} \label{current}
  J_i^a= (\tau_i)^I{}_J N\sqrt{\det q} \left( \delta_{I K} q^{a b}
      \phi^J \partial_b \phi^K - \phi^J \frac{N^a}{N} \frac{P_I}{\sqrt{\det
          q}} \right) \,.
\end{equation}
If we consider $H[N]+H_a[N^a]$ as a gauge transformations, $N$ and $N^a$
approach zero at any boundary, and therefore the smeared
$\int{\rm d}^3x G_i\alpha^i$ Poisson commutes with any gauge generator of
hypersurface deformations. The system is therefore first class. If $N$ or
$N^a$ do not approach zero at the boundary, they generate gravitational
symmetries, such as a time translation $H[1]$ in an asymptotically flat
space-time, that are not gauge.

The canonical equations of motion for the scalar field allow us to relate the
momenta $P_I$ in the spatial current (\ref{current}) to time derivatives
$\partial_t\phi^J$, and an emergent space-time metric $\tilde{g}_{\mu\nu}$
derived from the covariance condition of a modified gravitational theory
expresses $q^{ab}$ and $N^a$ through spatial and space-time components of the
metric. The components $g^{t\mu}$ of the inverse emergent space-time metric
then imply a unique expression for the time component $J^t_i$, and we have the
full 4-current with space-time density weight $N\sqrt{\det q}$ as
$\sqrt{-\det g}$ in Lorentzian signature.  The resulting covariant and
conserved current $J^\mu_i$ is of fundamental importance in quantum field
theory in curved space-time because it provides a well-defined inner product.
We will thus try to preserve the existence of a symmetry generator
(\ref{eq:Symmetry generator of scalar field}) or, equivalently, the
$\mathcal{G}$-invariance in the modified theory.

We illustrate this procedure for the case of a single complex scalar field
$\phi$, corresponding to the $\mathcal{G}={\rm U(1)}$ case.  Starting with the
Klein--Gordon action in curved space-time,
\begin{eqnarray}
    S_{\rm scalar} [\phi] = -\int {\rm d}^4 x \sqrt{-\det g} \left( g^{\mu \nu}
  (\nabla_\mu \phi^*) (\nabla_\nu \phi) + V (\phi^* \phi) \right) 
\end{eqnarray}
with a potential $V (\phi^*\phi)$.
Using the inverse metric
\begin{eqnarray}
    g^{\mu \nu} &=&
    q^{a b} s^\mu_a s^\nu_a
    - \frac{1}{N^2} \left(t^\mu - N^a s^\mu_a\right) \left(t^\nu - N^b s^\nu_b\right)
    \label{eq:Inverse metric}
\end{eqnarray}
in canonical form, the decomposition of the action is given by
\begin{eqnarray}
    S_{\rm scalar} [\phi] &=& \int {\rm d}^4 x N \sqrt{\det q} \bigg( - \frac{1}{N^2} \Dot{\phi}^*
    \Dot{\phi}
    + \frac{N^a}{N^2} \Dot{\phi}^*
    (\partial_a \phi)
    + \frac{N^a}{N^2} (\partial_a \phi^*) \Dot{\phi}
    \nonumber\\
    &&
    + \left( q^{a b} - \frac{N^a N^b}{N^2} \right) (\partial_a \phi^*)
    (\partial_b \phi)
    + V (\phi^* \phi) \bigg)
    \,.
\end{eqnarray}
The momenta are
\begin{eqnarray}
    P_\phi &=& \frac{\delta S_{\rm scalar} [\phi]}{\delta \Dot{\phi}}
    = - \frac{\sqrt{\det q}}{N} \left( \Dot{\phi}^*
    - N^a (\partial_a \phi^*) \right)
    \ , \\
    P_\phi^* &=& \frac{\delta S_{\rm scalar} [\phi]}{\delta \Dot{\phi}^*}
    =
    - \frac{\sqrt{\det q}}{N} \left( \Dot{\phi}
    - N^a (\partial_a \phi) \right)
    \label{eq:Momenta of complex KG field}
\end{eqnarray}
and therefore we can use
\begin{eqnarray}
    N^a P_\phi \partial_a \phi &=&
    - N^a \frac{\sqrt{\det q}}{N} \left( \Dot{\phi}^*
    - N^b (\partial_b \phi^*) \right) \partial_a \phi\\
    N^a P_\phi^* \partial_a \phi^* &=&
    - N^a \frac{\sqrt{\det q}}{N} \left( \Dot{\phi}
    - N^b (\partial_b \phi) \right) \partial_a \phi^*
\end{eqnarray}
and
\begin{eqnarray}
    \frac{N}{\sqrt{\det q}} P_\phi^* P_\phi &=&
    \frac{\sqrt{\det q}}{N} \left(
    \Dot{\phi}^* \Dot{\phi}
    - N^a \Dot{\phi}^* (\partial_a \phi)
    - N^a (\partial_a \phi^*) \Dot{\phi}
    + N^a N^b (\partial_a \phi) (\partial_b \phi^*) \right)
    \nonumber\\
    &=&
    \frac{\sqrt{\det q}}{N} \Dot{\phi}^* \Dot{\phi}
    - N^a \frac{\sqrt{\det q}}{N} \left(
    \Dot{\phi}^* (\partial_a \phi)
    + (\partial_a \phi^*) \Dot{\phi}
    - N^b (\partial_a \phi) (\partial_b \phi^*) \right)
    \nonumber\\
    &=&
    \frac{\sqrt{\det q}}{N} \left( \Dot{\phi}^* \Dot{\phi}
    - N^a N^b (\partial_a \phi) (\partial_b \phi^*) \right)
    + N^a \left( P_\phi \partial_a \phi + P_\phi^* \partial_a \phi^* \right)
\end{eqnarray}
in order to replace some of the time derivatives of $\phi$ and $\phi^*$ in the
action by momenta:
\begin{eqnarray}
    S_{\rm scalar} [\phi] &=& \int {\rm d}^4 x\
    \bigg[ P_\phi \Dot{\phi}
    + P_\phi^* \Dot{\phi}^*
    - N^a \left( P_\phi \partial_a \phi + P_\phi^* \partial_a \phi^* \right)
    \nonumber\\
    &&
    + N \left( \frac{P_\phi^* P_\phi}{\sqrt{\det q}}
    + \sqrt{\det q}\; q^{a b} (\partial_a \phi^*)
    (\partial_b \phi)
    + \sqrt{\det q}\;V (\phi^* \phi) \right) \bigg]
    \ .
\end{eqnarray}
In this form, we immediately read off the Hamiltonian and diffeomorphism constraints
\begin{eqnarray}
    H &=&
    \frac{P_\phi^* P_\phi}{\sqrt{\det q}}
    + \sqrt{\det q}\; q^{a b} (\partial_a \phi^*)
    (\partial_b \phi)
    + \sqrt{\det q}\; V (\phi^* \phi)
    \ , \\
    H_a &=& 
    P_\phi \partial_a \phi + P_\phi^* \partial_a \phi^*
\end{eqnarray}
which are both real.

The global symmetry transformation $\phi \to \phi e^{i \alpha}$ for constant
$\alpha$, which is manifest in the original action, is still pressent in
Hamiltonian form. It is completed to a canonical transformation
\begin{eqnarray}
    \phi \to \phi e^{i \alpha}
    \quad , \quad
    P_\phi \to P_\phi e^{- i \alpha}
\end{eqnarray}
by including the momenta, and an analogous version for their complex conjugate
counterparts.  The infinitesimal version
\begin{eqnarray}
    \phi \to \phi + i \alpha \phi
    \quad , \quad
    P_\phi \to P_\phi - i \alpha P_\phi
\end{eqnarray}
is generated by the phase-space function
\begin{eqnarray}
    G [\alpha] = \int {\rm d}^3 x\ \alpha i \left( \phi P_\phi - \phi^* P_\phi^* \right)
    \ ,
    \label{eq:Symmetry generator}
\end{eqnarray}
which we have smeared with the infinitesimal, real constant $\alpha$. (There
is a single global gauge generator $G[\alpha]$ rather than local transformations.)
This generator obeys the relations
\begin{eqnarray}
    \{ H[N] , G[\alpha] \} = \{ \Vec{H}[\Vec{N}] , G[\alpha] \} =  \{ G[\alpha_1] , G[\alpha_2] \} = 0
    \label{eq:Commutation Noether current and constraints}
\end{eqnarray}
to first order in the constant $\alpha$ and up to possible boundary terms
according to (\ref{currentboundary}). It therefore provides a global
first-class constraint in addition to the local ones, $H$ and $H_a$, and
implies ${\rm U}(1)$-invariance.

The physical meaning of this function  can be seen by replacing momenta with
time derivatives of the scalar field, using
(\ref{eq:Momenta of complex KG field}):
\begin{eqnarray}
    G &=&
    -  \frac{i}{N} \sqrt{\det q} \left( \phi \Dot{\phi}^* - \phi^* \Dot{\phi} + N^a \left( \phi^* (\partial_a \phi) - \phi (\partial_a \phi^*) \right) \right)
    \nonumber\\
    &=&
    - i \sqrt{- \det g} \left( g^{t t} \left( \phi^* \Dot{\phi} - \phi \Dot{\phi}^* \right) + g^{t a} \left( \phi^* (\partial_a \phi) - \phi (\partial_a \phi^*) \right) \right)
    \nonumber\\
    &=:& 
    g^{t t} J_t + g^{t a} J_a
    = J^t \label{GJt}
\end{eqnarray}
using standard expressions of the scalar-field current $J_{\mu}$. The metric
factors identify the global gauge generator $G=J^t$ with the time component of 
the densitized space-time current of the Klein--Gordon field.

The usual space-time formulation tells us that the current is conserved in the
sense that $ \partial_\mu J^\mu =  \nabla_\mu J^\mu = 0$, using
the space-time density weight of $J^\mu$. If we include boundary terms in
(\ref{eq:Commutation Noether current and constraints}) according to
(\ref{currentboundary}), we reproduce this conservation law of the space-time
current at the canonical level:
\begin{eqnarray}
    \partial_t J^t &=& \{ G , H [N] + H_a [N^a]\}
    \nonumber\\
    &=& - i \partial_a \left( \sqrt{-\det g} \left( q^{a b} \left( \phi \partial_b \phi^*
    - \phi^* \partial_b \phi \right)
    - \frac{N^a}{N} \left( \phi \frac{P_\phi}{\sqrt{\det q}} - \phi^* \frac{P_\phi^*}{\sqrt{\det q}} \right) \right) \right)
    \nonumber\\
    &=& - i \partial_a \left( \sqrt{-\det g} \left( g^{a b} \left( \phi \partial_b \phi^* - \phi^* \partial_b \phi \right)
    + g^{t a} \left( \phi \Dot{\phi}^* - \phi^* \Dot{\phi} \right) \right) \right)
    \nonumber\\
    &=&
    - \partial_a \left( g^{a b} J_b + g^{a t} J_t \right)
    = - \partial_a J^a
    \ ,
\end{eqnarray}
where we used (\ref{eq:Momenta of complex KG field}) in the third line.  This
result has several implications: (i) The spatial component of the
Klein--Gordon current with space-time density weight one is given by the
boundary terms of (\ref{eq:Commutation Noether current and constraints}),
derived after smearing $G$ with a constant $\alpha$. (ii) The unsmeared
symmetry generator equals the time component $J^t$ of the densitized
space-time current.  (iii) We need the space-time metric, including $g^{ta}$
and not just the spatial part, in order to combine the correct terms in
$\partial_tJ^t$ and derive the conservation law, equating this term to
$-\partial_aJ^a$. The conservation law is therefore related to covariance in
the sense that a well-defined emergent space-time metric must exist in the
modified case.

For the scalar field, our results show that the symmetry generator
(\ref{eq:Symmetry generator}) is the Noether charge density, the integration
of which is a conserved charge.  This may be generalized to other systems for
a Hamiltonian version of Noether's theorem, and also applied to a local
symmetry by the introduction of gauge fields. In canonical terms, the symmetry
generator (\ref{eq:Symmetry generator}) is a Dirac observable.

\subsection{Spherically symmetric sector}
\label{subsec:Spherically symmetric sector}

We will evaluate the full covariance conditions within a viewpoint of
effective field theory, starting with a generic Hamiltonian constraint with
terms up to a fixed number of spatial derivatives. It is easier to perform the
required calculations after a reduction to spherical symmetry, which is able
to provide new interesting models for non-vacuum black holes as well as
inhomogeneous cosmological models. 

\subsubsection{Classical theory}

Using spherical symmetry, the space-time line element can be written as
\begin{equation}\label{eq:ADM line element - spherical}
    {\rm d} s^2 = - N^2 {\rm d} t^2 + q_{x x} ( {\rm d} x + N^r {\rm d} t )^2
    + q_{\vartheta \vartheta} {\rm d} \Omega^2\,.
\end{equation}
As initially developed for models of loop quantum gravity
\cite{SymmRed,SphSymm,SphSymmHam}, it is convenient to parameterize the
metric components $q_{xx}$ and $q_{\vartheta\vartheta}$ as
\begin{equation}
  q_{x x} = \frac{(E^\varphi)^2}{E^x}
    \quad ,\quad
    q_{\vartheta \vartheta} = E^x
\end{equation}
where $E^x$ and $E^\varphi$ are the radial and angular densitized-triad
components, respectively. We assume $E^x>0$, fixing the orientation of space.

The canonical pairs for classical gravity are given by
$(K_\varphi , E^\varphi)$ and $(K_x , E^x)$ where $2K_x$ and
$K_{\varphi}$ are components of extrinsic curvature.  We have the canonical
pair $(\phi,P_\phi)$ for scalar matter. The basic Poisson brackets are given by
\begin{subequations}
\begin{align}
    \{ K_x (x) , E^x (y)\} = \{ K_\varphi(x) , E^\varphi (y) \} = \{ \phi(x) ,
  P_\phi (y) \} = \delta (x-y) . 
\end{align}
\end{subequations}
(Compared with other conventions, our
scalar phase-space variables are divided by $\sqrt{4\pi}$, absorbing the
remnant of a spherical integration. We use units in which Newton's constant
equals one.)

The Hamiltonian constraint has the vacuum  gravitational contribution
depending only on  $(K_\varphi , E^\varphi)$ and $(K_x , E^x)$, as well as a
matter contribution that depends also on $(\phi,P_\phi)$. To be specific,
we consider a minimally coupled scalar field in this section.
The Hamiltonian and diffeomorpshism constraints in the spherically symmetric
theory are then given by
\begin{eqnarray}
    H
    &=&
    - \frac{\sqrt{E^x}}{2} \Bigg[
    E^\varphi  \left( - V (\phi)
    + \frac{1}{E^x}
    + \frac{K_\varphi^2}{E^x}
    + 4 \frac{K_x}{E^\varphi} K_\varphi
    - \frac{1}{E^x} \frac{P_\phi^2}{(E^\varphi)^2} \right)
    \nonumber\\
    &&
    - E^x \frac{(\phi')^2}{E^\varphi}
    - \frac{1}{4 E^x} \frac{((E^x)')^2}{E^\varphi}
    + \frac{(E^x)' (E^\varphi)'}{(E^\varphi)^2}
    - \frac{(E^x)''}{E^\varphi}
    \Bigg]
    \ ,
    \label{eq:Hamiltonian constraint - spherical symmetry - Scalar field - Classical}
\end{eqnarray}
with a scalar potential $V(\phi)$ (or $\frac{1}{2}V(\phi)$, depending on conventions), and
\begin{eqnarray}
    H_r
    &=& E^\varphi K_\varphi' - K_x (E^x)'
    + P_\phi \phi'
    \,.
    \label{eq:Diffeomorphism constraint - spherical symmetry- Scalar field}
\end{eqnarray}
These constraints are first class and have Poisson brackets of
hypersurface-deformation form,
\begin{subequations}
\begin{align}
    \{ H_r [N^r] , H_r[M^r] \}=& H_r [N^r {M^r}' - {N^r}' M^r]
    \ , \\
    \{ H [N] , H_r [M^r] \}=& - H[M^r N'] 
    \ , \\
    \{ H [N] , H[M] \}=& H_r \left[ q^{x x} \left( N M' - N' M \right)\right]
\end{align}
\label{eq:Hypersurface deformation algebra - spherical - Scalar field1}
\end{subequations}
with the structure function $q^{x x} = E^x/(E^\varphi)^2$ equal to the
inverse radial component of the space-time metric.
The covariance conditions
\begin{subequations}
\begin{align}
    \frac{\partial \left(\{ q^{\theta \theta} , H[\epsilon^0] \}\right)}{\partial (\epsilon^0)'} \bigg|_{\text{O.S.}}
    = \frac{\partial \left(\{ q^{\theta \theta} , H[\epsilon^0] \}\right)}{\partial (\epsilon^0)''} \bigg|_{\text{O.S.}}
    = \cdots
    = 0
    \ ,
\end{align}
and
\begin{align}
    \frac{\partial \left(\{ q^{x x} , H[\epsilon^0] \}\right)}{\partial (\epsilon^0)'} \bigg|_{\text{O.S.}}
    = \frac{\partial \left(\{ q^{x x} , H[\epsilon^0] \}\right)}{\partial (\epsilon^0)''} \bigg|_{\text{O.S.}}
    = \cdots
    = 0
    \ ,
\end{align}
\label{eq:Covariance condition - spherical}
\end{subequations}
derived in \cite{HigherCov} are clearly satisfied.

The off-shell gauge transformations for lapse and shift,
\begin{subequations}
\begin{align}
    \delta_\epsilon N =& \dot{\epsilon}^0 + \epsilon^r N' - N^r (\epsilon^0)' ,\\
    \delta_\epsilon N^r =& \dot{\epsilon}^r + \epsilon^r (N^r)' - N^r (\epsilon^r)' + q^{x x} \left(\epsilon^0 N' - N (\epsilon^0)' \right) \,,
\end{align}
\label{eq:Off-shell gauge transformations for lapse and shift - spherical}
\end{subequations}
together with the realization of covariance conditions ensures that the line
element \eqref{eq:ADM line element - spherical} is invariant, with a covariant
metric tensor in the sense that its canonical gauge transformations reproduce
space-time diffeomorphisms on-shell: 
\begin{align}
    \delta_\epsilon g_{\mu \nu} \big|_{\text{O.S.}} =&
    \mathcal{L}_\xi g_{\mu \nu}
    \,.
\end{align}
The gauge functions $(\epsilon^0,\epsilon^r)$ on the left-hand side are related to
the 2-component vector generator $\xi^\mu = (\xi^t,\xi^r)$ of the
diffeomorphism on the right-hand side by
\begin{equation}
    \xi^\mu = \epsilon^0 n^\mu + \epsilon^r s^\mu = \xi^t t^\mu + \xi^r s^\mu
\end{equation}
with
\begin{equation}
    \xi^t = \frac{\epsilon^0}{N}
    \quad , \quad
    \xi^r = \epsilon^r - \frac{\epsilon^0}{N} N^r
    \ .
    \label{eq:Diffeomorphism generator projection - spherical}
\end{equation}

\subsubsection{Covariance in emergent  modified gravity}
\label{sec:Modified spherically symmetric theory}

We now consider modifications to the spherically symmetric theory with
canonical variables $(K_\varphi , E^\varphi)$ and $(K_x , E^x)$. Neither
$(E^{\varphi},E^x)$ nor $(K_{\varphi},K_x)$ then have a direct relationship
with a spatial metric or extrinsic curvature on spacelike hypersurfaces, but
we continue to use these symbols to denote the gravitational configuration
and momentum variables.

If we modify the Hamiltonian constraint such that the first-class nature is
maintained, the constraint brackets \eqref{eq:Hypersurface deformation algebra
  - spherical - Scalar field1} in general imply a modified structure function,
$\tilde{q}^{x x}\not=q^{xx}$.  There is no indication that the angular
component of the spatial metric should be modified because it does not appear
as a structure function in spherically symmetric hypersurface-deformation
brackets.  The emergent space-time metric then equals
\begin{subequations}
\begin{align}
    {\rm d} s^2 =& - N^2 {\rm d} t^2 + \tilde{q}_{x x} ( {\rm d} x + N^r {\rm
                   d} t )^2 + E^x {\rm d} \Omega^2
\end{align}
\label{eq:ADM line element - spherical - modified2a}
\end{subequations}
where $\tilde{q}_{x x} = 1 / \tilde{q}^{x x}$ (as long as $\tilde{q}^{xx}>0$).

The covariance condition \eqref{eq:Covariance condition - spherical} for the
angular component of the emergent spatial metric implies, using
$\delta_{\epsilon^0} E^x = - \delta \tilde{H} [\epsilon^0]/\delta K_x$,
\begin{align}
    \frac{\partial \tilde{H}}{\partial K_x'} \bigg|_{\text{O.S.}}
    = \frac{\partial \tilde{H}}{\partial K_x''} \bigg|_{\text{O.S.}}
    = \cdots
    = 0
    \ ,
    \label{eq:Covariance condition on K_x - modified - spherical}
\end{align}
which restricts possible modified Hamiltonian constraints to those that do
not contain radial derivatives of $K_x$.  The radial component of the
covariance condition becomes
\begin{align}
    \frac{\partial (\delta_{\epsilon^0} \tilde{q}^{x x})}{\partial (\epsilon^0)'} \bigg|_{\text{O.S.}}
    = \frac{\partial (\delta_{\epsilon^0} \tilde{q}^{x x})}{\partial (\epsilon^0)''} \bigg|_{\text{O.S.}}
    = \cdots
    = 0
    \label{eq:Covariance condition - modified - spherical}
\end{align}
and has important implications that cannot simply be summarized as
independence of the Hamiltonian constraint on certain spatial
derivatives. This condition will therefore be analyzed in more detail below.
The covariance condition  \eqref{eq:Spatial covariance condition - second
  reduced form} for the scalar field in spherical symmetry reduces to
\begin{align}
    \frac{\partial \tilde{H}}{\partial P_\phi'} \bigg|_{\text{O.S.}}
    = \frac{\partial \tilde{H}}{\partial P_\phi''} \bigg|_{\text{O.S.}}
    = \cdots
    = 0
    \ ,
    \label{eq:Covariance condition on phi - modified - spherical1}
\end{align}
which restricts the possible modified Hamiltonian constraints to those that do
not contain radial derivatives of $P_\phi$.

Given a modified structure function $\Tilde{q}^{xx}$ obtained from the vacuum
theory, and thus an emergent metric, one may postulate that a massive scalar field
obeys the Klein--Gordon equation
\begin{eqnarray}
    g^{\mu \nu} \nabla_\mu \nabla_\nu \phi - m^2 \phi = 0
    \ ,
\end{eqnarray}
where one uses the emergent metric instead of the classical one.
This equation of motion is derived from the invariant action functional
\begin{eqnarray} \label{action}
    S [\phi] =  \frac{1}{2} \int {\rm d}^4 x \sqrt{-\det g}\left( g^{\mu \nu}
  (\nabla_\mu \phi) (\nabla_\nu \phi) + m^2 \phi^2 \right) 
    \ ,
\end{eqnarray}
by varying with respect to the scalar field for a given background metric.
However, in the modified case, this proposal assumes that the emergent
space-time metric depends only on the gravitational matter variables and not
on the scalar field itself. The emergent nature of the space-time geometry
means that the canonical variables no longer have a close relationship with
their emergent geometrical roles, and any phase-space degree of freedom,
including a matter field, could possibly contribute to the geometry.
Moreover, previous studies of emergent modified gravity have shown that more
general equations of motion not necessarily derivable from an invariant action
functional such as (\ref{action}) may still be covariant. The following
sections extend this conclusion to gravity-scalar systems in spherical
symmetry, deriving several large classes of new models that go even beyond
non-minimal coupling terms in the standard action formalism. They also provide
explicit examples in which the emergent space-time metric depends on a scalar field.

Nevertheless, a specific and potentially interesting version of emergent
modified gravity coupled to a scalar field is given by minimal coupling of the
scalar field to an emergent space-time metric. Since the emergent space-time
metric is not one of the fundamental fields, it cannot be implemented by an
action principle of the form (\ref{action}) because $g_{\mu\nu}$ remains
unknown until the constraint brackets and equations of motion have been
analyzed. Such a theory requires a canonical formulation with due attention to
covariance conditions. The spatial part $\tilde{q}_{xx}$ of the emergent
space-time metric (in spherical symmetry) can then be used to replace the
classical $q_{xx}$ in the Hamiltonian constraint of the scalar field,
amounting to minimal coupling as suggested in \cite{SphSymmMinCoup}. Given the
non-fundamental nature of $\tilde{q}_{xx}$ and its potential dependence on
momentum variables, which can complicate constraint brackets, it is not
obvious that minimal coupling is always possible in emergent modified
gravity. The existence of such minimally coupled emergent gravity-scalar
theories therefore requires a proof, which we present here as a specific
application of our covariance conditions.

\subsubsection{Proof of minimal coupling in emergent modified gravity}

Minimal coupling of the scalar field, expressed in canonical form, amounts to
using a matter contribution to the constraint in which the phase-space
function $q_{xx}=(E^{\varphi})^2/E^x$ has been replaced by $\tilde{q}_{xx}$,
provided the latter depends only on the gravitational phase-space
variables. Otherwise, it would be impossible to have the correct
hypersurface-deformation terms for the gravitational contribution to the
Hamiltonian constraint,
$\{H_{\rm grav}[N_1],H_{\rm grav}[N_2]\}= H^{\rm grav}_x[\tilde{q}^{xx}(N_1'N_2-N_1N_2')]$
where all terms other than $\tilde{q}^{xx}$ by definition do not depend on
matter fields. As we will see later, polymerization of the scalar field, a
modification common in models of loop quantum gravity, requires a
scalar-dependent $\tilde{q}^{xx}$ and therefore cannot be minimally
coupled. Nevertheless, in cases of scalar independent $\tilde{q}^{xx}$,
minimal coupling might be a useful model to analyze certain matter properties.
Specifically, the matter contribution to the Hamiltonian constraint is then
given by
\begin{equation}
    H_{\rm matter}
    =
     E^x\sqrt{\tilde{q}_{xx}} \left(\frac{1}{2}
    \frac{P_{\phi}^2}{(E^x)^2 \tilde{q}_{xx}}   +\frac{1}{2} \frac{(\phi')^2}{\tilde{q}_{xx}}+ V(\phi)
     \right)
    \ ,
    \label{eq:HMin}
\end{equation}
generalizing the matter contribution in the classical (\ref{eq:Hamiltonian
  constraint - spherical symmetry - Scalar field - Classical}).  In this form,
the postulated emergent gravity-scalar theories with minimal coupling have been introduced in
\cite{SphSymmMinCoup}.

Anomaly-freedom of the vacuum constraints implies that $\tilde{q}_{xx}$
transforms just as the classical $q_{xx}$ under spatial coordinate changes,
such that these two expressions have the same Poisson bracket with the full
diffeomorphism constraint. Minimal coupling using the emergent metric is
therefore compatible with the Poisson bracket $\{H[N],D[M]\}$ where
$H[N]=H_{\rm grav}[N]+H_{\rm matter}[N]$ and
$D[M]=D_{\rm grav}[M]+D_{\rm matter}[M]$ contain both gravitational and scalar
contributions, the former with minimal coupling using $\tilde{q}_{xx}$.

The Poisson bracket of two Hamiltonian constraints, $\{H[N_1],H[N_2]\}$ is
more restrictive. If the vacuum theory is anomaly-free and covariant, and
$\tilde{q}^{xx}$ is independent of $\phi$, the
gravitational contribution $H_{\rm grav}$ by construction has a Poisson
bracket $\{H_{\rm grav}[N_1],H_{\rm grav}[N_2]\}$ of the correct form required
for hypersurface deformations, with structure function
$\tilde{q}^{xx}$. Similarly, $\{H_{\rm matter}[N_1],H_{\rm matter}[N_2]\}$ is
of the same form, with the same structure function, because antisymmetry of
the Poisson bracket implies that only derivative terms of momenta lead to
non-zero contributions to this bracket proportional to $N_1N_2'-N_1'N_2$ after
integrating by parts. The functional form of $\tilde{q}_{xx}$ does not matter
for this conclusion. For the gravitational variables, the terms in
(\ref{eq:HMin}) only depend on $E^x$ and $\tilde{q}_{xx}$, and since the
latter cannot depend on spatial derivatives of $K_x$ according to
(\ref{eq:Covariance condition on K_x - modified - spherical}), there are no
non-zero contributions to the matter Poisson bracket
$\{H_{\rm matter}[N_1],H_{\rm matter}[N_2]\}$ from the gravitational dependence. The
only non-zero contributions are from the $\phi'$-term with the $P_{\phi}$-term
using the Poisson bracket for matter variables, and these contributions
produce the correct diffeomorphism constraint with structure function
$\tilde{q}^{xx}$. Without the covariance condition, this part of the bracket
would not necessarily be correct. 

The gravity-matter cross-terms of the form
$\{\tilde{H}_{\rm grav}[N_1],\tilde{H}_{\rm matter}[N_2]\}$ in the Poisson
bracket of two full modified Hamiltonian constraints, given by
\begin{equation}
\{\tilde{H}[N_1],\tilde{H}[N_2]\}= \{\tilde{H}_{\rm grav}[N_1]+\tilde{H}_{\rm
  matter}[N_1], \tilde{H}_{\rm 
  grav}[N_2]+\tilde{H}_{\rm matter}[N_2]\}\,,
\end{equation}
are also non-trivial. They have
to vanish for an anomaly-free bracket of hypersurface-deformation
form. However, if $\tilde{q}_{xx}$ depends on $K_{\varphi}$, as it does in
many interesting examples of emergent modified gravity, there are non-trivial
Poisson brackets that result from $(E^{\varphi})'$-terms in $H_{\rm grav}$
with the $K_{\varphi}$-dependence of $\tilde{q}_{xx}$ in the minimally coupled
scalar Hamiltonian (\ref{eq:HMin}). Since there is a sum of two cross-terms,
$\{\tilde{H}_{\rm grav}[N_1],\tilde{H}_{\rm matter}[N_2]\}+ \{\tilde{H}_{\rm
  matter}[N_1],\tilde{H}_{\rm grav}[N_2]\}$, these contributions are still
antisymmetric under flipping $N_1$ and $N_2$. Any non-zero contribution must
therefore contain a derivative of one of the lapse functions obtained after
integrating by parts, resulting in the non-zero antisymmetric combination
$N_1N_2'-N_1'N_2$, as opposed to the vanishing $N_1N_2-N_2N_1$. Since the
gravitational Hamiltonian does not contain any matter variables, the only
relevant derivative terms are obtained from the Poisson bracket of
$H_{\rm grav}[N]$ with the emergent spatial metric $\tilde{q}_{xx}$ in the
minimally coupled scalar term. There are non-zero cross-terms, implying
anomalies in the constraint brackets, if and only if $\{\tilde{q}_{xx},H[N]\}$
depends on spatial derivatives of $N$. However, this possibility is ruled out
(on-shell) by the second gravitational covariance condition,
(\ref{eq:Covariance condition - modified - spherical}).

Minimal coupling of a scalar field is therefore consistent in spherically
symmetric emergent modified gravity, but only with a rather non-trivial
application of the covariance conditions. The arguments used rely on the
form of these conditions in spherical symmetry together with the assumption
that the structure function does not depend on the scalar field kinematically,
and they do not guarantee the consistency of minimal coupling beyond these
models.

\section{Conditions on the modified theory}
\label{sec:Conditions on the modified theory}

We are now ready to begin our systematic derivation of covariance and symmetry
conditions for scalar fields coupled to gravity. The resulting class of
allowed theories is vast and requires several restrictions not only from basic
physical principles but also to help organize different versions of these
theories. We therefore impose a variety of conditions, some of which are
necessary for consistency or based on fundamental principles, others are
useful for follow-up constructions, and there is yet another set that may be
used to classify different theories.

It is important to keep in mind that emergent modified gravity may be used in
different ways, and the necessity or desirability of some of our conditions
depends on the viewpoint taken toward this class of theories. One general
attitude toward modified gravity is as a collection of possible effective
theories that may be obtained in a semiclassical regime of quantum gravity. In
this case, we would only use the classical-type equations of a modified theory
for solutions, for instance in a phenomenological analysis, but we would not
use them as a starting point for quantization toward quantum gravity, or for
quantized matter fields on a curved classical background described by an
emergent space-time metric. Some of our conditions are then void.

However, since, as it turns out, there are non-trivial modifications of
general relativity within emergent modified gravity that retain the
second-order nature of field equations for both gravity and matter, emergent
modified gravity may well be an alternative to general relativity in a broader
sense. In particular, it would be meaningful to apply quantization procedures
to emergent modified gravity, both to the gravitational sector and to the
matter fields, the former resulting in a theory of quantum gravity and the
latter resulting in quantum field theory on a curved emergent
space-time. Since these may be viewed as fundamental constructions, we would
not be re-quantizing fields of an effective theory of some other fundamental
theory. The consistency of such quantization procedures then necessitates
additional conditions on allowed theories of emergent modified gravity.

\subsection{Required conditions}

Several conditions are necessary for the consistency of emergent modified
gravity itself and not just for possible quantizations, related mainly to their
gauge, symmetry and space-time structures.

\subsubsection{Anomaly-freedom}

Modifications to canonical gravity are usually encoded in a modified
Hamiltonian constraint, $\tilde{H}$.  A modified Hamiltonian constraint would
generally change the Poisson brackets with itself and with the diffeomorphism
constraint, risking a violation not only of covariance but also of their
consistency as gauge generators.  Thus, we need to restrict admissible
canonical theories to those given by modified constraints that preserve the
hypersurface deformation form (\ref{eq:Hypersurface deformation algebra -
  modified}) of their Poisson brackets,
\begin{subequations}
\begin{eqnarray}
    \{ \Vec{H} [ \Vec{N}] , \Vec{H} [ \Vec{M} ] \} &=& - \vec{H} [\mathcal{L}_{\Vec{M}} \Vec{N}]
    \ , \\
    \{ \tilde{H} [ N ] , \Vec{H} [ \Vec{M}]\} &=& - \tilde{H} [ M^b \partial_b N ]
    \ , \\
    \{ \tilde{H} [ N ] , \tilde{H} [ M ] \} &=& - \vec{H} [ \tilde{q}^{a b} ( M \partial_b N - N \partial_b M )]
    \ ,
\end{eqnarray}
\label{brackets}
\end{subequations}
where the structure function, $\tilde{q}^{a b}$, is modified and determined by
$\tilde{H}$. In an explicit calculation of Poisson bracket, this statement
contains several consistency conditions: The Poisson brackets must be closed
in the sense that they vanish when evaluated on the constraint surface
(anomaly-freedom as a gauge theory). And for a relationship between gauge
transformations and hypersurface deformations to be possible, they must
maintain the specific form (\ref{eq:Hypersurface deformation algebra -
  modified}) as seen in the classical theory where the structure function may
be modified in its dependence on phase-space degrees of freedom, but no
additional constraint terms appear such as a Hamiltonian constraint in the
Poisson bracket $\{ \tilde{H} [ N ] , \tilde{H} [ M ] \}$. If this condition
is satisfied, the theory has off-shell gauge transformations that may be
compared with hypersurface deformations. As already discussed, further
restrictions beyond anomaly-freedom are required for off-shell hypersurface
deformations to be equivalent to on-shell coordinate transformations in an
emergent space-time geometry, but anomaly-freedom is an important first step.

In their role as gauge functions labeling hypersurface-deformation generators
$\tilde{H}[N]$ and $\vec{H}[\vec{N}]$, the lapse function $N$ and shift vector
$\vec{N}$ are subject to gauge transformations that follow from consistency of
gauge transformations and evolution on phase space, generated by the same
constraints $\tilde{H}$ and $\vec{H}$. For constraint brackets of
hypersurface-deformation type, these gauge transformations are given by \cite{LapseGauge,CUP}
\begin{eqnarray}
    \delta_\epsilon N &=& \dot{\epsilon}^0 + \epsilon^a \partial_a N - N^a \partial_a \epsilon^0
    \ ,
    \label{eq:Off-shell gauge transformations for lapse - modified2}
    \\
    \delta_\epsilon N^a &=& \dot{\epsilon}^a + \epsilon^b \partial_b N^a - N^b \partial_b \epsilon^a + \tilde{q}^{a b} \left(\epsilon^0 \partial_b N - N \partial_b \epsilon^0 \right)
    \ ,
    \label{eq:Off-shell gauge transformations for shift - modified2}
\end{eqnarray}
where the only change with respect to the original theory is the use of the
modified structure function. If new terms would appear in modified constraint
brackets, such as a Hamiltonian constraint in the Poisson bracket of two
Hamiltonian constraints, there would also be extra terms in (\ref{eq:Off-shell
  gauge transformations for lapse - modified2}) and (\ref{eq:Off-shell gauge
  transformations for shift - modified2}) that could not be reconciled with
coordinate transformations of lapse and shift in a space-time line element. It
is therefore required that the constraints not only remain first class, with
Poisson brackets vanishing on the constraint surface, but also model the
classical form (\ref{brackets}) with the only option of having a modified
structure function. If the inverse of this modified function is used as the
spatial part of an emergent space-time metric, (\ref{eq:Off-shell gauge
  transformations for shift - modified2}) is compatible with coordinate
transformations as shown in \cite{EffLine}, provided that $\tilde{q}_{ab}$
indeed transforms like the spatial part of a space-time metric.

\subsubsection{Covariance}
\label{sec:Covariance in modified gravity Conditions}

A comparison between gauge transformations of lapse and shift with space-time
coordinate changes suggests that the lapse function and
shift vector may  play the role of time components of a space-time metric,
such that gauge transformations are on-shell equivalent to coordinate
transformations in space-time. If this step is still possible in the modified
theory, the corresponding space-time line element is given by
\begin{equation}
    {\rm d} s^2 = - N^2 {\rm d} t^2 + \tilde{q}_{a b} ( {\rm d} x^a + N^a {\rm d} t ) ( {\rm d} x^b + N^b {\rm d} t )
    \ ,
    \label{eq:ADM line element - Modified2}
\end{equation}
where the spatial metric, $\tilde{q}_{a b}$, is the inverse of the structure
function, $\tilde{q}^{a b}$. This conclusion is again obtained from the
geometrical behavior of hypersurface deformations, which have generators with
brackets (\ref{brackets}) provided
$\tilde{q}_{ab}$ is the induced metric on an embedded spacelike
hypersurface. In a modified theory, however, it is not guaranteed that the
inverse of the structure function (depending on the phase-space degrees of
freedom) indeed gauge transforms in a way equivalent to infinitesimal
coordinate changes of a spatial metric. The space-time interpretation
therefore implies a new consistency condition, in addition to anomaly freedom
of the underlying gauge theory.

We say that there is a covariant space-time with line element (\ref{eq:ADM
  line element - Modified2}) if
\begin{equation}
    \delta_\epsilon \tilde{g}_{\mu \nu} \big|_{\text{O.S.}} =
    \mathcal{L}_{\xi} \tilde{g}_{\mu \nu}
    \ ,
    \label{eq:Covariance condition of spacetime - modified2}
\end{equation}
that is, if the canonical gauge transformations with gauge functions
$(\epsilon^0, \epsilon^a)$ reproduce infinitesimal diffeomorphisms on-shell
with a space-time vector field $\xi$ related to the gauge functions by
\begin{subequations}
\begin{eqnarray}
    \xi^\mu &=& \epsilon^0 n^\mu + \epsilon^a s^\mu_a
    = \xi^t t^\mu + \xi^a s^\mu_a
    \ ,
    \\
    \xi^t &=& \frac{\epsilon^0}{N}
    \quad ,
    \quad
    \xi^a = \epsilon^a - \frac{\epsilon^0}{N} N^a
    \,.
\end{eqnarray}
\label{eq:Diffeomorphism generator projection2}
\end{subequations}
At this point, the on-shell condition requires that the constraints be solved
and equations of motion hold, which allows us to replace momenta with
time derivatives of the configuration variables on phase space.

The timelike components of the covariance condition are automatically
satisfied by virtue of the hypersurface-deformation brackets,
\eqref{eq:Hypersurface deformation algebra - modified}, via the gauge
transformation of the lapse function and shift vector, \eqref{eq:Off-shell
  gauge transformations for lapse - modified2} and \eqref{eq:Off-shell gauge
  transformations for shift - modified2}, provided the covariance condition of
the spatial metric,
$\delta_\epsilon q_{a b} |_{\text{O.S.}} = \mathcal{L}_{\xi} q_{a b}$, is
satisfied \cite{EffLine}.  The latter does not automatically hold for any
anomaly-free constraint algebra of hypersurface-deformation form. It can be
simplified to the conditions \cite{HigherCov}
\begin{equation}
    \frac{\partial (\delta_{\epsilon^0} \tilde{q}^{a b})}{\partial (\partial_c \epsilon^0)} \bigg|_{\text{O.S.}}
    = \frac{\partial (\delta_{\epsilon^0} \tilde{q}^{a b})}{\partial (\partial_c \partial_d \epsilon^0)} \bigg|_{\text{O.S.}}
    = \cdots
    = 0
\end{equation}
already shown in (\ref{eq:Covariance condition of 3-metric - modified - reduced}),
where $\delta_{\epsilon^0} \tilde{q}^{a b} = \{ \tilde{q}^{a b} ,
H[\epsilon^0]\}$ without a spatial shift.

We now extend the covariance condition to the scalar multiplet $\phi^I$. For a
canonical theory with hypersurface-deformation brackets \eqref{brackets}, we
say that the scalar field is covariant if its amplitude obeys
\begin{equation}
    \delta_\epsilon |\phi|^2 \big|_{\text{O.S.}} =
    \mathcal{L}_{\xi} |\phi|^2 \big|_{\text{O.S.}}
    \ .
    \label{eq:Covariance condition of phi - modified}
  \end{equation}
As in the case of a single-component scalar field, shown in Section~\ref{sec:Single},
this equation implies the conditions
\begin{equation}
    \phi^I \frac{\partial H}{\partial ( \partial_c P_I)}
    = \phi^I \frac{\partial H}{\partial ( \partial_c \partial_d P_I)}
    = \cdots
    = 0
    \ .
    \label{eq:Matter covariance condition - second reduced form}
\end{equation}
Unlike the single scalar field, the Hamiltonian constraint of a multiplet
allows derivatives of the conjugate momenta $P_I$ through the dependence
\begin{equation}
    H \left(\phi^{\bar{I}} \partial_c P_{\bar{I}} - \phi^{\bar{J}} \partial_c
      P_{\bar{J}} , \phi^{\bar{I}} \partial_{c_1} \partial_{c_2} P_{\bar{I}} -
      \phi^{\bar{J}} \partial_{c_1} \partial_{c_2} P_{\bar{J}} , \dots \right) 
    \ , \label{eq:H dependence on P derivatives - covcond}
\end{equation}
where $\bar{I}\neq \bar{J}$ are understood as non-contracted.

Anomaly-freedom of the constraints and general covariance of space-time as
well as matter are non-negotiable conditions to be placed on a modified theory
of space-time.  In the following, we formulate a series of further conditions
that we may require for a modified theory, but as we will find out, not all of
them are mutually inclusive.

\subsubsection{${\cal G}$-invariance and conservation of the scalar current}

In quantum field theory on a curved space-time, the generator
(\ref{eq:Symmetry generator of scalar field}) plays a role in the definition
of the Klein--Gordon inner product because of its many useful properties, in
particular its being preserved under time evolution.  If we require a
well-defined field quantization of matter in emergent modified gravity, we
should preserve the existence of the conserved current. The imposition of this
condition depends on the specific application of emergent modified gravity. If
it is used as an alternative to general relativity on which quantization may
be built, we must impose the condition of a conserved matter current. This
condition may be relaxed if emergent modified gravity is viewed as a possible
effective theory of some quantum theory of gravity constructed by other
means. If the underlying fundamental theory contains matter fields, it
provides quantized gravity and matter, and we do not need to re-quantize a
scalar field on an effective space-time geometry. The condition that there be
a conserved scalar current could then be relaxed. In practice, however, even
in this case one would usually desire an intermediate regime of quantized
matter coupled to classical gravity. If the gravitational sector of this
quantum-gravity theory is emergent, the intermediate regime would still need a
conserved scalar current for meaningful quantum fields on the emergent
background to exist. We are not re-quantizing the scalar field in this case,
but rather assume that it retains its quantum properties while gravity is
close to its classical limit.

The requirement that the theory is $\mathcal{G}$-invariant implies that the
brackets (\ref{eq:G-invariance}) hold, which in turn implies that
(\ref{eq:Symmetry generator of scalar field}) is a conserved charge associated
with a Noether current. The equivalence between $\mathcal{G}$-invariance and
the existence of a conserved current in general does not apply to the single
scalar field.  However, in the classical single-scalar theory, there is a
well-known conserved current for the {\em free} field, obtained when the
potential vanishes.  In what follows, we will assume that conservation of the
single-scalar current in the free limit (or, equivalently,
$\mathcal{G}$-invariance in the case of a scalar multiplet) is a necessary
condition because it covers a more useful set of interesting applications than
a non-conserved effective current.

Therefore, we postulate that the modified theory contains a conserved current.
In order to formulate this condition in a specific way, we make use of the
generator (\ref{eq:Symmetry generator of scalar field}), which does not depend
on the structure function or on any other phase-space function except for the scalar
field and its conjugate momentum, and demand that the Hamiltonian and
diffeomorphism constraints of the modified theory commute with it up to
possible boundary terms.  That is, given a lapse function $N$ and shift vector
$N^a$, the modified constraints $H$ and $H_a$ must commute with the generator
$G$ up to boundary terms such that, at least on-shell,
\begin{eqnarray}
    \{ G_i , H [N] + H [N^a] \} |_{\rm O.S.} = - \partial_a J^a_i |_{\rm O.S.}
    \,.
    \label{eq:Conserved current condition}
\end{eqnarray}
If this condition is satisfied, we identify $J^t_i=G_i$ as the charge density
associated to the $i$-th generator of the Lie algebra, as in (\ref{GJt}), and
the boundary term $J^a_i$ as the spatial current density associated with the
observer's frame (defined via $N$ and $N^a$ in the Hamiltonian and
diffeomorphism constraints).

Using the matter covariance condition (\ref{eq:Spatial covariance condition -
  second reduced form}), equation (\ref{eq:Conserved current condition}) can
be written as
\begin{eqnarray}\label{partialJ}
    &&
    \partial_a J_i^a
    =
    - (\tau_i)^I{}_J \bigg[
    \partial_{a} \left(
    \frac{\partial H}{\partial (\partial_a \phi^I)} \phi^J N
    + P_I \phi^J N^a \right) 
    \\
    &&- \left( 
    \frac{\partial H}{\partial \phi^I} \phi^J
    + \frac{\partial H}{\partial (\partial_{b} \phi^I)} \partial_{b} \phi^J
    - P_I \frac{\partial H}{\partial P_J}
    - \partial_b P_I \frac{\partial H}{\partial (\partial_b P_J)}
    - \partial_{b_1} \partial_{b_2} P_I \frac{\partial H}{\partial (\partial_{b_1} \partial_{b_2} P_J)} - \dotsi
    \right) N \bigg]
    \ , \nonumber
\end{eqnarray}
where we have assumed the constraints depend on derivatives of the field
up to first-order (while derivatives of the momentum are allowed to be of
higher finite order) and neglected boundary terms of the constraints. For the
symmetry generator to be preserved under time evolution, the second term in the
parenthesis on the right-hand side must vanish.  Using antisymmetry of the
Lie-algebra generators $\tau_i$, this condition implies the usual dependence
\begin{equation}
    H = H ( \delta^{I J} P_I P_J , \delta_{I J} \phi^I \phi^J , \delta^{I J} \partial_a P_I \partial_b P_J , \delta_{I J} \partial_a \phi^I \partial_b \phi^J , P_I \phi^I , P_I \partial_a \phi^I , \phi^I \partial_a P_I )
    \ ,
\end{equation}
of a possible modified Hamiltonian constraint on scalar fields and
momenta. (Higher-order spatial derivatives of $P_I$ are allowed as long as its
$\mathcal{G}$-index is contracted.)  Combining this dependence with the one allowed by
covariance, (\ref{eq:H dependence on P derivatives - covcond}), we conclude
that the Hamiltonian constraint cannot depend on derivatives of the momenta,
and is reduced to the dependence
\begin{equation}
    H = H ( P_I P^I , \phi_I \phi^I , \partial_a \phi_I \partial_b \phi^I , P_I \phi^I , P_I \partial_a \phi^I )
    \ .\label{eq:H dependence on scalar field - cov and gauge inv}
\end{equation}
This form
is compatible with, but is not limited to, the dependence of the classical
constraint (\ref{eq:Classical action for scalar field}).
The spatial component of the conserved current is then
\begin{eqnarray}
    J_i^a
    &=&
    - (\tau_i)^I{}_J \left( \frac{\partial H}{\partial (\partial_a \phi^I)} \phi^J N
    + P_I \phi^J N^a \right)
    \,,
\end{eqnarray}
read off from the boundary term in (\ref{partialJ}).

\subsubsection{Gravitational mass as an observable}

A Dirac observable is a phase-space function that weakly Poisson-commutes with
all the constraints, such that the Poisson brackets vanish when the
constraints are satisfied.  Dirac observables are thus preserved under time
evolution if the system is fully constrained.  The smeared symmetry generator
of the scalar field discussed above is an example of a Dirac observable
associated to the matter field. However, general relativity in its
4-dimensional form does not have such observables associated to the
gravitational field in any obvious way. The construction of gravitational
Dirac observables is simplified in the presence of boundaries or asymptotic
fall-off conditions, in which case boundary terms of the constraints can often
be related to Dirac observables with physical meaning
\cite{ReggeTeitelboim}. In vacuum spherical symmetry, which we will discuss in
detail in the next section, a Dirac observable exists which has the
physical meaning of mass.  The existence of such an observable is desirable
for various reasons, and therefore we postulate that the modified theory must
preserve the existence of a mass observable, at least in vacuum.  If this
condition is violated, there is no unambiguous definition of the gravitational
mass, a questionable outcome in a supposedly gravitational theory.

We conclude that the existence of both the matter and gravitational
observables is important.  They will play a crucial role
in restricting the class of anomaly-free, covariant constraints even further.

\subsubsection{Factoring out canonical transformations}

The canonical formulation of a specific theory in general is uniquely defined
only up to an application of canonical transformations. In a classification of
new versions of canonical theories it is therefore essential to eliminate the
freedom of performing canonical transformations by imposing suitable
relationships between the canonical variables or other phase-space
functions. If this step is omitted, a canonical transformation of the
classical theory might be misclassified as a new modified theory, even though
it would not imply new physics, or two equivalent modified theories might be
misclassified as different ones. A careful treatment of canonical
transformations also makes it possible to clarify whether specific
modifications are required by a certain quantization approach, such as
polymerization in models of loop quantum gravity, or merely appear because a
fixed set of canonical variables has been used.

Some canonical transformations can easily be eliminated because they would not
preserve the diffeomorphism constraint, which we always assume to be
unmodified. But the condition of preserving the diffeomorphism constraint
still leaves a large class of possible canonical transformations. We will
therefore impose additional conditions, guided for instance by how certain
modification terms appear in the Hamiltonian constraint that can sometimes be
eliminated by a canonical transformation, simplifying follow-up
calculations. While the general condition that canonical transformations be
factored out is essential, the specific implementation therefore depends on
detailed steps of our constructions and, to some degree, is subject to
preferences in the solution procedure. (For instance, the vacuum models of
\cite{SphSymmMinCoup} are based on different but canonically equivalent
choices compared with those of \cite{HigherCov}.)

\subsection{Desirable properties}

The structure of hypersurface-deformation brackets as well as general
properties of space-time solutions related to singularities suggest additional
conditions that may not be strictly necessary (as always, depending on how
emergent modified gravity is used) but are strongly desirable for common applications.

\subsubsection{Absence of singularities}

One motivation to pursue general physical theories beyond the standard model
and general relativity rests on the expectation that new physics may tame some
of the divergences present in standard dynamical solutions. In the case of
general relativity, the most well-known divergences are singularities at the
center of black holes and at the big bang. In some cases, coupling matter to
gravity can have a significant effect on the structure of singularities.

Emergent modified gravity may be viewed as a novel class of fundamental
theories that grant us access to new geometrical models of space-time beyond
general relativity. It is therefore important to ask what this class of
theories may tell us about divergences and singularities.  In vacuum, it has
been shown \cite{SphSymmEff,SphSymmEff2} that some of the modifications in
emergent modified gravity may resolve the classical singularity of a static
black hole.  In the presence of matter, the resolution of the singularity is
not guaranteed. For instance, by coupling a perfect fluid in a covariant way,
it was shown that the gravitational collapse of dust develops a singularity
once again, although in a more harmless way compared to the classical case
\cite{EmergentFluid}. (The case of a perfect fluid differs from the scalar
field in that the conditions of anomaly-freedom and covariance determine the
theory almost uniquely, except for a free function in the pressure term. A
perfect fluid is therefore always close to minimal coupling. Moreover, in this
case the structure function does not depend on the matter variables.)
Given this partial evidence, we expect
that a certain class of modified constraints coupled to the scalar field
within emergent modified gravity will still develop singularities, but there
is a chance that some modifications imply dynamical solutions free of this
dynamical divergence.

The matter case in spherical symmetry differs qualitatively from vacuum
solutions because the presence of scalar field implies a new local degree of
freedom. Different initial conditions chosen for the scalar field represent
different physical scenarios, which may have an effect on the nature of the
singularity (or its absence).  Furthermore, the equations of motion we will
obtain are complicated to work with analytically in general scenarios, for
instance because the matter field may contribute to the emergent space-time
metric. Emergent gravity-matter theories are usually more strongly coupled
than their classical counterparts.  In our explicit examples, we will focus on
a specific and simplified physical scenario, given by a spatially constant
scalar field on a collapsing homogeneous space with a topology suitable for a
region within spherically symmetric space-time. This scenario is intended to
model the interior of a black hole, which in the static vacuum case is indeed
homogeneous. While it is limited, it does allow us to observe interesting and
non-trivial distinctions between different outcomes, depending on the class of
scalar coupling in emergent modified gravity.

The resolution of singularities is a strongly desired fundamental property and
may therefore be used to rule out versions of emergent modified gravity that
do not lead to this outcome. However, emergent modified gravity presents a
classical setting of space-time physics, and additional quantum effects that
cannot be modeled by some of the modification functions in an effective way
could contribute to the resolution of singularities even if a version of
emergent modified gravity, by itself, does not do so. For this reason, we do
not consider singularity freedom as a strict condition to be imposed on
emergent modified gravity, in contrast to conditions such as covariance or
conservation laws that are required for internal consistency of a given
space-time theory coupled to matter.

\subsubsection{Absence of kinematical divergences in the Hamiltonian constraint}

As a phase-space function, the classical Hamiltonian constraint always takes
finite values provided it is evaluated for non-degenerate spatial metrics and
bounded extrinsic curvature as well as finite matter variables. The equations
of motion it generates then contain only finite terms under these conditions.
The non-degeneracy condition on the spatial metric and boundedness of
extrinsic curvature may not always be satisfied in certain regions of explicit
solutions of the equations of motion, for instance at a horizon or a physical
singularity. However, these divergences are properties of solutions in regions
where the canonical fields reach boundaries of phase space.

It turns out that some versions of emergent modified gravity imply stronger
divergences of the Hamiltonian constraint as a phase-space function, for
instance at values of some of the gravitational phase-space fields in the
interior of phase space, such as finite momenta with non-degenerate
configuration degrees of freedom. Such divergences then also appear in
equations of motion generated by the constraint, and not only in their
solutions. Similarly, it is possible to have divergences of the Hamiltonian
constraint at certain values of a matter field even if the gravitational
degrees of freedom are in well-defined interior regions of phase-space. When
this happens, the interpretation of the Hamiltonian constraint as a
well-defined generator of gauge transformations or evolution breaks down, even
though we have not reached a boundary of phase space where we may have to look
for a reparameterization of solutions for instance by a coordinate
transformation in the covariant space-time picture.

Depending on the solution procedures to be applied, it may therefore be
desirable to restrict modified theories to those cases in which the
Hamiltonian constraint does not have divergences in the phase-space
interior. This kind of divergences is sometimes related to, but usually not
identical with, the concept of space-time singularities. It does not make use
of the emergent space-time metric but only of the constraint functions and
their equations of motion.

\subsubsection{Partial Abelianization}

Partial Abelianization was proposed in \cite{LoopSchwarz,LoopSchwarz2} to
simplify common quantization procedures that are often untractable in the
presence of structure functions.  The general idea of this proposal is to
define a new phase-space function as a linear combination of the constraints,
$H^{\rm (A)}= B H + A^a H_a$, such that the Poisson brackets of $H^{\rm (A)}$
together with the classical diffeomorphism constraint take the form of
hypersurface-deformation brackets, but with a vanishing structure
function. The geometrical space-time interpretation is then lost because there
is no non-vanishing candidate for a spatial metric, but the resulting
partially Abelian algebra is free of structure functions and may be quantized
more easily through operator versions of the equations $H^{\rm (A)}=0$ and
$H_a=0$. The condition of partial Abelianization requires that 
$A^a$ and $B$ are phase-space functions, such that the off-shell behavior of the resulting
theory is different from the original version. For this reason, it may be
considered a modified theory, but not directly of space-time or gravity
because a compatible space-time geometry must be recovered in a more indirect
way than in emergent modified gravity.

A useful property of a partial Abelianization is that the phase-space
submanifold given by $H^{\rm (A)}=0$ and $H_a=0$ is identical with the
classical constraint surface. Classical solutions to the constraints can
therefore be used, but their gauge behavior and equations of motion are not
necessarily classical. Moreover, $H^{\rm (A)}$ is preserved under time
evolution in the normal direction in the absence of a spatial shift.

In \cite{LoopSchwarz,LoopSchwarz2}, a partial Abelianization for spherical
symmetry was constructed in two steps, first combining the classical
diffeomorphism and Hamiltonian constraints in order to remove $K_x$ from the
resulting expression, and then integrating by parts. The second step removes
spatial derivatives from the remaining terms of $K_{\varphi}$ and
$E^{\varphi}$, which implies a vanishing $\{H^{\rm (A)}[N],H^{\rm (A)}[M]\}$
based only on antisymmetry of the Poisson bracket: the bracket produces only
the vanishing $NM-MN$ and no term of the form $NM'-N'M$. However, integrating
the original local constraint functions turns them into global expressions
that require a careful analysis of boundary terms, and it obscures any
possible relationship of the resulting gauge theory with space-time geometry,
and the new lapse function becomes explicitly phase-space dependent. There is
an intrinsic consistency problem in this version of partial Abelianization
because boundary terms in a theory of gravitational variables require concepts
such as mass observables or asymptotically flat regions, but they are
available only if the theory has a consistent space-time interpretation. As
shown in \cite{HigherCov}, an emergent space-time metric does exist in some
partially Abelianized theories, but its spatial part is not necessarily
identical with the classical $(E^{\varphi})^2/E^x$ that had been implicitly
assumed in \cite{LoopSchwarz,LoopSchwarz2}.

A new method that manages to obtain a local off-shell partial Abelianization
with a compatible space-time interpretation has been given in
\cite{HigherCov}. Moreover, in vacuum spherical symmetry, it was shown that a
partial Abelianization of this kind, if it exists, is always unique up to an
overall factor multiplying the new constraint, both for the classical $H$ and
for a general covariantly modified $\tilde{H}$ as the initial expression of
the Hamiltonian constraint. In the latter case, partial Abelianization is
possible only if a specific modification function vanishes.  Therefore, the
possibility of a partial Abelianization can be used as another condition in a
classification of modified canonical theories.  The importance of this
condition depends on the specific application of emergent modified gravity,
viewing it as a potential effective description of some quantum theory of
gravity, or a new and more general starting point for a quantization of
gravity not necessarily based on general relativity. In the former case, the
existence of a partial Abelianization may simplify some calculations but is
not necessary because we would not re-quantize the underlying phase-space
degrees of freedom and constraints. In the latter case, the existence of a
partial Abelianization is strongly desired because it may help to construct
consistent quantizations of the constraints. As shown in \cite{HigherMOND}, a
fundamental origin of MOdified Newtonian Dynamics (MOND,
\cite{MOND1,MONDRotation,MOND2}) may then be obtained because the conditions
on partial Abelianization may require logarithmic terms in modification
functions that can be relevant on intermediate scales.

\subsubsection{Polymerization of the scalar  field}

An example of modified scalar theories is given by so-called polymerization,
motivated by mathematical constructions in loop quantum gravity.  An ongoing
challenge in this field is whether bounded phase-space functions such as
holonomies, used for a well-defined kinematical quantization scheme, can be
introduced and studied effectively as modifications of the constraints in a
way that preserves covariance.  In the example of a single real scalar field,
the general scheme requires that the Hamiltonian constraint be modified such
that it depends on the scalar field only via point holonomies
\cite{FermionHiggs}, defined as bounded and pointwise periodic functions
\begin{eqnarray}
    h_\phi (x) = \exp \left( i \nu \phi(x)\right)
\end{eqnarray}
with a constant $\nu$.  In what follows, we will refer to $\nu$ simply as the
holonomy parameter, which is usually considered a quantization ambiguity to be
fixed by phenomenological considerations. More generally in the context of
modified theories, $\nu$ may also be a phase-space function depending on the
gravitational variables of the canonical theory. A polymerized theory is a
modified scalar theory, possibly coupled to gravity, in which any
$\phi$-dependence of the constraints can be written through a dependence on
$h_{\phi}$.

Most scalar potentials of interest are not of a polymerized form and must
therefore be modified if a polymerized theory is desired. Moreover, the
spatial current (\ref{current}) is not of the required form and must be
adjusted to a polymerized theory, or be derived anew from a consistent
modified constraint if it is to comply with the principle of point
holonomies. Without systematic derivations, it is then unclear how a
compatible time component $J^t$ for a space-time current can be found.

Polymerization may also be applied to the gravitational dependence, in which
case it usually appears for $K_{\varphi}$ in spherically symmetric theories
because this component appears in the Ashtekar--Barbero connection
\cite{AshVar,AshVarReell} used in loop quantum gravity and, unlike $K_x$, has
spatial density weight zero and can therefore be exponentiated. If the
gravitational variables are polymerized in this way, general covariance is a
major question that can be addressed by emergent modified gravity. If gravity
is coupled to scalar matter, an important question is whether both kinds of
polymerization can be applied consistently, maintaining covariance and the
existence of a conserved current.

In this work we explore modified theories much more general than those
proposed by spherically symmetric models of loop quantum gravity.
Polymerization will therefore not play a central role in the modifications we
are seeking.  However, owing to the great interest enjoyed by polymerization
in loop quantum gravity and its critical covariance issues, which have rarely
been addressed in a successful manner, we will discuss possible ways in which
polymerization can be accommodated in emergent modified gravity. The consistent versions
turn out to be highly restricted and non-trivial, shedding light on the
important question of whether and how loop quantum gravity may be compatible
with space-time covariance even on a semiclassical level of effective
space-time line elements.

\subsection{Organizational principles}

Finally, we formulate several further conditions that may be used to classify
mutually distinct classes of emergent modified gravity. Most of these
conditions take the form of requiring the existence of certain limits, which
also help us to interpret possible physical effects in general terms.

\subsubsection{Classical constraint surface in a limit}

A modified constraint will inevitably change the dynamics of the system via
the equations of motion.  However, modified constraints may preserve the
classical constraint surface in some cases.  One example is given by the
spherically symmetric modified constraint first obtained in
\cite{SphSymmEff}. A simple way to arrive at a modified Hamiltonian constraint
obeying this condition is by postulating a new constraint as an invertible
linear combination of the classical Hamiltonian constraint and the
diffeomorphism constraint, $H^{\rm (new)}=B H^{\rm ({\rm cl})} + A H_x$, where
$A$ and $B\not=0$ are initially free phase-space functions.

One usually does not expect modifications of physical solutions if they are
derived from invertible linear combinations of the original constraints and
gauge generators.  However, while such linear combinations preserve the
constraint surface, they can change the off-shell behavior of gauge
transformations and, for the gravitational constraints, possibly the structure
function as well. These two ingredients are crucial in relating constraint
brackets first to hypersurface deformations, and then to a compatible
space-time geometry obeying general covariance such that infinitesimal
coordinate changes are equivalent to gauge transformations on-shell. If we
start with a modified canonical theory, a potential geometrical space-time
interpretation of its solutions is yet to be derived. Using the well-known
relationship $g^{\mu\nu}=q^{\mu\nu}-n^{\mu}n^{\nu}$ between the inverse
space-time metric $g^{\mu\nu}$, the inverse spatial metric $q^{\mu\nu}$, and
the unit normal $n^{\mu}$ on spacelike hypersurfaces of a foliation, we must
be able to identify both $q^{\mu\nu}$ and $n^{\mu}$ in order to find a
candidate for the space-time metric.

The inverse spatial metric is determined by the structure function of modified
but anomaly-free constraint brackets, using the known brackets of hypersurface
deformations.  The unit normal does not appear explicitly as another structure
function, but it is implicitly determined by what we consider the Hamiltonian
constraint to be among all the constraints. This property again follows from
the known brackets of hypersurface deformations, in which $H[N]$ is the
generator of normal deformations, singled out among the constraints by the
condition that it be the only one with a structure function in
$\{H[N],H[M]\}$. Replacing the classical constraint $H^{({\rm cl})}$ by a
linear combination $H^{({\rm new})}$ with the diffeomorphism constraint
changes the identification of the normal direction compared with the classical
theory, provided the linear combination is done in an anomaly-free way that
preserves the hypersurface-deformation property of
$\{H^{({\rm new})}[N],H^{({\rm new})}\}$ depending only on the diffeomorphism
constraint off-shell. Therefore, both $q^{\mu\nu}$ and $n^{\mu}$ can be
derived from anomaly-free constraint brackets, the former from the structure
function and the latter (implicitly) from how the Hamiltonian constraint is
singled out among all the constraints. The covariant space-time interpretation
of solutions of the theory is therefore not invariant under taking linear
combinations of the constraints. It also follows that we cannot replace the
diffeomorphism constraint by a linear combination with $H^{({\rm cl})}$
because doing so would introduce structure functions in Poisson brackets of
the diffeomorphism constraint, which is not compatible with
hypersurface-deformation commutators.

This discussion demonstrates the importance of two requirements to be imposed
on the new constraint $H^{({\rm new})}$.  Together with the classical
diffeomorphism constraint, the new Hamiltonian constraint must still satisfy
the hypersurface-deformation brackets, perhaps with a modified structure
function. And the emergent space-time metric obtained from this modified
structure function must be covariant according to the general conditions
derived in \cite{HigherCov} and reviewed earlier in the present paper.  In
vacuum, it turns out that these two requirements uniquely determine the
general form of the phase-space functions $A$ and $B$ (up to an overall factor
multiplying the new constraint), which in turn determine the general form of
the modified structure function.

If the new Hamiltonian constraint is a linear combination of the
classical constraints, the new constraint surface is the same as
determined by the classical constraints. However, gauge transformations
and the dynamics generated by the modified constraints are in general
non-classical, and so is the emergent space-time metric.  The most general
modified constraint in vacuum for spherically symmetric systems
\cite{HigherCov} allows further modifications that can make the modified
constraint surface non-classical. But given the existence of modified
theories with a classical constraint surface, there is a limit of any further
free functions in the general modification such that the classical constraint
surface is recovered, without having to take the full classical limit.
We refer to this non-trivial limit as a limit of reaching the classical
constraint surface. The dynamics and emergent space-time may remain modified
in this limit.

We are not aware of a fundamental argument that would require us to preserve
the classical constraint surface in a modified theory. However, one may use
this condition as a way of keeping the modifications as minimal as possible,
which is often useful in novel classes of theories that possess a large number
of free functions and possible modifications. For example, a standard
modification of general relativity may have an infinite number of independent
curvature scalars in the action. But the simplest non-trivial model is given
by the classical choice of simply using the Ricci scalar $R$, which can be
used to motivate $f(R)$ theories as a large class of tractable modifications.
In the same vein, we postulate the existence of a non-trivial limit of
reaching the classical constraint surface imposed as a condition on certain
modified theories as a principle that we can follow to differentiate between
two classes of modified constraints, those that do possess such a limit and
those that do not.  Unlike the conditions of anomaly-freedom and covariance,
or the existence of observables, we do not consider the existence of a
non-trivial limit of reaching the classical constraint surface as
non-negotiable or strongly desired. We use it only in order to define these
two distinct sets of modified constraints, thereby organizing a larger class
of possible modifications.

\subsubsection{Classical matter in a  limit}

We expect, and show below, that the coupling of matter to a modified theory
will allow modifications with additional free functions beyond those obtained
in vacuum, in particular functions depending on the canonical matter field.
With this result in mind, there is another limit of interest, which we call
the classical-matter limit. In this limit, by definition, the equations of
motion of the matter fields take their classical form, except for the
appearance of the emergent space-time metric instead of the classical
one. This limit is therefore closely related to a choice of minimal coupling.

Since we are focusing on the scalar field here, the classical-matter limit
will manifest itself as the condition that the Klein--Gordon equation be
reproduced in a curved, emergent spacetime.  This condition is similar to the
limit of reaching the classical constraint surface, in that it is neither
non-negotiable nor strongly desired, but it can be used to differentiate
between two classes of modified constraints, depending on whether the limit exists.

\subsubsection{Classical geometry in a limit}

We define the classical-geometry limit such that it leads to a space-time
picture of solutions with a classical, non-emergent space-time line element.
If the space-time is non-emergent, the spatial metric (or a triad) used as a configuration
degree of freedom on phase space is then equivalent to the gravitational
field, as in general relativity. However, equations of motion for the
gravitational field obtained in this limit may still be non-classical due to
residual freedom in modifications functions that do not affect the emergent
metric.

\subsubsection{Classical gravity in a limit}

Applying a further restriction or limit on the modification functions that
lead to the classical constraint surface, we may require that the equations of
motion have a limit equivalent to Einstein's equation with the classical
space-time metric. In the presence of matter, the stress-energy tensor may
retain non-classical featues in this limit, depending on some of the remaining
modification functions.

\subsubsection{Summary of classical limits}

After identifying the above conditions that may be imposed on a modified
theory, we conclude that there is more than one kind of limit that may be
considered classical.

\begin{itemize}
\item \textbf{Classical constraint surface in a limit:} Defined as the limit
  in which the modified constraints define the same constraint surface in
  phase space as the classical constraints, this property is possible even
  when the constraints and their emergent space-time are non-classical.
\item \textbf{Classical-matter limit:} This limit is defined such that the
  equations of motion of matter take the classical form, except for an
  appearance of the emergent rather than classical space-time metric. In the
  explicit example of the scalar field in spherical symmetry given below, the
  classical-matter limit means that the equation of motion for the matter
  field is the Klein--Gordon equation on a curved, emergent space-time.
    \item \textbf{Classical-geometry limit:}
    Defined as the limit in which the structure function in
    hypersurface-deformation brackets takes the classical form, 
    it retains a  possibility of modified dynamics on a space-time of
    classical type. 
  \item \textbf{Classical-gravity limit:} Defined as the limit in which the
    gravitational equations of motion take the form of Einstein's equation, it
    includes the classical-geometry limit but is more
    restrictive because the latter does not require classical equations of
    motion. While Einstein's equation is  recovered in this limit, the
    stress-energy tensor may be non-classical.
  \item \textbf{Vacuum limit:} One other limit we may be interested in is the
    vacuum limit, although it is not necessarily classical. From
    \cite{HigherCov}, we know the most general modified Hamiltonian constraint
    for the vacuum case in an expansion to second order in spatial derivative
    terms.  Thus, we can use this expression as a limiting case to be
    recovered when we remove the matter field.
\end{itemize}

\subsection{A-priori and a-posteriori principles}

We finish this section by noting the non-trivial nature of any application of
the conditions discussed above.  In applying these conditions we are
implicitly using them as guiding principles.  In particular, we will
distinguish between \emph{a-priori} and \emph{a-posteriori} principles based
on how they can be applied to restrict or classify the modified theories. This
dinstinction is different from the three sets of conditions, given by
necessary requirements, desirable properties, and the existence of certain
limits. In the following we classify the principles into a-priori and
a-posterior based on the procedures we followed for the spherically symmetric
system, the details of which are given in the following sections.

The a-priori principles are those that we can apply as conditions on the
modified theory before obtaining an explicit expression of the constraint.
The two archetypal a-priori principles here are anomaly-freedom and general
covariance. Because they are required for internal consistency of a space-time
theory, we must apply both conditions from the very beginning. They will
provide us with a system of differential equations that the constraints and
their modification functions must satisfy. However, the full system of
equations is complicated, and we will not be able to solve it exactly.  In
order to simplify these equations, we will apply a few additional conditions
in various combinations, which we will refer to as a-priori too.  One such
condition is the existence of the classical-matter limit and another is the
existence of a limit in which the classical constraint surface is reached. As
it will turn out, these two conditions are not mutually inclusive.  The
condition of the existence of the classical-matter limit will be restrictive
enough to simplify the conditions for anomaly-freedom and covariance such that
they can all be solved exactly. We then obtain an explicit expression of the
Hamiltonian constraint with some ambiguities in the modifications that
manifest themselves as undetermined functions of some of the phase-space variables.
On the other hand, the limit of a classical constraint surface, while
simplifying the anomaly-freedom and covariance conditions, is still too
complicated to be solved exactly.  We will find that a specific ambiguity in the
modification functions can be chosen in two distinct versions. The first one
complies with the classical constraint surface as a limit, and the other one
does not, giving rise to the two classes of modified theories.

The a-posteriori principles are the remaining ones listed in this section.
This includes the important ones given by conditions of being free of
singularities and divergences, as they cannot be checked until one has
obtained the dynamical solutions.  The conditions of the existence of the
matter and gravitational observables, and the partial Abelianization, as well
as the existence of the vacuum limit, and of scalar-field polymerization, can
be applied directly to the explicit expressions of the constraints obtained
from the a-priori principles, restricting (or classifying) their modification
ambiguities to comply with these conditions.

As an example, we may pick the simplest, but non-classical, constraint version
of each class, solve for the dynamical solutions it implies in the homogeneous
case, and check whether a singularity develops as expected classically.  The
outcome determines whether these constraints belong to the class of
singularity-free ones.  Surprisingly, we find that neither the class of
constraints compatible with the classical-matter limit nor with the limit of a
classical constraint surface are singularity-free. Singularity-freedom is
allowed only by the remaining class, following just the a-priori principles of
anomaly-freedom and covariance and some weaker conditions.  We also find that
scalar-field polymerization does not play a crucial role in the taming of a
spatially homogeneous singularity, but all classes can, in fact, be
polymerized.

\section{Spherically symmetric theory with a scalar field}
\label{sec:Spherically symmetric sector}

We now present detailed derivations of theories of emergent modified gravity subject to our
conditions from the preceding section.

\subsection{Classical theory}

From Section~\ref{subsec:Spherically symmetric sector} we recall the following
elements of the spherically symmetric classical theory in vacuum. The spacetime metric is
\begin{equation} \label{eq:ADM line element - spherical2}
    {\rm d} s^2 = - N^2 {\rm d} t^2 + q_{x x} ( {\rm d} x + N^r {\rm d} t )^2 + q_{\vartheta \vartheta} {\rm d} \Omega^2
\end{equation}
with
\begin{equation}
    q_{x x} = \frac{(E^\varphi)^2}{E^x}
    \quad , \quad
    q_{\vartheta \vartheta} = E^x
\end{equation}
where $E^x$ and $E^\varphi$ are the radial and angular components of a
densitized triad, respectively, assuming $E^x>0$ in order to fix spatial
parity.
The canonical pairs are $(K_\varphi , E^\varphi)$ and $(K_x , E^x)$ for
gravity and
$(\phi,P_\phi)$ for a single scalar field, such that
\begin{equation}
    \{ K_x (x) , E^x (y)\} = \{ K_\varphi , E^\varphi (y) \} = \{ \phi (x) ,
    P_\phi (y) \} = \delta (x-y)\,. 
\end{equation}
The diffeomorpshism and Hamiltonian constraints are given by
\begin{equation}
  H_x
    = E^\varphi K_\varphi' - K_x (E^x)'
    + P_\phi \phi'
    \label{eq:Diffeomorphism constraint - spherical symmetry - Scalar field}
  \end{equation}
  and
  \begin{equation}
        H = H_{\rm grav} + H_\phi
    \label{Hamiltonian constraint - spherical symmetry - Classical}
\end{equation}
where $H_{\rm grav}$ and $H_\phi$ are the gravitational and matter
contributions to the Hamiltonian constraint. In the classical theory with a
cosmological constant $\Lambda$ and minimal coupling of the scalar field, they
are given by
\begin{eqnarray}
    H_{\rm grav}
    &=&
    - \frac{\sqrt{E^x}}{2} \Bigg[
    E^\varphi  \left( - \Lambda
    + \frac{1}{E^x}
    + \frac{K_\varphi^2}{E^x}
    + 4 K_\varphi \frac{K_x}{E^\varphi} \right)
    \nonumber\\
    &&
    - \frac{1}{4 E^x} \frac{((E^x)')^2}{E^\varphi}
    + \frac{(E^x)' (E^\varphi)'}{(E^\varphi)^2}
    - \frac{(E^x)''}{E^\varphi}
    \Bigg]
    \label{eq:Hamiltonian constraint - spherical symmetry - Gravitational contribution - Classical}
\end{eqnarray}
and
\begin{equation}
    H_\phi =
    \frac{1}{2} \left(
    \frac{\sqrt{q^{xx}}}{E^x} P_\phi{}^2
    + E^x \sqrt{q^{xx}} (\phi')^2 + \sqrt{q_{xx}} E^x V (\phi) \right)
    \,.
    \label{Hamiltonian constraint - spherical symmetry - scalar field contribution - Classical}
  \end{equation}
  (A factor of $2$ may be absorbed in the scalar potential.)
These constraints have  Poisson brackets of hypersurface-deformation form,
\begin{subequations}
\begin{eqnarray}
    \{ H_x [N^x] , H_x[M^x] \} &=& - H_x [M^x (N^x)'-N^x(M^x)']
    \ , \label{eq:H_x,H_x bracket} \\
    \{ H [N] , H_x [M^x] \} &=& - H[M^x N'] 
    \ , \label{eq:H,H_x bracket} \\
    \{ H [N] , H[M] \} &=& - H_x \left[ q^{x x} \left( M N' - N M' \right)\right]
    \ , \label{eq:H,H bracket}
\end{eqnarray}
\label{eq:Hypersurface deformation algebra - spherical - Scalar field}
\end{subequations}
with the structure function $q^{x x} = E^x/(E^\varphi)^2$.

The off-shell gauge transformations for the lapse function and shift vector are
\begin{eqnarray}
    \delta_\epsilon N &=& \dot{\epsilon}^0 + \epsilon^x N' - N^x (\epsilon^0)' \ , \nonumber\\
    \delta_\epsilon N^x& =& \dot{\epsilon}^x + \epsilon^x (N^x)' - N^x
                            (\epsilon^x)' + q^{x x} \left(\epsilon^0 N' - N
                            (\epsilon^0)' \right) \ . 
    \label{eq:Off-shell gauge transformations for lapse and shift - spherical2}
\end{eqnarray}
The condition (\ref{eq:Covariance condition of 3-metric - modified - reduced})
for space-time covariance simplifies in spherical symmetry to two sets of equations,
\begin{subequations}
\begin{equation}
    \frac{\partial \left(\{ q^{\theta \theta} , H[\epsilon^0]
        \}\right)}{\partial (\epsilon^0)'} \bigg|_{\text{O.S.}} 
    = \frac{\partial \left(\{ q^{\theta \theta} , H[\epsilon^0]
        \}\right)}{\partial (\epsilon^0)''} \bigg|_{\text{O.S.}} 
    = \cdots
    = 0
\end{equation}
and
\begin{equation}
    \frac{\partial \left(\{ q^{x x} , H[\epsilon^0] \}\right)}{\partial
      (\epsilon^0)'} \bigg|_{\text{O.S.}} 
    = \frac{\partial \left(\{ q^{x x} , H[\epsilon^0] \}\right)}{\partial
      (\epsilon^0)''} \bigg|_{\text{O.S.}} 
    = \cdots
    = 0
    \,.
\end{equation}
\label{eq:Covariance condition - spherical2}
\end{subequations}
These conditions are clearly satisfied in the classical case because the
Hamiltonian constraint does not depend on spatial derivatives of the momenta
canonically conjugate to spatial metric components. The matter
covariance condition (\ref{eq:Matter covariance condition - second reduced
  form}) in spherical symmetry takes the simplified form
\begin{equation}
    \frac{\partial H}{\partial P_\phi'}
    = \frac{\partial H}{\partial P_\phi''}
    = \cdots
    = 0
    \ .
    \label{eq:Matter covariance condition - spherical}
\end{equation}
and is satisfied too.

The gauge transformations of the lapse function and shift vector,
\eqref{eq:Off-shell gauge transformations for lapse and shift - spherical2},
and the realization of the covariance condition \eqref{eq:Covariance condition
  - spherical2} ensure that the space-time metric \eqref{eq:ADM line element -
  spherical2} is covariant in the sense that canonical gauge transformations
applied to the metric reproduce diffeomorphisms when on-shell.  The gauge
functions $(\epsilon^0,\epsilon^x)$ are related to the 2-component vector
$\xi^\mu = (\xi^t,\xi^x)$ generating a radial space-time diffeomorphism by
\begin{eqnarray}
    \xi^\mu &=& \epsilon^0 n^\mu + \epsilon^x s^\mu = \xi^t t^\mu + \xi^x s^\mu
    \ ,
    \nonumber\\
    \xi^t &=& \frac{\epsilon^0}{N}
    \quad , \quad
    \xi^x = \epsilon^x - \frac{\epsilon^0}{N} N^x
    \ .
    \label{eq:Diffeomorphism generator projection - spherical2}
\end{eqnarray}

The global symmetry generator of the real scalar field is
\begin{eqnarray}
    G [\alpha] = \int {\rm d} x\ \alpha P_\phi
    \ ,
    \label{eq:Symmetry generator of real scalar field}
\end{eqnarray}
with constant $\alpha$.  However, unlike the scalar field multiplets with
values in some Lie group, the symmetry of the real scalar field holds only for
the free field, $V=0$. Its Poisson brackets with the constraints is given by
\begin{eqnarray}
    \{ G , H_x [N^x] \} = \left(N^x G\right)'
    \quad,\quad
    \{ G , H [N] \} = 0
    \ ,
    \label{eq:Poisson bracket symmetry generator w constraints - classical - spherical2}
\end{eqnarray}
which is a boundary term.
This gives rise to the conserved current with components
\begin{equation}
    J^t = P_\phi
    \ , \hspace{1cm}
    J^x = - N^x P_\phi
    \ .
\end{equation}

The gravitational mass observable is
\begin{equation}
    m =
    \frac{\sqrt{E^x}}{2} \left(
    1 + K_{\varphi}^2
    - \left(\frac{(E^x)'}{2 E^\varphi}\right)^2 - \frac{\Lambda}{3} E^x \right)
    \ .
    \label{eq:Gravitational observable - Classical}
\end{equation}

\subsection{Covariance in the modified theory}

We consider modifications of the spherically symmetric theory with canonical
variables $(K_\varphi , E^\varphi)$ and $(K_x , E^x)$.  If we modify the
Hamiltonian constraint, then the constraint brackets \eqref{eq:Hypersurface
  deformation algebra - spherical - Scalar field} imply a modified structure
function, $\tilde{q}^{x x}$, which then determines the emergent spatial
metric. The angular component of the metric, which does not independently
appear in the structure functions, remains
unmodified.
The emergent space-time metric is then
\begin{subequations}
\begin{align}
    {\rm d} s^2 =& - N^2 {\rm d} t^2 + \tilde{q}_{x x} ( {\rm d} x + N^x {\rm
                   d} t )^2 + E^x {\rm d} \Omega^2
    \ ,
\end{align}
\label{eq:ADM line element - spherical - modified2}
\end{subequations}
where $\tilde{q}_{x x} = 1 / \tilde{q}^{x x}$, provided
$\tilde{q}^{xx}>0$. (More generally, we can allow for a modified angular
component $\tilde{E}^x\not=E^x$, but we will show that it can always be mapped
back to $E^x$ by a canonical transformation. This function does not affect the
covariance condition.) There is no direct correspondence between the
phase-space variable $E^{\varphi}$ and the spatial metric or a densitized
triad. And since modified constraints generate non-classical equations of
motion, $K_x$ and $K_{\varphi}$ do not have a direct relationship with
extrinsic curvature of spacelike hypersurfaces in the emergent space-time. We
will therefore refer to $E^{\varphi}$ and $E^x$ simply as the gravitational
configuration variables, and to $K_{\varphi}$ and $K_x$ as the gravitational
momenta. (As usual, the roles of configuration variables and momenta could be
reversed.)

The space-time covariance
condition (\ref{eq:Covariance condition of 3-metric - modified - reduced}) for
the angular component of the emergent spatial metric implies, using
$\delta_{\epsilon^0} E^x = - \delta \tilde{H} [\epsilon^0]/\delta K_x$,
\begin{align}
    \frac{\partial \tilde{H}}{\partial K_x'} \bigg|_{\text{O.S.}}
    = \frac{\partial \tilde{H}}{\partial K_x''} \bigg|_{\text{O.S.}}
    = \cdots
    = 0
    \ ,
    \label{eq:Covariance condition on K_x - modified - spherical2}
\end{align}
which restricts the possible modified Hamiltonian constraints to those that do
not contain radial derivatives of $K_x$.  The radial component of the
space-time covariance condition becomes
\begin{align}
    \frac{\partial (\delta_{\epsilon^0} \tilde{q}^{x x})}{\partial (\epsilon^0)'} \bigg|_{\text{O.S.}}
    = \frac{\partial (\delta_{\epsilon^0} \tilde{q}^{x x})}{\partial (\epsilon^0)''} \bigg|_{\text{O.S.}}
    = \cdots
    = 0
    \label{eq:Covariance condition - modified - spherical2}
\end{align}
and does not have simple solutions.
The covariance condition for the scalar field, \eqref{eq:Matter covariance
  condition - second reduced form}, is reduced in spherical symmetry to
\begin{align}
    \frac{\partial \tilde{H}}{\partial P_\phi'} \bigg|_{\text{O.S.}}
    = \frac{\partial \tilde{H}}{\partial P_\phi''} \bigg|_{\text{O.S.}}
    = \cdots
    = 0
    \ ,
    \label{eq:Covariance condition on phi - modified - spherical}
\end{align}
which restricts possible modified Hamiltonian constraints to those that do not
contain radial derivatives of $P_\phi$.

\subsection{Linear combinations of the constraints and  the limit of reaching the classical constraint
  surface}
\label{sec:Linear combination}

The aim of this section is to obtain a covariant modified constraint from a
linear combination of the classical constraints.  We will use these results
later on when we compute more general modified constraints because the class of
modified constraints that comply with the limit of reaching the  classical
constraint surface is closely related to modified theories obtained from
linear combinations of the classical constraints. Such linear combinations
also provide a useful and tractable example of the general analysis.

\subsubsection{Anomaly-free linear combination}

Consider the following linear combination of the classical constraints,
\begin{eqnarray}
    H^{\text{(new)}} = B H^{\text{(old)}} + A H_x
    \ ,
    \label{eq:Linear combinations of constraints - generic}
\end{eqnarray}
where $A$ and $B\not=0$ are, at this point, undetermined phase-space
functions.  We restrict ourselves to the dependence
$B= B(K_\varphi , E^x , \phi)$ including only phase-space fields of spatial
density weight zero. Unlike $B$, the function $A$ must have density weight
minus one and may therefore depend on the remaining fields as well, for
instance through $(E^x)'/(E^{\varphi})^2$. Given these density weights, the
bracket $\{H^{({\rm new})}[N],H_x[M]\}$ is then of the required form, and only
the bracket of two new Hamiltonian constraints must be checked. The derivation
here follows the method of \cite{HigherCov} almost line by line, with the only
major difference given by the inclusion of a scalar field.

We begin by defining the quantities ${\cal B}$ and ${\cal B}^x$ according to
\begin{eqnarray}
    \{ B , H^{({\rm old})} [\epsilon^0] \} 
    \big|_{\text{O.S.}}
    &=:&\left(
    {\cal B} \epsilon^0
    + {\cal B}^x (\epsilon^0)'
    \right)\big|_{\text{O.S.}}\,.
    \label{eq:Transformation of B - Geometric condition - Linear combination}
\end{eqnarray}
In this equation, no second-order derivative of $\epsilon^0$ can appear
because we assumed that $B$ does not depend on the momentum $K_x$ conjugate to the
only variable, $E^x$, that appears with a second-order derivative in the
Hamiltonian constraint. An explicit application of the classical Hamiltonian
constraint $H^{({\rm old})}$ shows that
\begin{equation}
  {\cal B}^x
    =  \sqrt{E^x} \frac{(E^x)'}{2 (E^\varphi)^2} \frac{\partial B}{\partial
      K_\varphi}\,.
\end{equation}
Anomaly-freedom of
\begin{eqnarray} \label{BB}
  &&\{H^{({\rm new})}[N_1],H^{({\rm new})}[N_2]\}
  =  - H_x[B^2 q^{xx}  (N_2N_1'-N_1N_2')]\\
  &&\qquad+  H^{({\rm old})}[BN_2\{H^{({\rm old})}[N_1],B\}]
     -H^{({\rm old})}[BN_1\{H^{({\rm old})}[N_2],B\}]\nonumber\\
  &&\qquad+ H_x[BN_2\{H^{({\rm
     old})}[N_1],A\}]- H_x[BN_1\{BH^{({\rm   old})}[N_2],A\}]\nonumber\\
  &&\qquad- H^{({\rm
     old})}[ABN_2N_1']+ H^{({\rm   old})}[AB N_1'N_2]\nonumber\\
  &&\qquad
- H_x[A^2(N_2N_1'-N_1N_2')]+ H_x[A(N_1A)'N_2]-H_x[AN_1(N_2A)']\nonumber
\end{eqnarray}
in hypersurface-deformation form
implies that all terms proportional to $H^{({\rm old})}$ must cancel out. (We
have used the density weight minus one of $A$ in the last line, which then
vanishes identically.)
This is the case only if
\begin{eqnarray}
    A  &=& - {\cal B}^x
    = - \sqrt{E^x} \frac{(E^x)'}{2 (E^\varphi)^2} \frac{\partial B}{\partial K_\varphi}
    \label{eq:A coefficient - Linear combination}
\end{eqnarray}
from the second and fourth line, which indeed has density weight minus
one. (As usual, antisymmetry means that only terms with derivatives of $N$ 
need be checked.)

Given this expression for $A$, we now write
\begin{equation}
    \{ A , H^{({\rm old})} [\epsilon^0] \}
    =:
    {\cal A} \epsilon^0
    + {\cal A}^x (\epsilon^0)'
    \ ,
    \label{eq:Transformation of A - Geometric condition - Linear combination}
\end{equation}
where
\begin{equation}
    {\cal A}^x
    =
    - \frac{E^x}{(E^\varphi)^2} \left( K_\varphi \frac{\partial B}{\partial K_\varphi}
    + \left(\frac{(E^x)'}{2 E^\varphi}\right)^2 \frac{\partial^2 B}{(\partial K_\varphi)^2} \right)
    \ .
    \label{eq:Transformation of Ax - Geometric condition - Linear combination}
  \end{equation}
  This Poisson bracket, together with
  $B^2\{H^{({\rm old})}[N_1],H^{({\rm old})}[N_2]\}$ in (\ref{BB}),
  contributes a term proportional to the diffeomorphism constraint which is
  allowed for brackets in hypersurface-deformation form.
The combined coefficient of all terms of this form determines the new structure function
\begin{equation}
    \tilde{q}^{xx} =
    B^2 q^{xx} + B {\cal A}^x
    \,,
    \label{eq:New structure function - Linear combination}
\end{equation}
implementing anomaly-freedom.

\subsubsection{Covariant modified theory}

In order to impose the covariance condition (\ref{eq:Covariance condition - spherical2}), applied
to the new structure function (\ref{eq:New structure function - Linear
  combination}) and using the new constraint (\ref{eq:Linear combinations of
  constraints - generic}), we now write
\begin{equation}
    \{ \mathcal{A}^x , H [\bar{\epsilon}^0] \} =:
    \Lambda^0 \bar{\epsilon}^0
    + \Lambda^x (\bar{\epsilon}^0)'
    \,,
\end{equation}
defining $\Lambda^0$ and $\Lambda^x$.
The covariance condition then implies that
\begin{eqnarray}
    0
    =
    \left(\Lambda^x
    - B^{-1} {\cal B}^x {\cal A}^x
    \right)\big|_{\text{O.S.}}
    =\mathcal{C}=
    \mathcal{C}_{\varepsilon} (E^x)'
    + \mathcal{C}_{\varepsilon \varepsilon \varepsilon} ((E^x)')^3
    \label{eq:Covariance condition - Linear combination}
\end{eqnarray}
must vanish, defining two new coefficients $\mathcal{C}_{\varepsilon}$ and
$\mathcal{C}_{\varepsilon\varepsilon\varepsilon}$ which must vanish
independently if $\mathcal{C}$ is to vanish for all functions $E^x(x)$.  The equation
$\mathcal{C}_{\varepsilon} = 0$ implies
\begin{equation}
    K_{\varphi} \left( \frac{\partial B}{\partial K_\varphi} \right)^2 + B \left(K_{\varphi}
  \frac{\partial^2 B}{(\partial K_\varphi)^2} - \frac{\partial B}{\partial
  K_\varphi} \right) = 0
\end{equation}
solved by
\begin{equation}
    B = c_1 \sqrt{c_2 \pm K_\varphi^2}
\end{equation}
where $c_1$ and $c_2$ are free functions of $E^x$ and $\phi$.
The equation
$\mathcal{C}_{\varepsilon\varepsilon\varepsilon}=0$ implies
\begin{equation}
    B \frac{\partial^3 B}{(\partial K_\varphi)^3} + 3 \frac{\partial B}{\partial K_\varphi} \frac{\partial^2 B}{(\partial K_\varphi)^2}
    = 0
\end{equation}
solved by
\begin{equation}
    B= \tilde{c}_1 \sqrt{ \tilde{c}_2 \pm K_\varphi^2 + \tilde{c}_3 K_\varphi}
\end{equation}
with additional free functions of $E^x$ and $\phi$, $\tilde{c}_1$,
$\tilde{c}_2$ and $\tilde{c}_3$.
Consistency between the two solutions requires $\tilde{c}_3=0$ while
$\tilde{c}_1=c_1$ and $\tilde{c}_2=c_2$, leaving two free functions of $E^x$
and $\phi$ which we write in a form such that
\begin{equation}
    B_s (K_\varphi , E^x , \phi) = \lambda_0 \sqrt{ 1 - s \lambda^2 K_\varphi^2}
\end{equation}
where $\lambda_0 = \lambda_0 (E^x , \phi) , \lambda = \lambda(E^x , \phi)$,
and we have split off an explicit sign choice by $s = \pm 1$. For non-zero
$\lambda$, this solution restricts the phase space to a range of $K_{\varphi}$
such that $1 - s \lambda^2 K_\varphi^2 \geq 0$, which is a non-trivial
condition only if $s=+1$.

Inserting this solution in (\ref{eq:A coefficient - Linear combination}), we derive
\begin{equation}
    A_s =
    \lambda_0 \frac{\sqrt{E^x}}{2} \frac{(E^x)'}{(E^\varphi)^2} \frac{s \lambda^2 K_\varphi}{\sqrt{ 1 - s \lambda^2 K_\varphi^2}}
\end{equation}
and the new structure function
\begin{equation}\label{eq:Covariant linear combination}
    q^{x x}_{\text{(new)}} = 
    \lambda_0^2 \left( 1 + \frac{s \lambda^2}{1- s \lambda^2 K_\varphi^2} \left(\frac{(E^x)'}{2 E^\varphi}\right)^2\right) \frac{E^x}{(E^\varphi)^2}
\end{equation}
from (\ref{eq:New structure function - Linear combination}).
With these results, the modified Hamiltonian constraint is
\begin{eqnarray}
    H^{\rm (new)}
    &=&
    - \lambda_0 \frac{\sqrt{E^x}}{2} \sqrt{ 1 - s \lambda^2 K_\varphi^2} \Bigg[E^\varphi  \left( - V (\phi)
    + \frac{1}{E^x}
    + \frac{K_\varphi^2}{E^x}
    + 4 K_x \frac{K_\varphi}{E^\varphi}
    - \frac{1}{E^x} \frac{P_\phi^2}{(E^\varphi)^2} \right)
    \nonumber\\
    &&
    - E^x \frac{(\phi')^2}{E^\varphi}
    - \frac{1}{4 E^x} \frac{((E^x)')^2}{E^\varphi}
    + \frac{(E^x)' (E^\varphi)'}{(E^\varphi)^2}
    - \frac{(E^x)''}{E^\varphi}
    \nonumber\\
    &&
    - \frac{(E^x)'}{(E^\varphi)^2} \frac{s \lambda^2 K_\varphi}{1 - s \lambda^2 K_\varphi^2} \left( E^\varphi K_\varphi' + P_\phi \phi'
    - K_x (E^x)' \right)
    \Bigg]
    \,,
    \label{eq:Covariant Hamiltonian - Linear combination}
\end{eqnarray}
parameterized by the same two functions, $\lambda_0$ and $\lambda$, and the
sign parameter $s$.

It is interesting to note that the two sign choices for $s$ suggest physically
distinct new phenomena. The case $s=1$, together with a reality condition
imposed on the
constraint, implies a curvature bound $K_\varphi<1 / \lambda$. In this case,
$q^{xx}_{({\rm new})}$ is guaranteed to be positive within the allowed range
of $K_{\varphi}$.
The case $s=-1$ is compatible with the classical range of $K_{\varphi}$, but
the structure function $q^{xx}_{({\rm new})}$ may become negative. In this
case, as discussed in more detail in \cite{HigherCov}, we have to separate
the sign of this function before we can define the spatial metric. The
emergent space-time line element then reads
\begin{equation}
 {\rm d}s^2= -{\rm sgn}(q^{xx}_{({\rm new})}) N^2{\rm d}t^2+
 \frac{1}{|q^{xx}_{({\rm new})}|} ({\rm d}x+N^x{\rm d}t)^2+ E^x{\rm d}\Omega^2\,.
\end{equation}
This
case therefore  implies a possibility of signature change.

In the case of $s = 1$, a natural canonical transformation is given by
\begin{eqnarray}
    K_\varphi &\to&
    \frac{\sin (\lambda K_\varphi)}{\lambda}
    \ ,
    \hspace{1cm}
    E^\varphi \to \frac{E^\varphi}{\cos (\lambda K_\varphi)}
    \ ,
    \nonumber\\
    \phi &\to&
    \phi
    \ ,
    \hspace{2.4cm}
    P_\phi \to P_\phi
    - \frac{E^\varphi}{\cos (\lambda K_\varphi)} \frac{\partial}{\partial \phi} \left(\frac{\sin (\lambda K_\varphi)}{\lambda}\right)
    \ ,
    \nonumber\\
    E^x &\to& E^x
    \ ,
    \hspace{2.1cm}
    K_x \to
    K_x
    + \frac{E^\varphi}{\cos (\lambda K_\varphi)} \frac{\partial}{\partial E^x} \left(\frac{\sin (\lambda K_\varphi)}{\lambda}\right)
    \label{eq:Canonical transformation to holonomy variables}
\end{eqnarray}
which makes the bound on $K_{\varphi}$ explicit by replacing this variable
with the bounded sine function. (When checking the canonical transformation,
note that $\lambda$ is a function only of $E^x$ and $\phi$, but not of
$E^{\varphi}$.)  After the canonical transformation, the previous modified
Hamiltonian constraint becomes
\begin{eqnarray}
    H^{\rm (c)}
    &=&
    - \lambda_0 \frac{\sqrt{E^x}}{2} \Bigg[ E^\varphi  \Bigg( - V (\phi)
    + \frac{1}{E^x}
    + \frac{1}{E^x} \frac{\sin^2 (\lambda K_\varphi)}{\lambda^2}
    \nonumber\\
    &&
    + 4 \left(K_x
    + E^\varphi \left(
    K_\varphi
    - \frac{\tan (\lambda K_\varphi)}{\lambda} \right) \frac{\partial \ln \lambda}{\partial E^x}\right) \frac{1}{E^\varphi} \frac{\sin (2 \lambda K_\varphi)}{2 \lambda}
    \nonumber\\
    &&
    - \frac{1}{E^x} \frac{\cos^2 (\lambda K_\varphi)}{(E^\varphi)^2} \left( P_\phi
    - E^\varphi \left(
    K_\varphi
    - \frac{\tan (\lambda K_\varphi)}{\lambda} \right) \frac{\partial \ln \lambda}{\partial \phi} \right)^2 \Bigg)
    \nonumber\\
    &&
    - \left( \frac{\cos^2 (\lambda K_\varphi)}{4 E^x}
    - \lambda^2 \frac{\sin (2 \lambda K_\varphi)}{2 \lambda} \frac{1}{E^\varphi} \left( K_x + E^\varphi K_\varphi \frac{\partial \ln \lambda}{\partial E^x} \right) \right) \frac{((E^x)')^2}{E^\varphi}
    \nonumber\\
    &&
    - \lambda^2 \frac{\sin (2 \lambda K_\varphi)}{2 \lambda} \left( P_\phi + E^\varphi K_\varphi \frac{\partial \ln \lambda}{\partial \phi} \right) \frac{(E^x)' (\phi)'}{(E^\varphi)^2}
    \nonumber\\
    &&
    - E^x \cos^2 (\lambda K_\varphi) \frac{(\phi')^2}{E^\varphi}
    + \left( \frac{(E^x)' (E^\varphi)'}{(E^\varphi)^2}
    - \frac{(E^x)''}{E^\varphi} \right) \cos^2 (\lambda K_\varphi)
    \Bigg]
    \label{eq:Covariant Hamiltonian - Linear combination - holonomy variables}
\end{eqnarray}
with structure function
\begin{eqnarray}
    q^{x x}_{c} &=& 
    \lambda_0^2 \cos^2 (\lambda K_\varphi) \left( 1 + \lambda^2 \left(\frac{(E^x)'}{2 E^\varphi}\right)^2\right) \frac{E^x}{(E^\varphi)^2}
    \ .
    \label{eq:Structure function - Linear combination - holonomy variables}
\end{eqnarray}
A second canonical transformation
\begin{eqnarray}
    K_\varphi &\to&
    \frac{\bar{\lambda}}{\lambda} K_\varphi
    \ ,
    \hspace{1cm}
    E^\varphi \to \frac{\lambda}{\bar{\lambda}} E^\varphi
    \ ,
    \nonumber\\
    \phi &\to&
    \phi
    \ ,
    \hspace{1.5cm}
    P_\phi \to P_\phi
    + E^\varphi K_\varphi \frac{\partial \ln \lambda}{\partial \phi}
    \ ,
    \nonumber\\
    E^x &\to& E^x
    \ ,
    \hspace{3cm}
    K_x \to
    K_x
    - E^\varphi K_\varphi \frac{\partial \ln \lambda}{\partial E^x}
    \ ,
    \label{eq:Canonical transformation to periodic variables}
\end{eqnarray}
with constant $\bar{\lambda}$ renders the modified Hamiltonian constraint periodic in $K_\varphi$:
\begin{eqnarray}
    H^{\rm (cc)}
    &=&
    - \frac{\bar{\lambda}}{\lambda} \lambda_0 \frac{\sqrt{E^x}}{2} \Bigg[ E^\varphi  \Bigg( \frac{\lambda^2}{\bar{\lambda}^2} \left( - V (\phi)
    + \frac{1}{E^x} \right)
    + \frac{1}{E^x} \frac{\sin^2 (\bar{\lambda} K_\varphi)}{\bar{\lambda}^2}
    \nonumber\\
    &&
    + 4 \left(\frac{K_x}{E^\varphi}
    - \frac{\tan (\bar{\lambda} K_\varphi)}{\bar{\lambda}} \frac{\partial \ln \lambda}{\partial E^x}\right) \frac{\sin (2 \bar{\lambda} K_\varphi)}{2 \bar{\lambda}}\nonumber\\
&&    - \frac{\cos^2 (\bar{\lambda} K_\varphi)}{E^x} \left( \frac{P_\phi}{E^\varphi}
    + \frac{\tan (\bar{\lambda} K_\varphi)}{\bar{\lambda}} \frac{\partial \ln \lambda}{\partial \phi} \right)^2 \Bigg)
    \nonumber\\
    &&
    + \left( 
    \left( \frac{\partial \ln \lambda}{\partial E^x}
    - \frac{1}{4 E^x} \right)  \cos^2 (\bar{\lambda} K_\varphi)
    + \bar{\lambda}^2 \frac{\sin (2 \bar{\lambda} K_\varphi)}{2 \bar{\lambda}} \frac{K_x}{E^\varphi} \right) \frac{((E^x)')^2}{E^\varphi}
    \nonumber\\
    &&
    + \left( 
    \frac{\partial \ln \lambda}{\partial \phi} \cos^2 (\bar{\lambda} K_\varphi)
    - \bar{\lambda}^2 \frac{\sin (2 \bar{\lambda} K_\varphi)}{2 \bar{\lambda}} \frac{P_\phi}{E^\varphi} \right) \frac{(E^x)' \phi'}{E^\varphi}
    \nonumber\\
    &&
    - E^x \cos^2 (\bar{\lambda} K_\varphi) \frac{(\phi')^2}{E^\varphi}
    + \left( \frac{(E^x)' (E^\varphi)'}{(E^\varphi)^2}
    - \frac{(E^x)''}{E^\varphi} \right) \cos^2 (\bar{\lambda} K_\varphi)
    \Bigg]
    \ ,
    \label{eq:Covariant Hamiltonian - Linear combination - periodic variables}
\end{eqnarray}
with structure function
\begin{eqnarray}
    q^{x x}_{(cc)} &=& 
    \frac{\bar{\lambda}^2}{\lambda^2} \lambda_0^2 \cos^2 (\bar{\lambda} K_\varphi) \left( 1 + \bar{\lambda}^2 \left(\frac{(E^x)'}{2 E^\varphi}\right)^2\right) \frac{E^x}{(E^\varphi)^2}
    \ .
    \label{eq:Structure function - Linear combination - periodic variables}
\end{eqnarray}
One can then redefine $\lambda_0 \to \lambda_0 \lambda / \bar{\lambda}$ to
absorb the overall factor in the Hamiltonian constraint and in the structure
function.  Unlike the expression in (\ref{eq:Covariant Hamiltonian - Linear
  combination}), which contains a term of $1/\sqrt{1-s\lambda^2K_{\varphi}^2}$
that diverges at maximum $K_{\varphi}$ for $s=1$, the holonomy-like
coordinates of (\ref{eq:Covariant Hamiltonian - Linear combination - periodic
  variables}) maintain a finite constraint even at the curvature bound. (There
are two coefficients of $\tan(\bar{\lambda}K_{\varphi})$ in the latter
expression, but they are both multiplied by at least one factor of
$\cos(\bar{\lambda}K_{\varphi})$ which removes the divergence.) The
divergence-free version, which in the vacuum case had been obtained by
different means in \cite{SphSymmEff,SphSymmEff2}, allows crossing this
maximum-curvature hypersurface at least in the absence of matter, as
explicitly shown in these papers.

The modified constraint (\ref{eq:Covariant Hamiltonian - Linear combination -
  periodic variables}) represents the non-trivial limit of reaching the
classical constraint surface, to be used for the more general modified
constraints we will obtain in the next subsections.

\subsubsection{Matter and gravitational observables}

The system with modified Hamiltonian constraint (\ref{eq:Covariant Hamiltonian
  - Linear combination}), obtained from a linear combination of the classical
constraints, retains the global symmetry generated by (\ref{eq:Symmetry
  generator of real scalar field}) on-shell when $V=0$.  Thus, the constraints
(\ref{eq:Covariant Hamiltonian - Linear combination - holonomy variables}) and
(\ref{eq:Covariant Hamiltonian - Linear combination - periodic variables})
retain the same symmetry generator, but only if the proper canonical
transformations (\ref{eq:Canonical transformation to holonomy variables}) and
(\ref{eq:Canonical transformation to periodic variables}), respectively, are
applied to the symmetry generator.  Therefore, the symmetry generator of
(\ref{eq:Covariant Hamiltonian - Linear combination - holonomy variables}) is
given by
\begin{eqnarray}
    G [\alpha] = \int {\rm d} x\ \alpha \left( P_\phi
    - E^\varphi \left( K_\varphi - \frac{\tan (\lambda K_\varphi)}{\lambda} \right) \frac{\partial \ln \lambda}{\partial \phi} \right)
    \ .
\end{eqnarray}
while the symmetry generator of (\ref{eq:Covariant Hamiltonian - Linear
  combination - periodic variables}) is given by
\begin{eqnarray}
    G [\alpha] = \int {\rm d} x\ \alpha \left( P_\phi
    + E^\varphi \frac{\tan (\bar{\lambda} K_\varphi)}{\bar{\lambda}} \frac{\partial \ln \lambda}{\partial \phi} \right)
    \ .
    \label{eq:Symmetry generator of real scalar field - Linear combination - periodic variables}
\end{eqnarray}

Similarly, the gravitational observable (\ref{eq:Gravitational observable -
  Classical}) is also preserved by the new constraint (\ref{eq:Covariant
  Hamiltonian - Linear combination}), but only in the vacuum limit where
$\phi , P_\phi \to 0$.  Also in this case, its form changes because of the
application of canonical transformations. In particular, the observable
associated with (\ref{eq:Covariant Hamiltonian - Linear combination - holonomy
  variables}) is given by
\begin{equation}
    m =
    \frac{\sqrt{E^x}}{2} \left(
    1 + \frac{\sin^2 (\lambda K_{\varphi})}{\lambda^2}
    - \left(\frac{(E^x)'}{2 E^\varphi}\right)^2 \cos^2 (\lambda K_{\varphi}) - \frac{\Lambda}{3} E^x \right)
    \ ,
\end{equation}
while that of (\ref{eq:Canonical transformation to periodic variables}) is given by
\begin{equation}
    m =
    \frac{\sqrt{E^x}}{2} \left(
    1 + \frac{\bar{\lambda}^2}{\lambda^2} \left( \frac{\sin^2 (\bar{\lambda} K_{\varphi})}{\bar{\lambda}^2}
    - \left(\frac{(E^x)'}{2 E^\varphi}\right)^2 \cos^2 (\bar{\lambda} K_{\varphi}) \right) - \frac{\Lambda}{3} E^x \right)
    \ .
    \label{eq:Gravitational observable - Linear combination - periodic variables}
\end{equation}

\subsection{General modified constraint}
\label{sec:Modified constraint algebra}

We will now derive a general modified constraint in spherical symmetry that
depends on the canonical fields for gravity and scalar matter with up to
second order in spatial derivatives. There are no additional phase-space
degrees of freedom that could represent higher time derivatives. We are
therefore working at the classical order of derivatives, seen from a viewpoint
of effective field theory, and yet we will find that general relativity
minimally coupled to a scalar field is not the only solution of our
conditions. Within spherical symmetry, there is therefore a difference between
manifestly covariant space-time actions of gravity and scalar matter, and the
larger class of covariant canonical theories. Moreover, the uniqueness results
of \cite{Regained} in vacuum and their extensions to matter fields in
\cite{LagrangianRegained,KucharHypI,KucharHypII,KucharHypIII}, derived like
ours in a Hamiltonian formulation, are based on implicit assumptions, in
particular that the spatial part of a space-time metric is one of the
canonical configuration fields. In our analysis, we have eliminated these
assumptions and obtain a larger class of admissible theories.

\subsubsection{Second-order constraints}

Based on past models considered for instance in
\cite{SphSymmMatter,SphSymmMatter2,HigherCov}, we consider the following
ansatz for a Hamiltonian constraint that, together with the classical
diffeomorphism constraint, (\ref{eq:Diffeomorphism constraint - spherical
  symmetry - Scalar field}), has anomaly-free hypersurface-deformation
brackets for the spherically symmetric theory with scalar field coupling:
\begin{eqnarray}
    H
    &=&
    a
    + ((E^x)')^2 e_{x x}
    + (E^x)' (E^\varphi)' e_{x \varphi}
    + (E^x)'' e_{2 x}
    + (E^x)' K_\varphi' c_{x \varphi}
    \nonumber\\
    &&
    + (E^\varphi)' K_\varphi' c_{\varphi \varphi}
    + ((E^\varphi)')^2 e_{\varphi\varphi}
    + e_{2 \varphi} (E^\varphi)''
    + c_{2 \varphi} K_\varphi''
    \nonumber\\
    &&
    + (\phi')^2 f_{\phi \phi}
    + (E^x)' \phi' f_{x \phi}
    + (E^\varphi)' \phi' f_{\varphi \phi}
    \,.
    \label{eq:Hamiltonian constraint ansatz}
\end{eqnarray}
The free functions $a$, $e_{i}$, $c_{i}$, and $f_i$ have spatial density
weight zero and depend on the basic phase-space degrees of freedom. The
covariance conditions (\ref{eq:Covariance condition on K_x - modified -
  spherical2}) and (\ref{eq:Covariance condition on phi - modified -
  spherical}) have already ruled out spatial derivatives of $K_x$ and
$P_{\phi}$ in the Hamiltonian constraint.  For second-order field equations,
the constraint must be linear in any second-order derivative terms of the
remaining fields, $E^x$, $E^{\varphi}$, $K_{\varphi}$.

We do not include a term with $\phi''$, also here modeling the classical
derivative order of standard scalar field theories, and we do not include
coupling terms between spatial derivatives of the scalar field and those of
$K_{\varphi}$. There may be covariant theories that include such terms, but
the relevant equations become rather intractable. One conceptual difficulty of
including a $\phi''$-term is that this variable then becomes indistinguishable
from $E^x$ in the general second-order constraint and in the covariance
conditions that prohibit derivatives of their momenta (in contrast to
$K_{\varphi}$). Omitting this term allows us to have a well-defined distinction
between gravitational and matter degrees of freedom in a modified theory.

The dependence on first-order derivatives may in principle be higher-order or
even non-polynomial, but the specific form is restricted by the condition that
$H$ have density weight zero. The main higher-order or non-polynomial
dependence on spatial derivatives to be expected is a dependence on the ratio
$(E^x)'/E^{\varphi}$, which has density weight zero. The free functions in
(\ref{eq:Hamiltonian constraint ansatz}) may therefore depend on this
ratio. Since previous results in vacuum showed that the modified structure
function in non-classical models depends on this expression, it turns out to
be convenient to parameterize a dependence on $(E^x)'/E^{\varphi}$ as a
dependence on the future structure function $q^{xx}$, or alternatively as a
dependence on the quantity $\sqrt{q_{xx}}=1/\sqrt{q^{xx}}$ which is required
to have spatial density weight one. We will therefore include $\sqrt{q_{xx}}$
among the canonical fields $K_x$, $E^{\varphi}$ and $P_{\phi}$ of density
weight one.  For now, a dependence on $\sqrt{q_{xx}}$ parameterizes a
dependence on first-order spatial derivatives, but by evaluating the
consistency conditions we will simultaneously be solving for $\sqrt{q_{xx}}$
as a phase-space function.

\subsubsection{Anomaly-freedom of the bracket $\{H,H_x\}$}

We first compute the bracket $\{ H[N] , H_x [N^x] \}$ where, as we recall,
\begin{equation}
  H_x
    = E^\varphi K_\varphi' - K_x (E^x)'
    + P_\phi \phi'
\end{equation}
remains classical,  and put it in the form
\begin{eqnarray}
    \{ H[N] , H_x [N^x] \}
    &=& \int {\rm d} x \ N^x \bigg[ N \mathcal{F}_0 + N' \mathcal{F}_1 +
        N'' \mathcal{F}_2 + N''' \mathcal{F}_3 \bigg] \,,
\end{eqnarray}
using integration by parts to avoid derivatives of $N^x$.  For this expression
to match (\ref{eq:H,H_x bracket}) we set
$\mathcal{F}_0=\mathcal{F}_2=\mathcal{F}_3=0$ and $\mathcal{F}_1 + H = 0$.
Since all the functions in the Hamiltonian constraint (\ref{eq:Hamiltonian
  constraint ansatz}) are independent of spatial derivatives of the phase-space
variables, each term in these equations multiplying derivatives must vanish
independently.

The term $\mathcal{F}_3 = 0$ implies
\begin{eqnarray}
    e_{2 \varphi} = 0
    \ .
\end{eqnarray}
The term $\mathcal{F}_2 = 0$ can be separated into the following derivative
terms, which must vanish independently:
\begin{eqnarray}
    K_\varphi' :&&
    c_{2\varepsilon} = 0
    \ , \\
    \phi' : &&
    f_{\varphi \phi} = 0
    \ , \\
    (E^x)' && e_{2 x} = - E^\varphi e_{x \varphi}
    \ , \\
    (E^\varphi)': && e_{\varphi \varphi} = 0
    \ .
\end{eqnarray}
Using these results, the term $\mathcal{F}_1+H = 0$ can be separated into
derivatives, each of which must again vanish independently (where $0^{\rm th}$
means no derivatives):
\begin{eqnarray}
    0^{\rm th} : \hspace{1cm}
    a &=&
    \sqrt{q_{x x}} \frac{\partial a}{\partial \sqrt{q_{x x}}}
    + P_\phi \frac{\partial a}{\partial P_\phi}
    + K_x \frac{\partial a}{\partial K_x}
    + E^\varphi \frac{\partial a}{\partial E^\varphi}
    \ , \\
    (E^x)' K_\varphi' : \hspace{1cm}
    c_{x \varphi} &=&
    - \sqrt{q_{x x}}\frac{\partial c_{x \varphi}}{\partial \sqrt{q_{x x}}}
    - P_\phi \frac{\partial c_{x \varphi}}{\partial P_\phi}
    - K_x \frac{\partial c_{x \varphi}}{\partial K_x}
    - E^\varphi \frac{\partial c_{x \varphi}}{\partial E^\varphi}
    \ , \\
    ((E^x)')^2 : \hspace{1cm}
    e_{x x} &=&
    - \sqrt{q_{x x}}\frac{\partial e_{x x}}{\partial \sqrt{q_{x x}}}
    - P_\phi \frac{\partial e_{x x}}{\partial P_\phi}
    - K_x \frac{\partial e_{x x}}{\partial K_x}
    - E^\varphi \frac{\partial e_{x x}}{\partial E^\varphi}
    \ , \\
    (E^x)' (E^\varphi)' : \hspace{1cm}
    2 e_{x \varphi} &=&
    - \sqrt{q_{x x}}\frac{\partial c_{x \varphi}}{\partial \sqrt{q_{x x}}}
    - P_\phi \frac{\partial c_{x \varphi}}{\partial P_\phi}
    - K_x \frac{\partial c_{x \varphi}}{\partial K_x}
    - E^\varphi \frac{\partial c_{x \varphi}}{\partial E^\varphi}
    \ , \\
    (E^x)' \phi' : \hspace{1cm}
    f_{x \phi} &=&
    - \sqrt{q_{x x}}\frac{\partial f_{x \phi}}{\partial \sqrt{q_{x x}}}
    - P_\phi \frac{\partial f_{x \phi}}{\partial P_\phi}
    - K_x \frac{\partial f_{x \phi}}{\partial K_x}
    - E^\varphi \frac{\partial f_{x \phi}}{\partial E^\varphi}
    \ , \\
    (E^x)' \phi' : \hspace{1cm}
    f_{\phi \phi} &=&
    - \sqrt{q_{x x}}\frac{\partial f_{\phi \phi}}{\partial \sqrt{q_{x x}}}
    - P_\phi \frac{\partial f_{\phi \phi}}{\partial P_\phi}
    - K_x \frac{\partial f_{\phi \phi}}{\partial K_x}
    - E^\varphi \frac{\partial f_{\phi \phi}}{\partial E^\varphi}
    \ , \\
    K_\varphi'' : \hspace{1cm}
    2 c_{\varphi \varphi} &=&
    - \sqrt{q_{x x}}\frac{\partial c_{\varphi \varphi}}{\partial \sqrt{q_{x x}}}
    - P_\phi \frac{\partial c_{\varphi \varphi}}{\partial P_\phi}
    - K_x \frac{\partial c_{\varphi \varphi}}{\partial K_x}
    - E^\varphi \frac{\partial c_{\varphi \varphi}}{\partial E^\varphi}
    \ .
\end{eqnarray}
These equations are derived from Poisson brackets, separating terms according
to derivative orders. We do not know the phase-space function $\sqrt{q_{xx}}$
at this point, and it does not have an obvious momentum. Derivatives by
$\sqrt{q_{xx}}$ therefore do not follow from basic Poisson brackets, but they
are nevertheless uniquely determined because we are computing a Poisson
bracket with the diffeomorphism constraint. The fact that $\sqrt{q_{xx}}$ has
spatial density weight one, as determined by its geometrical role in
hypersurface deformations, then implies the Poisson-bracket terms as used
here, where $\sqrt{q_{xx}}$ appears in the same way as the basic phase-space
variables of density weight one.

Thus,
\begin{eqnarray}
    a &=& - \frac{\sqrt{E^x}}{2} E^\varphi g A \left( \frac{\sqrt{E^xq_{x x}}}{E^\varphi} , \frac{P_\phi}{E^\varphi} , \frac{K_x}{E^\varphi} \right)
    \ , \\
    e_{x x} &=& - \frac{\sqrt{E^x}}{2} \frac{1}{E^\varphi} g E_{x x} \left( \frac{\sqrt{E^xq_{x x}}}{E^\varphi} , \frac{P_\phi}{E^\varphi} , \frac{K_x}{E^\varphi} \right)
    \ , \\
    e_{x \varphi} &=& - \frac{\sqrt{E^x}}{2} \frac{1}{(E^\varphi)^2} g \left( \frac{\sqrt{E^xq_{x x}}}{E^\varphi} , \frac{P_\phi}{E^\varphi} , \frac{K_x}{E^\varphi} \right)
    \ , \\
    c_{x \varphi} &=& - \frac{\sqrt{E^x}}{2} \frac{1}{E^\varphi} g C_{x \varphi} \left( \frac{\sqrt{E^xq_{x x}}}{E^\varphi} , \frac{P_\phi}{E^\varphi} , \frac{K_x}{E^\varphi} \right)
    \ , \\
    f_{\phi \phi} &=& - \frac{\sqrt{E^x}}{2} \frac{1}{E^\varphi} g F_{\phi \phi} \left( \frac{\sqrt{E^xq_{x x}}}{E^\varphi} , \frac{P_\phi}{E^\varphi} , \frac{K_x}{E^\varphi}  \right)
    \ , \\
    f_{x \phi} &=& - \frac{\sqrt{E^x}}{2} \frac{1}{E^\varphi} g F_{x \phi} \left( \frac{\sqrt{E^xq_{x x}}}{E^\varphi} , \frac{P_\phi^*}{E^\varphi} , \frac{K_x}{E^\varphi} \right)
    \ , \\
    c_{\varphi \varphi} &=& - \frac{\sqrt{E^x}}{2} \frac{1}{(E^\varphi)^2} g C_{\varphi \varphi} \left( \frac{\sqrt{E^xq_{x x}}}{E^\varphi} , \frac{P_\phi}{E^\varphi} , \frac{K_x}{E^\varphi} \right)
    \ .
\end{eqnarray}
For later convenience, it turns out to be useful to include a factor of the
function $g$ from $e_{x\varphi}$ in the remaining functions, which all have
the same general dependence. In addition to the dependence on phase-space
variables with density weight one, all free functions are at this point
allowed to have an unrestricted dependence on the variables $E^x$,
$K_{\varphi}$ and $\phi$ with density weight zero. A factor of
$-\sqrt{E^x}/2$, matching the classical limit, has been extracted in each
function for later convenience.
Using this, the term $\mathcal{F}_0 = 0$ is satisfied automatically.

\subsubsection{Anomaly-freedom of the bracket  $\{H,H\}$}

The analysis of the bracket of two Hamiltonian constraints can be split into
parts, first removing any term that does not obey the hypersurface-deformation
form and would therefore be anomalous, and then analyzing the remaining terms
in order to derive the structure function. We begin with the removal of
anomalous terms, but already at this stage the covariance condition is useful
because it implies that $\{q_{xx},H[N]\}$ does not depend on derivatives of
$N$ and therefore does not contrinute to $\{H[N],H[M]\}$ thanks to
antisymmetry in $(N,M)$. (The combination $\sqrt{E^xq_{xx}}/E^{\varphi}$ in
some of our modification functions does contribute to the Poisson bracket, but
only because it depends on $E^{\varphi}$ and there may be terms in the modified
Hamiltonian with derivatives of $K_{\varphi}$.)

Computing the bracket $\{ H[N] , H [M] \}$, it can be  put in the form
\begin{eqnarray}
    \{ H[N] , H [M] \}
    &=& \int {\rm d} x \ \bigg[ (N M' - M N') \mathcal{G}_0 + (N M'' - M N'') \mathcal{G}_1 + (N M''' - M N''') \mathcal{G}_2 \bigg]
    \nonumber\\
    &=& \int {\rm d} x \ \bigg[ (N M' - M N') \left( \mathcal{G}_0 - \mathcal{G}_1' + (N M''' - M N''') \mathcal{G}_2 \right) \bigg]
    \ ,
\end{eqnarray}
where we used several integrations by parts.  For this to match \eqref{eq:H,H
  bracket} we must set $\mathcal{G}_2=0$ and
$\mathcal{G}\equiv \mathcal{G}_0 - \mathcal{G}_1' = H_r q^{x x}$ for some
function $q^{x x}$ of density weight $-2$.

The equation $\mathcal{G}_2 = 0$ implies
\begin{eqnarray}
    C_{\varphi \varphi} = 0
    \ .
\end{eqnarray}
Any terms in $\mathcal{G}$ that do not contain $K_\varphi'$, $(E^x)'$, or
$\phi'$ cannot contribute to reproducing $H_x$. These terms are
\begin{eqnarray}
    \mathcal{G} &\supset&
    G_0
    + G^\phi P_\phi'
    + G^x K_x'
    + G_\varphi (E^\varphi)'
    + G_{2 x} (E^x)''
    + G_{3 x} (E^x)'''
    + G_{2 x \varphi} (E^x)'' (E^\varphi)'
    \nonumber\\
    &&
    + G_{2 x}^\phi (E^x)'' P_\phi'
    + G_{2 x}^x (E^x)'' K_x'
    \ ,
\end{eqnarray}
which must all vanish in order to obtain an anomaly-free bracket of
hypersurface-deformation form.  It turns out that the equations implied by
each of these terms being zero are not all independent, and only four
independent ones remain:
\begin{eqnarray}
    G_{3 x}=0 \ &:& \hspace{0.5cm}
    \frac{\partial g}{\partial (K_x / E^\varphi)} = 0
    \ , \\
    G^x = 0 \ &:& \hspace{0.5cm}
    - \frac{\partial A}{\partial (K_x / E^\varphi)} \frac{\partial g}{\partial (K_x / E^\varphi)}
    + g \frac{\partial^2 A}{(\partial (K_x / E^\varphi) )^2} = 0
    \ , \\
    G^\phi = 0 \ &:& \hspace{0.5cm}
    - \frac{\partial A}{\partial (K_x / E^\varphi)} \frac{\partial g}{\partial (P_\phi / E^\varphi)}
    + \frac{\partial A^x}{\partial (P_\phi / E^\varphi)} = 0
    \ , \\
    G_\varphi = 0 \ &:& \hspace{0.5cm}
    \frac{\partial A^x}{\partial (\sqrt{q_{x x} E^x} / E^\varphi)} = 0
    \ .
\end{eqnarray}
Their solution is given by
\begin{eqnarray}
    g &=& g \left( E^x , \phi , K_\varphi , \frac{\sqrt{E^x}}{E^\varphi} \sqrt{q_{x x}} , \frac{P_\phi}{E^\varphi} \right)
    \ , \\
    A &=& A_0 \left( E^x , \phi , K_\varphi , \frac{\sqrt{E^x}}{E^\varphi}
          \sqrt{q_{x x}} , \frac{P_\phi}{E^\varphi} \right) +
          \frac{K_x}{E^\varphi} f_1 \left( E^x , \phi , K_\varphi \right) \label{Af1}
\end{eqnarray}
with a new function $f_1 ( E^x , \phi , K_\varphi )$ defined as the
coefficient of $K_x/E^{\varphi}$ in $A$.
With these results, the remaining anomalous terms vanish automatically, and we
can continue with the analysis of structure-function terms.

The remaining non-zero terms in $\mathcal{G}$ contain either $K_\varphi'$,
$(E^x)'$, or $\phi'$, but they must be of the right form in order to
contribute to reproducing the diffeomorphism constraint.
They are:
\begin{eqnarray}
    \mathcal{G} &=&
    \left( G^\varphi E^\varphi K_\varphi' - G_x K_x (E^x)' + G_\phi P_\phi \phi' \right) \frac{1}{(E^\varphi)^2}
    \nonumber\\
    &&
    + \left( G_{(x x)}{}^{\varphi} E^\varphi K_\varphi'
    - G_{(x x)}{}_x K_x (E^x)'
    + G_{(x x)}{}^{\phi} P_\phi \phi' \right) \frac{((E^x)')^2}{(E^\varphi)^4}
    \nonumber\\
    &&
    + \left( G_{(\phi \phi)}{}^{\varphi} E^\varphi K_\varphi'
    - G_{(\phi \phi)}{}_x K_x (E^x)'
    + G_{(\phi \phi)}{}^{\phi} P_\phi \phi' \right) \frac{(\phi')^2}{(E^\varphi)^4}
    \nonumber\\
    &&+ \left( G_{{(2x)}}{}^{\varphi} E^\varphi (K_\varphi)' - G_{{(2x)}}{}_x K_x (E^x)'
    + G_{{(2x)}}{}_{\phi} P_\phi \phi' \right) \frac{(E^x)''}{(E^\varphi)^4}
    \nonumber\\
    &&+ \left( - G_{{(2\phi)}}{}_x K_x (E^x)'
    + G_{{(2\phi)}}{}_{\phi} P_\phi \phi' \right) \frac{\phi''}{(E^\varphi)^4}
    \nonumber\\
    &&
    - G^{{(2\varphi)}}{}_x K_x (E^x)' \frac{K_\varphi''}{(E^\varphi)^4}
    \nonumber\\
    &&
    + \left( G^{(\varphi \varphi)}{}_x E^\varphi K_\varphi' + G^\varphi_{x \phi} P_\phi \phi' \right) \frac{(E^x)' K_\varphi'}{(E^\varphi)^6}
    \nonumber\\
    &&+ \left[ \left( G^{(\phi)}{}_{\varphi x} K_\varphi' + G^{(\phi)}{}_{\phi
       x} \phi' + G^{(\phi)}{}_{xx} (E^x)' \right) (E^x)'
    + G^{(\phi)}{}_{\phi\phi} (\phi')^2
    \right] \frac{P_\phi'}{(E^\varphi)^6}
    \nonumber\\
    &&+ \left[ \left( G_{(\varphi)}{}^\varphi_x K_\varphi' +
       G_{(\varphi)}{}_{\phi x} \phi' + G_{(\varphi)}{}_{xx} (E^x)' \right) (E^x)'
    + G_{(\varphi)}{}_{\phi\phi} (\phi')^2
    \right] \frac{(E^\varphi)'}{(E^\varphi)^6}
    \nonumber\\
    &&+ \left[ \left( G^{(x)}{}_{\varphi x} K_\varphi' + G^{(x)}{}_{\phi x} \phi' + G^{(x)}{}_{xx} (E^x)' \right) (E^x)'
    + G^{(x)}{}_{\phi\phi} (\phi')^2
    \right] \frac{K_x'}{(E^\varphi)^6}
    \nonumber\\
    &&
    + \left( \left( G_{(q)}{}^\varphi_x K_\varphi' + G_{(q)}{}_{\phi x} \phi'
       + G_{(q)}{}_{x x} (E^x)' \right) (E^x)'
    + G_{(q)}{}_{\phi\phi} (\phi')^2
    \right) (\sqrt{q_{x x}})'
    \ .
    \label{eq:G function - Generalized vacuum - Extended}
\end{eqnarray}

Any terms multiplying $K_\varphi''$ and $\phi''$, given by
\begin{eqnarray}
    G^{(2\varphi)}{}_x &=& \frac{E^x}{4} g^2 \frac{E^\varphi}{K_x} \frac{\partial C_{x \varphi}}{\partial (K_x/E^\varphi)}
    \ , \\
    G_{(2\phi)}{}_x &=& \frac{E^x}{4} g^2 \frac{E^\varphi}{K_x} \frac{\partial F_{x \phi}}{\partial (K_x/E^\varphi)}
    \ , \\
    G_{(2\phi)}{}_{\phi} &=& - \frac{E^x}{4} g^2 \frac{E^\varphi}{P_\phi} \frac{\partial F_{\phi \phi}}{\partial (K_x/E^\varphi)}\,,
\end{eqnarray}
must each vanish independently, implying that $C_{x \varphi}$, $F_{x \phi}$
and $F_{\phi \phi}$ are independent of $K_x/E^{\varphi}$.
The only non-trivial term multiplying
$P_\phi'$ is then $G^{(\phi)}{}_{xx}$, and it must vanish:
\begin{eqnarray}
    \frac{\partial C_{x\varphi}}{\partial (P_\phi/E^\varphi)}
    + \frac{\partial^2 E_{xx}}{\partial (K_x/E^\varphi)\partial (P_\phi/E^\varphi)} &=& 0
    \ . \label{eq:Cxphi/Exx Pphi}
\end{eqnarray}

Using all the above results, we obtain the following conditions: All the terms
multiplying $K_x'$ trivialize, except for $G^{(x)}{}_{xx}$ which must vanish
and implies
\begin{eqnarray}
    \frac{\partial^2 E_{xx}}{\partial (K_x/E^\varphi)^2} &=& 0
    \ . \label{eq:AF Exx/kx^2}
\end{eqnarray}
All the terms multiplying $(E^\varphi)'$ then trivialize, except for
$G_{(\varphi)}{}_{x\phi}$ and $G_{(\varphi)}{}_{xx}$, which now must vanish.
The former implies
\begin{eqnarray}
    0 &=& \frac{\partial g}{\partial (P_\phi/E^\varphi)}
\end{eqnarray}
and we will soon return to the latter.  The term
$G^{(\varphi \varphi)}{}_{\phi}$ then trivializes, while the non-trivial
equation $G^{(\varphi \varphi)}{}_x=0$ implies
\begin{eqnarray}
    \frac{\partial^2 E_{xx}}{\partial (K_x/E^\varphi)\partial (P_\phi/E^\varphi)}= 0
    \ . \label{eq:AF Exx/kxPphi}
\end{eqnarray}
Using this result in (\ref{eq:Cxphi/Exx Pphi}) we obtain that $C_{x \varphi}$
is independent of $P_\phi$.

Continuing using these results in the remaining equations, all terms
multiplying $(\sqrt{q_{xx}})'$ trivialize, except for $G_{(q)}{}_{xx}$ which
must vanish and implies
\begin{eqnarray}
    \frac{\partial C_{x\varphi}}{\partial (\sqrt{q_{xx} E^x}/E^\varphi)}
    + \frac{\partial^2 E_{xx}}{\partial (K_x/E^\varphi)\partial (\sqrt{q_{xx} E^x}/E^\varphi)} &=& 0
    \ . \label{eq:AF Cxphi/Exx qxx}
\end{eqnarray}
All terms multiplying $(E^x)''$ and $(E^\varphi)'$ then trivialize, except for
$G_{{(2x)}}{}_{x}$ and $G_{(\varphi)}{}_{xx}$ which must both vanish but imply
the same equation,
\begin{eqnarray}
    C_{x \varphi} \frac{\sqrt{q_{xx} E^x}}{E^\varphi} \frac{\partial \ln
  g}{\partial (\sqrt{q_{xx}} E^x/E^\varphi)} 
    + C_{x \varphi}
    + \frac{\partial \ln g}{\partial K_\varphi}
    + 2 \frac{\partial E_{xx}}{\partial (K_x/E^\varphi)} &=& 0
    \ . \label{eq:AF Cxphi/Exx/g}
\end{eqnarray}
Finally, the term $G_{(\phi \phi)}{}^{\varphi}$ trivializes, while the
non-trivial equations from $G_{(\phi \phi)}{}^{\phi}$ and $G_{(\phi \phi)}{}_x$ now
imply
\begin{eqnarray}
    \frac{\partial F_{\phi\phi}}{\partial (P_\phi/E^\varphi)} = 0
    \ , \label{eq:AF Fphiphi / Pphi}
\end{eqnarray}
and
\begin{eqnarray}
    0 &=&
    C_{x \varphi} \frac{\sqrt{q_{xx} E^x}}{E^\varphi} \frac{\partial F_{\phi \phi}}{\partial (\sqrt{q_{xx}} E^x/E^\varphi)}
    - 2 F_{\phi\phi} \frac{\partial F_{x\phi}}{\partial (P_\phi/E^\varphi)}
    - 2 F_{\phi\phi} \frac{\partial E_{xx}}{\partial (K_x/E^\varphi)}
    + \frac{\partial F_{\phi \phi}}{\partial K_{\varphi}}
    \,, \label{eq:AF Cxphi/Fphiphi}
\end{eqnarray}
respectively.

The structure function can now be obtained from
\begin{eqnarray}
    q^{xx} &=&
    \frac{G^\varphi}{(E^\varphi)^2} + G_{(\phi \phi)}{}^{\varphi} \frac{((E^x)')^2}{(E^\varphi)^4}
    \label{qxxfirst}\\
    &=&
    \frac{E^x}{4 (E^\varphi)^2} g^2 \Bigg(
    \frac{\partial f_1}{\partial K_\varphi} - C_{x \varphi} f_1
    + \frac{((E^x)')^2}{(E^\varphi)^2} \Bigg(
    - C_{x \varphi}{}^2 \left( 1
    + \sqrt{q_{xx}} \frac{\sqrt{E^x}}{E^\varphi} \frac{\partial \ln g}{\partial (\sqrt{q_{xx} E^x}/E^\varphi)} \right)
    \nonumber\\
    &&
    - C_{x \varphi} \left( \sqrt{q_{xx}} \frac{\sqrt{E^x}}{E^\varphi} \frac{\partial C_{x \varphi}}{\partial (\sqrt{q_{xx} E^x}/E^\varphi)}
    + \frac{\partial \ln g}{\partial K_\varphi}
    + \frac{\partial E_{xx}}{\partial (K_x / E^\varphi)} \right)
    + \frac{\partial^2 E_{xx}}{\partial K_\varphi\partial (K_x / E^\varphi)}
    \Bigg)
    \Bigg)
    \ .\nonumber
\end{eqnarray}
This function is composed of the free functions $g$, $C_{x\varphi}$, $f_1$,
and $\partial E_{xx} / \partial (K_x/E^\varphi)$. Its inverse $q_{xx}$
appeared in some of the original dependences allowed for free functions,
except for $f_1$ which was introduced in (\ref{Af1}) as a function independent
of $q_{xx}$ as a consequence of anomaly-freedom. For the sake of simplicity,
we will assume that the remaining functions that determine $q^{xx}$, given by
$g$, $C_{x\varphi}$ and $\partial E_{xx} / \partial (K_x/E^\varphi)$, cannot
independently depend on the structure function itself or its inverse.

This asumption turns (\ref{qxxfirst}) into an explicit equation for the
structure function, which simplifies to
\begin{eqnarray}
    q^{xx} &=&
    \frac{G^\varphi}{(E^\varphi)^2} + G_{(\phi \phi)}{}^{\varphi} \frac{((E^x)')^2}{(E^\varphi)^4}
    \\
    &=&
    \frac{E^xg^2}{4 (E^\varphi)^2}  \Bigg(
    \frac{\partial f_1}{\partial K_\varphi} - C_{x \varphi} f_1\nonumber\\
&& \qquad   + \frac{((E^x)')^2}{(E^\varphi)^2} \Bigg(
    \frac{\partial^2 E_{xx}}{\partial K_\varphi\partial (K_x / E^\varphi)}
    - C_{x \varphi} \left( C_{x \varphi}
    + \frac{\partial \ln g}{\partial K_\varphi}
    + \frac{\partial E_{xx}}{\partial (K_x / E^\varphi)} \right)
    \Bigg)
    \Bigg)
    \ .\nonumber
\end{eqnarray}
Equation~(\ref{eq:AF Cxphi/Exx qxx}) for anomaly-freedom now
trivializes, while (\ref{eq:AF Cxphi/Exx/g}) simplifies to
\begin{eqnarray}
    0 &=& C_{x \varphi}
    + \frac{\partial \ln g}{\partial K_\varphi}
    + 2 \frac{\partial E_{xx}}{\partial (K_x/E^\varphi)}
    \ . \label{eq:AF Cxphi/Exx/g - Simplification qxx}
\end{eqnarray}
Combining the latter with (\ref{eq:AF Exx/kxPphi}) and (\ref{eq:AF Fphiphi / Pphi}), we find that 
\begin{eqnarray}
    \frac{\partial^2 F_{x\phi}}{\partial (P_\phi/E^\varphi)^2} = 0
    \ .
\end{eqnarray}

As a summary so far, the Hamiltonian constraint is of the form
\begin{eqnarray}
    H
    &=&
    - \frac{\sqrt{E^x}}{2} \bar{g} \Bigg[
    E^\varphi \left( A_0 + \frac{K_x}{E^\varphi} \bar{f}_1 \right)
    + \frac{((E^x)')^2}{E^\varphi} E_{xx}
    + \frac{(E^x)' (E^\varphi)'}{(E^\varphi)^2} - \frac{(E^x)''}{E^\varphi}
    \nonumber\\
    &&\qquad
    + \frac{(E^x)' K_\varphi'}{E^\varphi} \bar{C}_{x \varphi}
    + \frac{(\phi')^2}{E^\varphi} F_{\phi \phi}
    + \frac{(E^x)' \phi'}{E^\varphi} F_{x \phi}
    \Bigg]
    \label{eq:Hamiltonian constraint ansatz - Simplification I}
\end{eqnarray}
where we use a bar on $\bar{g}=g$, $\bar{f}_1=f_1$, $\bar{E}_{xx}=E_{xx}$ and
$\bar{C}_{x\varphi}=C_{x\varphi}$ in order to indicate that these functions
(and any other free function with a bar) may depend on $E^x$, $\phi$ and
$K_\varphi$ but not on $\sqrt{q_{x x} E^x}/E^\varphi$. Any unbarred free
function is allowed to depend also on $\sqrt{q_{x x} E^x}/E^\varphi$.
We use the same convention in the expansions
\begin{eqnarray}
    A_0 &=&
    f_0
    + \frac{P_\phi}{E^\varphi} h_0
    + \frac{P_\phi{}^2}{(E^\varphi)^2} f_3
    \ , \\
    E_{xx} &=&
    f_2
    + \frac{K_x}{E^\varphi} \bar{h}
    + \frac{P_\phi}{E^\varphi} h_1
    + \frac{P_\phi{}^2}{(E^\varphi)^2} h_2
    \ , \\
    F_{\phi\phi} &=&
    f_4
    \ , \\
    F_{x\phi} &=&
    h_3
    + \frac{P_\phi}{E^\varphi} h_4
\end{eqnarray}
of some of the coefficient functions in the Hamiltonian constraint, observing
conditions implied by anomaly-freedom.  (While $A_0$ and $E_{xx}$ are so far
allowed to have higher powers of $P_\phi/E^\varphi$, the remaining equations
for anomaly-freedom, to be analyzed in the following section, imply that they
must vanish.) With these expansions, we have four functions, $\bar{g}$,
$\bar{C}_{x\varphi}$, $\bar{f}_1$ and $\bar{h}$, depending only on $E^x$,
$\phi$ and $K_\varphi$, and nine functions, $f_0$, $f_2$, $f_3$, $f_4$, and
$h_i$, $i=0,\ldots,4$, depending on $E^x$, $\phi$, $K_\varphi$ and
$\sqrt{q_{x x} E^x}/E^\varphi$.
Anomaly-freedom requires that these functions satisfy the equations
\begin{eqnarray}
    0 &=&
    \bar{C}_{x \varphi} \frac{\sqrt{E^x}}{E^\varphi} \sqrt{q_{xx}} \frac{\partial f_4}{\partial (\sqrt{q_{xx} E^x}/E^\varphi)}
    - 2 f_4 (h_4 + \bar{h}) 
    + \frac{\partial f_4}{\partial K_\varphi}
    \ , \label{eq:AF f4,h4,h}
    \\
    0 &=& \bar{C}_{x \varphi}
    + \frac{\partial \ln \bar{g}}{\partial K_\varphi}
    + 2 \bar{h}
    \ . \label{eq:AF h,g}
\end{eqnarray}

In order to proceed, it is convenient to apply suitable canonical
transformations in order to eliminate some of the free functions.

\subsubsection{Canonical transformations I}

The constraint (\ref{eq:Hamiltonian constraint ansatz - Simplification I}) was
shown in \cite{HigherCov} to fully determine the vacuum theory by completely
factoring out canonical transformations that preserve the diffeomorphism
constraint.  Here, we generalize the set of diffeomorphism-preserving canonical
transformations to include the scalar field:
\begin{subequations}
\begin{eqnarray}
    &&\phi = f_c^\phi (E^x , \tilde{\phi})
    \quad,\quad
    P_\phi = \tilde{P}_\phi \left( \frac{\partial f_c^\phi}{\partial \tilde{\phi}} \right)^{-1}
    - \tilde{E}^\varphi \frac{\partial f_c^\varphi}{\partial \tilde{\phi}} \left( \frac{\partial f_c^\varphi}{\partial \tilde{K}_\varphi} \right)^{-1}
    \ ,
    \label{eq:Diffeomorphism-constraint-preserving canonical transformations - phi - Spherical - general}
    \\
    &&K_\varphi = f_c^\varphi (E^x , \tilde{\phi}, \tilde{K}_\varphi)
    \quad,\quad
    E^\varphi = \tilde{E}^\varphi \left( \frac{\partial f_c^\varphi}{\partial \tilde{K}_\varphi} \right)^{-1}
    \ ,
    \label{eq:Diffeomorphism-constraint-preserving canonical transformations - K_phi - Spherical - general}
    \\
   && K_x = \frac{\partial (\alpha_c^2 E^x)}{\partial E^x} \tilde{K}_x
    + \tilde{E}^\varphi \frac{\partial f_c^\varphi}{\partial E^x} \left( \frac{\partial f_c^\varphi}{\partial \tilde{K}_\varphi} \right)^{-1}\!\!\!\!\!
    + \tilde{P}_\phi \frac{\partial f_c^\phi}{\partial E^x} \left( \frac{\partial f_c^\phi}{\partial \tilde{\phi}} \right)^{-1}\!\!
    ,\;
    \tilde{E}^x = \alpha_c^2 (E^x) E^x
    \label{eq:Diffeomorphism-constraint-preserving canonical transformations - K_x - Spherical - general}
\end{eqnarray}
    \label{eq:Diffeomorphism-constraint-preserving canonical transformations - Spherical - general}
\end{subequations}
where the new phase-space variables are written with a tilde.  A
transformation with $f_c^\phi = \tilde{\phi}$,
$f_c^\varphi=\tilde{K}_\varphi$, and $\alpha_c = \alpha_c (E^x)$ can always be
used to transform the angular component of the metric from a potentially
modified $q_{\vartheta \vartheta} (E^x)$ to its classical expression
$\tilde{E}^x$.  If we fix the classical form for this component, the residual
canonical transformations are given by
(\ref{eq:Diffeomorphism-constraint-preserving canonical transformations -
  Spherical - general}) with $\alpha_c=1$.  Following \cite{HigherCov}, we can
use a canonical transformation with $f_c^\phi = \tilde{\phi}$, and a function
$f_c^\varphi (E^x, \tilde{\phi}, \tilde{K}_\varphi)$ such that the transformed
$\bar{C}_{x\varphi} (E^x, \tilde{\phi}, \tilde{K}_\varphi)$ vanishes.  In the
following, we will assume that we have applied this canonical transformation,
setting $\bar{C}_{x\varphi}=0$. Equations (\ref{eq:AF
  f4,h4,h}) and (\ref{eq:AF h,g}) for anomaly-freedom then simplify to
\begin{eqnarray}
    \frac{\partial f_4}{\partial K_\varphi} &=&
    - 2 f_4 (h_4 + \bar{h}) 
    \ , \label{eq:AF f4,h4,h - CT}
    \\
    \bar{h} &=& - \frac{1}{2} \frac{\partial \ln \bar{g}}{\partial K_\varphi}
    \,, \label{eq:AF h,g - CT}
\end{eqnarray}
and the structure function turns into
\begin{eqnarray}
    q^{xx}
    &=&
    \Bigg(
    \frac{1}{4}\frac{\partial \bar{f}_1}{\partial K_\varphi} - \frac{1}{2} \left(\frac{(E^x)'}{2 E^\varphi}\right)^2 \frac{\partial \ln \bar{g}}{\partial K_\varphi^2}
    \Bigg) \bar{g}^2 \frac{E^x}{(E^\varphi)^2}
    \ . \label{eq:Anomaly-free structure function}
\end{eqnarray}

The residual canonical transformations that preserve both $q_{\vartheta
  \vartheta}=E^x$ and $\bar{C}_{x\varphi}=0$ are
\begin{subequations}
\begin{eqnarray}
    &&\phi =f_c^\phi (E^x , \tilde{\phi})
    \quad,\quad
    P_\phi = \tilde{P}_\phi \left( \frac{\partial f_c^\phi}{\partial \tilde{\phi}} \right)^{-1}
    - \frac{\tilde{E}^\varphi}{f_x^\varphi} \left( \frac{\partial f_x^\varphi}{\partial \Tilde{\phi}} \tilde{K}_\varphi - \frac{\partial \Tilde{\mu}_\varphi}{\partial \Tilde{\phi}} \right)
    \ ,
    \\
    &&K_\varphi = f_x^\varphi (E^x , \phi) \tilde{K}_\varphi - \Tilde{\mu}_\varphi (E^x , \phi)
    \quad,\quad
    E^\varphi = \tilde{E}^\varphi / f_x^\varphi
    \ ,
    \\
    &&K_x = \tilde{K}_x
    + \frac{\tilde{E}^\varphi}{f_x^\varphi} \left( \frac{\partial f_x^\varphi}{\partial E^x} \tilde{K}_\varphi - \frac{\partial \Tilde{\mu}_\varphi}{\partial E^x} \right)
    + \tilde{P}_\phi \frac{\partial f_c^\phi}{\partial E^x} \left( \frac{\partial f_c^\phi}{\partial \tilde{\phi}} \right)^{-1}
    \quad, \quad
    \tilde{E}^x = E^x
    \ .
\end{eqnarray}
    \label{eq:Diffeomorphism-constraint-preserving canonical transformations - First residue}
\end{subequations}

\subsubsection{Expansion by the scalar momentum}

In order to complete the conditions for anomaly-freedom, the remaining
undetermined functions must reproduce all terms in the diffeomorphism constraint,
\begin{eqnarray}
    \mathcal{G} &=&
    \left( G^\varphi + G_{(x x)}{}^{\varphi} ((E^x)')^2 \right) K_\varphi'
    + \left( G_x + G_{(x x)}{}_x ((E^x)')^2 \right) (E^x)'
    + \left( G_\phi + G_{(x x)}{}_{\phi} ((E^x)')^2 \right) \phi'
    \nonumber\\
    &=:&
    \mathcal{G}^\varphi E^\varphi K_\varphi'
    - \mathcal{G}_x K_x (E^x)'
    + \mathcal{G}_\phi P_\phi \phi'
    \nonumber\\
    &=:&
    q^{x x} H_r
    \ .
\end{eqnarray}
Thus, the undetermined functions must satisfy the equations
$\mathcal{G}^\varphi = \mathcal{G}_x$ and
$\mathcal{G}^\varphi = \mathcal{G}_\phi$.  Because they are all independent of
$P_\phi$, an expansion of the relevant equations in this variable will be
useful, in which each power $P_\phi$ must vanish independently.

We first note that the non-trivial terms in the expansion are
$\mathcal{G} = Q^0 + Q^\phi P_\phi + Q^{\phi\phi}
P_\phi{}^2/(E^\varphi)^2$, where
\begin{eqnarray}
    P_\phi{}^2 Q^{\phi\phi} &=&
    P_\phi{}^2 \frac{E^x}{2} \frac{(E^x)'}{E^\varphi}
    \Bigg[
    \bar{g} \frac{\partial (\bar{g} f_3)}{\partial K_\varphi}
    - 2 \bar{g} f_3 \bar{g} h_4
    + 2 \bar{g} \bar{f}_1 \bar{g} h_2
    \nonumber\\
    &&
    + \frac{((E^x)')^2}{(E^\varphi)^2}
    \left(
    \bar{g} \frac{\partial (\bar{g} h_2)}{\partial K_\varphi}
    - \bar{g} h_2 \left( 2 \bar{g} h_4 + \frac{\partial \bar{g}}{\partial K_\varphi} \right)
    \right)
    \Bigg]
    = 0
    \ ,
    \label{eq:AFeq momenta expansion}
\end{eqnarray}
must vanish, as it cannot contribute to the diffeomorphism constraint.

The expansion of the condition $\mathcal{G}^\varphi = \mathcal{G}_x$ gives the equations
\begin{eqnarray}
    0 &=&
    \bar{g} h_0 \bar{g} h_3
    - \bar{g} \bar{f}_1 \left( 2 \bar{g} f_2 + \frac{\partial \bar{g}}{\partial E^x} \right)
    + g \left( \frac{\partial (\bar{g} \bar{f}_1)}{\partial E^x}
    - \frac{\partial (\bar{g} f_0)}{\partial K_\varphi} \right)
    \nonumber\\
    &&
    + \frac{((E^x)')^2}{(E^\varphi)^2}
    \left(
    - \bar{g} \frac{\partial (\bar{g} f_2)}{\partial K_\varphi}
    + \bar{g} f_2 \frac{\partial \bar{g}}{\partial K_\varphi}
    + \bar{g} h_1 \bar{g} h_3
    - \bar{g}^2 \frac{1}{2} \frac{\partial}{\partial E^x} \frac{\partial \ln \bar{g}}{\partial K_\varphi}
    \right)    \label{eq:AFeq Gphi-Gx}
\end{eqnarray}
and
\begin{eqnarray}
0 &=& 
    P_\phi \Bigg[
    - \bar{g} \frac{\partial (\bar{g} h_0)}{\partial K_\varphi}
    - 2 \bar{g} \bar{f}_1 \bar{g} h_1
    + 2 \bar{g} f_3 \bar{g} h_3
    + \bar{g} h_0 \bar{g} h_4
    \nonumber\\
    &&
    + \frac{((E^x)')^2}{(E^\varphi)^2}
    \left(
    - \bar{g} \frac{\partial (\bar{g} h_1)}{\partial K_\varphi}
    + 2 \bar{g} h_2 \bar{g} h_3
    + \bar{g} h_1 \left( \bar{g} h_4 + \frac{\partial \bar{g}}{\partial K_\varphi}\right)
    \right)
    \Bigg]
    \ .
    \label{eq:AFeq Gphi-Gx Pphi}
\end{eqnarray}
The expansion of the condition $\mathcal{G}^\varphi =
\mathcal{G}_\phi$ gives the equations
\begin{eqnarray}
    0 &=&
    2 \bar{g} f_4 \bar{g} h_0
    - \bar{g} \bar{f}_1 \left( \bar{g} h_3 + \frac{\partial \bar{g}}{\partial \phi} \right)
    + \bar{g} \frac{\partial (\bar{g} \bar{f}_1)}{\partial \phi}
    \nonumber\\
    &&
    + \frac{((E^x)')^2}{(E^\varphi)^2}
    \left(
    - \bar{g} \frac{\partial (\bar{g} h_3)}{\partial K_\varphi}
    + 2 \bar{g} f_4 \bar{g} h_1
    + \bar{g} h_3 \left(
    \bar{g} h_4 + \frac{1}{2} \frac{\partial \bar{g}}{\partial K_\varphi}
    \right)
    - \bar{g}^2 \frac{1}{2} \frac{\partial^2 \ln \bar{g}}{\partial \phi \partial K_\varphi}
    \right)   \label{eq:AFeq Gphi-Gphi}
\end{eqnarray}
and
\begin{eqnarray}
  0 &=& 
    P_\phi \Bigg[
    - \bar{g} \frac{\partial (\bar{g} \bar{f}_1)}{\partial K_\varphi}
    + 4 \bar{g} f_3 \bar{g} f_4
    + \bar{g} \bar{f}_1 \left(
    - \bar{g} h_4 + \frac{\partial \bar{g}}{\partial K_\varphi}
    \right)
    \nonumber\\
    &&
    + \frac{((E^x)')^2}{(E^\varphi)^2}
    \left(
    - \bar{g} \frac{\partial (\bar{g} h_4)}{\partial K_\varphi}
    + \bar{g} h_4 \left( \bar{g} h_4
    + \frac{1}{2} \frac{\partial \bar{g}}{\partial K_\varphi} \right)
    + 4 \bar{g} f_4 \bar{g} h_2
    + \frac{1}{2} \bar{g}^2 \frac{\partial^2 \ln \bar{g}}{\partial K_\varphi^2}
    \right)
    \Bigg]
    \ .
    \label{eq:AFeq Gphi-Gphi Pphi}
\end{eqnarray}

We keep the $P_\phi$ factor in the previous equations because it helps us to
identify which equations must be neglected in the vacuum limit,
$\phi , P_\phi \to 0$.  Anomaly-freedom is then ensured by the equations
(\ref{eq:AF f4,h4,h - CT}), (\ref{eq:AF h,g - CT}), (\ref{eq:AFeq momenta
  expansion})--(\ref{eq:AFeq Gphi-Gphi Pphi}).

\subsubsection{Expansion by the structure function}

In order to implement the classical-matter limit in the modified constraint,
we will use an expansion by the structure function in
the coefficients $f_3$ and $f_4$ relevant to the scalar equations of motion,
\begin{eqnarray}
    f_3 &=&
    \bar{f}_3
    + E^\varphi \frac{\sqrt{q^{xx}}}{\sqrt{E^x}} \bar{f}_3^q
    \ , \\
    f_4 &=& 
    \bar{f}_4
    + E^\varphi \frac{\sqrt{q^{xx}}}{\sqrt{E^x}} \bar{f}_4^q
    \ ,
\end{eqnarray}
where, as before, we write a bar on some functions to indicate that they are
independent of $q^{xx}$. We use the structure function rather than its inverse
that appeared in previous equations, such that $E^{\varphi}\sqrt{q^{xx}}$ has
spatial density weight zero. This expansion is useful because $\bar{f}_3^q$
and $\bar{f}_4^q$ will be responsible for obtaining the classical-matter
limit, while $\bar{f}_3$ and $\bar{f}_4$ allow us to explore alternative
theories. There can be no higher-order terms in $\sqrt{q^{xx}}$ because the
product of two such functions, one from $H[N]$ and one from $H[M]$ in a
Poisson bracket of two Hamiltonian constraints, must give us a single factor
of $q^{xx}$ in the hypersurface-deformation bracket.

We perform the same expansion by the structure function for the remaining functions,
\begin{eqnarray}
    f_0 &=&
    \frac{\sqrt{q_{xx} E^x}}{E^\varphi} \bar{f}_{0q}
    + \bar{f}_0
    + E^\varphi \frac{\sqrt{q^{xx} }}{\sqrt{E^x}} \bar{f}_0^q
    \ , \\
    f_2 &=&
    \bar{f}_2
    + E^\varphi \frac{\sqrt{q^{xx} }}{\sqrt{E^x}} \bar{f}_2^q
    \ , \\
    h_0 &=&
    \bar{h}_0
    + E^\varphi \frac{\sqrt{q^{xx} }}{\sqrt{E^x}} \bar{h}_0^q
    \ , \\
    h_1 &=&
    \bar{h}_1
    + E^\varphi \frac{\sqrt{q^{xx} }}{\sqrt{E^x}} \bar{h}_1^q
    \ , \\
    h_2 &=&
    \bar{h}_2
    + E^\varphi \frac{\sqrt{q^{xx} }}{\sqrt{E^x}} \bar{h}_2^q
    \ , \\
    h_3 &=&
    \bar{h}_3
    + E^\varphi \frac{\sqrt{q^{xx}}}{\sqrt{E^x}} \bar{h}_3^q
    \ , \\
    h_4 &=&
    \bar{h}_4
    \ ,
\end{eqnarray}
where we have chosen the expansion coefficients according to what they
multiply in the constraints.  The function $f_0$ is the only one with a
$\sqrt{q_{xx}}$ term, suitable for a measure of radial integration, because no
other function can reproduce the potential term of the Klein--Gordon
constraint in the classical-matter limit.  The function $h_2$ contains a
$\sqrt{q^{xx}}$ term because it multiplies $P_\phi^2$, just as $f_3$ and
$f_4$.  Equation (\ref{eq:AF f4,h4,h - CT}) implies that $h_4$ cannot have any
structure-function term, hence $h_4 = \bar{h}_4$: The left-hand side is linear
in a derivative of $f_4$ by $K_{\varphi}$ and is therefore at most linear in
$\sqrt{q^{xx}}$. The right-hand side multiplies $f_4$ by $h_4$, which can be
at most linear in $\sqrt{q^{xx}}$ only if $h_4$ does not depend on
$\sqrt{q^{xx}}$. (The same equation shows that $\bar{f}_4^q$ must depend on
$K_{\varphi}$ if it is non-zero.)  For the sake of generality, we have
expanded the remaining functions $f_2$, $h_0$, $h_1$, and $h_3$ by including a
$0^{\rm th}$-order term and a linear term in $\sqrt{q^{xx}}$.

We will proceed by substituting these expansions into the conditions
(\ref{eq:AF f4,h4,h - CT}), (\ref{eq:AFeq momenta expansion})--(\ref{eq:AFeq
  Gphi-Gphi Pphi}) for anomaly-freedom, taking into account that cross-terms
multiplying $q^{x x}$ may mix the $0^{\rm th}$-order terms and linear terms in
$((E^x)')^2$, and that the functions $f_i$ must be non-zero because they are
responsible for reproducing the complete classical limit.

Condition (\ref{eq:AF f4,h4,h - CT}) can be rewritten as a combination of two equations,
\begin{eqnarray}
    \frac{\partial (\bar{g} \bar{f}_4)}{\partial K_\varphi} &=& 2 \bar{g} \bar{f}_4 \bar{h}_4
    \ , \label{eq:AF f4}
    \\
    \frac{\partial (\bar{g} \bar{f}_4^q)}{\partial K_\varphi} &=& 2 \bar{g} \bar{f}_4^q \bar{h}_4
    \ . \label{eq:AF f4q}
\end{eqnarray}
Condition (\ref{eq:AFeq momenta expansion}) becomes
\begin{eqnarray}
    0 &=&
    \bar{g} \left( \frac{\partial (\bar{g} \bar{f}_3)}{\partial K_\varphi}
    - 2 \bar{g} \bar{f}_3 \bar{h}_4
    + 2 \bar{g} \bar{f}_1 \bar{h}_2 \right)
    + \bar{g} E^\varphi \frac{\sqrt{q^{xx}}}{\sqrt{E^x}} \left( \frac{\partial (\bar{g} \bar{f}_3^q)}{\partial K_\varphi}
    - 2 \bar{g} \bar{f}_3^q \bar{h}_4
    + 2 \bar{g} \bar{f}_1 \bar{h}_2^q \right)
    \nonumber\\
    &&
    + \bar{g}^2 \frac{((E^x)')^2}{(E^\varphi)^2}
    \left(
    \frac{\partial \bar{h}_2}{\partial K_\varphi}
    - 2 \bar{h}_2 \bar{h}_4
    + E^\varphi \frac{\sqrt{q^{xx}}}{\sqrt{E^x}} \left( \frac{\partial \bar{h}_2^q}{\partial K_\varphi}
    - 2 \bar{g} \bar{h}_2^q \bar{h}_4 \right)
    \right)
    \label{eq:AF f3,f3q,h2,h2q}
\end{eqnarray}
and equation (\ref{eq:AFeq Gphi-Gx}) is turned into
\begin{eqnarray}
    0 &=&
    - \bar{g} \frac{\sqrt{q_{xx} E^x}}{E^\varphi} \frac{\partial (\bar{g} \bar{f}_{0q})}{\partial K_\varphi}
    - \bar{g} \frac{\partial (\bar{g} \bar{f}_0)}{\partial K_\varphi}
    + \bar{g} \bar{h}_0 \bar{g} \bar{h}_3
    - 2 \bar{g} \bar{f}_1 \bar{g} \bar{f}_2
    - \bar{g} \bar{f}_1  \frac{\partial \bar{g}}{\partial E^x}\nonumber\\
&&    + \bar{g} \frac{\partial (\bar{g} \bar{f}_1)}{\partial E^x}
    + \frac{1}{4} \bar{g}^2 \frac{\partial \bar{f}_1}{\partial K_\varphi} \bar{g} \bar{h}_0^q \bar{g} \bar{h}_3^q
    \nonumber\\
    &&
    + E^\varphi \frac{\sqrt{q^{xx}}}{\sqrt{E^x}} \left(
    - \bar{g} \frac{\partial (\bar{g} \bar{f}_0^q)}{\partial K_\varphi}
    - 2 \bar{g} \bar{f}_1 \bar{g}\bar{f}_2^q
    + \bar{g}\bar{h}_0^q \bar{g} \bar{h}_3
    + \bar{g} \bar{h}_0 \bar{g} \bar{h}_3^q \right)
    \nonumber\\
    &&
    + \frac{((E^x)')^2}{(E^\varphi)^2}
    \Bigg(
    - \bar{g} \frac{\partial (\bar{g} \bar{f}_2)}{\partial K_\varphi}
    + \bar{g} \bar{f}_2 \frac{\partial \bar{g}}{\partial K_\varphi}
    - \bar{g}^2 \frac{1}{2} \frac{\partial}{\partial E^x} \frac{\partial \ln \bar{g}}{\partial K_\varphi}
    + \bar{g} \bar{h}_1 \bar{g} \bar{h}_3
    - \frac{1}{8} \bar{g}^2 \frac{\partial^2 \ln \bar{g}}{\partial K_\varphi^2} \bar{g} \bar{h}_0^q \bar{g} \bar{h}_3^q
    \nonumber\\
    &&
    + E^\varphi \frac{\sqrt{q^{xx}}}{\sqrt{E^x}} \left(
    - \bar{g} \frac{\partial (\bar{g} \bar{f}_2^q)}{\partial K_\varphi}
    + \bar{g} \bar{f}_2^q \frac{\partial \bar{g}}{\partial K_\varphi}
    + \bar{g} \bar{h}_1^q \bar{g} \bar{h}_3
    + \bar{g} \bar{h}_1 \bar{g} \bar{h}_3^q \right)
    + \frac{(E^\varphi)^2}{E^x} q^{xx} \bar{g} \bar{h}_1^q \bar{g} \bar{h}_3^q
    \Bigg)
    \ ,
    \label{eq:AF f0,f2,f2q}
\end{eqnarray}
where we have used (\ref{eq:Anomaly-free structure function})

Equation (\ref{eq:AFeq Gphi-Gx Pphi}) now reads
\begin{eqnarray}
    0 &=& 
    P_\phi \Bigg[
    - \bar{g} \frac{\partial (\bar{g} \bar{h}_0)}{\partial K_\varphi}
    - 2 \bar{g} \bar{f}_1 \bar{g} \bar{h}_1
    + \bar{g} \bar{h}_0 \bar{g} \bar{h}_4
    + 2 \bar{g} \bar{f}_3 \bar{g} \bar{h}_3
    + \frac{1}{2} \bar{g}^2 \frac{\partial \bar{f}_1}{\partial K_\varphi} \bar{g} \bar{f}_3^q \bar{g} \bar{h}_3^q
    \nonumber\\
    &&
    + E^\varphi \frac{\sqrt{q^{xx}}}{\sqrt{E^x}} \left(
    - \bar{g} \frac{\partial (\bar{g} \bar{h}_0^q)}{\partial K_\varphi}
    - 2 \bar{g} \bar{f}_1 \bar{g}\bar{h}_1^q
    + \bar{g} \bar{h}_0^q \bar{g} \bar{h}_4
    + 2 \bar{g} \bar{f}_3^q \bar{g} \bar{h}_3
    + 2 \bar{g} \bar{f}_3 \bar{g} \bar{h}_3^q \right)
    \nonumber\\
    &&
    + \frac{((E^x)')^2}{(E^\varphi)^2}
    \Bigg(
    - \bar{g} \frac{\partial (\bar{g} \bar{h}_1)}{\partial K_\varphi}
    + 2 \bar{g} \bar{h}_2 \bar{g} \bar{h}_3
    + \bar{g} \bar{h}_1 \left( \bar{g} \bar{h}_4 + \frac{\partial \bar{g}}{\partial K_\varphi}\right)
    - \frac{1}{4} \bar{g}^2 \frac{\partial^2 \ln \bar{g}}{\partial K_\varphi^2} \bar{g}\bar{f}_3^q \bar{g} \bar{h}_3^q
    \nonumber\\
    &&
    + E^\varphi \frac{\sqrt{q^{xx}}}{\sqrt{E^x}} \left( 
    - \bar{g} \frac{\partial (\bar{g} \bar{h}_1^q)}{\partial K_\varphi}
    + 2 \bar{g} \bar{h}_2^q \bar{g} \bar{h}_3
    + 2 \bar{g} \bar{h}_2 \bar{g} \bar{h}_3^q
    + \bar{g} \bar{h}_1^q \left( \bar{g} \bar{h}_4 + \frac{\partial \bar{g}}{\partial K_\varphi}\right)
       \right)\nonumber\\
  &&
    + \frac{(E^\varphi)^2}{E^x} q^{xx} 2 \bar{g} \bar{h}_2^q \bar{g} \bar{h}_3^q
    \Bigg)
    \Bigg]
    \ , \label{eq:AF h0,h0q,h1,h1q}
\end{eqnarray}
where we have used (\ref{eq:Anomaly-free structure function}), and
equation (\ref{eq:AFeq Gphi-Gphi}) becomes
\begin{eqnarray}
    0 &=&
    \bar{g} \frac{\partial (\bar{g} \bar{f}_1)}{\partial \phi}
    - \bar{g} \bar{f}_1 \left( \bar{g} \bar{h}_3 + \frac{\partial \bar{g}}{\partial \phi} \right)
    + 2 \bar{g} \bar{h}_0 \bar{g} \bar{f}_4 
    + \frac{1}{2} \bar{g}^2 \frac{\partial \bar{f}_1}{\partial K_\varphi} \bar{g} \bar{h}_0^q \bar{g} \bar{f}_4^q
    \nonumber\\
    &&
    + E^\varphi \frac{\sqrt{q^{xx}}}{\sqrt{E^x}} \left(
    2 \bar{g} \bar{h}_0 \bar{g} \bar{f}_4^q
    + 2 \bar{g} \bar{h}_0^q \bar{g} \bar{f}_4
    - \bar{g} \bar{f}_1 \bar{g} \bar{h}_3^q
    \right)
    \nonumber\\
    &&
    + \frac{((E^x)')^2}{(E^\varphi)^2}
    \Bigg(
    - \bar{g} \frac{\partial (\bar{g} \bar{h}_3)}{\partial K_\varphi}
    + \bar{g} \bar{h}_3 \left(
    \bar{g} \bar{h}_4 + \frac{1}{2} \frac{\partial \bar{g}}{\partial K_\varphi}
    \right)
       + 2 \bar{g} \bar{h}_1 \bar{g} \bar{f}_4\nonumber\\
  &&\qquad
    - \bar{g}^2 \frac{1}{2} \frac{\partial^2 \ln \bar{g}}{\partial \phi \partial K_\varphi}
    - \frac{1}{4} \bar{g}^2 \frac{\partial^2 \ln \bar{g}}{\partial K_\varphi^2} \bar{g} \bar{h}_0^q \bar{g} \bar{f}_4^q
    \nonumber\\
    &&\qquad
    + E^\varphi \frac{\sqrt{q^{xx}}}{\sqrt{E^x}} \left(
    - \bar{g} \frac{\partial (\bar{g} \bar{h}_3^q)}{\partial K_\varphi}
    + \bar{g} \bar{h}_3^q \left(
    \bar{g} \bar{h}_4 + \frac{1}{2} \frac{\partial \bar{g}}{\partial K_\varphi}
    \right)
    + 2 \bar{g} \bar{h}_1 \bar{g}\bar{f}_4^q
    + 2 \bar{g} \bar{h}_1^q \bar{g} \bar{f}_4
       \right)\nonumber\\
  &&\qquad
    + (E^\varphi)^2 \frac{q^{xx}}{E^x} 2 \bar{g} \bar{h}_1^q \bar{g} \bar{f}_4^q
    \Bigg)
    \ ,
    \label{eq:AF f1,h3}
\end{eqnarray}
where we have used (\ref{eq:Anomaly-free structure function}).
Finally, Equation~(\ref{eq:AFeq Gphi-Gphi Pphi}) appears as
\begin{eqnarray}
    0 &=& 
    P_\phi \Bigg[
    - \bar{g}^2 \frac{\partial \bar{f}_1}{\partial K_\varphi} \left( 1 - \bar{g} \bar{f}_3^q \bar{g} \bar{f}_4^q \right)
    + 4 \bar{g} \bar{f}_3 \bar{g} \bar{f}_4
    - \bar{g} \bar{f}_1 \bar{g} \bar{h}_4
    + 4 E^\varphi \frac{\sqrt{q^{xx}}}{\sqrt{E^x}} \left( \bar{g} \bar{f}_3 \bar{g} \bar{f}_4^q
    + \bar{g} \bar{f}_3^q \bar{g} \bar{f}_4 \right)
    \nonumber\\
    &&
    + \frac{((E^x)')^2}{(E^\varphi)^2}
    \Bigg(
    - \bar{g} \frac{\partial (\bar{g} \bar{h}_4)}{\partial K_\varphi}
    + \bar{g} \bar{h}_4 \left( \bar{g} \bar{h}_4
    + \frac{1}{2} \frac{\partial \bar{g}}{\partial K_\varphi} \right)
    + 4 \bar{g} \bar{h}_2 \bar{g} \bar{f}_4
    + \frac{1}{2} \bar{g}^2 \frac{\partial^2 \ln \bar{g}}{\partial K_\varphi^2} \left( 1 - \bar{g} \bar{f}_3^q \bar{g} \bar{f}_4^q \right)
    \nonumber\\
    &&
    + 4 E^\varphi \frac{\sqrt{q^{xx}}}{\sqrt{E^x}} \left( \bar{g} \bar{h}_2 \bar{g} \bar{f}_4^q
    + \bar{g} \bar{h}_2^q \bar{g} \bar{f}_4 \right)
    + 4 \frac{(E^\varphi)^2}{E^x} q^{xx} \bar{g}\bar{h}_2^q \bar{g} \bar{f}_4^q
    \Bigg)
    \Bigg]
    \ ,
    \label{eq:AF f3f4,h4}
\end{eqnarray}
where we have used (\ref{eq:Anomaly-free structure function}).

Only the $\sqrt{q_{xx}E^x}$-term in (\ref{eq:AF f0,f2,f2q}) can be readily solved at this stage,
\begin{eqnarray}
    \bar{g} \bar{f}_{0q} = - \lambda_0^2 V_q
    \ ,
    \label{eq:AF solution to f0_q}
\end{eqnarray}
where $V_q$ and $\lambda_0$ are undetermined functions of $E^x$ and $\phi$.
This function represents the freedom to choose a potential for a scalar field
on an emergent spacetime.

Equations (\ref{eq:AF f4})--(\ref{eq:AF f3f4,h4}) are the whole anomaly-freedom
equations left, and their $0^{\rm th}$ orders as well as linear orders in
$\sqrt{q^{xx}}$, $((E^x)')^2$, $((E^x)')^2 \sqrt{q^{xx}}$, and
$((E^x)')^2 q^{xx}$ must all vanish separately.  Due to the complexity of
these equations, they cannot be solved exactly, and we must rely on a number
of principles to simplify them further.  The first and primary such principle
is covariance.

\subsubsection{Covariance}

The covariance condition imposed on the structure function
(\ref{eq:Anomaly-free structure function}) is trivial, except for the
first-order derivative term in the gauge function.  This condition has one
term independent of spatial derivatives of the phase-space variables, and
another term multiplying $((E^x)')^2$.  Because the on-shell condition cannot
mix these two terms, they must vanish independently, such that the covariance
condition is satisfied off the constraint surface (while the equations of
motion are still being used in order to compare time-derivative terms).  The
two equations implied by the covariance condition are
\begin{equation}
    0 =
    \frac{\partial^3 \ln \bar{g}}{\partial K_\varphi^3}
    + \frac{\partial \ln \bar{g}}{\partial K_\varphi} \frac{\partial^2 \ln \bar{g}}{\partial K_\varphi^2}
          \end{equation}
          and
          \begin{equation}
    0 =
    \frac{\partial^2 \ln g}{\partial K_\varphi^2} f_1
    - 2 \frac{\partial \ln g}{\partial K_\varphi} \frac{\partial f_1}{\partial K_\varphi}
    - \frac{\partial^2 f_1}{\partial K_\varphi^2}
    \ .
\end{equation}
They have the general solutions
\begin{eqnarray}
    \bar{g} &=&
    \lambda_0 \cos^2 ( \lambda (K_\varphi + \mu_\varphi) )
    \ , \nonumber\\
    \bar{g} \bar{f}_1 &=&
    4 \lambda_0 \left(c_f \frac{\sin (2 \lambda (K_\varphi + \mu_\varphi))}{2 \lambda}
    + q \cos(2 \lambda (K_\varphi + \mu_\varphi))\right)
    \ ,
    \label{eq:Covariance condition solution g,f1}
\end{eqnarray}
recovering the classical limit for $\lambda , \mu_\varphi , q \to 0$ and $\lambda_0 , c_f \to 1$.

\subsubsection{Canonical transformations II}

It is now convenient to employ the residual canonical transformation
(\ref{eq:Diffeomorphism-constraint-preserving canonical transformations -
  First residue}).
To simplify the anomaly-freedom equations we will perform the canonical transformation
\begin{eqnarray}
    &&\phi \to \phi
    \quad ,\quad
    P_\phi \to P_\phi
    + E^\varphi \left( \frac{\partial \ln \lambda}{\partial \phi} K_\varphi + \frac{\lambda}{\bar{\lambda}} \frac{\partial \mu_\varphi}{\partial \Tilde{\phi}} \right)
    \ ,
    \nonumber\\
    &&K_\varphi \to \frac{\bar{\lambda}}{\lambda} K_\varphi - \mu_\varphi
    \quad ,\quad
    E^\varphi \to \frac{\lambda}{\bar{\lambda}} E^\varphi
    \ ,
    \nonumber\\
    &&K_x \to K_x
    - E^\varphi \left( \frac{\partial \ln \lambda}{\partial E^x} K_\varphi + \frac{\lambda}{\bar{\lambda}} \frac{\partial \mu_\varphi}{\partial E^x} \right)
    \quad , \quad
    E^x \to E^x
    \label{eq:Canonical transformations to periodic phase-space}
\end{eqnarray}
with constant $\bar{\lambda}$, and we redefine the undetermined functions so
as to absorb the $\lambda$ and $\mu_\varphi$ factors.  As was shown in
\cite{HigherCov}, this particular canonical transformations renders the
constraint periodic in $K_\varphi$. Any non-periodic modification and the
freedom of non-constant $\lambda$ as coefficient of $K_{\varphi}$ in
trigonometric functions can then be recovered by inverting the canonical
transformation once the anomaly-freedom equations have been solved.

Thus, in
the new phase-space coordinates we have
\begin{eqnarray}
    \bar{g} &=&
    \lambda_0 \cos^2 (\bar{\lambda} K_\varphi)
    \ , \\
    \bar{g} \bar{f}_1 &=&
    4 \lambda_0 \left(c_f \frac{\sin (2 \bar{\lambda} K_\varphi)}{2 \bar{\lambda}}
    + q \cos(2 \bar{\lambda} K_\varphi)\right)
    \ ,
    \label{eq:Covariance condition solution g,f1 - periodic}
\end{eqnarray}
recovering the classical limit for $\bar{\lambda} , q \to 0$ and $\lambda_0 , c_f \to 1$.
The structure function (\ref{eq:Anomaly-free structure function}) can now be explicitly obtained,
\begin{eqnarray}
    q^{xx}
    &=&
    \left(
    \left( c_{f}
    + \left(\frac{\bar{\lambda} (E^x)'}{2 E^\varphi} \right)^2 \right) \cos^2 \left(\bar{\lambda} K_\varphi\right)
    - 2 q \bar{\lambda}^2 \frac{\sin \left(2 \bar{\lambda} K_\varphi\right)}{2 \bar{\lambda}}\right)
    \lambda_0^2 \frac{E^x}{(E^\varphi)^2}
    \ . \label{eq:Covariant structure function}
\end{eqnarray}

This still leaves the freedom of a final residual canonical transformation
preserving periodicity, which takes the form 
\begin{eqnarray}
    &&\phi = f_c^\phi (E^x , \tilde{\phi})
    \quad ,\quad
    P_\phi = \tilde{P}_\phi \left( \frac{\partial f_c^\phi}{\partial \tilde{\phi}} \right)^{-1}
    \ ,
    \nonumber\\
    &&K_\varphi = \tilde{K}_\varphi
    \quad ,\quad
    E^\varphi = \tilde{E}^\varphi
    \ ,
    \nonumber\\
    &&K_x =\tilde{K}_x
    + \tilde{P}_\phi \frac{\partial f_c^\phi}{\partial E^x} \left( \frac{\partial f_c^\phi}{\partial \tilde{\phi}} \right)^{-1}
    \quad , \quad
    \tilde{E}^x = E^x
    \ .
    \label{eq:Residual canonical transformation phi,ex}
\end{eqnarray}
Unlike the effects of the previous canonical transformations, which had
already been understood in the vacuum case \cite{HigherCov}, this last
residual canonical transformation of the matter variable remains to be
factored out and fully interpreted.  We will do so after solving the
anomaly-freedom equations in the remainder of this section.

Equations (\ref{eq:AF f4})--(\ref{eq:AF f3f4,h4}) for anomaly-freedom are
hard to solve exactly.  We will thus rely on the principles
described in Section~\ref{sec:Conditions on the modified theory} to simplify
their solutions.  These principles will differentiate between three classes of
constraints which we will obtain in the following subsections.  The first
class of constraints is given by those compatible with the classical-matter
limit, the second class by those compatible with the limit of reaching the
classical constraint surface, and the third one by having a dynamical solution
free of singularities. For now, we will look for possible
restrictions from the remaining principles.

\subsubsection{Vacuum limit}

The vacuum limit is given by $\phi , P_\phi \to 0$, and in the anomaly-freedom
equations one has to further take
$\partial_\phi g , \partial_\phi f_1 , h_i , f_3 , f_4 \to 0$ wherever such
terms survive. The equations (\ref{eq:AF f4})--(\ref{eq:AF f3f4,h4}) for
anomaly-freedom then reduce to
\begin{eqnarray}
    \bar{g} \frac{\partial (\bar{g} \bar{f}_0)}{\partial K_\varphi} &=&
    - 2 \bar{g} \bar{f}_1 \bar{g} \bar{f}_2
    - \bar{g} \bar{f}_1  \frac{\partial \bar{g}}{\partial E^x}
    + \bar{g} \frac{\partial (\bar{g} \bar{f}_1)}{\partial E^x}
    \ , \\
    \bar{g} \frac{\partial (\bar{g} \bar{f}_0^q)}{\partial K_\varphi} &=&
    - 2 \bar{g} \bar{f}_1 \bar{g} \bar{f}_2^q
    \ , \\
    \bar{g} \frac{\partial (\bar{g} \bar{f}_2)}{\partial K_\varphi} &=&
    \bar{g} \bar{f}_2 \frac{\partial \bar{g}}{\partial K_\varphi}
    - \bar{g}^2 \frac{1}{2} \frac{\partial}{\partial E^x} \frac{\partial \ln \bar{g}}{\partial K_\varphi}
    \ , \\
    \bar{g} \frac{\partial (\bar{g} \bar{f}_2^q)}{\partial K_\varphi} &=&
    \bar{g} \bar{f}_2^q \frac{\partial \bar{g}}{\partial K_\varphi}
    \ ,
    \label{eq:AF f0,f2,f2q - vacuum}
\end{eqnarray}
which can all be solved exactly,
\begin{eqnarray}
    \bar{g} \bar{f}_0 &=&
    \lambda_0 \left( \frac{\alpha_{0}}{E^x}
    + \frac{\sin^2 (\bar{\lambda} K_\varphi)}{\bar{\lambda}^2} \left( c_f \frac{\alpha_2}{E^x} + 2 \frac{\partial c_f}{\partial E^x}\right)
    + 2 \bar{\lambda}^2 \frac{\sin (2\bar{\lambda} K_\varphi)}{2 \bar{\lambda}} \left( q \frac{\alpha_2}{E^x} + 2 \frac{\partial q}{\partial E^x}\right)
    \right)
    \ , \\
    \bar{g} \bar{f}_0^q &=&
    \frac{\alpha_{0q}}{E^x}
    + \frac{\alpha_{2q}}{E^x} \left( c_f \frac{\sin^2 (\bar{\lambda} K_\varphi)}{\bar{\lambda}^2} + 2 q \frac{\sin(2\bar{\lambda} K_\varphi)}{2 \bar{\lambda}} \right)
    \ , \\
    \bar{g} \bar{f}_2 &=&
    - \frac{\alpha_2}{4 E^x} \lambda_0 \cos^2 (\bar{\lambda} K_\varphi)
    \ , \\
    \bar{g} \bar{f}_2^q &=&
    - \frac{\alpha_{2q}}{4 E^x} \cos^2 (\bar{\lambda} K_\varphi)
    \ .
\end{eqnarray}

The general vacuum Hamiltonian constraint is
\begin{eqnarray}
    H &=&
    \lambda_0^2 \frac{E^x}{2} \sqrt{q_{xx}} V_q
    - \frac{E^\varphi}{2} \sqrt{q^{xx}} \Bigg[
    E^\varphi \left(
    \frac{\alpha_{0q}}{E^x}
    + \frac{\alpha_{2q}}{E^x} \left( c_f \frac{\sin^2 (\bar{\lambda} K_\varphi)}{\bar{\lambda}^2} + 2 q \frac{\sin(2\bar{\lambda} K_\varphi)}{2 \bar{\lambda}} \right) \right)
    \nonumber\\
    &&
    - \frac{((E^x)')^2}{E^\varphi} \frac{\alpha_{2q}}{4 E^x} \cos^2 (\bar{\lambda} K_\varphi)
    \Bigg]
    \nonumber\\
    &&
    - \lambda_0 \frac{\sqrt{E^x}}{2} \Bigg[
    E^\varphi \Bigg(
    \frac{\alpha_0}{E^x}
    + \left( c_f \frac{\alpha_2}{E^x} + 2 \frac{\partial c_{f}}{\partial E^x} \right) \frac{\sin^2 \left(\bar{\lambda} K_\varphi\right)}{\bar{\lambda}^2}
     \nonumber\\
    &&
   + 2 \left( q \frac{\alpha_2}{E^x} + 2 \frac{\partial q}{\partial E^x} \right) \frac{\sin \left(2 \bar{\lambda} K_\varphi\right)}{2 \bar{\lambda}}
    + 4 \frac{K_x}{E^\varphi} \left(c_f \frac{\sin (2 \bar{\lambda} K_\varphi)}{2 \bar{\lambda}}
    + q \cos(2 \bar{\lambda} K_\varphi)\right) \Bigg)
    \nonumber\\
    &&
    + \frac{((E^x)')^2}{E^\varphi} \left( - \frac{\alpha_2}{4 E^x} \cos^2 (\bar{\lambda} K_{\varphi})
    + \bar{\lambda}^2 \frac{K_x}{E^\varphi} \frac{\sin (2 \bar{\lambda} K_\varphi)}{2 \bar{\lambda}} \right)\nonumber\\
&&    + \left( \frac{(E^x)' (E^\varphi)'}{(E^\varphi)^2} - \frac{(E^x)''}{E^\varphi} \right) \cos^2 (\bar{\lambda} K_\varphi)
    \Bigg]
    \ ,
    \label{eq:Hamiltonian constraint - vacuum}
\end{eqnarray}
with structure function 
\begin{eqnarray}
    q^{x x}
    =
    \left(
    \left( c_{f}
    + \left(\frac{\bar{\lambda} (E^x)'}{2 E^\varphi} \right)^2 \right) \cos^2 \left(\bar{\lambda} K_\varphi\right)
    - 2 q \bar{\lambda}^2 \frac{\sin \left(2 \bar{\lambda} K_\varphi\right)}{2 \bar{\lambda}}\right)
    \lambda_0^2 \frac{E^x}{(E^\varphi)^2}
    \ ,
    \label{eq:Structure function - periodic - vacuum}
\end{eqnarray}
where $V_q$, $\alpha_i$, and $\alpha_{iq}$ are undetermined functions of $E^x$.

\subsubsection{Existence of a gravitational observable}

The vacuum constraint (\ref{eq:Hamiltonian constraint - vacuum}) admits a
Dirac observable $\mathcal{D}$ only if
$\delta_\epsilon \mathcal{D} = \mathcal{D}_H H + \mathcal{D}_x H_x$, where
$\mathcal{D}_H$ and $\mathcal{D}_x$ depend on the phase-space variables and on
the gauge function $\epsilon$.  We consider the dependence
$\mathcal{D} (E^x , K_\varphi , (E^x)'/E^\varphi , (E^\varphi)^2 q^{xx} /
E^x)$ and require that this expression has the classical mass observable as a limit.

The condition for a Dirac observable can then be rewritten as
\begin{eqnarray}
    \mathcal{O}&=&
    \frac{\partial \mathcal{D}}{\partial E^x} \delta_\epsilon E^x
    + \frac{\partial \mathcal{D}}{\partial K_\varphi} \delta_\epsilon K_\varphi
    + \frac{\partial \mathcal{D}}{\partial z} \delta_\epsilon z
    + \frac{\partial \mathcal{D}}{\partial \beta} \delta_\epsilon \beta
    - \mathcal{D}_H H
    - \mathcal{D}_x H_x
    \nonumber\\
    &=&
    \left( \frac{\partial \mathcal{D}}{\partial E^x} 
    + \frac{\partial \mathcal{D}}{\partial \beta}
    \frac{\partial \beta}{\partial E^x} \right) \delta_\epsilon E^x
    + \left( \frac{\partial \mathcal{D}}{\partial K_\varphi} + \frac{\partial \mathcal{D}}{\partial \beta}
    \frac{\partial \beta}{\partial K_\varphi} \right) \delta_\epsilon K_\varphi
    \nonumber\\
    &&
    + \left( \frac{\partial \mathcal{D}}{\partial z}
    + \frac{\partial \mathcal{D}}{\partial \beta} \frac{\partial \beta}{\partial z} \right) \delta_\epsilon z
    - \mathcal{D}_H H
    - \mathcal{D}_x H_x
    = 0
    \ ,
    \label{eq:Weak observable condition}
\end{eqnarray}
where $z = (E^x)'/E^\varphi$ and $\beta = (E^\varphi)^2 q^{xx} / E^x$, and
each partial derivative is taken by leaving the rest of the variables
constant. (Thus, $\partial \mathcal{D} / \partial E^x$ does not act on the
dependence of $\mathcal{D}$ on $z$ and $\beta$.)

The condition $\mathcal{O}=0$ can be analyzed by derivative conditions.  For example,
the derivative terms
$\partial \mathcal{O}/\partial (E^\varphi)' = \partial \mathcal{O}/\partial (E^x)'' = 0$,
which are both proportional to the overall factor in the Hamiltonian
constraint, determine the coefficient $\mathcal{D}_H$ in terms of the
observable $\mathcal{D}$:
\begin{eqnarray}
    \mathcal{D}_H = \frac{2 \epsilon^x}{\sqrt{E^x} \lambda_0 \cos^2(\lambda K_\varphi)} \left( \frac{\partial \mathcal{D}}{\partial z}
    + \frac{\partial \mathcal{D}}{\partial \beta} \frac{\partial \beta}{\partial z} \right)
    \ .
    \label{eq:DH - scalar weak observable}
\end{eqnarray}

The derivative term $\partial \mathcal{O}/\partial (\epsilon^0)'$ does not have the
necessary phase-space dependence to contribute to either $H$ or $H_x$, so it
must vanish independently and implies
\begin{eqnarray}
    0 &=&
    \left( \frac{\partial \mathcal{D}}{\partial K_\varphi} + \frac{\partial \mathcal{D}}{\partial \beta}
    \frac{\partial \beta}{\partial K_\varphi} \right) \cos^2 (\bar{\lambda} K_\varphi) z
    \nonumber\\
    &&
    + \left( \frac{\partial \mathcal{D}}{\partial z} + \frac{\partial \mathcal{D}}{\partial \beta}
    \frac{\partial \beta}{\partial z} \right) \left( \left( 4 c_f + \bar{\lambda}^2 z^2 \right) \frac{\sin (2 \bar{\lambda} K_\varphi)}{2 \bar{\lambda}}
    + 4 q \cos (2 \bar{\lambda} K_\varphi) \right)
    \ .
    \label{eq:Condition (epsilon^0)' - scalar weak observable}
\end{eqnarray}
Using this condition, we can obtain the coefficient $\mathcal{D}_x$ from the
derivative term $\partial \mathcal{O}/\partial K_\varphi'=0$:
\begin{eqnarray}
    \mathcal{D}_x &=&
    \left( \frac{\partial \mathcal{D}}{\partial K_\varphi} + \frac{\partial \mathcal{D}}{\partial \beta}
    \frac{\partial \beta}{\partial K_\varphi} \right)
    \left(
    \frac{\epsilon^x}{E^\varphi} - \lambda_0 \epsilon^0 \frac{\sqrt{E^x}}{(E^\varphi)^2} \bar{\lambda}^2 \frac{\sin (2 \bar{\lambda} K_\varphi)}{2 \bar{\lambda}} z
    \right)
    \nonumber\\
    &&
    + \left( \frac{\partial \mathcal{D}}{\partial z} + \frac{\partial \mathcal{D}}{\partial \beta}
    \frac{\partial \beta}{\partial z} \right) \lambda_0 \epsilon^0 \frac{\sqrt{E^x}}{2 (E^\varphi)^2}
    \left( \left( 4 c_f + \bar{\lambda}^2 z^2 \right) \cos (2 \bar{\lambda} K_\varphi)
    - 16 q \bar{\lambda}^2 \frac{\sin (2 \bar{\lambda} K_\varphi)}{2 \bar{\lambda}} \right)
    \nonumber\\
    &=&
    \frac{1}{z} \left( \frac{\partial \mathcal{D}}{\partial z} + \frac{\partial \mathcal{D}}{\partial \beta}
    \frac{\partial \beta}{\partial z} \right) \Bigg[ \lambda_0 \epsilon^0 \frac{\sqrt{E^x}}{2 (E^\varphi)^2}
    z \left( \left( 4 c_f + \bar{\lambda}^2 z^2 \right) \cos (2 \bar{\lambda} K_\varphi)
    - 16 q \bar{\lambda}^2 \frac{\sin (2 \bar{\lambda} K_\varphi)}{2 \bar{\lambda}} \right)
    \nonumber\\
    &&
    - \sec^2 (\bar{\lambda} K_\varphi) \left( \left( 4 c_f + \bar{\lambda}^2 z^2 \right) \frac{\sin (2 \bar{\lambda} K_\varphi)}{2 \bar{\lambda}}
       + 4 q \cos (2 \bar{\lambda} K_\varphi) \right)\nonumber\\
  &&\qquad\qquad\qquad\times
    \left(
    \frac{\epsilon^x}{E^\varphi} - \lambda_0 \epsilon^0 \frac{\sqrt{E^x}}{(E^\varphi)^2} \bar{\lambda}^2 \frac{\sin (2 \bar{\lambda} K_\varphi)}{2 \bar{\lambda}} z
    \right)
    \Bigg]
    \ .
    \label{eq:Dx - scalar weak observable}
\end{eqnarray}
With these results, the dependence of the condition $\mathcal{O}=0$ on $z$
(independently of the intrinsic dependence of $\mathcal{D}$, $\mathcal{D}_H$,
$\mathcal{D}_x$, and $\sqrt{q^{xx}}$ on $z$) is polynomial up to order $z^2$.
Therefore, we consider an expression for the observable of the form
$\mathcal{D} = \mathcal{D}_0 + \mathcal{D}_1 z + \mathcal{D}_2 z^2$, with
$\mathcal{D}_i = \mathcal{D}_i (E^x,K_\varphi)$ for $i=0,1,2$, and then expand
$\mathcal{O}=0$ in $z$ with highest order $z^3$.  In addition, $\mathcal{O}=0$ can be
expanded in powers of $\sqrt{\beta}$ and $K_x$, which should vanish
independently.  The terms in $\mathcal{O}=0$ proportional to $K_x$ are
\begin{eqnarray}
    0 &=&
    \left( \frac{\partial \mathcal{D}}{\partial K_\varphi} + \frac{\partial \mathcal{D}}{\partial \beta}
    \frac{\partial \beta}{\partial K_\varphi} \right)
    \lambda_0 \epsilon^0 \frac{\sqrt{E^x}}{E^\varphi} \bar{\lambda}^2 \frac{\sin (2 \bar{\lambda} K_\varphi)}{2 \bar{\lambda}} z^2
    \nonumber\\
    &&
    - \left( \frac{\partial \mathcal{D}}{\partial z}
    + \frac{\partial \mathcal{D}}{\partial \beta} \frac{\partial \beta}{\partial z} \right)
    \lambda_0 \epsilon^0 \frac{\sqrt{E^x}}{2 E^\varphi} z
    \left( \left( 4 c_f + \bar{\lambda}^2 z^2 \right) \cos (2 \bar{\lambda} K_\varphi)
    - 16 q \bar{\lambda}^2 \frac{\sin (2 \bar{\lambda} K_\varphi)}{2 \bar{\lambda}} \right)
    \nonumber\\
    &&
    + \mathcal{D}_H \lambda_0 \frac{\sqrt{E^x}}{2} \left( \left( 4 c_f + \bar{\lambda}^2 z^2 \right) \frac{\sin (2 \bar{\lambda} K_\varphi)}{2 \bar{\lambda}}
    + 4 q \cos (2 \bar{\lambda} K_\varphi) \right)
    + z \mathcal{D}_x E^\varphi
    \ ,
\end{eqnarray}
the vanishing of which is implied by (\ref{eq:DH - scalar weak observable}),
(\ref{eq:Dx - scalar weak observable}) and (\ref{eq:Condition (epsilon^0)' -
  scalar weak observable}).

We next expand $\mathcal{O}$ in powers of $\beta$, independently of the intrinsic
dependence of $\mathcal{D}$ on $\beta$, giving
\begin{eqnarray}
    \mathcal{O} &=& \frac{\mathcal{D}^\beta}{\sqrt{\beta}}
    + \sqrt{\beta} \mathcal{D}_\beta
    + \mathcal{D}_0
    \ ,
\end{eqnarray}
where
\begin{eqnarray}
    \mathcal{D}^\beta &\propto& \left( \frac{\partial \mathcal{D}}{\partial z}
    + \frac{\partial \mathcal{D}}{\partial \beta} \frac{\partial \beta}{\partial z} \right) \Bigg[
    \frac{\epsilon^0}{z} A_{-1} + \epsilon^x A_0 + \epsilon^0 A_1 z + \epsilon^0 A_3 z^3 + \epsilon^0 A_5 z^5
    \Bigg]
    \ ,
\end{eqnarray}
\begin{eqnarray}
    \mathcal{D}_\beta &\propto& \left( \frac{\partial \mathcal{D}}{\partial z}
    + \frac{\partial \mathcal{D}}{\partial \beta} \frac{\partial \beta}{\partial z} \right) \Bigg[ \frac{\epsilon^0}{z} B_{-1} + \epsilon^0 B_1 z \Bigg]
    \ ,
\end{eqnarray}
and
\begin{eqnarray}
    \mathcal{D}_0 &\propto&
    \frac{\epsilon^0}{z} C_{-1} + \left( \frac{\partial \mathcal{D}}{\partial E^x} \epsilon^0 C_0^x + \left( \frac{\partial \mathcal{D}}{\partial z}
    + \frac{\partial \mathcal{D}}{\partial \beta} \frac{\partial \beta}{\partial z} \right) \epsilon^x C_0^z \right)
    \nonumber\\
    &&
    + \left( \frac{\partial \mathcal{D}}{\partial E^x} \epsilon^x C_1^x + \left( \frac{\partial \mathcal{D}}{\partial z}
    + \frac{\partial \mathcal{D}}{\partial \beta} \frac{\partial \beta}{\partial z} \right) \epsilon^0 C_1^z \right) z
    \nonumber\\
    &&
    + \left( \frac{\partial \mathcal{D}}{\partial E^x} \epsilon^0 C_2^x + \left( \frac{\partial \mathcal{D}}{\partial z}
    + \frac{\partial \mathcal{D}}{\partial \beta} \frac{\partial \beta}{\partial z} \right) \epsilon^x C_2^z \right) z^2
    + \left( \frac{\partial \mathcal{D}}{\partial z}
    + \frac{\partial \mathcal{D}}{\partial \beta} \frac{\partial \beta}{\partial z} \right) \epsilon^0 z^3
    \,.
\end{eqnarray}
The functions $A_i$ are in general complicated expressions of the undetermined
functions of the phase-space variables.  We find that no $\beta$-dependence of
$\mathcal{D}$ can properly mix the $\mathcal{D}^\beta$, $\mathcal{D}_\beta$,
and $\mathcal{D}_0$ such that $\mathcal{D}^\beta$ and $\mathcal{D}_\beta$ are
non-trivial.  Therefore, and since the classical limit requires
${\rm d} \mathcal{D} / {\rm d} z \neq 0$, we must have $A_i = B_i = 0$ and
take $\partial \mathcal{D} / \partial \beta = 0$.  We start with the simplest
of these expressions.

Since the phase-space variables are non-vanishing off-shell, we are interested
only in the dependence of $A_i$ and $B_i$ on the undetermined functions in
the constraint, with the condition that the undetermined functions with
non-vanishing classical limit cannot be trivial.  We then have
\begin{eqnarray}
    A_0 &\propto& V_q = 0
    \ , \\
    A_5 &\propto& \alpha_{2q} = 0
    \ .
\end{eqnarray}
Using this, the other terms simplify to
\begin{eqnarray}
    A_{-1} &\propto& A_1 \propto A_3 \propto \alpha_{0q} = 0
\end{eqnarray}
which implies that  the $B_i$ terms automatically vanish.

With $V_q = \alpha_{0q} = \alpha_{2q} = 0$, the Hamiltonian constraint takes
the form of the expression previously obtained for vacuum in \cite{HigherCov}.
The solution to $\mathcal{O}=0$ is then straightforward, giving
\begin{eqnarray}
    \mathcal{D}
    &=&
    d_0
    + \frac{d_2}{2} \left(\exp \int {\rm d} E^x \ \frac{\alpha_2}{2 E^x}\right)
    \left(
    c_f \frac{\sin^2\left(\bar{\lambda} K_{\varphi}\right)}{\bar{\lambda}^2}
    + 2 q \frac{\sin \left(2 \bar{\lambda}  K_{\varphi}\right)}{2 \bar{\lambda}}
    - \cos^2 (\bar{\lambda} K_\varphi) \left(\frac{(E^x)'}{2 E^\varphi}\right)^2
    \right)
    \notag\\
    &&
    + \frac{d_2}{4} \int {\rm d} E^x \ \left( \left(
    \Lambda_0
    + \frac{\alpha_0}{E^x}
    \right) \exp \int {\rm d} E^x \ \frac{\alpha_2}{2 E^x} \right)
    \ ,
    \label{eq:Gravitational weak observable}
\end{eqnarray}
where $d_0$ and $d_2$ are constants with classical limit $d_0 \to 0$ and $d_2 \to 1$.

In what follows, we impose the condition that this observable be preserved in the vacuum
limit, thus restricting (\ref{eq:Hamiltonian constraint - vacuum}) to the
case where
\begin{eqnarray}
    V_q , \bar{f}_0^q , \bar{f}_2^q \xrightarrow[\phi \to 0]{\empty} 0
    \ .
    \label{eq:Gravitational vacuum observable condition}
\end{eqnarray}
(This restriction eliminates the possibility of using $V_q$ to introduce a
cosmological constant coupled to the emergent space-time metric.)

\subsubsection{Existence of a matter observable,  and residual canonical transformation}

The general Hamiltonian constraint takes the form
\begin{eqnarray}
    H
    &=&
    - \frac{\sqrt{E^x}}{2} \lambda_0 \not{\bar{g}} \Bigg[
    E^\varphi \bar{f}_0
    + K_x \bar{f}_1 
    + P_\phi \bar{h}_0
    + \frac{P_\phi{}^2}{E^\varphi} \bar{f}_3\nonumber\\
&&\qquad    + \frac{((E^x)')^2}{E^\varphi} \left( \bar{f}_2 + \frac{K_x}{E^\varphi} \bar{h} 
    + \frac{P_\phi}{E^\varphi} \bar{h}_1
    + \frac{P_\phi{}^2}{(E^\varphi)^2} \bar{h}_2 \right)
    \nonumber\\
    &&\qquad
    + \frac{(\phi')^2}{E^\varphi} \bar{f}_4
    + \frac{(E^x)' \phi'}{E^\varphi} \left( \bar{h}_3 + \frac{P_\phi}{E^\varphi} \bar{h}_4 \right)
    + \frac{(E^x)' (E^\varphi)'}{(E^\varphi)^2} - \frac{(E^x)''}{E^\varphi}
    \Bigg]
    \nonumber\\
    &&
    - \frac{\sqrt{E^x}}{2} \lambda_0 \sqrt{\beta} \Bigg[
    E^\varphi \bar{g}\bar{f}_0^q
    + P_\phi \bar{g}\bar{h}_0^q
    + \frac{P_\phi{}^2}{E^\varphi} \bar{g}\bar{f}_3^q
    + \frac{((E^x)')^2}{E^\varphi} \left( \bar{g}\bar{f}_2^q
    + \frac{P_\phi}{E^\varphi} \bar{g}\bar{h}_1^q
    + \frac{P_\phi{}^2}{(E^\varphi)^2} \bar{g}\bar{h}_2^q \right)
    \nonumber\\
    &&\qquad
    + \frac{(\phi')^2}{E^\varphi} \bar{g}\bar{f}_4^q
    + \frac{(E^x)' \phi'}{E^\varphi} \bar{g}\bar{h}_3^q
    \Bigg]
    + \frac{\lambda_0^2}{2} E^x \sqrt{q_{xx}} V_q
    \ ,
    \label{eq:Hamiltonian constraint ansatz - Matter observable}
\end{eqnarray}
if we write $\bar{g} = \lambda_0 \not{\bar{g}}$ and
$q^{xx} = \beta \lambda_0^2 E^x/ E^\varphi$. We now consider a slight
generalization of the symmetry generator (\ref{eq:Symmetry generator of real
  scalar field - Linear combination - periodic variables})
\begin{eqnarray}
    G [\alpha] = \int {\rm d} x\ \alpha \left( P_\phi \frac{\partial f_\phi}{\partial \phi}
    + E^\varphi \frac{\tan (\bar{\lambda} K_\varphi)}{\bar{\lambda}} f_\varphi \right)
    \ ,
    \label{eq:Symmetry generator of real scalar field - general}
\end{eqnarray}
for undetermined functions $f_\phi (\phi , E^x)$ and $f_\varphi (\phi , E^x)$,
and a constant $\alpha$.  A canonical transformation of the form
(\ref{eq:Residual canonical transformation phi,ex}) such that
$f_c^\phi = f_\phi$ simplifies the symmetry generator:
\begin{eqnarray}
    G [\alpha] = \int {\rm d} x\ \alpha \left( P_\phi
    + E^\varphi \frac{\tan (\bar{\lambda} K_\varphi)}{\bar{\lambda}} \frac{\partial \ln \lambda}{\partial \phi} \right)
    \ ,
    \label{eq:Symmetry generator of real scalar field - CT modded}
\end{eqnarray}
where we have rewritten the transformed $f_\varphi$ as
$\partial \ln \lambda / \partial \phi$ for some undetermined function
$\lambda (\phi , E^x)$. We may also redefine the functions in the constraint
(\ref{eq:Hamiltonian constraint ansatz - Matter observable}) such that they
now depend on these new versions.  This step completes factoring out the
diffeomorphism-preserving canonical transformations.

We now require that the smeared phase-space function (\ref{eq:Symmetry
  generator of real scalar field - CT modded}) Poisson-commutes with the
Hamiltonian constraint (\ref{eq:Hamiltonian constraint ansatz - Matter
  observable}) on-shell when $V_q=0$.  Defining
$\delta_{G[\alpha]} \cdot \equiv \{ \cdot , G[\alpha] \}$, we find that the
bracket $\delta_{G[\alpha]} H[\epsilon^0]$ contains an $(E^x)''$ term.
The rest of the terms can then be rearranged to complement this term into
reproducing the Hamiltonian constraint, which vanishes on-shell and need not
be considered for the existence of a global matter symmetry.  In practice, it is easier to
subtract such a term and require that the rest vanish.  We do this together with
the expansion
\begin{eqnarray} \label{exp}
    \delta_{G[\alpha]} H[\epsilon^0]
    - H[\epsilon^0] \delta_{G[\alpha]} \ln \left( \frac{\not{\bar{g}}}{E^\varphi} \right)
    &=&
    H^{(-1)} / \sqrt{\beta} + H^{(0)} + H^{(1)} \sqrt{\beta}
    \ ,
\end{eqnarray}
where each term must vanish independently.  Further using the fact that all the
undetermined functions are independent of the phase-space variables $K_x$,
$E^\varphi$, and $P_\phi$ with density weight one and of derivatives, all the
sub-terms obtained from the expansion (\ref{exp}) must vanish
independently.  Thus, the first term being zero implies the equations
\begin{eqnarray}
    0 &=&
    B^{(-1)} \bar{f}_0^q
    \ , \label{eq:f0q - Matter observable - H(-1)} \\
    0 &=&
    B^{(-1)} \bar{f}_2^q
    \ , \label{eq:f2q - Matter observable - H(-1)} \\
    0 &=&
    B^{(-1)} \bar{f}_3^q
    \ , \label{eq:f3q - Matter observable - H(-1)} \\
    0 &=&
    B^{(-1)}\bar{f}_4^q
    \ , \label{eq:f4q - Matter observable - H(-1)} \\
    0 &=&
    B^{(-1)} \bar{h}_0^q
    \ , \label{eq:h0q - Matter observable - H(-1)} \\
    0 &=&
    B^{(-1)} \bar{h}_1^q
    \ , \label{eq:h1q - Matter observable - H(-1)} \\
    0 &=&
    B^{(-1)}\bar{h}_2^q
    \ , \label{eq:h2q - Matter observable - H(-1)} \\
    0 &=&
    B^{(-1)} \bar{h}_3^q
    \ , \label{eq:h3q - Matter observable - H(-1)}
\end{eqnarray}
where
\begin{eqnarray}
    B^{(-1)} 
    &=& \frac{\partial \beta}{\partial \phi}
    + \frac{\partial \ln \lambda}{\partial \phi} \left( \frac{\tan (\bar{\lambda} K_\varphi)}{\bar{\lambda}} \frac{\partial \beta}{\partial K_\varphi}
    - E^\varphi \sec^2 (\bar{\lambda} K_\varphi) \frac{\partial \beta}{\partial E^\varphi} \right)
    \nonumber\\
    &=&
    \left( \frac{\partial c_{f}}{\partial \phi} \cos^2 \left(\bar{\lambda} K_\varphi\right)
    - 2 \frac{\partial q}{\partial \phi} \bar{\lambda}^2 \frac{\sin \left(2 \bar{\lambda} K_\varphi\right)}{2 \bar{\lambda}}\right)
    \nonumber\\
    &&
    - \frac{\partial \ln \lambda^2}{\partial \phi} \bar{\lambda}^2 \frac{\tan (\bar{\lambda} K_\varphi)}{\bar{\lambda}} \left( c_{f} \frac{\sin \left(2 \bar{\lambda} K_\varphi\right)}{2 \bar{\lambda}}
    + q \cos \left(2 \bar{\lambda} K_\varphi\right) \right)
    \nonumber\\
    &&
    + \frac{\partial \ln \lambda^2}{\partial \phi} \left(\frac{\bar{\lambda} (E^x)'}{2 E^\varphi} \right)^2 \cos^2 \left(\bar{\lambda} K_\varphi\right)
    \ . \label{eq:B factor - Matter observable - H(-1)}
\end{eqnarray}

The last term $H^{(1)}=0$ being zero implies the equations
\begin{eqnarray}
    \frac{\partial (\bar{g} \bar{f}_0^q)}{\partial \phi} &=&
    \left( 2 \bar{g} \bar{f}_0^q
    - \frac{\tan (\bar{\lambda} K_\varphi)}{\bar{\lambda}} \frac{\partial (\bar{g} \bar{f}_0^q)}{\partial K_\varphi} \right) \frac{\partial \ln \lambda}{\partial \phi}
    + \frac{\tan (\bar{\lambda} K_\varphi)}{\bar{\lambda}} \bar{g} \bar{h}_0^q \frac{\partial^2 \ln \lambda}{\partial \phi^2}
    \ , \label{eq:f0q - Matter observable - H(1)} \\
    \frac{\partial (\bar{g} \bar{f}_2^q)}{\partial \phi} &=&
    - \left( 2 \bar{\lambda}^2 \frac{\tan^2(\bar{\lambda} K_\varphi)}{\bar{\lambda}^2} \bar{g}\bar{f}_2^q
    + \frac{\tan (\bar{\lambda} K_\varphi)}{\bar{\lambda}} \frac{\partial (\bar{g} \bar{f}_2^q)}{\partial K_\varphi} \right) \frac{\partial \ln \lambda}{\partial \phi}
    \nonumber\\
    &&
    + \frac{\tan (\bar{\lambda} K_\varphi)}{\bar{\lambda}} \bar{g} \bar{h}_1^q \frac{\partial^2 \ln \lambda}{\partial \phi^2}
    \ , \label{eq:f2q - Matter observable - H(1)} \\
    \frac{\partial (\bar{g}\bar{f}_3^q)}{\partial \phi} &=&
    - \left( 2 \bar{\lambda}^2 \frac{\tan^2(\bar{\lambda} K_\varphi)}{\bar{\lambda}^2} \bar{g} \bar{f}_3^q
    + \frac{\tan (\bar{\lambda} K_\varphi)}{\bar{\lambda}} \frac{\partial (\bar{g} \bar{f}_3^q)}{\partial K_\varphi} \right) \frac{\partial \ln \lambda}{\partial \phi}
    \ , \label{eq:f3q - Matter observable - H(1)} \\
    \frac{\partial (\bar{g} \bar{f}_4^q)}{\partial \phi} &=&
    - \left( 2 \bar{\lambda}^2 \frac{\tan^2(\bar{\lambda} K_\varphi)}{\bar{\lambda}^2} \bar{g} \bar{f}_4^q
    + \frac{\tan (\bar{\lambda} K_\varphi)}{\bar{\lambda}} \frac{\partial (\bar{g} \bar{f}_4^q)}{\partial K_\varphi} \right) \frac{\partial \ln \lambda}{\partial \phi}
    \ , \label{eq:f4q - Matter observable - H(1)} \\
    \frac{\partial (\bar{g}\bar{h}_0^q)}{\partial \phi} &=&
    \left( \frac{\cos(2 \bar{\lambda} K_\varphi)}{\cos^2 (\bar{\lambda} K_\varphi)} \bar{g} \bar{h}_0^q
    - \frac{\tan (\bar{\lambda} K_\varphi)}{\bar{\lambda}} \frac{\partial (\bar{g} \bar{h}_0^q)}{\partial K_\varphi} \right) \frac{\partial \ln \lambda}{\partial \phi}
    \nonumber\\
    &&
    + 2 \frac{\tan (\bar{\lambda} K_\varphi)}{\bar{\lambda}} \bar{g}\bar{f}_3^q \frac{\partial^2 \ln \lambda}{\partial \phi^2}
    \ , \label{eq:h0q - Matter observable - H(1)} \\
    \frac{\partial (\bar{g} \bar{h}_1^q)}{\partial \phi} &=&
    - \left( \frac{1+2\sin^2(\bar{\lambda} K_\varphi)}{\cos^2 (\bar{\lambda} K_\varphi)} \bar{g} \bar{h}_1^q
    + \frac{\tan (\bar{\lambda} K_\varphi)}{\bar{\lambda}} \frac{\partial (\bar{g} \bar{h}_1^q)}{\partial K_\varphi} \right) \frac{\partial \ln \lambda}{\partial \phi}
    \nonumber\\
    &&
    + 2 \frac{\tan (\bar{\lambda} K_\varphi)}{\bar{\lambda}} \bar{g} \bar{h}_2^q \frac{\partial^2 \ln \lambda}{\partial \phi^2}
    \ , \label{eq:h1q - Matter observable - H(1)} \\
    \frac{\partial (\bar{g} \bar{h}_2^q)}{\partial \phi} &=&
    - \left( 2\frac{1+\sin^2(\bar{\lambda} K_\varphi)}{\cos^2 (\bar{\lambda} K_\varphi)} \bar{g} \bar{h}_2^q
    + \frac{\tan (\bar{\lambda} K_\varphi)}{\bar{\lambda}} \frac{\partial (\bar{g} \bar{h}_2^q)}{\partial K_\varphi} \right) \frac{\partial \ln \lambda}{\partial \phi}
    \ , \label{eq:h2q - Matter observable - H(1)} \\
    \frac{\partial (\bar{g} \bar{h}_3^q)}{\partial \phi} &=&
    - \left( 2 \bar{\lambda}^2 \frac{\tan^2(\bar{\lambda} K_\varphi)}{\bar{\lambda}^2} \bar{g}\bar{h}_3^q
    + \frac{\tan (\bar{\lambda} K_\varphi)}{\bar{\lambda}} \frac{\partial (\bar{g}\bar{h}_3^q)}{\partial K_\varphi} \right) \frac{\partial \ln \lambda}{\partial \phi}
    \ . \label{eq:h3q - Matter observable - H(1)}
\end{eqnarray}

The last term $H^{(0)}=0$ being zero implies the equations
\begin{eqnarray}
    \frac{\partial (\not{\bar{g}} \bar{f}_0)}{\partial \phi} &=&
    \left( 2 \not{\bar{g}} \bar{f}_0
    - \frac{\tan (\bar{\lambda} K_\varphi)}{\bar{\lambda}} \frac{\partial (\not{\bar{g}} \bar{f}_0)}{\partial K_\varphi} \right) \frac{\partial \ln \lambda}{\partial \phi}
    + \frac{\tan (\bar{\lambda} K_\varphi)}{\bar{\lambda}} \not{\bar{g}} \bar{h}_0 \frac{\partial^2 \ln \lambda}{\partial \phi^2}
    \nonumber\\
    &&
    - \frac{\tan (\bar{\lambda} K_\varphi)}{\bar{\lambda}} \not{\bar{g}} \bar{f}_1 \frac{\partial^2 \ln \lambda}{\partial \phi\partial E^x}
    \ , \label{eq:f0 - Matter observable - H(0)} \\
    \frac{\partial (\not{\bar{g}} \bar{f}_1)}{\partial \phi} &=&
    \left( \frac{\cos(2 \bar{\lambda} K_\varphi)}{\cos^2(\bar{\lambda} K_\varphi)} \not{\bar{g}} \bar{f}_1
    - \frac{\tan (\bar{\lambda} K_\varphi)}{\bar{\lambda}} \frac{\partial (\not{\bar{g}} \bar{f}_1)}{\partial K_\varphi} \right) \frac{\partial \ln \lambda}{\partial \phi}
    \ , \label{eq:f1 - Matter observable - H(0)} \\
    \frac{\partial (\not{\bar{g}} \bar{f}_2)}{\partial \phi} &=&
    - \left( 2 \bar{\lambda}^2 \frac{\tan^2(\bar{\lambda} K_\varphi)}{\bar{\lambda}^2} \not{\bar{g}} \bar{f}_2
    + \frac{\tan (\bar{\lambda} K_\varphi)}{\bar{\lambda}} \frac{\partial (\not{\bar{g}} \bar{f}_2)}{\partial K_\varphi} \right) \frac{\partial \ln \lambda}{\partial \phi}
    \nonumber\\
    &&
    + \frac{\tan (\bar{\lambda} K_\varphi)}{\bar{\lambda}} \not{\bar{g}} \bar{h}_1 \frac{\partial^2 \ln \lambda}{\partial \phi^2}
    + \cos^2 (\bar{\lambda} K_\varphi) \frac{\partial^2 \ln \lambda}{\partial \phi\partial E^x}
    \ , \label{eq:f2 - Matter observable - H(0)} \\
    \frac{\partial (\not{\bar{g}} \bar{f}_3)}{\partial \phi} &=&
    - \left( 2 \bar{\lambda}^2 \frac{\tan^2(\bar{\lambda} K_\varphi)}{\bar{\lambda}^2} \not{\bar{g}} \bar{f}_3
    + \frac{\tan (\bar{\lambda} K_\varphi)}{\bar{\lambda}} \frac{\partial (\not{\bar{g}} \bar{f}_3)}{\partial K_\varphi} \right) \frac{\partial \ln \lambda}{\partial \phi}
    \ , \label{eq:f3 - Matter observable - H(0)} \\
    \frac{\partial (\not{\bar{g}} \bar{f}_4)}{\partial \phi} &=&
    - \left( 2 \bar{\lambda}^2 \frac{\tan^2(\bar{\lambda} K_\varphi)}{\bar{\lambda}^2} \not{\bar{g}} \bar{f}_4
    + \frac{\tan (\bar{\lambda} K_\varphi)}{\bar{\lambda}} \frac{\partial (\not{\bar{g}} \bar{f}_4)}{\partial K_\varphi} \right) \frac{\partial \ln \lambda}{\partial \phi}
    \ , \label{eq:f4 - Matter observable - H(0)} \\
    \frac{\partial (\not{\bar{g}} \bar{h}_0)}{\partial \phi} &=&
    \left( \frac{\cos(2 \bar{\lambda} K_\varphi)}{\cos^2 (\bar{\lambda} K_\varphi)} \not{\bar{g}} \bar{h}_0
    - \frac{\tan (\bar{\lambda} K_\varphi)}{\bar{\lambda}} \frac{\partial (\not{\bar{g}} \bar{h}_0)}{\partial K_\varphi} \right) \frac{\partial \ln \lambda}{\partial \phi}
    \nonumber\\
    &&
    + 2 \frac{\tan (\bar{\lambda} K_\varphi)}{\bar{\lambda}} \not{\bar{g}} \bar{f}_3 \frac{\partial^2 \ln \lambda}{\partial \phi^2}
    \ , \label{eq:h0 - Matter observable - H(0)} \\
    \frac{\partial (\not{\bar{g}} \bar{h}_1)}{\partial \phi} &=&
    - \left( \frac{1+2\sin^2(\bar{\lambda} K_\varphi)}{\cos^2 (\bar{\lambda} K_\varphi)} \not{\bar{g}} \bar{h}_1
    + \frac{\tan (\bar{\lambda} K_\varphi)}{\bar{\lambda}} \frac{\partial (\not{\bar{g}} \bar{h}_1)}{\partial K_\varphi} \right) \frac{\partial \ln \lambda}{\partial \phi}
    \nonumber\\
    &&
    + 2 \frac{\tan (\bar{\lambda} K_\varphi)}{\bar{\lambda}} \not{\bar{g}} \bar{h}_2 \frac{\partial^2 \ln \lambda}{\partial \phi^2}
    \ , \label{eq:h1 - Matter observable - H(0)} \\
    \frac{\partial (\not{\bar{g}} \bar{h}_2)}{\partial \phi} &=&
    - \left( 2 \frac{1+\sin^2(\bar{\lambda} K_\varphi)}{\cos^2 (\bar{\lambda} K_\varphi)} \not{\bar{g}} \bar{h}_2
    + \frac{\tan (\bar{\lambda} K_\varphi)}{\bar{\lambda}} \frac{\partial (\not{\bar{g}} \bar{h}_2)}{\partial K_\varphi} \right) \frac{\partial \ln \lambda}{\partial \phi}
    \ , \label{eq:h2 - Matter observable - H(0)} \\
    \frac{\partial (\not{\bar{g}} \bar{h}_3)}{\partial \phi} &=&
    - \left( 2 \bar{\lambda}^2 \frac{\tan^2(\bar{\lambda} K_\varphi)}{\bar{\lambda}^2} \not{\bar{g}} \bar{h}_3
    + \frac{\tan (\bar{\lambda} K_\varphi)}{\bar{\lambda}} \frac{\partial (\not{\bar{g}} \bar{h}_3)}{\partial K_\varphi} \right) \frac{\partial \ln \lambda}{\partial \phi}
    \nonumber\\
    &&
    + \left( 1 + \frac{\tan (\bar{\lambda} K_\varphi)}{\bar{\lambda}} \not{\bar{g}} \bar{h}_4 \right) \frac{\partial^2 \ln \lambda}{\partial \phi^2}
    \ , \label{eq:h3 - Matter observable - H(0)} \\
    \frac{\partial (\not{\bar{g}} \bar{h}_4)}{\partial \phi} &=&
    \left( - 2 \frac{1 + 2 \sin^2 (\bar{\lambda} K_\varphi)}{\cos^2 (\bar{\lambda} K_\varphi)} \not{\bar{g}} \bar{h}_4
    - \frac{\tan (\bar{\lambda} K_\varphi)}{\bar{\lambda}} \left( \frac{\partial (\not{\bar{g}} \bar{h}_4)}{\partial K_\varphi}
    + 2 \bar{\lambda}^2 \right) \right) \frac{\partial \ln \lambda}{\partial \phi}
    \ . \label{eq:h4 - Matter observable - H(0)}
\end{eqnarray}

The condition for the existence of the matter observable is thus highly
non-trivial, leading to the set of equations (\ref{eq:f0q - Matter observable
  - H(-1)})--(\ref{eq:h3q - Matter observable - H(-1)}) and (\ref{eq:f0q -
  Matter observable - H(1)})--(\ref{eq:h4 - Matter observable - H(0)}).  These
are, however, too complicated to be solved completely, and yet not sufficient
to fully restrict the form of the Hamiltonian constraint. In order to make
progress, we continue to impose additional conditions.

\subsubsection{Partial Abelianization}

We apply the generalized techniques for partial Abelianization developed in
\cite{HigherCov} by simply including the new degree of freedom given by the
scalar field. The procedure is identical to that of the earlier 
Section~\ref{sec:Linear combination} up to the definition of the new structure
function and using the modified constraint instead of the classical one.

We consider the following linear combination
\begin{eqnarray}
    H^{(A)} = B H + A H_x
    \label{eq:Linear combinations of constraints - generic2}
\end{eqnarray}
of the constraints, where $A$ and $B\not=0$ are so far undetermined
phase-space functions, and $H$ is the previous modified constraint.  Reusing
the definitions (\ref{eq:Transformation of B - Geometric condition - Linear
  combination})--(\ref{eq:Transformation of Ax - Geometric condition - Linear
  combination}) for coefficients such as ${\cal A}^x$, now applied to the
modified constraint, the structure function in the bracket of two $H^{(A)}$ is
given by
\begin{equation}
    q^{(A)} =
    B^2 q^{xx} + B {\cal A}^x
    \ .
    \label{eq:To-be abelianized structure function}
\end{equation}

Partial Abelianization is achieved by setting $q^{(A)}=0$.  Assuming the
dependence $B=B(K_\varphi,E^x,\phi)$ and a constraint of the general form
(\ref{eq:Hamiltonian constraint ansatz - Matter observable}), we obtain
\begin{eqnarray}
    A &=& \frac{\partial B}{\partial K_\varphi} \frac{\partial H}{\partial (E^\varphi)'}
    \nonumber\\
    &=&
    - \frac{\sqrt{E^x}}{2} \lambda_0 \cos^2 (\bar{\lambda} K_\varphi) \frac{\partial B}{\partial K_\varphi} \frac{(E^x)'}{(E^\varphi)^2}
\end{eqnarray}
and
\begin{eqnarray}
    {\cal A}^x &=& - \frac{\partial A}{\partial K_\varphi} \frac{\partial H}{\partial (E^\varphi)'}
    - \frac{\partial A}{\partial (E^x)'} \frac{\partial H}{\partial K_x}
    \nonumber\\
    &=&
    - \lambda_0^2 \frac{E^x}{4} \frac{\cos^4 (\bar{\lambda} K_\varphi)}{(E^\varphi)^2} \Bigg[ - \frac{\partial B}{\partial K_\varphi} \bar{f}_1
    + \left(\frac{(E^x)'}{E^\varphi}\right)^2 \left(
    \frac{\partial^2 B}{\partial K_\varphi^2} + \frac{3}{2} \frac{\partial B}{\partial K_\varphi} \frac{\partial \ln \bar{g}}{\partial K_\varphi} \right)
    \Bigg]
    \nonumber\\
    &=&
    - \lambda_0^2 E^x \frac{\cos^2 (\bar{\lambda} K_\varphi)}{(E^\varphi)^2} \Bigg[ - \frac{\partial B}{\partial K_\varphi} \left(c_f \frac{\sin (2 \bar{\lambda} K_\varphi)}{2 \bar{\lambda}}
    + q \cos(2 \bar{\lambda} K_\varphi)\right)
    \nonumber\\
    &&
    + \left(\frac{(E^x)'}{2 E^\varphi}\right)^2 \left(
    \frac{\partial^2 B}{\partial K_\varphi^2} - 3 \bar{\lambda}^2 \frac{\partial B}{\partial K_\varphi} \frac{\tan (\bar{\lambda} K_\varphi)}{\bar{\lambda}} \right) \cos^2 (\bar{\lambda} K_\varphi)
    \Bigg]
    \,.
\end{eqnarray}
Condition (\ref{eq:To-be abelianized structure function}) for partial
Abelianization, such that $q^{(A)}=0$, then implies
\begin{eqnarray}
    0 &=&
    \Bigg[ B \left( c_{f}
    - 2 q \bar{\lambda}^2 \frac{\tan \left(\bar{\lambda} K_\varphi\right)}{\bar{\lambda}} \right)
    - \frac{\partial B}{\partial K_\varphi} \left(c_f \frac{\sin (2 \bar{\lambda} K_\varphi)}{2 \bar{\lambda}}
    + q \cos(2 \bar{\lambda} K_\varphi)\right) \Bigg] \cos^2 (\bar{\lambda} K_\varphi)
    \nonumber\\
    &&
    - \left(\frac{(E^x)'}{2 E^\varphi}\right)^2 \left(
    \frac{\partial^2 B}{\partial K_\varphi^2}\cos^2 (\bar{\lambda} K_\varphi)
    + \bar{\lambda}^2 B - 3 \bar{\lambda}^2 \frac{\partial B}{\partial K_\varphi} \frac{\sin (2 \bar{\lambda} K_\varphi)}{2 \bar{\lambda}} \right) \cos^2 (\bar{\lambda} K_\varphi)
    \Bigg]
    \ .
    \label{eq:To-be abelianized structure function2}
\end{eqnarray}
Since $B$ is independent of $(E^x)'$, the two lines in this equation must vanish independently.
The first line implies
\begin{eqnarray}
    B = \frac{\bar{B}}{\cos^2 (\bar{\lambda} K_\varphi)} \left( c_f \frac{\sin (2 \bar{\lambda} K_\varphi)}{2 \bar{\lambda}} + q \cos (2 \bar{\lambda} K_\varphi) \right)
\end{eqnarray}
where $\bar{B}$ is an undetermined function of $E^x$ and $\phi$.
Substituting this result in the second line and demanding that it  vanish, we obtain
\begin{eqnarray}
    \bar{\lambda} \bar{B}  q \sec (\bar{\lambda} K_\varphi) = 0
    \ .
    \label{eq:Partial Abelianization condition on q}
\end{eqnarray}
For a non-trivial Abelianization with non-zero $\bar{\lambda}$, this equation determines $q=0$.
The Abelianization coefficients are then
\begin{equation}
    B^{(A)} = \bar{B} c_f \frac{\tan (\bar{\lambda} K_\varphi)}{\bar{\lambda}}
    \quad,\quad
    A^{(A)} = - \bar{B} \lambda_0 c_f \frac{\sqrt{E^x}}{2} \frac{(E^x)'}{(E^\varphi)^2}
    \ ,
    \label{eq:Abelianization coefficients}
\end{equation}
where we have included the superscript $(A)$ in order to distinguish them from the
previous coefficients in linear combinations.  The coefficients
(\ref{eq:Abelianization coefficients}) together with the condition that $q=0$,
implied by (\ref{eq:Partial Abelianization condition on q}), Abelianize any
constraint of the general form (\ref{eq:Hamiltonian constraint ansatz - Matter
  observable}).

\section{Classes of constraints}
\label{sec:Classes}

There is a large number of free functions in the generic modified Hamiltonian
constraint, subject to conditions that include coupled non-linear differential
equations. It is hard to solve these equations in complete generality, but
several physically motivated conditions impose additional equations that can
be used to simplify and solve the original restrictions on modification functions.

\subsection{Constraints compatible with the classical-matter limit}

A special class of modified constraints is given by those that recover
classical matter behavior (on a modified background) in a certain limit. This
requirement imposes additional conditions that can be used in order to solve
for some of the free functions.

\subsubsection{Anomaly-freedom}

In order to recover the Klein--Gordon Hamiltonian on a curved, emergent
space-time, we must impose $\bar{f}_3^q \neq 0$ and $\bar{f}_4^q \neq
0$. Equations (\ref{eq:f3q - Matter observable - H(-1)}) and (\ref{eq:f4q -
  Matter observable - H(-1)}) then imply that the $B^{(-1)}$ factor (\ref{eq:B
  factor - Matter observable - H(-1)}) must vanish, which in turn implies that
\begin{eqnarray}
    \frac{\partial c_{f}}{\partial \phi} = \frac{\partial q}{\partial \phi}
    = \frac{\partial \lambda}{\partial \phi} &=& 0
    \,.
\end{eqnarray}
Hence, $c_f$, $q$, and $\lambda$ can only depend on $E^x$.  With these results,
equations~(\ref{eq:f0q - Matter observable - H(-1)})--(\ref{eq:h3q - Matter
  observable - H(-1)}) are trivially satisfied, while equations~(\ref{eq:f0q -
  Matter observable - H(1)})--(\ref{eq:h4 - Matter observable - H(0)})
imply that $\bar{g} \bar{f}_0^q$ as well as $\bar{g} \bar{h}_j^q$, $\bar{f}_j$
and $\bar{h}_j$ for $j=0,1,2,3,4$ must be independent of $\phi$. Considering
these conditions, the only undetermined function that is allowed to depend on
$\phi$ is the global factor $\lambda_0$. Combined with the results for the
existence of a gravitational vacuum observable, (\ref{eq:Gravitational vacuum
  observable condition}), we obtain
\begin{eqnarray}
    \bar{f}_{0q} = \bar{f}_0^q = \bar{f}_2^q = 0
    \ .
\end{eqnarray}

At this point, only the conditions for anomaly-freedom  remain to be solved.  The
vanishing of the $((E^x)')^2 q^{xx}$ term in equations (\ref{eq:AF f1,h3}) and
(\ref{eq:AF f3f4,h4}) implies $\bar{h}_1^q = 0$ and $\bar{h}_2^q = 0$,
respectively.  The $((E^x)')^2 \sqrt{q^{xx}}$ term in (\ref{eq:AF f3f4,h4})
then implies that $\bar{h}_2 = 0$.
Five additional equations are derived from (\ref{eq:AF f3,f3q,h2,h2q}), and
(\ref{eq:AF f3f4,h4}), implementing anomaly-freedom: 
\begin{eqnarray}
    \frac{\partial (\bar{g} \bar{f}_3)}{\partial K_\varphi} &=&
    2 \bar{g} \bar{f}_3 \bar{h}_4
    \ , \label{eq:AF f3 - CML}
    \\
    \frac{\partial (\bar{g} \bar{f}_3^q)}{\partial K_\varphi} &=&
    2 \bar{g} \bar{f}_3^q \bar{h}_4
    \ .
    \label{eq:AF f3q - CML}
\end{eqnarray}
as well as
\begin{eqnarray}
     0 &=& 
    P_\phi \Bigg[- \bar{g}^2 \frac{\partial \bar{f}_1}{\partial K_\varphi} \left( 1 - \bar{g} \bar{f}_3^q \bar{g} \bar{f}_4^q \right)
    + 4 \bar{g} \bar{f}_3 \bar{g} \bar{f}_4
    - \bar{g} \bar{f}_1 \bar{g} \bar{h}_4
    \Bigg]
    \ , \label{eq:AF f3q,f4q - CML}
    \\
     0 &=& P_\phi \Bigg[\bar{g} \bar{f}_3 \bar{g} \bar{f}_4^q
    + \bar{g}\bar{f}_3^q \bar{g} \bar{f}_4
    \Bigg]
    \ , \label{eq:AF f3,f4q - CML}
    \\
    P_\phi \Bigg[ \bar{g} \frac{\partial (\bar{g} \bar{h}_4)}{\partial K_\varphi} \Bigg]&=&
    P_\phi \Bigg[ \bar{g} \bar{h}_4 \left( \bar{g} \bar{h}_4
    + \frac{1}{2} \frac{\partial \bar{g}}{\partial K_\varphi} \right)
    + \frac{1}{2} \bar{g} \frac{\partial^2 \ln \bar{g}}{\partial K_\varphi^2} \left( 1 -\bar{f}_3^q \bar{g}\bar{f}_4^q \right)
    \Bigg]
    \ .
    \label{eq:AF h4 - CML}
\end{eqnarray}

One can solve (\ref{eq:AF f4q}) and (\ref{eq:AF f3q - CML}) for $\bar{f}_4^q$
and $\bar{f}_3^q$ in terms of $\bar{h}_4$, substitute in (\ref{eq:AF h4 -
  CML}), and solve for $\bar{h}_4$, which has the rather lengthy solution
\begin{eqnarray}
    \bar{h}_4 &=&
    q \bar{\lambda}^2 \Bigg(
    c_f \cos^2 (\bar{\lambda} K_\varphi))
    - 2 q \bar{\lambda}^2 \frac{\sin (2 \bar{\lambda} K_\varphi)}{2 \bar{\lambda}}
    \nonumber\\
    &&
    + c_{h4} q \bar{\lambda}^2 |\cos (\lambda (K_\varphi + \mu_\varphi))|
    \sqrt{c_f \cos^2 (\bar{\lambda} K_\varphi)
    - 2 q \bar{\lambda}^2 \frac{\sin (2 \bar{\lambda} K_\varphi)}{2 \bar{\lambda}}}
    \Bigg)^{-1}
\end{eqnarray}
where $c_{h4}$ is an undetermined function of $E^x$.  Upon substitution in
(\ref{eq:AF f3,f4q - CML}) and solving for $\bar{f}_3$ and $\bar{f}_4$ using
(\ref{eq:AF f4}) and (\ref{eq:AF f3 - CML}), and substituting all the results in
(\ref{eq:AF f3q,f4q - CML}), consistency forces us to take the limit
$c_{h4} \to \infty$, that is, the function involved must have the form
\begin{eqnarray}
    \bar{h}_4 &=& \bar{f}_3  = \bar{f}_4 =  0
    \ , \\
    \bar{g} \bar{f}_3^q &=& - \frac{\alpha_3}{E^x}
    \ , \\
    \bar{g} \bar{f}_4^q &=& - \frac{E^x}{\alpha_3}
    \ ,
\end{eqnarray}
where $\alpha_3$ is an undetermined function of $E^x$ with classical limit $\alpha_3\to 1$.

The remaining equations for anomaly-freedom then simplify to
\begin{eqnarray}
    \bar{g} \frac{\partial (\bar{g} \bar{f}_0)}{\partial K_\varphi} &=&
    \bar{g} \bar{h}_0 \bar{g} \bar{h}_3
    - 2 \bar{g} \bar{f}_1 \bar{g} \bar{f}_2
    - \bar{g} \bar{f}_1  \frac{\partial \bar{g}}{\partial E^x}
    + \bar{g} \frac{\partial (\bar{g} \bar{f}_1)}{\partial E^x}
    + \frac{1}{4} \bar{g}^2 \frac{\partial \bar{f}_1}{\partial K_\varphi} \bar{g} \bar{h}_0^q \bar{g} \bar{h}_3^q
    \\
    0 &=&
    \bar{h}_0^q \bar{h}_3
    + \bar{h}_0 \bar{h}_3^q
    \ , \label{eq:AF f2q - CML}
    \\
    \frac{\partial \bar{f}_2}{\partial K_\varphi} &=&
    - \frac{1}{2} \frac{\partial}{\partial K_\varphi} \frac{\partial \ln \bar{g}}{\partial E^x}
    + \bar{h}_1 \bar{h}_3
    - \frac{1}{8} \frac{\partial^2 \ln \bar{g}}{\partial K_\varphi^2} \bar{g} \bar{h}_0^q \bar{g} \bar{h}_3^q
    \ , \label{eq:AF f2 - CML}
    \\
    \bar{h}_1 \bar{h}_3^q &=&
    0
    \ , \label{eq:AF f2q2 - CML}
\end{eqnarray}
as well as
\begin{eqnarray}
    P_\phi \Bigg[ \bar{g} \frac{\partial (\bar{g} \bar{h}_0)}{\partial K_\varphi}\Bigg] &=& 
     P_\phi \Bigg[ - 2 \bar{g} \bar{f}_1 \bar{g} \bar{h}_1
    + \frac{1}{2} \bar{g}^2 \frac{\partial \bar{f}_1}{\partial K_\varphi} \bar{g} \bar{f}_3^q \bar{g} \bar{h}_3^q
    \Bigg]
    \ , \label{eq:AF h0 - CML}
    \\
    P_\phi \Bigg[ \bar{g} \frac{\partial (\bar{g} \bar{h}_0^q)}{\partial K_\varphi}\Bigg] &=&
     P_\phi \Bigg[ 2 \bar{g} \bar{f}_3^q \bar{g} \bar{h}_3
    \Bigg]
    \ , \label{eq:AF h3 - CML}
    \\
    P_\phi \Bigg[ \frac{\partial \bar{h}_1}{\partial K_\varphi}\Bigg] &=& 
     P_\phi \Bigg[ - \frac{1}{4} \frac{\partial^2 \ln \bar{g}}{\partial K_\varphi^2} \bar{g} \bar{f}_3^q \bar{g} \bar{h}_3^q
    \Bigg]
    \ , \label{eq:AF h1 - CML}
\end{eqnarray}
and
\begin{eqnarray}
    \bar{h}_3 &=& \frac{1}{2} \frac{\partial \ln \bar{f}_1}{\partial K_\varphi} \bar{g} \bar{h}_0^q \bar{g} \bar{f}_4^q
    \ , \label{eq:AF f1/phi - CML}
    \\
    2 \bar{h}_0 \bar{f}_4^q &=&
    \bar{f}_1 \bar{h}_3^q
    \ , \label{eq:AF h0,h3q - CML}
    \\
    \bar{g} \frac{\partial (\bar{g} \bar{h}_3)}{\partial K_\varphi} &=&
    \bar{g} \bar{h}_3 \frac{1}{2} \frac{\partial \bar{g}}{\partial K_\varphi}
    - \frac{1}{4} \bar{g}^2 \frac{\partial^2 \ln \bar{g}}{\partial K_\varphi^2} \bar{g} \bar{h}_0^q \bar{g} \bar{f}_4^q
    \ , \label{eq:AF g/phi - CML}
    \\
    \frac{\partial (\bar{g} \bar{h}_3^q)}{\partial K_\varphi} &=&
    \bar{g} \bar{h}_3^q \frac{1}{2} \frac{\partial \ln \bar{g}}{\partial K_\varphi}
    + 2 \bar{h}_1 \bar{g} \bar{f}_4^q
    \ , \label{eq:AF h3q - CML}
\end{eqnarray}

Combining equations (\ref{eq:AF f2q2 - CML}), (\ref{eq:AF h1 - CML}), and
(\ref{eq:AF h3q - CML}),  we conclude that
\begin{eqnarray}
    \bar{h}_1 = \bar{h}_3^q = 0
    \ ,
\end{eqnarray}
such that these three equations are now satisfied.  Using these results,
equations (\ref{eq:AF f2q - CML}), (\ref{eq:AF h3 - CML}), (\ref{eq:AF f1/phi
  - CML})--(\ref{eq:AF g/phi - CML}) can all be solved, concluding that
\begin{eqnarray}
    \bar{h}_0^q = \bar{h}_0 = \bar{h}_3 = 0
    \,.
\end{eqnarray}

The remaining equations for anomaly-freedom now greatly simplify to
\begin{eqnarray}
    \bar{g} \frac{\partial (\bar{g} \bar{f}_0)}{\partial K_\varphi} &=&
    - 2 \bar{g} \bar{f}_1 \bar{g} \bar{f}_2
    - \bar{g} \bar{f}_1  \frac{\partial \bar{g}}{\partial E^x}
    + \bar{g} \frac{\partial (\bar{g} \bar{f}_1)}{\partial E^x}
    \\
    \frac{\partial \bar{f}_2}{\partial K_\varphi} &=&
    - \frac{1}{2} \frac{\partial}{\partial K_\varphi} \frac{\partial \ln \bar{g}}{\partial E^x}
    \ ,
\end{eqnarray}
with the general solutions
\begin{eqnarray}
    \bar{g} \bar{f}_0 &=&
    \lambda_0 \left( - \Lambda_0 + \frac{\alpha_{0}}{E^x}
    + \frac{\sin^2 (\bar{\lambda} K_\varphi)}{\bar{\lambda}^2} \left( c_f \frac{\alpha_2}{E^x} + 2 \frac{\partial c_f}{\partial E^x}\right)
    + 2 \bar{\lambda}^2 \frac{\sin (2\bar{\lambda} K_\varphi)}{2 \bar{\lambda}} \left( q \frac{\alpha_2}{E^x} + 2 \frac{\partial q}{\partial E^x}\right)
    \right)
    \ , \nonumber\\
    \bar{g} \bar{f}_2 &=&
    - \frac{\alpha_2}{4 E^x} \lambda_0 \cos^2 (\bar{\lambda} K_\varphi)
    \ ,
\end{eqnarray}
where $\Lambda_0$, $\alpha_i$, and $\alpha_{iq}$ are undetermined functions of
$E^x$.  This exhausts all the anomaly-freedom equations.  Here, $\Lambda_0$
and $\alpha_0$, are not independent functions, but we keep them separate
because of their physical significance in the classical limit, which will be
explained below.

The general Hamiltonian constraint obtained from the assumed conditions is
\begin{eqnarray}
    H &=&
    - \lambda_0 \frac{\sqrt{E^x}}{2} \Bigg[
    E^\varphi \left(
    - \Lambda_0
    + \frac{\alpha_0}{E^x}
    + \left( c_f \frac{\alpha_2}{E^x} + 2 \frac{\partial c_{f}}{\partial E^x} \right) \frac{\sin^2 \left(\bar{\lambda} K_\varphi\right)}{\bar{\lambda}^2} \right)
    \nonumber\\
    &&\qquad
    + 2 E^\varphi \left( q \frac{\alpha_2}{E^x} + 2 \frac{\partial q}{\partial E^x} \right) \frac{\sin \left(2 \bar{\lambda} K_\varphi\right)}{2 \bar{\lambda}}
    + 4 K_x \left(c_f \frac{\sin (2 \bar{\lambda} K_\varphi)}{2 \bar{\lambda}}
    + q \cos(2 \bar{\lambda} K_\varphi)\right)
    \nonumber\\
    &&\qquad
    + \frac{((E^x)')^2}{E^\varphi} \left( - \frac{\alpha_2}{4 E^x} \cos^2 (\bar{\lambda} K_{\varphi})
    + \bar{\lambda}^2 \frac{K_x}{E^\varphi} \frac{\sin (2 \bar{\lambda}
       K_\varphi)}{2 \bar{\lambda}} \right)\nonumber\\
  &&\qquad
    + \left( \frac{(E^x)' (E^\varphi)'}{(E^\varphi)^2} - \frac{(E^x)''}{E^\varphi} \right) \cos^2 (\bar{\lambda} K_\varphi)
    \Bigg]
    \nonumber\\
    &&
    + \frac{E^\varphi}{2} \sqrt{q^{xx}} \left[
    \frac{P_\phi{}^2}{E^\varphi} \frac{\alpha_3}{E^x}
    + \frac{(\phi')^2}{E^\varphi} \frac{E^x}{\alpha_3}
    \right]
    + \lambda_0^2 \frac{E^x}{2} \sqrt{q_{xx}} V_q
    \ ,
    \label{eq:Hamiltonian constraint - CML}
\end{eqnarray}
with structure function 
\begin{eqnarray}
    q^{x x}
    =
    \left(
    \left( c_{f}
    + \left(\frac{\bar{\lambda} (E^x)'}{2 E^\varphi} \right)^2 \right) \cos^2 \left(\bar{\lambda} K_\varphi\right)
    - 2 q \bar{\lambda}^2 \frac{\sin \left(2 \bar{\lambda} K_\varphi\right)}{2 \bar{\lambda}}\right)
    \lambda_0^2 \frac{E^x}{(E^\varphi)^2}
    \ .
    \label{eq:Structure function - periodic - general}
\end{eqnarray}
All free functions, except for the constant $\bar{\lambda}$, may depend on
$E^x$, and only $\lambda_0$ may also depend on $\phi$.  We will discuss the
different classical limits below.

\subsubsection{Recovery of a non-constant holonomy parameter}

We have used canonical transformations in order to restrict the dependence of
the Hamiltonian constraint and make it more manageable, in particular by
setting $\bar{\lambda}$ equal to a constant. By undoing some of the canonical
transformations, it is possible to replace $\bar{\lambda}$ with a function, at
the expense of introducing additional terms in the constraint.

Our discussion of the symmetry generator implies that a non-constant holonomy
parameter $\lambda$ replacing $\bar{\lambda}$ in (\ref{eq:Hamiltonian
  constraint - CML}) cannot depend on $\phi$, but it may depend on $E^x$.  In
order to recover such non-constant holonomy effects, we simply have to invert
some of our canonical transformations and redefine the rest of the parameters
accordingly. Redefining
\begin{eqnarray}
    &&\lambda_0 \to \lambda_0 \frac{\bar{\lambda}}{\lambda}
    \ ,\ 
    q \to q \frac{\lambda}{\bar{\lambda}}
    \ , \
    \Lambda_0 \to \Lambda_0 \frac{\lambda^2}{\bar{\lambda}^2}
    \ , \
    \alpha_0 \to \alpha_0 \frac{\lambda^2}{\bar{\lambda}^2}
    \ , \nonumber\\
    &&
    \alpha_2 \to \alpha_2 - 4 E^x \frac{\partial \ln \lambda}{\partial E^x}
    \ , \
    V_q \to V_q \frac{\lambda^2}{\bar{\lambda}^2}
    \label{eq:Redefinition for non-constant holonomy - CML}
\end{eqnarray}
implies that the general constraint and structure function now resemble
(\ref{eq:Covariant Hamiltonian - Linear combination - periodic variables}) and
(\ref{eq:Structure function - Linear combination - periodic variables}):
\begin{eqnarray}
    H &=&
    - \frac{\bar{\lambda}}{\lambda} \lambda_0 \frac{\sqrt{E^x}}{2} \Bigg[
    E^\varphi \left(
    \frac{\lambda^2}{\bar{\lambda}^2} \left( - \Lambda_0
    + \frac{\alpha_0}{E^x} \right)
    + \left( c_f \left(\frac{\alpha_2}{E^x} - 4 \frac{\partial \ln \lambda}{\partial E^x}\right) + 2 \frac{\partial c_{f}}{\partial E^x} \right) \frac{\sin^2 \left(\bar{\lambda} K_\varphi\right)}{\bar{\lambda}^2} \right)
    \nonumber\\
    &&\qquad
    + 2 E^\varphi \left( q \left(\frac{\alpha_2}{E^x} - 2 \frac{\partial \ln \lambda}{\partial E^x}\right) + 2 \frac{\lambda}{\bar{\lambda}} \frac{\partial q}{\partial E^x} \right) \frac{\sin \left(2 \bar{\lambda} K_\varphi\right)}{2 \bar{\lambda}}\nonumber\\
&& \qquad   + 4 K_x \left(c_f \frac{\sin (2 \bar{\lambda} K_\varphi)}{2 \bar{\lambda}}
    + \frac{\lambda}{\bar{\lambda}} q \cos(2 \bar{\lambda} K_\varphi)\right)
    \nonumber\\
    &&\qquad
    + \frac{((E^x)')^2}{E^\varphi} \left( \left(\frac{\partial \ln \lambda}{\partial E^x}-\frac{\alpha_2}{4 E^x}\right) \cos^2 (\bar{\lambda} K_{\varphi})
    + \bar{\lambda}^2 \frac{K_x}{E^\varphi} \frac{\sin (2 \bar{\lambda} K_\varphi)}{2 \bar{\lambda}} \right)
    \nonumber\\
    &&\qquad
    + \left( \frac{(E^x)' (E^\varphi)'}{(E^\varphi)^2} - \frac{(E^x)''}{E^\varphi} \right) \cos^2 (\bar{\lambda} K_\varphi)
    \Bigg]\nonumber\\
  &&
    + \frac{E^\varphi}{2} \sqrt{q^{xx}} \left[
    \frac{P_\phi{}^2}{E^\varphi} \frac{\alpha_3}{E^x}
    + \frac{(\phi')^2}{E^\varphi} \frac{E^x}{\alpha_3}
       \right]    + \lambda_0^2 \frac{E^x}{2} \sqrt{q_{xx}} V_q
    \ ,
    \label{eq:Hamiltonian constraint - periodic - CML - nonconstant holonomy}
\end{eqnarray}
with structure function 
\begin{eqnarray}
    q^{x x}
    =
    \left(
    \left( c_{f}
    + \left(\frac{\bar{\lambda} (E^x)'}{2 E^\varphi} \right)^2 \right) \cos^2 \left(\bar{\lambda} K_\varphi\right)
    - 2 q \frac{\lambda}{\bar{\lambda}} \bar{\lambda}^2 \frac{\sin \left(2 \bar{\lambda} K_\varphi\right)}{2 \bar{\lambda}}\right)
    \frac{\bar{\lambda}^2}{\lambda^2} \lambda_0^2 \frac{E^x}{(E^\varphi)^2}
    \ .
    \label{eq:Structure function - periodic - general - nonconstant holonomy}
\end{eqnarray}

A canonical transformation of the form
(\ref{eq:Diffeomorphism-constraint-preserving canonical transformations -
  First residue}) with $f_x^\varphi=\lambda/\bar{\lambda}$,
$\tilde{\mu}_\varphi=0$, $f_c^\phi=\phi$ then eliminates all traces of
$\bar{\lambda}$:
\begin{eqnarray}
    H &=&
    - \lambda_0 \frac{\sqrt{E^x}}{2} \Bigg[
    E^\varphi \left(
    - \Lambda_0
    + \frac{\alpha_0}{E^x}
    + \left( c_f \left(\frac{\alpha_2}{E^x} - 4 \frac{\partial \ln \lambda}{\partial E^x}\right) + 2 \frac{\partial c_{f}}{\partial E^x} \right) \frac{\sin^2 \left(\lambda K_\varphi\right)}{\lambda^2} \right)
    \nonumber\\
    &&\qquad
    + 2 E^\varphi \left( q \left(\frac{\alpha_2}{E^x} - 2 \frac{\partial \ln \lambda}{\partial E^x}\right) + 2 \frac{\partial q}{\partial E^x} \right) \frac{\sin \left(2 \lambda K_\varphi\right)}{2 \lambda}
    \nonumber\\
    &&\qquad
    + 4 \left( K_x + E^\varphi K_\varphi \frac{\partial \ln \lambda}{\partial E^x} \right) \left(c_f \frac{\sin (2 \lambda K_\varphi)}{2 \lambda}
    + q \cos(2 \lambda K_\varphi)\right)
    \nonumber\\
    &&\qquad
    + \frac{((E^x)')^2}{E^\varphi} \left( - \frac{\alpha_2}{4 E^x} \cos^2 (\lambda K_{\varphi})
    + \lambda^2 \left(\frac{K_x}{E^\varphi} + K_\varphi \frac{\partial \ln \lambda}{\partial E^x} \right) \frac{\sin (2 \lambda K_\varphi)}{2 \lambda} \right)
    \nonumber\\
    &&\qquad
    + \left( \frac{(E^x)' (E^\varphi)'}{(E^\varphi)^2}
    - \frac{(E^x)''}{E^\varphi} \right) \cos^2 (\lambda K_\varphi)
       \Bigg]\nonumber\\
  &&
    + \frac{E^\varphi}{2} \sqrt{q^{xx}} \left[
    \frac{P_\phi{}^2}{E^\varphi} \frac{\alpha_3}{E^x}
    + \frac{(\phi')^2}{E^\varphi} \frac{E^x}{\alpha_3}
    \right]
    + \lambda_0^2 \frac{E^x}{2} \sqrt{q_{xx}} V_q
    \ ,
    \label{eq:Hamiltonian constraint - nonperiodic - CML - nonconstant holonomy}
\end{eqnarray}
and
\begin{eqnarray}
    q^{x x}
    =
    \left(
    \left( c_{f}
    + \left(\frac{\lambda (E^x)'}{2 E^\varphi} \right)^2 \right) \cos^2 \left(\lambda K_\varphi\right)
    - 2 q \lambda^2 \frac{\sin \left(2 \lambda K_\varphi\right)}{2 \lambda}\right) \lambda_0^2 \frac{E^x}{(E^\varphi)^2}
    \ .
    \label{eq:Structure function - nonperiodic - general - nonconstant holonomy}
\end{eqnarray}
now resemble (\ref{eq:Covariant Hamiltonian - Linear combination - holonomy
  variables}) and (\ref{eq:Structure function - Linear combination - holonomy
  variables}).  In these phase-space coordinates, the constraint is no
longer periodic in $K_\varphi$ (see the third and fourth lines of $H$), but
the classical limit is now direct.  This shows that we have properly taken
into account all effects of a non-constant $\lambda$.

\subsubsection{Polymerization of the scalar field}

The quantization strategy of loop quantum gravity requires a ``polymerization''
of the scalar field, usually done by replacing $\phi$ with
$\sin (\bar{\nu} \phi) / \bar{\nu}$ in the Hamiltonian constraint, where
$\bar{\nu}$ is a constant and the classical limit is obtained for
$\bar{\nu}\to 0$. Such a replacement might be performed in a version of the
constraint that is to be turned into an operator, in which case the
boundedness of $\sin (\bar{\nu} \phi)$ may be beneficial, or it could be used
as an effective constraint that is supposed to mimick some of the effects of
loop quantization in an analysis of classical type, revealing potential
space-time effects.

However, this replacement is not compatible with the general constraint
(\ref{eq:Hamiltonian constraint - nonperiodic - CML - nonconstant holonomy}),
where the classical $(\phi')^2$-term can only be multiplied by $E^x$-dependent
functions while a loop quantization would require a version of the form
$\sin(\bar{\nu}\phi)' = \bar{\nu} \cos(\bar{\nu}\phi) \phi'$ with a
$\phi$-dependent multiplier. This version of polymerization is therefore not a
covariant modification that preserves the classical-matter limit.

In fact, there is no room for any modification involving the scalar matter
field except for one undetermined function that can be used to this end: the
overall factor $\lambda_0$.  The remaining freedom in applying the
polymerization procedure is non-unique, but it can be further restricted and
completed by taking inspiration from how a polymerization of the gravitational
variable $K_\varphi$ emerges without the need of a canonical transformation.

Physically, polymerization of the scalar field should imply boundedness
effects from the field dependence since the field $\phi$ itself appears, by
definition of polymerization, as an argument of a trigonometric function.
More generally, we may want to allow polymerization to have an $E^x$-dependent
point-holonomy parameter $\nu (E^x)$, such that it is sensitive to distance
and energy scales and automatically implies the classical limit $\nu\to0$ for
large spherical areas $E^x$ if $\nu$ is a decreasing function. If possible, a
substitution of the form $\phi \to \sin (\nu \phi) / \nu$ is preferable
because it has been most commonly used, which is bounded by
$|\sin (\nu \phi) / \nu|<1/\nu$. Since the relationship between $\phi$ and
$\sin(\nu\phi)/\nu$ is not one-to-one, we have to limit the range of $\phi$
after a canonical transformation to polymerized form such that
$|\phi| \le 1/ \nu$ (in an $E^x$-dependent way) if the replacement
$\phi \to \sin (\nu \phi) / \nu$ is to be implemented by a well-defined canonical
transformation.

In order to have a dynamically stable range limited in this way, we compute
the evolution equations of the scalar field and require that
$\dot{\phi}|_{\phi \to 1/\nu}\to0$. There is then no evolution
transversal to the surface $\phi = 1 / \nu$ in phase space, and it is
consistent to assume that the value of the scalar field does not increase
beyond this limit. Geometrically, this condition means that whenever there is
a point or a region on where $\phi = 1 / \nu$, we must have
$\dot\phi=0$ at this place. As a specific case, we assume that this condition
is obtained in an extended spatial region that defines part of a hypersurface
of a canonical foliation. Since we need to limit only the normal component of
evolution on the hypersurface, we may assume $N^x=0$. The condition then leads
to the equations
\begin{eqnarray}
    \dot{\phi} |_{\phi \to 1/\nu}
    &=& 
    \left\{ \phi , H [N] \right\}
    \\
    &=& 
    \frac{\bar{\lambda}}{\lambda} \lambda_0 N \sqrt{\left( c_{f}
    + \left(\frac{\bar{\lambda} (E^x)'}{2 E^\varphi} \right)^2 \right) \cos^2 \left(\bar{\lambda} K_\varphi\right)
    - 2 q \frac{\lambda}{\bar{\lambda}} \bar{\lambda}^2 \frac{\sin \left(2 \bar{\lambda} K_\varphi\right)}{2 \bar{\lambda}}} \frac{P_\phi}{E^\varphi} \frac{\alpha_3}{\sqrt{E^x}}
    \ ,\nonumber \\
    \dot{P}_\phi |_{\phi \to 1/\nu}
    &=& 
    \frac{\bar{\lambda}\lambda_0}{2\lambda}
        \frac{N\sqrt{E^x} E^\varphi\partial V_q/\partial \phi}{\sqrt{\left( c_{f} 
    + \left(\bar{\lambda} (E^x)'/(2 E^\varphi) \right)^2 \right) \cos^2 \left(\bar{\lambda} K_\varphi\right)
    - 2 q \lambda \bar{\lambda} \sin \left(2
        \bar{\lambda} K_\varphi\right)/(2 \bar{\lambda})}} 
\end{eqnarray}
using (\ref{eq:Hamiltonian constraint - periodic - CML - nonconstant
  holonomy}), the second equation at spatial points where $\phi'=0$ according
to our assumption that the maximum $\phi$ is reached in a subset of a
hypersurface where, if it is sufficiently small, $E^x$ and therefore $\nu$ can
be assumed to be nearly constant. In $\dot{P}_\phi$, we omitted the term
implied by $\partial \lambda_0 / \partial \phi$ because it vanishes on-shell.

We need both expressions to vanish because $\lambda_0$ is always positive.
The sign of $P_\phi$ must therefore change in order to start decreasing the
value of $\phi$ past the hypersurface.  The solution to this problem is non-unique
because we are not restricting the rate at which $P_{\phi}$ approaches
zero. Based on how boundedness comes about in the gravitational case where the
limiting value of $\bar{\lambda}K_{\varphi}$ implies a similar transition
hypersurface studied for instance in \cite{SphSymmEff,SphSymmEff2}, we
redefine the overall factor by
\begin{eqnarray}
    \lambda_0 (E^x,\phi) \to \lambda_0 (E^x,\phi) \left(1 - \nu^2 \phi^2\right)
    \ ,
    \label{eq:Global factor boundedness redefinition}
\end{eqnarray}
where the residual dependence of the redefined $\lambda_0$ on $E^x$ and $\phi$
is required to be non-zero if $\phi=1/\nu$ so as not to interfere with the
bound.

After this preparation, we perform a canonical transformation of the form
(\ref{eq:Residual canonical transformation phi,ex}) with
$f_c^\phi=\sin (\nu \phi) / \nu$.  This transformation turns the right-hand
side of the redefinition (\ref{eq:Global factor boundedness redefinition})
into $\lambda_0^2 \cos^2 (\nu \phi)$, and the Hamiltonian constraint
(\ref{eq:Hamiltonian constraint - periodic - CML - nonconstant holonomy})
into
\begin{eqnarray}
    H &=&
    - \frac{\bar{\lambda}}{\lambda} \lambda_0 \cos^2 (\nu \phi) \frac{\sqrt{E^x}}{2} \Bigg[
    E^\varphi \left(
    \frac{\lambda^2}{\bar{\lambda}^2} \left( - \Lambda_0
          + \frac{\alpha_0}{E^x} \right)\right.\nonumber\\
  &&\qquad \left.
    + \left( c_f \left(\frac{\alpha_2}{E^x} - 4 \frac{\partial \ln \lambda}{\partial E^x}\right) + 2 \frac{\partial c_{f}}{\partial E^x} \right) \frac{\sin^2 \left(\bar{\lambda} K_\varphi\right)}{\bar{\lambda}^2} \right)
    \nonumber\\
    &&\qquad
    + 2 E^\varphi \left( q \left(\frac{\alpha_2}{E^x} - 2 \frac{\partial \ln \lambda}{\partial E^x}\right) + 2 \frac{\lambda}{\bar{\lambda}} \frac{\partial q}{\partial E^x} \right) \frac{\sin \left(2 \bar{\lambda} K_\varphi\right)}{2 \bar{\lambda}}
    \nonumber\\
    &&\qquad
    + 4 \left(K_x + P_\phi \left( \phi - \frac{\tan (\nu \phi)}{\nu} \right) \frac{\partial \ln \nu}{\partial E^x}\right) \left(c_f \frac{\sin (2 \bar{\lambda} K_\varphi)}{2 \bar{\lambda}}
    + \frac{\lambda}{\bar{\lambda}} q \cos(2 \bar{\lambda} K_\varphi)\right)
    \nonumber\\
    &&\qquad
    + \frac{((E^x)')^2}{E^\varphi} \left( \left(\frac{\partial \ln
       \lambda}{\partial E^x}-\frac{\alpha_2}{4 E^x}\right) \cos^2
       (\bar{\lambda} K_{\varphi})\right.\nonumber\\
  &&\qquad \left.
    + \bar{\lambda}^2 \left(\frac{K_x}{E^\varphi} + \frac{P_\phi}{E^\varphi} \left( \phi - \frac{\tan (\nu \phi)}{\nu} \right) \frac{\partial \ln \nu}{\partial E^x}\right) \frac{\sin (2 \bar{\lambda} K_\varphi)}{2 \bar{\lambda}} \right)
    \nonumber\\
    &&\qquad
    + \left( \frac{(E^x)' (E^\varphi)'}{(E^\varphi)^2} - \frac{(E^x)''}{E^\varphi} \right) \cos^2 (\bar{\lambda} K_\varphi)
    \Bigg]
    \nonumber\\
    &&
    + \frac{E^\varphi}{2} \sqrt{q^{xx}} \left[
    \frac{P_\phi{}^2}{E^\varphi \cos^2 (\nu \phi)} \frac{\alpha_3}{E^x}
     \right.  \nonumber\\
  &&\qquad \left.
    + \frac{1}{E^\varphi} \frac{E^x}{\alpha_3} \left( \phi' \cos (\nu \phi)
    + \left(\phi \cos (\nu \phi)
    - \frac{\sin (\nu \phi)}{\nu}\right) \frac{\partial\ln \nu}{\partial E^x} (E^x)' \right)^2
    \right]
    \nonumber\\
    &&
    + \lambda_0^2 \cos^4 (\nu \phi) \frac{E^x}{2} \sqrt{q_{xx}} V_q
    \ ,
    \label{eq:Hamiltonian constraint - CML - scalar polymerization}
\end{eqnarray}
with structure function
\begin{eqnarray}
    q^{x x}
    =
    \left(
    \left( c_{f}
    + \left(\frac{\bar{\lambda} (E^x)'}{2 E^\varphi} \right)^2 \right) \cos^2 \left(\bar{\lambda} K_\varphi\right)
    - 2 q \frac{\lambda}{\bar{\lambda}} \bar{\lambda}^2 \frac{\sin \left(2 \bar{\lambda} K_\varphi\right)}{2 \bar{\lambda}}\right)
    \frac{\bar{\lambda}^2}{\lambda^2}
    \lambda_0^2 \cos^4 (\nu \phi) \frac{E^x}{(E^\varphi)^2}
    \ .
    \label{eq:Structure function - periodic - general - Scalar field polymerization}
\end{eqnarray}

For non-constant $\nu$, this constraint is not periodic in $\phi$.  Performing
a second canonical transformation of the form (\ref{eq:Residual canonical
  transformation phi,ex}) with $f_c^\phi = (\bar{\nu}/\nu) \phi$ and a
constant $\bar{\nu}$, the constraint is rendered periodic,
\begin{eqnarray}
    H &=&
    - \frac{\bar{\lambda}}{\lambda} \lambda_0 \cos^2 (\bar{\nu} \phi) \frac{\sqrt{E^x}}{2} \Bigg[
    E^\varphi \left(
    \frac{\lambda^2}{\bar{\lambda}^2} \left( - \Lambda_0
          + \frac{\alpha_0}{E^x} \right)\right.\nonumber\\
  &&\left.\qquad
    + \left( c_f \left(\frac{\alpha_2}{E^x} - 4 \frac{\partial \ln \lambda}{\partial E^x}\right) + 2 \frac{\partial c_{f}}{\partial E^x} \right) \frac{\sin^2 \left(\bar{\lambda} K_\varphi\right)}{\bar{\lambda}^2} \right)
    \nonumber\\
    && \qquad
    + 2 E^\varphi \left( q \left(\frac{\alpha_2}{E^x} - 2 \frac{\partial \ln \lambda}{\partial E^x}\right) + 2 \frac{\lambda}{\bar{\lambda}} \frac{\partial q}{\partial E^x} \right) \frac{\sin \left(2 \bar{\lambda} K_\varphi\right)}{2 \bar{\lambda}}
    \nonumber\\
    &&\qquad
    + 4 \left( K_x  - P_\phi\frac{\tan (\bar{\nu} \phi)}{\bar{\nu}} \frac{\partial \ln \nu}{\partial E^x}\right) \left(c_f \frac{\sin (2 \bar{\lambda} K_\varphi)}{2 \bar{\lambda}}
    + \frac{\lambda}{\bar{\lambda}} q \cos(2 \bar{\lambda} K_\varphi)\right)
    \nonumber\\
    &&\qquad
    + \frac{((E^x)')^2}{E^\varphi} \left( \left(\frac{\partial \ln
       \lambda}{\partial E^x}-\frac{\alpha_2}{4 E^x}\right) \cos^2
       (\bar{\lambda} K_{\varphi})\right.\nonumber\\
  &&\left.\qquad
    + \bar{\lambda}^2 \left(\frac{K_x}{E^\varphi} - \frac{P_\phi}{E^\varphi} \frac{\tan (\bar{\nu} \phi)}{\bar{\nu}} \frac{\partial \ln \nu}{\partial E^x}\right) \frac{\sin (2 \bar{\lambda} K_\varphi)}{2 \bar{\lambda}} \right)
    \nonumber\\
    &&\qquad
    + \left( \frac{(E^x)' (E^\varphi)'}{(E^\varphi)^2} - \frac{(E^x)''}{E^\varphi} \right) \cos^2 (\bar{\lambda} K_\varphi)
    \Bigg]
    \nonumber\\
    &&
    + \frac{\bar{\nu}^2}{\nu^2} \frac{\sqrt{q^{xx}}}{2} \left[
    \frac{P_\phi{}^2}{\cos^2 (\bar{\nu} \phi)} \frac{\alpha_3}{E^x}
    + \frac{E^x}{\alpha_3} \left( \left(\frac{\sin (\bar{\nu} \phi)}{\bar{\nu}}\right)'
    - \frac{\sin (\bar{\nu} \phi)}{\bar{\nu}} \frac{\partial\ln \nu}{\partial E^x} (E^x)' \right)^2
       \right] \nonumber\\
  &&
    + \lambda_0^2 \lambda_0^2 \cos^4 (\bar{\nu} \phi) \frac{E^x}{2} \sqrt{q_{xx}} V_q
    \ ,
    \label{eq:Hamiltonian constraint - CML - scalar polymerization - periodic}
\end{eqnarray}
and the structure function  becomes
\begin{eqnarray}
    q^{x x}
    =
    \left(
    \left( c_{f}
    + \left(\frac{\bar{\lambda} (E^x)'}{2 E^\varphi} \right)^2 \right) \cos^2 \left(\bar{\lambda} K_\varphi\right)
    - 2 q \frac{\lambda}{\bar{\lambda}} \bar{\lambda}^2 \frac{\sin \left(2 \bar{\lambda} K_\varphi\right)}{2 \bar{\lambda}}\right)
    \frac{\bar{\lambda}^2}{\lambda^2}
    \lambda_0^2 \cos^4 (\bar{\nu} \phi) \frac{E^x}{(E^\varphi)^2}
    \ .
    \label{eq:Structure function - periodic - general - Scalar field polymerization - periodic}
\end{eqnarray}

We note that effects implied by boundedness of the scalar field in the
polymerized constraint (\ref{eq:Hamiltonian constraint - CML - scalar
  polymerization}) are not due to the canonical transformations, which cannot
change physical implications, but rather a consequence of the non-classical
overall factor $\lambda_0$ and its $\phi$-dependence. The result has two
general implications of importance for discussions of polymerization in models
of loop quantum gravity. First, while the $P_{\phi}^2$-term and the new
$(\phi')^2$-term may look like something one may have chosen with standard
polymerization, there are additional terms in the consistent Hamiltonian
constraint depending on $\phi$ and $P_{\phi}$. In particular, there is a
coupling term between $\phi$ and the spatial derivative $(E^x)'$ in the last
line, as well as a terms linear in $P_{\phi}$ in the third and fourth lines.
Such terms are not part of standard polymerization procedures.

Secondly, the
structure function necessarily depends on the scalar field even at the
kinematical level, after redefining the overall factor in order to comply with
a bounded $\phi$-dependent function in the constraint.  This result is
physically meaningful only within our new viewpoint of emergent modified
gravity, in which space-time geometry is not described directly by a
fundamental field, but rather an emergent object composed of the truly
fundamental fields, in this case both the gravitational degree of freedom
\emph{and} the scalar matter field.

The Hamiltonian constraint (\ref{eq:Hamiltonian constraint - CML - scalar
  polymerization - periodic}) is then periodic in both $K_\varphi$ and $\phi$
as a modified constraint with holonomy or polymerization effects. It includes
the option of polymerization functions with non-constant parameters $\lambda$
and $\nu$ upon using canonical transformations.  The vacuum mass observable
associated to (\ref{eq:Hamiltonian constraint - CML - scalar polymerization -
  periodic}) is given by
\begin{eqnarray}
    \mathcal{M}
    &=&
    d_0
    + \frac{d_2}{2} \left(\exp \int {\rm d} E^x \ \left(\frac{\alpha_2}{2 E^x} - \frac{\partial \ln \lambda^2}{\partial E^x}\right)\right)\notag\\
&&\qquad\quad\times    \left(
    \frac{\sin^2\left(\bar{\lambda} K_{\varphi}\right)}{\bar{\lambda}^2}
    + 2 \frac{\lambda}{\bar{\lambda}} q \frac{\sin \left(2 \bar{\lambda}  K_{\varphi}\right)}{2 \bar{\lambda}}
    - \cos^2 (\bar{\lambda} K_\varphi) \left(\frac{(E^x)'}{2 E^\varphi}\right)^2
    \right)
    \notag\\
    &&
    + \frac{d_2}{4} \int {\rm d} E^x \ \left(\frac{\lambda^2}{\bar{\lambda}^2}\left( \Lambda_0
    + \frac{\alpha_0}{E^x} \right) \exp \int {\rm d} E^x \ \left(\frac{\alpha_2}{2 E^x} - \frac{\partial \ln \lambda^2}{\partial E^x}\right)\right)
    \ ,
    \label{eq:Gravitational weak observable - CML - scalar polymerization - periodic}
\end{eqnarray}
and, when $V=V_q=V^q=0$, its scalar-field observable by
\begin{eqnarray}
    G [\alpha] &=& \int {\rm d}^3 x\ \alpha \frac{\nu}{\bar{\nu}} \frac{P_\phi}{\cos(\bar{\nu}\phi)}
    \label{eq:Scalar field symmetry generator - CML - scalar polymerization - periodic}
\end{eqnarray}
where $\alpha$, $d_0$, and $d_2$ are constants.
The associated conserved matter current $J^\mu$ has the components
\begin{eqnarray}
    J^t &=& \frac{\nu}{\bar{\nu}} \frac{P_\phi}{\cos(\bar{\nu}\phi)}
    \ , \\
    J^x &=& \frac{\partial G}{\partial P_\phi} \frac{\partial H}{\partial \phi'}
    = \frac{\bar{\nu}}{\nu} \sqrt{q^{xx}}
    \frac{E^x}{\alpha_3} \left( \left(\frac{\sin (\bar{\nu} \phi)}{\bar{\nu}}\right)'
    - (E^x)' \frac{\sin (\bar{\nu} \phi)}{\bar{\nu}} \frac{\partial\ln \nu}{\partial E^x} \right)
    \ .
    \label{eq:Conserved matter current - CML}
\end{eqnarray}

\subsubsection{Partial Abelianization}

The partial Abelianization of the constraint (\ref{eq:Hamiltonian constraint - CML - scalar polymerization - periodic}) is easily achieved by using the coefficients (\ref{eq:Abelianization coefficients}) under the redefinitions (\ref{eq:Redefinition for non-constant holonomy - CML}) and (\ref{eq:Global factor boundedness redefinition}) and taking $q=0$ according to the condition (\ref{eq:Partial Abelianization condition on q}).
The resulting Abelianized constraint is given by
\begin{eqnarray}
    \frac{H^{(A)}}{\bar{B} c_f} &=& - \frac{\bar{\lambda}}{\lambda} \lambda_0 \cos^2 (\Bar{\nu} \phi) \frac{\sqrt{E^x}}{2} \frac{\tan (\bar{\lambda} K_\varphi)}{\bar{\lambda}} \Bigg[
    E^\varphi \left(
    \frac{\lambda^2}{\bar{\lambda}^2} \left( - \Lambda_0  +
                                    \frac{\alpha_0}{E^x}
                                    \right)\right. \nonumber\\ 
&&\qquad  \left.
    + \left( c_f \left(\frac{\alpha_2}{E^x} - 4 \frac{\partial \ln \lambda}{\partial E^x}\right) + 2 \frac{\partial c_{f}}{\partial E^x} \right) \frac{\sin^2 \left(\bar{\lambda} K_\varphi\right)}{\bar{\lambda}^2} \right)
    \nonumber\\
    && \qquad
    + 4 \left( K_x  - P_\phi\frac{\tan (\bar{\nu} \phi)}{\bar{\nu}} \frac{\partial \ln \nu}{\partial E^x}\right) c_f \frac{\sin (2 \Bar{\lambda} K_\varphi)}{2 \Bar{\lambda}}
    \nonumber\\
    &&\qquad
    + \frac{((E^x)')^2}{E^\varphi} \left( \left(\frac{\partial \ln \lambda}{\partial E^x}-\frac{\alpha_2}{4 E^x}\right) \cos^2 (\Bar{\lambda} K_{\varphi})\right.\nonumber\\
&&\qquad\left.    + \Bar{\lambda}^2 \left(\frac{K_x}{E^\varphi} - \frac{P_\phi}{E^\varphi} \frac{\tan (\bar{\nu} \phi)}{\bar{\nu}} \frac{\partial \ln \nu}{\partial E^x}\right) \frac{\sin (2 \Bar{\lambda} K_\varphi)}{2 \Bar{\lambda}} \right)
    \nonumber\\
    && \qquad
    + \left( \frac{(E^x)' (E^\varphi)'}{(E^\varphi)^2} - \frac{(E^x)''}{E^\varphi} \right) \cos^2 (\Bar{\lambda} K_\varphi)
    \Bigg]
    \nonumber\\
    &&
    - \frac{\bar{\lambda}}{\lambda} \lambda_0 \cos^2 (\Bar{\nu} \phi) \frac{\sqrt{E^x}}{2} \frac{(E^x)'}{E^\varphi} \left( K_\varphi' - \frac{K_x}{E^\varphi} (E^x)'
    + \frac{P_\phi}{E^\varphi} \phi'\right)
    \nonumber\\
    &&
    + \frac{\tan (\bar{\lambda} K_\varphi)}{\bar{\lambda}} \frac{\bar{\nu}^2}{\nu^2} \frac{\sqrt{q^{xx}}}{2} \left(
    \frac{P_\phi{}^2}{\cos^2 (\bar{\nu} \phi)} \frac{\alpha_3}{E^x}
    + \frac{E^x}{\alpha_3} \left( \left(\frac{\sin (\bar{\nu} \phi)}{\bar{\nu}}\right)'
    - \frac{\sin (\bar{\nu} \phi)}{\bar{\nu}} \frac{\partial\ln \nu}{\partial E^x} (E^x)' \right)^2
    \right)
    \nonumber\\
    &&
    + \frac{\tan (\bar{\lambda} K_\varphi)}{\bar{\lambda}} \frac{\bar{\lambda}^2}{\lambda^2} \lambda_0^2 \cos^4 (\Bar{\nu} \phi) \frac{E^x}{2} \sqrt{q_{xx}} V_q
    \ .
    \label{eq:Abelian CML constraint}
\end{eqnarray}
This Abelian constraint has kinematical divergences at
$K_\varphi= \pm \pi / (2 \bar{\lambda})$ in from the first line and last line.
The latter can easily be resolved by simply restricting the constraint to the
free scalar case, $V_q=0$, while the divergence of the first line can be
eliminated if the equation
\begin{equation}
    \lambda^2\left( - \Lambda_0
    + \frac{\alpha_0}{E^x} \right)
    + 2 \frac{\partial c_f}{\partial E^x}
    + \left(\frac{\alpha_2}{E^x} - 4 \frac{\partial \ln \lambda}{\partial E^x}\right) c_f = 0
    \ ,
\end{equation}
holds.  If this equation is interpreted as a condition on $c_f$, its general
solution is not compatible with the classical limit.  However, if we exclude
the last term of this equation, it reduces to
\begin{equation}
    \frac{\partial c_f}{\partial E^x}
    = \frac{\lambda^2}{2} \left( \Lambda_0
    - \frac{\alpha_0}{E^x} \right)
    \label{eq:Partial Abelianization divergence resolution c_f}
\end{equation}
which can be directly integrated to obtain a non-classical $c_f$ compatible
with the classical limit.  For instance, if we choose the classical values
$\Lambda_0 = \Lambda$ and $\alpha_0=1$, and a constant
$\lambda = \bar{\lambda}$, we obtain
\begin{equation}
    c_f = 1 + \frac{\bar{\lambda}^2}{2} \left( \Lambda E^x - \ln \left(\frac{E^x}{c_0}\right) \right)
    \ ,
    \label{eq:MONDian cf - constant holonomy}
\end{equation}
where $c_0$ is the constant of integration.  If one instead chooses
$\lambda^2 = \Delta / E^x$, motivated for instance by loop quantum gravity,
one obtains
\begin{equation}
    c_f = 1 + \frac{\Delta}{2} \left( \Lambda \ln \left(\frac{E^x}{c_0}\right) + \frac{1}{E^x} \right)
    \ .
    \label{eq:MONDian cf - mu-scheme}
\end{equation}
In \cite{HigherMOND} it was shown that a non-classical function
of the form (\ref{eq:MONDian cf - constant holonomy}) can be related to MOND via the
logarithmic term.  Similarly, the version (\ref{eq:MONDian cf - mu-scheme})
can be related to MOND effects, too, since it has a logarithmic term.  With
this procedure, only the term multiplying $c_f$ in (\ref{eq:Abelian CML
  constraint}) retains its kinematical divergence.

\subsubsection{Classical limits and conditions}

We have different types of classical limits that can be demonstrated explicitly
for the polymerized Hamiltonian constraint (\ref{eq:Hamiltonian constraint -
  CML - scalar polymerization}), on which the conditions for gravitational and
matter observables have been imposed. The polymerization can always be undone
by setting $\nu\to\bar{\nu}$ followed by $\bar{\nu}\to0$.  The following two
canonical transformations will also be useful for the discussion of classical limits:
\begin{eqnarray}
    &&\phi \to \phi
    \quad ,\quad
    P_\phi \to P_\phi
    - E^\varphi \frac{\partial \ln \lambda}{\partial \phi} K_\varphi
    \ ,
    \nonumber\\
    &&K_\varphi \to \frac{\lambda}{\bar{\lambda}} K_\varphi
    \quad ,\quad
    E^\varphi \to \frac{\bar{\lambda}}{\lambda} E^\varphi
    \ ,
    \nonumber\\
    &&K_x \to K_x
    + E^\varphi \frac{\partial \ln \lambda}{\partial E^x} K_\varphi
    \quad ,\quad
    E^x \to E^x
    \ , \label{eq:Inversion of gravitational periodic CT}
\end{eqnarray}
where $\lambda$ may depend on $E^x$ and $\phi$, and
\begin{eqnarray}
    &&\phi\to \frac{\nu}{\bar{\nu}} \phi
    \quad ,\quad
    P_\phi \to \frac{\bar{\nu}}{\nu} P_\phi
    \ ,
    \nonumber\\
    &&K_\varphi \to K_\varphi
    \quad ,\quad
    E^\varphi \to E^\varphi
    \ ,
    \nonumber\\
    &&K_x \to K_x
    + P_\phi \frac{\partial \ln \nu}{\partial E^x}
    \quad , \quad
    E^x \to E^x
    \ , \label{eq:Inversion of scalar periodic CT}
\end{eqnarray}
where $\nu$ may depend on $E^x$.
We have the following limits:
\begin{itemize}
  \item The classical-matter limit is given by first performing the canonical
transformation (\ref{eq:Inversion of scalar periodic CT}), turning the
constraint (\ref{eq:Hamiltonian constraint - CML - scalar polymerization -
  periodic}) into (\ref{eq:Hamiltonian constraint - CML - scalar
  polymerization}) where $\bar{\nu}$ no longer appears, followed by
$\alpha_3 \to 1$, $\nu \to 0$, and $V_q$ becoming the classical potential of
the scalar field.  The resulting Hamiltonian constraint implies the
Klein--Gordon equation on a curved, emergent spacetime.
\item
The classical-geometry limit is given by first performing the canonical
transformations (\ref{eq:Inversion of scalar periodic CT}) and
(\ref{eq:Inversion of gravitational periodic CT}), eliminating $\bar{\nu}$ and
$\bar{\lambda}$ respectively, followed by $\lambda_0 , c_f \to 1$ and
$\lambda,\nu \to 0$.  In this limit, we recover  residual canonical
transformations linear in $K_\varphi$ which can be used to eliminate $q$ by
absorbing it into $\Lambda_0$.
\item
The classical-gravity limit is given by the classical-geometry limit together
with $\Lambda_0 \to - \Lambda$, $\alpha_0 , \alpha_2 \to 1$, $\alpha_3$
becoming a constant, and $V_q$ becoming a free function of $\phi$
only.
\item
A comparison with the constraint (\ref{eq:Covariant Hamiltonian - Linear
  combination - periodic variables}), obtained from a linear combination of
the classical constraints and subsequent canonical transformations, shows that
the modified constraint under consideration cannot reproduce the limit of reaching
the classical constraint surface unless we take the classical-geometry limit
since $q^{xx}$ appears explicitly in (\ref{eq:Hamiltonian constraint - CML -
  scalar polymerization - periodic}), but not in (\ref{eq:Covariant
  Hamiltonian - Linear combination - periodic variables}).  However, the
constraint (\ref{eq:Covariant Hamiltonian - Linear combination - periodic
  variables}) is incompatible with the classical-geometry limit. Thus, the
limit of reaching the classical constraint surface for (\ref{eq:Hamiltonian constraint - CML
  - scalar polymerization - periodic}) is trivial, as it exists only in the
full classical limit.
\item
The full classical limit is given by the classical-gravity limit together with
the classical-matter limit.
\item
The vacuum limit is given by $P_\phi , \phi , V_q \to 0$, recovering the
vacuum constraint (\ref{eq:Hamiltonian constraint - vacuum}).
\end{itemize}
Moreover, the constraint (\ref{eq:Hamiltonian constraint - CML - scalar
  polymerization - periodic}) can easily be Abelianized by imposing the
condition (\ref{eq:Partial Abelianization condition on q}), which simply
requires that we set $q=0$.

As will be shown in Section~\ref{sec:Homogeneous spacetime}, the constraint
(\ref{eq:Covariant Hamiltonian - Linear combination - periodic variables})
implies a physical singularity at the maximum-curvature surface in spatially
homogeneous dynamical solutions.

\subsection{Constraints compatible with the classical constraint surface as a limit}

A second class of tractable conditions is obtained by requiring that the
modified constraint has a limit in which the classical constraint surface is recovered.

\subsubsection{Anomaly-freedom}

We just found that the modified constraint compatible with the
classical-matter limit is not compatible with the limit of reaching the
classical constraint surface unless we take the full classical limit.  Since
the limit in which the classical constraint surface is reached is given by
(\ref{eq:Covariant Hamiltonian - Linear combination - periodic variables}),
the existence of this limit requires a modified constraint that can be
reduced to this version.

By inspection of the constraint (\ref{eq:Covariant Hamiltonian - Linear
  combination - periodic variables}), we require that the functions
$\bar{f}_0$, $\bar{f}_2$, $\bar{f}_3$, $\bar{f}_4$, $\bar{h}_0$, $\bar{h}_3$,
$\bar{h}_4$ are non-vanishing such that it can match (\ref{eq:Symmetry
  generator of real scalar field - Linear combination - periodic variables})
in some limit.  Direct substitution of the covariance solution
(\ref{eq:Covariance condition solution g,f1 - periodic}) in the condition
(\ref{eq:f1 - Matter observable - H(0)}) for the existence of a matter
observable implies the restriction
\begin{eqnarray}
    \frac{\tan (\bar{\lambda} K_\varphi)}{\bar{\lambda}} \frac{\partial c_f}{\partial \phi}
    + q \frac{\partial \ln \lambda}{\partial \phi} = 0
    \ .
\end{eqnarray}
For a non-zero $\bar{\lambda}$ to be possible, $c_f$ must be independent of
$\phi$ because the second term does not depend on $K_{\varphi}$, unlike the
first one. The second term then leaves us with two mutually exclusive options,
a $\phi$-dependent $\lambda$ or a non-vanishing $q$.  It turns out that
anomaly-freedom restricts us to the first option, as we will show now.

First, if we assume $\bar{f}_4^q \neq 0$, equation~(\ref{eq:AF f3f4,h4})
implies that $\bar{f}_4^q=\bar{h}_2=0$ from its $((E^x)')^2 \sqrt{q^{xx}}$ and
$((E^x)')^2 q^{xx}$ terms.  This turns equation (\ref{eq:AF f3f4,h4}) into the
same set of equations (\ref{eq:AF f3q - CML})--(\ref{eq:AF h4 - CML}) used in
the previous section, which allowed us to conclude that $\bar{f}_3$,
$\bar{f}_4$, and $\bar{h}_4$ must vanish for non-zero $\bar{f}_4^q$.  This
contradiction with our opening conditions shows that we must instead choose
$\bar{f}_4^q=0$.

With $\bar{f}_4^q=0$, the $\sqrt{q^{xx}}$ and $((E^x)')^2 \sqrt{q^{xx}}$ terms
of (\ref{eq:AF f3f4,h4}) imply $\bar{f}_3^q=0$ and $\bar{h}_2^q=0$,
respectively.  One can then show that the consistency between the
$((E^x)')^2 q^{xx}$ term of (\ref{eq:AF f0,f2,f2q}), the $((E^x)')^2$ term of
(\ref{eq:AF f3,f3q,h2,h2q}), the $\sqrt{q^{xx}}$ and
$((E^x)')^2 \sqrt{q^{xx}}$ terms of (\ref{eq:AF h0,h0q,h1,h1q}) and
(\ref{eq:AF f1,h3}), and the $0^{th}$-order and $((E^x)')^2$ terms of (\ref{eq:AF
  f3f4,h4}) determine
\begin{eqnarray}
    \bar{h}_0^q=\bar{h}_1^q=\bar{h}_3^q= 0
    \ .
\end{eqnarray}
With this, the $\sqrt{q^{xx}}$, $((E^x)')^2\sqrt{q^{xx}}$ terms of (\ref{eq:AF f0,f2,f2q}) can be solved for
\begin{eqnarray}
    \bar{g} \bar{f}_2^q &=& - \frac{\alpha_{2 q}}{4 E^x} \cos^2 (\bar{\lambda} K_\varphi)
    \ , \\
    \bar{g} \bar{f}_0^q &=& \frac{\alpha_{0 q}}{E^x}
    + \frac{\alpha_{2 q}}{E^x} \left( c_f \cos^2 (\bar{\lambda} K_\varphi) + 2 q \frac{ \sin (2 \bar{\lambda} K_\varphi)}{2 \bar{\lambda}}\right)
    \ ,
\end{eqnarray}
where $\alpha_{0q}$ and $\alpha_{2 q}$ are undetermined functions of $E^x$ and $\phi$.

The non-trivial equations for anomaly-freedom become
\begin{eqnarray}
    \bar{g} \frac{\partial (\bar{g} \bar{f}_0)}{\partial K_\varphi} &=&
    \bar{g} \bar{h}_0 \bar{g} \bar{h}_3
    - 2 \bar{g} \bar{f}_1 \bar{g} \bar{f}_2
    - \bar{g} \bar{f}_1  \frac{\partial \bar{g}}{\partial E^x}
    + \bar{g} \frac{\partial (\bar{g} \bar{f}_1)}{\partial E^x}
    \ , \label{eq:AF f0 - CCSL}
    \\
    \bar{g} \frac{\partial (\bar{g} \bar{f}_2)}{\partial K_\varphi} &=&
    \bar{g} \bar{f}_2 \frac{\partial \bar{g}}{\partial K_\varphi}
    - \bar{g}^2 \frac{1}{2} \frac{\partial}{\partial E^x} \frac{\partial \ln \bar{g}}{\partial K_\varphi}
    + \bar{g} \bar{h}_1 \bar{g} \bar{h}_3
    \ , \label{eq:AF f2 - CCSL}
    \\
    \bar{g} \frac{\partial (\bar{g} \bar{h}_0)}{\partial K_\varphi} &=&
    - 2 \bar{g} \bar{f}_1 \bar{g} \bar{h}_1
    + \bar{g} \bar{h}_0 \bar{g} \bar{h}_4
    + 2 \bar{g} \bar{f}_3 \bar{g} \bar{h}_3
    \ , \label{eq:AF h0 - CCSL}
    \\
    \frac{\partial (\bar{g} \bar{h}_1)}{\partial K_\varphi} &=&
    2 \bar{h}_2 \bar{g} \bar{h}_3
    + \bar{g} \bar{h}_1 \left( \bar{h}_4 + \frac{\partial \ln \bar{g}}{\partial K_\varphi}\right)
    \ , \label{eq:AF h1 - CCSL}
    \\
    \bar{g} \frac{\partial (\bar{g} \bar{h}_3)}{\partial K_\varphi} &=&
    \bar{g} \bar{h}_3 \left(
    \bar{g} \bar{h}_4 + \frac{1}{2} \frac{\partial \bar{g}}{\partial K_\varphi}
    \right)
    + 2 \bar{g} \bar{h}_1 \bar{g} \bar{f}_4
    - \bar{g}^2 \frac{1}{2} \frac{\partial^2 \ln \bar{g}}{\partial \phi \partial K_\varphi}
    \ , \label{eq:AF h3 - CCSL}
    \\
    \bar{g}^2 \frac{\partial \bar{f}_1}{\partial \phi} &=&
    \bar{g} \bar{f}_1 \bar{g} \bar{h}_3
    - 2 \bar{g} \bar{h}_0 \bar{g} \bar{f}_4
    \ , \label{eq:AF f1/phi - CCSL}
\end{eqnarray}
and
\begin{eqnarray}
    \frac{\partial \ln (\bar{g} \bar{f}_4)}{\partial K_\varphi} &=& 2 \bar{h}_4
    \ , \label{eq:AF f4 - CCSL}
    \\
    \frac{\partial (\bar{g} \bar{f}_3)}{\partial K_\varphi} &=&
    2 \bar{g} \bar{f}_3 \bar{h}_4
    - 2 \bar{g} \bar{f}_1 \bar{h}_2
    \ , \label{eq:AF f3 - CCSL}
    \\
    \frac{\partial \bar{h}_2}{\partial K_\varphi} &=&
    2 \bar{h}_2 \bar{h}_4
    \ , \label{eq:AF h2 - CCSL}
    \\
    0 &=& 
    P_\phi \Bigg[- \bar{g}^2 \frac{\partial \bar{f}_1}{\partial K_\varphi}
    + 4 \bar{g} \bar{f}_3 \bar{g} \bar{f}_4
    - \bar{g} \bar{f}_1 \bar{g} \bar{h}_4
    \Bigg]
    \ , \label{eq:AF f3,f4 - CCSL}
    \\
    P_\phi \Bigg[
    \bar{g}^2 \frac{\partial \bar{h}_4}{\partial K_\varphi} \Bigg] &=&
    P_\phi \Bigg[\bar{g}^2 \bar{h}_4 \left( \bar{h}_4
    - \frac{1}{2} \frac{\partial \ln \bar{g}}{\partial K_\varphi} \right)
    + 4 \bar{g} \bar{h}_2 \bar{g} \bar{f}_4
    + \frac{1}{2} \bar{g}^2 \frac{\partial^2 \ln \bar{g}}{\partial K_\varphi^2}
    \Bigg]
    \ . \label{eq:AF h4 - CCSL}
\end{eqnarray}

The last five equations form an over-determined system of equations for
$\bar{f}_3$, $\bar{f}_4$, $\bar{h}_2$, and $\bar{h}_4$, since we already know
$\bar{g}$ and $\bar{f}_1$.  This system is hard, if not impossible, to solve
exactly, and we will split our analysis into two versions with additional
assumptions.  In the first version we assume $\bar{h}_2=0$, and in the second
one $\bar{h}_4=0$.  The former is compatible with the limit of reaching the
classical constraint surface considered in this subsection, while the latter will
be analyzed in the next subsection.

With $\bar{h}_2=0$, the general solution of equation (\ref{eq:AF h4 - CCSL}) is
\begin{eqnarray}
    \bar{h}_4 &=&
    - \bar{\lambda}^2 \sec (\bar{\lambda} K_\varphi) \frac{ \sin (\bar{\lambda} K_\varphi)/\bar{\lambda} + c_{h4}}{ c_{h4} \bar{\lambda} \sin (\bar{\lambda} K_\varphi) + 1}
\end{eqnarray}
with an integration function $c_{h4}$ independent of $K_{\varphi}$.
Equations (\ref{eq:AF f4 -
  CCSL}) and (\ref{eq:AF f3 - CCSL}) can now be directly integrated, yielding
\begin{eqnarray}
    \bar{g} \bar{f}_3 &=& - \lambda_0 \frac{\alpha_3}{E^x} \cos^2 (\bar{\lambda} K_\varphi) \left( 1 + c_{h4} \bar{\lambda}^2 \frac{\sin (\bar{\lambda} K_\varphi)}{\bar{\lambda}}\right)^{-2}
    \ , \\
    \bar{g} \bar{f}_4 &=& - \lambda_0 \frac{E^x}{\alpha_4} \cos^2 (\bar{\lambda} K_\varphi) \left( 1 + c_{h4} \bar{\lambda}^2 \frac{\sin (\bar{\lambda} K_\varphi)}{\bar{\lambda}}\right)^{-2}
    \ ,
\end{eqnarray}
where $\alpha_3$ and $\alpha_4$ are undetermined functions of $E^x$ and $\phi$
that cannot identically vanish under the current assumptions.
Inserting these results in (\ref{eq:AF f3,f4 - CCSL}), we obtain
\begin{eqnarray}
    0 &=&
    \cos (\bar{\lambda} K_\varphi) \left( -4 c_f \cos (2 \bar{\lambda} K_\varphi)
    + 16 q \bar{\lambda}^2 \frac{\sin (2 \bar{\lambda} K_\varphi)}{2 \bar{\lambda}} \right)
    + 4 \frac{\alpha_3}{\alpha_4} \frac{\cos^3 (\bar{\lambda} K_\varphi)}{\left( 1 + c_{h4} \bar{\lambda} \sin (\bar{\lambda} K_\varphi)\right)^4}
    \nonumber\\
    &&
    + 4 \bar{\lambda} \frac{c_{h4} \bar{\lambda} \cos (2 \bar{\lambda} K_\varphi) - \sin (\bar{\lambda} K_\varphi)}{1 + c_{h4} \bar{\lambda} \sin (\bar{\lambda} K_\varphi)} \left( c_f \frac{\sin (2 \bar{\lambda} K_\varphi)}{2 \bar{\lambda}} + q \cos (\bar{\lambda} K_\varphi) \right)
\end{eqnarray}
which must hold for all $K_{\varphi}$. If $c_{h4}\not=0$, the dependence of
this equation on $K_{\varphi}$ is such that it can be valid only if
$\alpha_3=0$, which is not allowed by the classical limit. Therefore, we have 
$c_{h4}=0$ and the equation simplifies to
\begin{equation}
0=    \left(c_f - \frac{\alpha_3}{\alpha_4}\right) 4 \cos^3 (\bar{\lambda} K_\varphi)
    - 2 q \bar{\lambda} \left( 3 \sin (\bar{\lambda} K_\varphi) + 3 \cos^2 (\bar{\lambda} K_\varphi) \sin (\bar{\lambda} K_\varphi)- \sin^3 (\bar{\lambda} K_\varphi) \right)
\end{equation}
again for all $K_{\varphi}$.
This equation  restricts the values of the following free functions:
\begin{eqnarray}
    c_{h4} &=& q = 0
    \ , \\
    \alpha_3 &=& c_f \alpha_4
    \ .
\end{eqnarray}

We can now solve the remaining equations (\ref{eq:AF f0 - CCSL})--(\ref{eq:AF
  f1/phi - CCSL}) for anomaly-freedom, which again form an over-determined
system of equations.  Equation (\ref{eq:AF h1 - CCSL}), assuming $\bar{h}_2=0$
in the present case, has the general solution
\begin{eqnarray}
    \bar{g} \bar{h}_1 &=& \lambda_0 c_{h1} \frac{\cos^3 (\bar{\lambda}
                          K_\varphi)}{1+c_{h4} \bar{\lambda} \sin
                          (\bar{\lambda} K_\varphi)}= \lambda_0 c_{h1} \cos^3 (\bar{\lambda}
                          K_\varphi)
    \ ,
\end{eqnarray}
where $c_{h1}$ is an undetermined function of $E^x$ and $\phi$.
Equation (\ref{eq:AF h3 - CCSL}) can now be solved by
\begin{eqnarray}
    \bar{g} \bar{h}_3 &=&
    \lambda_0 \cos^2 (\bar{\lambda} K_\varphi) \left( c_{h3}
    - 2 c_{h1} \frac{E^x}{\alpha_4} \frac{\sin (\bar{\lambda} K_\varphi)}{\bar{\lambda}}
    \right)
    \ ,
\end{eqnarray}
where $c_{h3}$ is an undetermined function of $E^x$ and $\phi$.
Next, we solve equation (\ref{eq:AF h0 - CCSL}) by
\begin{eqnarray}
    \bar{g} \bar{h}_0 &=&
    \lambda_0 \left( c_{h0} \cos (\bar{\lambda} K_\varphi)
    - 2 \left( c_{h3} \frac{\alpha_3}{E^x}
    + c_{h1} c_f \frac{\sin (\bar{\lambda} K_\varphi)}{\bar{\lambda}} \right) \frac{\sin (2\bar{\lambda} K_\varphi)}{2\bar{\lambda}}
    \right)
    \ ,
\end{eqnarray}
where $c_{h0}$ is an undetermined function of $E^x$ and $\phi$.  Inserting
these results into (\ref{eq:AF f1/phi - CCSL}), we obtain the
condition
\begin{eqnarray}
    0 &=&
    \bar{\lambda}^2 c_{h0}
    - c_{h1} c_f \left( c_{h1} + 2 \right)
    + 3 c_{h1} c_f \cos(2 \bar{\lambda} K_\varphi)
\end{eqnarray}
for all $K_{\varphi}$, 
which determines
\begin{eqnarray}
    c_{h0} &=& c_{h1} = 0
\end{eqnarray}
since the classical limit requires $c_f$ and $\alpha_3$ to be non-zero.

Using all the results obtained so far, we solve the last two equations
(\ref{eq:AF f0 - CCSL}) and (\ref{eq:AF f2 - CCSL}):
\begin{eqnarray}
    \bar{g} \bar{f}_2 &=& - \lambda_0 \frac{\alpha_2}{4 E^x} \cos^2 (\bar{\lambda} K_\varphi)
    \ , \\
    \bar{g} \bar{f}_0 &=&
    \lambda_0 \left( - \Lambda_0
    + \frac{\alpha_0}{E^x} - V
    + \left( \frac{\alpha_2}{E^x} c_f
    + 2 \frac{\partial c_f}{\partial E^x} - c_{h3}{}^2 \frac{\alpha_3}{E^x} \right) \frac{\sin^2 (\bar{\lambda} K_\varphi)}{\bar{\lambda^2}} \right)
    \ ,
\end{eqnarray}
where $\alpha_0$, $V$, and $\alpha_2$ are undetermined functions of $E^x$ and
$\phi$.  Here, as in a similar case before, $V$ and $\Lambda_0$ are not
independent of $\alpha_0$, but we keep them separate so as to be able to
define a scalar-field potential independent of the gravitational terms.  This
exhausts all the equations for anomaly-freedom.

We now go back to the condition that there be a matter observable, since the
choice $q=0$ has been forced upon us by the conditions for anomaly-freedom.
For $\partial_\phi \lambda \neq 0$, conditions (\ref{eq:f0q - Matter
  observable - H(-1)}) and (\ref{eq:h3q - Matter observable - H(-1)}) for the
existence of a matter observable imply that $\bar{f}_i^q$ and $\bar{h}_i^q$,
for all $i$, must vanish, thus determining $\alpha_{0 q} = \alpha_{2 q} = 0$.
With this, the conditions (\ref{eq:f0q - Matter observable -
  H(1)})--(\ref{eq:h3q - Matter observable - H(1)}) for the existence of a
matter observable and condition (\ref{eq:Gravitational vacuum observable
  condition}) for the existence of a gravitational observable are trivially
satisfied.  The last step is to substitute all our solutions into the
conditions (\ref{eq:f0 - Matter observable - H(0)})--(\ref{eq:h4 - Matter
  observable - H(0)}), which then become constraining equations for the
undetermined functions.  Equation (\ref{eq:f3 - Matter observable - H(0)})
requires that $\partial \alpha_3 / \partial \phi = 0$, and, similarly,
equation (\ref{eq:f4 - Matter observable - H(0)}) requires
$\partial \alpha_4 / \partial \phi = 0$.  Using this, equation~(\ref{eq:h0 -
  Matter observable - H(0)}) implies
\begin{equation}
    \frac{\partial c_{h3}}{\partial \phi} = \frac{\partial^2 \ln \lambda}{\partial \phi^2}
    \ ,
\end{equation}
and thus
\begin{eqnarray}
    c_{h3} = c_{h3}^x + \frac{\partial \ln \lambda}{\partial \phi}
    \ ,
\end{eqnarray}
where $c_{h3}^x$ is an undetermined function of $E^x$.
Substitution in (\ref{eq:h3 - Matter observable - H(0)}) shows $c_{h3}^x=0$.
Equation (\ref{eq:f2 - Matter observable - H(0)}) implies
\begin{equation}
    \frac{\partial \alpha_2}{\partial \phi} = - 2 E^x \frac{\partial^2 \ln \lambda^2}{\partial \phi \partial E^x}
    \ ,
\end{equation}
and thus
\begin{eqnarray}
    \alpha_2 = \alpha_2^x - 4 E^x \frac{\partial \ln \lambda}{\partial E^x}
    \ ,
\end{eqnarray}
where $\alpha_2^x$ is an undetermined function of $E^x$.
Substituting all the above into (\ref{eq:f0 - Matter observable - H(0)}) with $V=0$ we obtain the equation
\begin{eqnarray}
    \frac{\partial \ln \left( \alpha_0 - E^x \Lambda_0 \right)}{\partial \phi}
    = \frac{\partial \ln \lambda^2}{\partial \phi}
    \ ,
\end{eqnarray}
and thus
\begin{eqnarray}
    \alpha_0 - E^x \Lambda_0 = \frac{\lambda^2}{\bar{\lambda}} \left(\alpha_0^x - E^x \Lambda_0^x\right)
    \ ,
\end{eqnarray}
where $\alpha_0^x$ and $\Lambda_0^x$ are undetermined functions of $E^x$.
Comparison with (\ref{eq:Covariant Hamiltonian - Linear combination - periodic
  variables}) suggests that we multiply $V$ by the same factor, and we do so
in what follows.  This exhausts all the conditions for the existence of matter
and gravitational observables.

\subsubsection{General Hamiltonian constraint}

In order to recover the full effects of non-constant $\lambda$, it suffices to redefine 
\begin{eqnarray}
    \lambda_0 \to \lambda_0 \frac{\bar{\lambda}}{\lambda}
    \ ,
    \hspace{1cm}
    V_q \to \frac{\lambda^2}{\bar{\lambda}^2} V_q
    \ , \label{eq:Redefinition for non-constant holonomy - CCSL}
\end{eqnarray}
such that the general constraint now resembles (\ref{eq:Covariant Hamiltonian
  - Linear combination - periodic variables}).  The Hamiltonian constraint is
then
\begin{eqnarray}
    H &=&
    - \frac{\bar{\lambda}}{\lambda} \lambda_0 \frac{\sqrt{E^x}}{2} \Bigg[
    E^\varphi \bigg( \frac{\lambda^2}{\bar{\lambda}} \left( - \Lambda_0^x
    + \frac{\alpha_0^x}{E^x} - V \right) 
    + \left( \frac{\alpha_2^x}{E^x} c_f
    + 2 \frac{\partial c_f}{\partial E^x} \right) \frac{\sin^2 (\bar{\lambda} K_\varphi)}{\bar{\lambda^2}}
    \nonumber\\
    &&
    + \left( \frac{K_x}{E^\varphi}
    - \frac{\tan (\bar{\lambda} K_\varphi)}{\bar{\lambda}} \frac{\partial \ln \lambda}{\partial E^x} \right) 4 c_f \frac{\sin (2\bar{\lambda} K_\varphi)}{2\bar{\lambda}}
    \nonumber\\
    &&
    - \left( \frac{P_\phi}{E^\varphi} + \frac{\tan (\bar{\lambda} K_\varphi)}{\bar{\lambda}} \left(c_{h3}^x + \frac{\partial \ln \lambda}{\partial \phi}\right) \right)^2 \frac{\alpha_4}{E^x} c_f \cos^2 (\bar{\lambda} K_\varphi)
    \bigg)
    \nonumber\\
    &&
    + \frac{((E^x)')^2}{E^\varphi} \left( \left( \frac{\partial \ln \lambda}{\partial E^x} - \frac{\alpha_2^x}{4 E^x} \right) \cos^2 (\bar{\lambda} K_{\varphi})
    + \bar{\lambda}^2 \frac{K_x}{E^\varphi} \frac{\sin (2 \bar{\lambda} K_\varphi)}{2 \bar{\lambda}} \right)
    \nonumber\\
    &&
    + \left( \frac{(E^x)' (E^\varphi)'}{(E^\varphi)^2} - \frac{(E^x)''}{E^\varphi} \right) \cos^2 (\bar{\lambda} K_\varphi)
    \nonumber\\
    &&
    + \cos^2 (\bar{\lambda} K_\varphi) \left( 
    - \frac{(\phi')^2}{E^\varphi} \frac{E^x}{\alpha_4}
    + \frac{(E^x)' \phi'}{E^\varphi} \left( c_{h3}^x + \frac{\partial \ln \lambda}{\partial \phi} - \frac{P_\phi}{E^\varphi} \bar{\lambda}^2 \frac{\tan (\bar{\lambda} K_\varphi)}{\bar{\lambda}} \right) \right)
       \Bigg]\nonumber\\
  &&
    + \lambda_0^2 \frac{E^x}{2} \sqrt{q_{xx}} V_q
    \label{eq:Hamiltonian constraint - periodic - CCSL}
\end{eqnarray}
with structure function
\begin{eqnarray}
    q^{x x}
    =
    \left( c_f
    + \left(\frac{\bar{\lambda} (E^x)'}{2 E^\varphi} \right)^2 \right) \cos^2 \left(\bar{\lambda} K_\varphi\right)
    \frac{\bar{\lambda}^2}{\lambda^2} \lambda_0^2 \frac{E^x}{(E^\varphi)^2}
    \ ,
    \label{eq:Structure function - periodic - CCSL}
\end{eqnarray}
where all parameters are undetermined functions of $E^x$, except for
$\lambda_0$, $\lambda$, $V$, and $V_q$ which depend on both $E^x$ and $\phi$,
and $\bar{\lambda}$ is a constant.  For the constraint to be invariant under
the transformation generated by (\ref{eq:Symmetry generator of real scalar
  field - CT modded}), one must take $V=V_q=0$.  The classical limit can be
obtained in different ways, as discussed below.

A canonical transformation of the form
$K_\varphi \to (\lambda/\bar{\lambda}) K_\varphi$ eliminates all traces of
$\bar{\lambda}$, but the constraint becomes non-periodic in $K_\varphi$.  This
shows that we have properly taken into account all effects of the non-constant
$\lambda$.

\subsubsection{Polymerization of the scalar field}

Following the discussion of the previous section, we place an upper bound on the
absolute value of the scalar matter field of the constraint (\ref{eq:Hamiltonian
  constraint - periodic - CCSL}) by using the redefinition (\ref{eq:Global
  factor boundedness redefinition}).  We can then take two consecutive
canonical transformations of the form (\ref{eq:Residual canonical
  transformation phi,ex}), the first one with $f_c^\phi=\sin (\nu \phi) / \nu$,
and the second one with $f_c^\phi = (\bar{\nu}/\nu) \phi$.  Applying these two
transformations is equivalent to using the single canonical
transformation
\begin{eqnarray}
    && \phi \to \frac{\sin (\bar{\nu}\phi)}{\nu}
    \quad , \quad
    P_\phi \to \frac{\nu}{\bar{\nu}} \frac{P_\phi}{\cos (\bar{\nu}\phi)}
    \nonumber\\
    &&
    K_x \to K_x - P_\phi \frac{\tan (\bar{\nu}\phi)}{\bar{\nu}} \frac{\partial \ln \nu}{\partial E^x}
    \label{eq:Canonical transformation - Scalar polymerization}
\end{eqnarray}
while all other phase-space variables remain unchanged.

The constraint (\ref{eq:Hamiltonian constraint - periodic - CCSL}) then becomes
\begin{eqnarray}
    H &=&
    - \frac{\bar{\lambda}}{\lambda} \lambda_0 \cos^2 (\bar{\nu} \phi) \frac{\sqrt{E^x}}{2} \Bigg[
    E^\varphi \bigg( \frac{\lambda^2}{\bar{\lambda}} \left( - \Lambda_0^x
    + \frac{\alpha_0^x}{E^x} - V \right) 
    + \left( \frac{\alpha_2^x}{E^x} c_f
    + 2 \frac{\partial c_f}{\partial E^x} \right) \frac{\sin^2 (\bar{\lambda} K_\varphi)}{\bar{\lambda^2}}
    \nonumber\\
    &&
    + \left( \frac{K_x}{E^\varphi} - \frac{P_\phi}{E^\varphi} \frac{\tan (\bar{\nu}\phi)}{\bar{\nu}} \frac{\partial \ln \nu}{\partial E^x}
    - \frac{\tan (\bar{\lambda} K_\varphi)}{\bar{\lambda}} \frac{\partial \ln \lambda}{\partial E^x} \right) 4 c_f \frac{\sin (2\bar{\lambda} K_\varphi)}{2\bar{\lambda}}
    \nonumber\\
    &&
    - \left( \frac{P_\phi}{E^\varphi \cos (\bar{\nu} \phi)} + \frac{\tan (\bar{\lambda} K_\varphi)}{\bar{\lambda}} \left(\frac{\bar{\nu}}{\nu} c_{h3}^x + \frac{\partial \ln \lambda}{\partial \phi}\right) \right)^2 \frac{\alpha_4}{E^x} \frac{\nu^2}{\bar{\nu}^2} c_f \cos^2 (\bar{\lambda} K_\varphi)
    \bigg)
    \nonumber\\
    &&
    + \frac{((E^x)')^2}{E^\varphi} \Bigg( 
    \bar{\lambda}^2 \frac{K_x}{E^\varphi} \frac{\sin (2 \bar{\lambda} K_\varphi)}{2 \bar{\lambda}}
    \nonumber\\
    &&
    + \cos^2 (\bar{\lambda} K_\varphi) \left( \frac{\partial \ln \lambda}{\partial E^x} - \frac{\alpha_2^x}{4 E^x} - \frac{\sin (\bar{\nu}\phi)}{\bar{\nu}} \frac{\partial \ln \nu}{\partial E^x} \left( \frac{\bar{\nu}}{\nu} c_{h3}^x + \frac{\partial \ln \lambda}{\partial \phi}
    + \frac{\sin (\bar{\nu}\phi)}{\bar{\nu}} \frac{\partial \ln \nu}{\partial E^x} \frac{E^x}{\alpha_4} \right) \right)
    \Bigg)
    \nonumber\\
    &&
    + \left( \frac{(E^x)' (E^\varphi)'}{(E^\varphi)^2} - \frac{(E^x)''}{E^\varphi} \right) \cos^2 (\bar{\lambda} K_\varphi)
    \nonumber\\
    &&
    + \cos^2 (\bar{\lambda} K_\varphi) \Bigg( 
    - \frac{1}{E^\varphi} \left(\left(\frac{\sin (\bar{\nu}\phi)}{\bar{\nu}}\right)'\right)^2 \frac{E^x}{\alpha_4}
    + \frac{(E^x)'}{E^\varphi} \left(\frac{\sin (\bar{\nu}\phi)}{\bar{\nu}}\right)' \Bigg( \frac{2 E^x}{\alpha_4} \frac{\sin (\bar{\nu}\phi)}{\bar{\nu}} \frac{\partial \ln \nu}{\partial E^x}
    \nonumber\\
    &&
    + \frac{\bar{\nu}}{\nu} c_{h3}^x
    + \frac{\partial \ln \lambda}{\partial \phi}
    - \frac{P_\phi}{E^\varphi \cos (\bar{\nu} \phi)} \bar{\lambda}^2 \frac{\tan (\bar{\lambda} K_\varphi)}{\bar{\lambda}}
    \Bigg)\Bigg)
    \Bigg]
    + \lambda_0^2 \cos^4 (\bar{\nu} \phi) \frac{E^x}{2} \sqrt{q_{xx}} V_q
    \ ,
    \label{eq:Hamiltonian constraint - periodic - CCSL - scalar polymerized}
\end{eqnarray}
with structure function
\begin{eqnarray}
    q^{x x}
    =
    \left( c_f
    + \left(\frac{\bar{\lambda} (E^x)'}{2 E^\varphi} \right)^2 \right) \cos^2 \left(\bar{\lambda} K_\varphi\right)
    \frac{\bar{\lambda}^2}{\lambda^2} \lambda_0^2 \cos^4 (\bar{\nu} \phi) \frac{E^x}{(E^\varphi)^2}
    \ .
    \label{eq:Structure function - periodic - CCSL - scalar polymerized}
\end{eqnarray}
The constraint (\ref{eq:Hamiltonian constraint - periodic - CCSL - scalar
  polymerized}) has been succesfully polymerized: it is periodic in both
$K_\varphi$ and $\phi$ and allows for remnants of non-constant holonomy
parameters $\lambda$ and $\nu$.

The vacuum mass observable associated to (\ref{eq:Hamiltonian constraint -
  periodic - CCSL - scalar polymerized}) is given by
\begin{eqnarray}
    \mathcal{M}
    &=&
    d_0
    + \frac{d_2}{2} \left(\exp \int {\rm d} E^x \ \left(\frac{\alpha_2^x}{2 E^x} - \frac{\partial \ln \lambda^2}{\partial E^x}\right)\right)
    \left(
    \frac{\sin^2\left(\bar{\lambda} K_{\varphi}\right)}{\bar{\lambda}^2}
    - \cos^2 (\bar{\lambda} K_\varphi) \left(\frac{(E^x)'}{2 E^\varphi}\right)^2
    \right)
    \notag\\
    &&
    + \frac{d_2}{4} \int {\rm d} E^x \ \left(\frac{\lambda^2}{\bar{\lambda}^2}\left( \Lambda_0^x
    + \frac{\alpha_0^x}{E^x} \right) \exp \int {\rm d} E^x \ \left(\frac{\alpha_2^x}{2 E^x} - \frac{\partial \ln \lambda^2}{\partial E^x}\right)\right)
    \ ,
    \label{eq:Gravitational weak observable - periodic - CCSL - scalar polymerized}
\end{eqnarray}
and, when $V=V_q=V^q=0$, its scalar-field observable by
\begin{eqnarray}
    G [\alpha] &=& \int {\rm d}^3 x\ \alpha \frac{\nu}{\bar{\nu}} \left( \frac{P_\phi}{\cos(\bar{\nu}\phi)}
    + E^\varphi \frac{\tan (\bar{\lambda} K_\varphi)}{\bar{\lambda}} \frac{\partial \ln \lambda}{\partial \phi} \right)
    \ ,
    \label{eq:Scalar field symmetry generator - periodic - CCSL - scalar polymerized}
\end{eqnarray}
where $\alpha$, $d_0$, and $d_2$ are constants.
The associated conserved matter current $J^\mu$ has the components
\begin{eqnarray}
    J^t &=& \frac{\nu}{\bar{\nu}} \left( \frac{P_\phi}{\cos(\bar{\nu}\phi)}
    + E^\varphi \frac{\tan (\bar{\lambda} K_\varphi)}{\bar{\lambda}} \frac{\partial \ln \lambda}{\partial \phi} \right)
    \ , \\
    J^x &=& \frac{\partial G}{\partial P_\phi} \frac{\partial H}{\partial \phi'}
    = - \frac{\nu}{\bar{\nu}}
    \frac{\bar{\lambda}}{\lambda} \lambda_0 \frac{\sqrt{E^x}}{2}
    \cos^2 (\bar{\nu} \phi) \cos^2 (\bar{\lambda} K_\varphi) \Bigg( 
    - \frac{2}{E^\varphi} \left(\frac{\sin (\bar{\nu}\phi)}{\bar{\nu}}\right)' \frac{E^x}{\alpha_4}
    \label{eq:Conserved matter current - CCSL}\\
    &&
    + \frac{(E^x)'}{E^\varphi} \Bigg( \frac{2 E^x}{\alpha_4} \frac{\sin (\bar{\nu}\phi)}{\bar{\nu}} \frac{\partial \ln \nu}{\partial E^x}
    + \frac{\bar{\nu}}{\nu} c_{h3}^x
    + \frac{\partial \ln \lambda}{\partial \phi}
    - \frac{P_\phi}{E^\varphi \cos (\bar{\nu} \phi)} \bar{\lambda}^2 \frac{\tan (\bar{\lambda} K_\varphi)}{\bar{\lambda}}
    \Bigg)\Bigg)
    \ .\nonumber
\end{eqnarray}

\subsubsection{Partial Abelianization}

The partial Abelianization of the constraint (\ref{eq:Hamiltonian constraint - periodic - CCSL}) is easily achieved by using the coefficients (\ref{eq:Abelianization coefficients}) under the redefinitions (\ref{eq:Redefinition for non-constant holonomy - CCSL}) and (\ref{eq:Global factor boundedness redefinition}).
The resulting Abelianized constraint is given by
\begin{eqnarray}
    \frac{H^{(A)}}{\bar{B} c_f} &=& - \frac{\bar{\lambda}}{\lambda} \lambda_0 \cos^2 (\Bar{\nu} \phi) \frac{\sqrt{E^x}}{2} \frac{\tan (\bar{\lambda} K_\varphi)}{\bar{\lambda}} \Bigg[
    E^\varphi \bigg( \frac{\lambda^2}{\bar{\lambda}} \left( - V - \Lambda_0^x
    + \frac{\alpha_0^x}{E^x} \right) \nonumber\\
&&\qquad    + \left( \frac{\alpha_2^x}{E^x} c_f
    + 2 \frac{\partial c_f}{\partial E^x} \right) \frac{\sin^2 (\Bar{\lambda} K_\varphi)}{\Bar{\lambda^2}}
    \nonumber\\
    &&\qquad
    + \left( \frac{K_x}{E^\varphi} - \frac{P_\phi}{E^\varphi} \frac{\tan (\bar{\nu}\phi)}{\bar{\nu}} \frac{\partial \ln \nu}{\partial E^x}
    - \frac{\tan (\Bar{\lambda} K_\varphi)}{\Bar{\lambda}} \frac{\partial \ln \lambda}{\partial E^x} \right) 4 c_f \frac{\sin (2\Bar{\lambda} K_\varphi)}{2\Bar{\lambda}}
    \nonumber\\
    &&\qquad
    - \left( \frac{P_\phi}{E^\varphi \cos (\bar{\nu} \phi)} + \frac{\tan (\Bar{\lambda} K_\varphi)}{\Bar{\lambda}} \left(\frac{\bar{\nu}}{\nu} c_{h3}^x + \frac{\partial \ln \lambda}{\partial \phi}\right) \right)^2 \frac{\alpha_4}{E^x} \frac{\nu^2}{\bar{\nu}^2} c_f \cos^2 (\Bar{\lambda} K_\varphi)
    \bigg)
    \nonumber\\
    &&\qquad
    + \frac{((E^x)')^2}{E^\varphi} \Bigg( 
    \Bar{\lambda}^2 \frac{K_x}{E^\varphi} \frac{\sin (2 \Bar{\lambda} K_\varphi)}{2 \Bar{\lambda}}
    \nonumber\\
    && \qquad
    + \cos^2 (\Bar{\lambda} K_\varphi) \left( \frac{\partial \ln
       \lambda}{\partial E^x} - \frac{\alpha_2^x}{4 E^x}\right.\nonumber\\
  &&\qquad\left. - \frac{\sin (\bar{\nu}\phi)}{\bar{\nu}} \frac{\partial \ln \nu}{\partial E^x} \left( \frac{\bar{\nu}}{\nu} c_{h3}^x + \frac{\partial \ln \lambda}{\partial \phi}
    + \frac{\sin (\bar{\nu}\phi)}{\bar{\nu}} \frac{\partial \ln \nu}{\partial E^x} \frac{E^x}{\alpha_4} \right) \right)
    \Bigg)
    \nonumber\\
    &&\qquad
    + \left( \frac{(E^x)' (E^\varphi)'}{(E^\varphi)^2} - \frac{(E^x)''}{E^\varphi} \right) \cos^2 (\Bar{\lambda} K_\varphi)
    \nonumber\\
    && \qquad
    + \cos^2 (\Bar{\lambda} K_\varphi) \Bigg( 
    - \frac{1}{E^\varphi} \left(\left(\frac{\sin
       (\bar{\nu}\phi)}{\bar{\nu}}\right)'\right)^2 \frac{E^x}{\alpha_4}
       \nonumber\\
  &&\qquad 
    + \frac{(E^x)'}{E^\varphi} \left(\frac{\sin (\bar{\nu}\phi)}{\bar{\nu}}\right)' \Bigg( \frac{2 E^x}{\alpha_4} \frac{\sin (\bar{\nu}\phi)}{\bar{\nu}} \frac{\partial \ln \nu}{\partial E^x}
     + \frac{\bar{\nu}}{\nu} c_{h3}^x
    + \frac{\partial \ln \lambda}{\partial \phi}
   \nonumber\\
    && \qquad
    - \frac{P_\phi}{E^\varphi \cos (\bar{\nu} \phi)} \bar{\lambda}^2 \frac{\tan (\bar{\lambda} K_\varphi)}{\bar{\lambda}}
    \Bigg)\Bigg)
    \Bigg]
    \nonumber\\
    &&
    - \frac{\bar{\lambda}}{\lambda} \lambda_0 \cos^2 (\Bar{\nu} \phi) \frac{\sqrt{E^x}}{2} \frac{(E^x)'}{E^\varphi} \left( K_\varphi' - \frac{K_x}{E^\varphi} (E^x)'
    + \frac{P_\phi}{E^\varphi} \phi'\right)
    \nonumber\\
    &&
    + \frac{\tan (\bar{\lambda} K_\varphi)}{\bar{\lambda}} \lambda_0^2 \cos^4 (\bar{\nu} \phi) \frac{E^x}{2} \sqrt{q_{xx}} V_q
    \ .
    \label{eq:Abelian CCSL constraint}
\end{eqnarray}
This Abelian constraint has kinematical divergences at
$K_\varphi= \pm \pi / (2 \bar{\lambda})$ in the first line and last line.  The
latter can be easily resolved by simply restricting the constraint to the case
$V_q=0$.  While the divergence of the first line can be treated as in the past
section, such that it can be partially resolved if the equation
\begin{equation}
    \frac{\partial c_f}{\partial E^x}
    = \frac{\lambda^2}{2} \left( V + \Lambda_0^x
    - \frac{\alpha_0^x}{E^x} \right)
    \ .
\end{equation}
is solved for $c_f$.  However, the difference between this equation and the
one of the past section (\ref{eq:Partial Abelianization divergence resolution
  c_f}) is that the former involves the potential $V$.  Recalling that $c_f$
cannot depend on $\phi$, we must either exclude the $V$-term from the equation
(leaving it as a divergent term) or restrict the constraint to the free scalar
case $V=0$.  Doing this, we recover the equation (\ref{eq:Partial
  Abelianization divergence resolution c_f}).

\subsubsection{Classical limits and conditions}

The constraint (\ref{eq:Hamiltonian constraint - periodic - CCSL - scalar
  polymerized}) cannot reproduce the classical-matter limit because it does
not have the necessary structure-function terms. It has the following limits:
\begin{itemize}
\item The classical-geometry limit is given by first performing the canonical
  transformations (\ref{eq:Inversion of scalar periodic CT}) and
  (\ref{eq:Inversion of gravitational periodic CT}), which
  eliminate $\bar{\nu}$ and $\bar{\lambda}$, respectively,
  followed by $\lambda_0 , c_f \to 1$ and $\lambda,\nu \to 0$.
In this limit, we can absorb $q$ into $\Lambda_0$ via a canonical transformation.
\item The classical-gravity limit is given by the classical-geometry limit
  together with $\Lambda_0^x \to - \Lambda$, $\alpha_0^x , \alpha_2^x \to 1$,
  $\alpha_4$ becoming a constant, and $V(E^x,\phi)\to V(\phi)$ becoming a
  free function of $\phi$ only.
\item Unlike the constraint of the previous section, (\ref{eq:Hamiltonian
    constraint - periodic - CCSL - scalar polymerized}) has a non-trivial
  limit of reaching the classical constraint surface.  This is given by taking
  the classical values for all the undetermined functions except $\lambda_0$
  and $\bar{\lambda}$.  The limit  correspond precisely to
  (\ref{eq:Covariant Hamiltonian - Linear combination - periodic variables}).
\item The vacuum limit is given by $P_\phi , \phi , V_q \to 0$.  However, the
  constraint (\ref{eq:Hamiltonian constraint - periodic - CCSL - scalar
    polymerized}) does not match the vacuum constraint (\ref{eq:Hamiltonian
    constraint - vacuum}) because it lacks the $q$-function.
\end{itemize}
The constraint (\ref{eq:Hamiltonian constraint - periodic - CCSL - scalar
  polymerized}) can easily be Abelianized by imposing the condition
(\ref{eq:Partial Abelianization condition on q}), which simply requires that
we set $q=0$.

As will be shown in Section~\ref{sec:Homogeneous spacetime}, the constraint
(\ref{eq:Hamiltonian constraint - periodic - CCSL - scalar polymerized})
develops a singularity at a maximum-curvature surface of spatially
homogeneous dynamical solution.

\subsection{Singularity-free constraints}

Our final class of examples is given by constraints that have non-singular
space-time solutions at least for homogeneous spatial slices. While this
statement does not guarantee complete removal of singularities, it sets this
set of modified theories apart from the previous two classes.

We start with assumptions on some of the free functions that are apparently
unrelated to the existence of non-singular solutions. The next section will
demonstrate the existence of solutions free of space-time singularities.

\subsubsection{Anomaly-freedom}

We use the initial steps of the preceding subsection, but instead of using
zero $\bar{h}_2$, we now assume $\bar{h}_2$ to be non-zero and $\bar{h}_4=0$.
Equations~(\ref{eq:AF f4 - CCSL}) and (\ref{eq:AF h2 - CCSL}) then imply that
$\bar{h}_2$ and $\bar{g} \bar{f}_4$ are independent of
$K_\varphi$. Equation~(\ref{eq:AF h4 - CCSL}) then has the solution
\begin{eqnarray}
    \bar{g} \bar{f}_4 &=& - \lambda_0 \frac{E^x}{\alpha_3}
    \ , \\
    \bar{h}_2 
    &=&
    - \bar{\lambda}^2 \frac{\alpha_3}{4 E^x}
    \ ,
\end{eqnarray}
where $\alpha_3$ is an undetermined non-vanishing function of $E^x$ and $\phi$.
We can now solve (\ref{eq:AF f3 - CCSL}) by
\begin{eqnarray}
    \bar{g} \bar{f}_3 &=& \lambda_0 \left( c_{f3}
    - \frac{\alpha_3}{E^x} \left( c_f \cos^2 (\bar{\lambda} K_\varphi)
    - 2 q \bar{\lambda}^2 \frac{\sin (2 \bar{\lambda} K_\varphi)}{2 \bar{\lambda}} \right) \right)
    \ ,
\end{eqnarray}
where $c_{f3}$ is an undetermined function of $E^x$ and $\phi$.
Equation (\ref{eq:AF f3,f4 - CCSL}) then requires
\begin{eqnarray}
    c_{f3} &=& 0
    \ .
\end{eqnarray}

We next solve the system of equations (\ref{eq:AF f0 - CCSL})--(\ref{eq:AF
  f1/phi - CCSL}), solving (\ref{eq:AF h1 - CCSL}) for $\bar{h}_3$ in terms of
$\bar{h}_1$ according to
$\bar{g} \bar{h}_3 = (\bar{g}/2 \bar{h}_2) \partial \bar{h}_1/\partial
K_\varphi$. This solution together with (\ref{eq:AF h3 - CCSL}) implies
\begin{eqnarray}
    \frac{\bar{g}}{2 \bar{h}_2} \frac{\partial^2 \bar{h}_1}{\partial K_\varphi^2} &=&
    - \frac{1}{2 \bar{h}_2} \frac{\partial \bar{h}_1}{\partial K_\varphi} \frac{1}{2} \frac{\partial \bar{g}}{\partial K_\varphi}
    + 2 \bar{h}_1 \bar{g} \bar{f}_4
    - \bar{g} \frac{1}{2} \frac{\partial^2 \ln \bar{g}}{\partial \phi \partial K_\varphi}
    \ ,
\end{eqnarray}
with general solution
\begin{eqnarray}
    \bar{h}_1 &=& c_{h1} \sec (\bar{\lambda} K_\varphi)
    + \bar{\lambda}^2 c_{h3} \frac{\tan (\bar{\lambda} K_\varphi)}{\bar{\lambda}}
\end{eqnarray}
where $c_{h1}$ and $c_{h3}$ are undetermined functions of $E^x$ and $\phi$.
Inserting this solution back into the previous expression, we obtain
\begin{eqnarray}
    \bar{g} \bar{h}_3 &=&
    - \lambda_0 \frac{2 E^x}{\alpha_3} \left( c_{h3} + c_{h1} \frac{\sin (\bar{\lambda} K_\varphi)}{\bar{\lambda}} \right)
    \ .
\end{eqnarray}

We now solve (\ref{eq:AF h0 - CCSL}) by
\begin{eqnarray}
    \bar{g} \bar{h}_0 &=&
    \lambda_0 \bigg(
    c_{h0}
    + 4 c_f \cos (\bar{\lambda} K_\varphi) \left( \frac{c_{h1}}{\bar{\lambda}^2} + c_{h3} \frac{\sin (\bar{\lambda} K_\varphi)}{\bar{\lambda}} \right)\nonumber\\
&&\qquad    + 8 q \left( - c_{h1} \frac{\sin (\bar{\lambda} K_\varphi)}{\bar{\lambda}} + c_{h3} \cos^2 (\bar{\lambda} K_\varphi) \right)
    \bigg)
    \ .
\end{eqnarray}
Using all these results in (\ref{eq:AF f1/phi - CCSL}), we find the condition
\begin{eqnarray}
    0 &=&
    \bar{\lambda}^2 c_{h0}
    + 2 q \bar{\lambda}^2 \left( 2 c_{h3}
    - 3 c_{h1} \left( \frac{\sin (\bar{\lambda} K_\varphi)}{\bar{\lambda}}
    + \frac{\sin (3 \bar{\lambda} K_\varphi)}{3 \bar{\lambda}} \right) \right)
    + 4 c_{h1} c_f \cos^3 (\bar{\lambda} K_\varphi)
    \nonumber\\
    &&
    - \frac{\alpha_3}{E^x} \cos^2 (\bar{\lambda} K_\varphi) \left( \sin (2 \bar{\lambda} K_\varphi) \frac{\partial c_f}{\partial \phi}
    + 2 \bar{\lambda} \cos (2 \bar{\lambda} K_\varphi) \frac{\partial q}{\partial \phi} \right)
\end{eqnarray}
that must be valid for all $K_{\varphi}$, and therefore
determines
\begin{eqnarray}
    \frac{\partial c_f}{\partial \phi} &=& \frac{\partial q}{\partial \phi} = c_{h1} = 0
    \ , \\
    c_{h0} &=& - 4 q c_{h3}
    \ .
\end{eqnarray}

Finally, we solve the last two remaining equations for anomaly-freedom,
(\ref{eq:AF f0 - CCSL}) and (\ref{eq:AF f2 - CCSL}) by
\begin{eqnarray}
    \bar{g} \bar{f}_0 &=&
    \lambda_0 \left( - \Lambda_0
    + \frac{\alpha_0}{E^x}
    + \frac{\sin^2 (\bar{\lambda} K_\varphi)}{\bar{\lambda}^2} \left( \frac{\alpha_2}{E^x} c_f + 2 \frac{\partial c_f}{\partial E^x} \right)
    + 2 \frac{\sin (2 \bar{\lambda} K_\varphi)}{2 \bar{\lambda}} \left( \frac{\alpha_2}{E^x} q + 2 \frac{\partial q}{\partial E^x} \right)
    \right)
    \ , \nonumber\\
    \bar{g} \bar{f}_2 &=&
    \lambda_0 \left( - V - \frac{\alpha_2}{4 E^x} \cos^2 (\bar{\lambda} K_\varphi) - c_{h3}{}^2 \frac{E^x}{\alpha_3} \right)
    \ ,
\end{eqnarray}
where $\alpha_0$, $\alpha_2$, $\Lambda_0$, and $V$ are undetermined functions of $E^x$ and $\phi$.

We now use these results in the case of $V=V_q=0$ in order to to address
equations (\ref{eq:f0q - Matter observable - H(-1)})--(\ref{eq:h3q - Matter
  observable - H(-1)}) and (\ref{eq:f0q - Matter observable -
  H(1)})--(\ref{eq:h4 - Matter observable - H(0)}) for the existence of a
matter observable.  Equation (\ref{eq:h2 - Matter observable - H(0)}) is
turned into the condition
\begin{eqnarray}
    - \cos^2 (\bar{\lambda} K_\varphi) \frac{\partial \ln \alpha_3}{\partial\phi}
    = \frac{\partial \ln \lambda^2}{\partial \phi}
    \ ,
\end{eqnarray}
which implies that both $\alpha_3$ and $\lambda$ must be independent of $\phi$.

The independence of $\lambda$ on $\phi$ (as well as that of $c_f$ and $q$)
implies that equations (\ref{eq:f0q - Matter observable -
  H(-1)})--(\ref{eq:h3q - Matter observable - H(-1)}) are trivially satisfied
because the $B^{-1}$ factor (\ref{eq:B factor - Matter observable - H(-1)})
vanishes.  Equation~(\ref{eq:f2q - Matter observable - H(1)}) requires that
$\bar{g} \bar{f}_2^q$ is independent of $\phi$ which, together with condition
(\ref{eq:Gravitational vacuum observable condition}) for the existence of a
gravitational observable, implies that $\bar{f}_2^q=0$.  Similarly, 
equation~(\ref{eq:f0q - Matter observable - H(-1)}) requires that
$\bar{g} \bar{f}_2^q$ is independent of $\phi$ and must thus vanish.

In this case, we can introduce a new potential term
$\bar{g} \bar{f}_2^q = V^q$, such that we recover the matter symmetry when
$V^q=V_q=V=0$.  Finally, the right-hand sides of equations~(\ref{eq:f0 - Matter
  observable - H(0)})--(\ref{eq:h4 - Matter observable - H(0)}) vanish,
implying that all the remaining undetermined functions, except for
$\lambda_0$, must be independent of $\phi$.

\subsubsection{General Hamiltonian constraint}

As in the previous sections, we redefine 
\begin{eqnarray}
    &&\lambda_0 \to \lambda_0 \frac{\bar{\lambda}}{\lambda}
    \ ,\ 
    q \to q \frac{\lambda}{\bar{\lambda}}
    \ , \
    \Lambda_0 \to \Lambda_0 \frac{\lambda^2}{\bar{\lambda}^2}
    \ , \
    \alpha_0 \to \alpha_0 \frac{\lambda^2}{\bar{\lambda}^2}
    \ , \
    \alpha_2 \to \alpha_2 - 4 E^x \frac{\partial \ln \lambda}{\partial E^x}
    \ , \nonumber\\
    &&
    V \to V \frac{\lambda^2}{\bar{\lambda}^2}
    \ , \
    V_q \to V_q \frac{\lambda^2}{\bar{\lambda}^2}
    \ , \
    V^q \to V_q \frac{\lambda^2}{\bar{\lambda}^2}
    \ , \label{eq:Redefinition for non-constant holonomy - SF}
\end{eqnarray}
in order to recover all allowed effects of a non-constant holonomy parameter $\lambda$.
The Hamiltonian constraint is then
\begin{eqnarray}
    H &=&
    - \frac{\bar{\lambda}}{\lambda} \lambda_0 \frac{\sqrt{E^x}}{2} \Bigg[
    E^\varphi \left(
    \frac{\lambda^2}{\bar{\lambda}^2}\left( - \Lambda_0
    + \frac{\alpha_0}{E^x} \right)
    + \frac{\sin^2 (\bar{\lambda} K_\varphi)}{\bar{\lambda}^2} \left( \left(\frac{\alpha_2}{E^x} - 4 \frac{\partial \ln \lambda}{\partial E^x}\right) c_f + 2 \frac{\partial c_f}{\partial E^x} \right)\right)
    \nonumber\\
    &&\qquad
    + 2E^\varphi \frac{\sin (2 \bar{\lambda} K_\varphi)}{2 \bar{\lambda}} \left( \left(\frac{\alpha_2}{E^x} - 2 \frac{\partial \ln \lambda}{\partial E^x}\right) \frac{\lambda}{\bar{\lambda}} q
    + 2 \frac{\lambda}{\bar{\lambda}} \frac{\partial q}{\partial E^x} \right)
    \nonumber\\
    &&\qquad
    + 4 \left( K_x + P_\phi c_{h3} \right) \left(c_f \frac{\sin (2 \bar{\lambda} K_\varphi)}{2 \bar{\lambda}}
    + \frac{\lambda}{\bar{\lambda}} q \cos(2 \bar{\lambda} K_\varphi)\right)\nonumber\\
&&\qquad    - \frac{P_\phi{}^2}{E^\varphi} \frac{\alpha_3}{E^x} \left( c_f \cos^2 (\bar{\lambda} K_\varphi)
    - 2 \frac{\lambda}{\bar{\lambda}} q \bar{\lambda}^2 \frac{\sin (2 \bar{\lambda} K_\varphi)}{2 \bar{\lambda}} \right)
    \nonumber\\
    &&\qquad
    - \frac{((E^x)')^2}{E^\varphi} \left( \left(\frac{\alpha_2}{4 E^x} - \frac{\partial \ln \lambda}{\partial E^x}\right) \cos^2 (\bar{\lambda} K_{\varphi})
    - \left( \frac{K_x}{E^\varphi}
    + \frac{P_\phi}{E^\varphi} c_{h3} \right) \bar{\lambda}^2 \frac{\sin (2
       \bar{\lambda} K_\varphi)}{2 \bar{\lambda}}\right.\nonumber\\
  &&\qquad  + \left.\frac{P_\phi{}^2}{(E^\varphi)^2} \bar{\lambda}^2
     \frac{\alpha_3}{4 E^x} \cos^2 (\bar{\lambda} K_{\varphi}) \right) 
    \nonumber\\
    && \qquad
    + \left( \frac{(E^x)' (E^\varphi)'}{(E^\varphi)^2} - \frac{(E^x)''}{E^\varphi} \right) \cos^2 (\bar{\lambda} K_\varphi)
     - \frac{\left( \phi'
    + c_{h3} (E^x)' \right)^2}{E^\varphi} \frac{E^x}{\alpha_3}
    - \frac{\lambda^2}{\bar{\lambda}^2} E^\varphi V
    \Bigg]
    \nonumber\\
    &&
   + \lambda_0^2 \frac{E^x}{2} \sqrt{q_{xx}} V_q
    + \frac{\lambda^2}{\bar{\lambda}^2} \frac{(E^\varphi)^2}{2} \sqrt{q^{xx}} V^q
    \ ,
    \label{eq:Hamiltonian constraint - DF}
\end{eqnarray}
with structure function 
\begin{eqnarray}
    q^{x x}
    =
    \frac{\bar{\lambda}^2}{\lambda^2} \lambda_0^2 \left( \left( c_f
    + \left(\frac{\bar{\lambda} (E^x)'}{2 E^\varphi}\right)^2 \right) \cos^2\left(\bar{\lambda} K_{\varphi}\right)
    - 2 \bar{\lambda}^2 \frac{\lambda}{\bar{\lambda}} q \frac{\sin \left(2 \bar{\lambda} K_{\varphi}\right)}{2 \bar{\lambda}} \right) \frac{E^x}{(E^\varphi)^2}
    \ ,
    \label{eq:Structure function - periodic - DF}
\end{eqnarray}
where all parameters are free functions of $E^x$, except for $\lambda_0$,
$V$, and $V_q$, and $V^q$ which may depend on both $E^x$ and $\phi$, while
$\bar{\lambda}$ is a constant.  The classical limit can be obtained in
different ways, as discussed below.  The matter symmetry is recovered
for $V=V_q=V^q=0$.

\subsubsection{Partial Abelianization}

The partial Abelianization of the constraint (\ref{eq:Hamiltonian constraint - DF - scalar polymerization}) is easily achieved by using the coefficients (\ref{eq:Abelianization coefficients}) under the redefinitions (\ref{eq:Global factor boundedness redefinition}) and (\ref{eq:Redefinition for non-constant holonomy - SF}) and taking $q=0$ according to the condition (\ref{eq:Partial Abelianization condition on q}).
The resulting Abelianized constraint is given by
\begin{eqnarray}
    \frac{H^{(A)}}{\bar{B} c_f} &=& - \frac{\bar{\lambda}}{\lambda} \lambda_0 \cos^2 (\Bar{\nu} \phi) \frac{\sqrt{E^x}}{2} \frac{\tan (\bar{\lambda} K_\varphi)}{\bar{\lambda}} \Bigg[
    E^\varphi \left(
    \frac{\lambda^2}{\bar{\lambda}^2}\left( - V - \Lambda_0
    + \frac{\alpha_0}{E^x} \right)\right.\nonumber\\
&&\qquad\left.    + \frac{\sin^2 (\Bar{\lambda} K_\varphi)}{\Bar{\lambda}^2} \left( \left(\frac{\alpha_2}{E^x} - 4 \frac{\partial \ln \lambda}{\partial E^x}\right) c_f + 2 \frac{\partial c_f}{\partial E^x} \right)\right)
    \nonumber\\
    &&\qquad
    + 4 \left( K_x + \frac{P_\phi}{\cos (\bar{\nu}\phi)} \left( \frac{\nu}{\bar{\nu}} c_{h3}
    - \frac{\sin (\bar{\nu}\phi)}{\bar{\nu}} \frac{\partial \ln \nu}{\partial E^x} \right) \right) c_f \frac{\sin (2 \Bar{\lambda} K_\varphi)}{2 \Bar{\lambda}}
    \nonumber\\
    &&\qquad
    - \frac{\nu^2}{\bar{\nu}^2} \frac{P_\phi{}^2}{E^\varphi \cos^2 (\bar{\nu}\phi)} \frac{\alpha_3}{E^x} c_f \cos^2 (\Bar{\lambda} K_\varphi)
    + \left( \frac{(E^x)' (E^\varphi)'}{(E^\varphi)^2} - \frac{(E^x)''}{E^\varphi} \right) \cos^2 (\Bar{\lambda} K_\varphi)
    \nonumber\\
    &&\qquad
    - \frac{((E^x)')^2}{E^\varphi} \Bigg( \left(\frac{\alpha_2}{4 E^x} - \frac{\partial \ln \lambda}{\partial E^x}
    + \frac{\nu^2}{\bar{\nu}^2} \frac{P_\phi{}^2}{(E^\varphi)^2 \cos^2 (\bar{\nu}\phi)} \Bar{\lambda}^2 \frac{\alpha_3}{4 E^x} \right) \cos^2 (\Bar{\lambda} K_{\varphi})
    \nonumber\\
    &&\qquad
    - \left( \frac{K_x}{E^\varphi}
    + \frac{P_\phi}{E^\varphi \cos (\bar{\nu}\phi)} \left( \frac{\nu}{\bar{\nu}} c_{h3}
    - \frac{\sin (\bar{\nu}\phi)}{\bar{\nu}} \frac{\partial \ln \nu}{\partial E^x} \right) \right) \Bar{\lambda}^2 \frac{\sin (2 \Bar{\lambda} K_\varphi)}{2 \Bar{\lambda}} \Bigg)
    \nonumber\\
    &&\qquad
    - \frac{1}{E^\varphi} \frac{\bar{\nu}^2}{\nu^2} \left(  \left(\frac{\sin (\bar{\nu}\phi)}{\bar{\nu}}\right)'
    + (E^x)' \left( \frac{\nu}{\bar{\nu}} c_{h3}
    - \frac{\sin (\bar{\nu}\phi)}{\bar{\nu}} \frac{\partial \ln \nu}{\partial E^x} \right) \right)^2 \frac{E^x}{\alpha_3}
    \Bigg]
    \nonumber\\
    &&
    - \frac{\bar{\lambda}}{\lambda} \lambda_0 \cos^2 (\Bar{\nu} \phi) \frac{\sqrt{E^x}}{2} \frac{(E^x)'}{E^\varphi} \left( K_\varphi' - \frac{K_x}{E^\varphi} (E^x)'
    + \frac{P_\phi}{E^\varphi} \phi'\right)
    \nonumber\\
    &&
    + \frac{\tan (\bar{\lambda} K_\varphi)}{\bar{\lambda}} \left( \lambda_0^2 \cos^4 (\Bar{\nu} \phi) \frac{E^x}{2} \sqrt{q_{xx}} V_q
    + \frac{\lambda^2}{\bar{\lambda}^2} \frac{(E^\varphi)^2}{2} \sqrt{q^{xx}} V^q\right)
    \ .
    \label{eq:Abelian SF constraint}
\end{eqnarray}
This Abelian constraint has some kinematical divergences at
$K_\varphi= \pm \pi / (2 \bar{\lambda})$ coming from the first line and the
first term of the last line.  The latter can be easily resolved by simply
restricting the constraint to the case $V_q=0$.  While the divergence of the
first line can be treated as in the past section, such that it can be
partially resolved if the equation (\ref{eq:Partial Abelianization divergence
  resolution c_f}) is solved for $c_f$ in the free scalar case $V=0$.

\subsubsection{Polymerization of the scalar field}

As discussed before, we place an upper bound on the absolute value of the
scalar field in the constraint (\ref{eq:Hamiltonian constraint -
  periodic - CCSL}) by using the redefinition (\ref{eq:Global factor
  boundedness redefinition}).  We can then apply the canonical transformation
(\ref{eq:Canonical transformation - Scalar polymerization}), resulting
in
\begin{eqnarray}
    H &=&
    - \frac{\bar{\lambda}}{\lambda} \lambda_0 \cos^2 (\bar{\nu} \phi) \frac{\sqrt{E^x}}{2} \Bigg[
    E^\varphi \left(
    \frac{\lambda^2}{\bar{\lambda}^2}\left( - \Lambda_0
    + \frac{\alpha_0}{E^x} \right)\right.\nonumber\\
&&\qquad    \left.+ \frac{\sin^2 (\bar{\lambda} K_\varphi)}{\bar{\lambda}^2} \left( \left(\frac{\alpha_2}{E^x} - 4 \frac{\partial \ln \lambda}{\partial E^x}\right) c_f + 2 \frac{\partial c_f}{\partial E^x} \right)\right)
    \nonumber\\
    &&\qquad
    + 2 E^\varphi \frac{\sin (2 \bar{\lambda} K_\varphi)}{2 \bar{\lambda}} \left( \left(\frac{\alpha_2}{E^x} - 2 \frac{\partial \ln \lambda}{\partial E^x}\right) \frac{\lambda}{\bar{\lambda}} q
    + 2 \frac{\lambda}{\bar{\lambda}} \frac{\partial q}{\partial E^x} \right)
    \nonumber\\
    &&\qquad
    + 4 \left( K_x + \frac{P_\phi}{\cos (\bar{\nu}\phi)} \left( \frac{\nu}{\bar{\nu}} c_{h3}
    - \frac{\sin (\bar{\nu}\phi)}{\bar{\nu}} \frac{\partial \ln \nu}{\partial E^x} \right) \right) \left(c_f \frac{\sin (2 \bar{\lambda} K_\varphi)}{2 \bar{\lambda}}
    + \frac{\lambda}{\bar{\lambda}} q \cos(2 \bar{\lambda} K_\varphi)\right)
    \nonumber\\
    &&\qquad
    - \frac{\nu^2}{\bar{\nu}^2} \frac{P_\phi{}^2}{E^\varphi \cos^2 (\bar{\nu}\phi)} \frac{\alpha_3}{E^x} \left( c_f \cos^2 (\bar{\lambda} K_\varphi)
    - 2 \frac{\lambda}{\bar{\lambda}} q \bar{\lambda}^2 \frac{\sin (2
       \bar{\lambda} K_\varphi)}{2 \bar{\lambda}} \right) \nonumber\\
&&\qquad    + \left( \frac{(E^x)' (E^\varphi)'}{(E^\varphi)^2} - \frac{(E^x)''}{E^\varphi} \right) \cos^2 (\bar{\lambda} K_\varphi)
    \nonumber\\
    &&\qquad
    - \frac{((E^x)')^2}{E^\varphi} \Bigg( \left(\frac{\alpha_2}{4 E^x} - \frac{\partial \ln \lambda}{\partial E^x}\right) \cos^2 (\bar{\lambda} K_{\varphi})\nonumber\\
&&\qquad\qquad    - \left( \frac{K_x}{E^\varphi}
    + \frac{P_\phi}{E^\varphi \cos (\bar{\nu}\phi)} \left( \frac{\nu}{\bar{\nu}} c_{h3}
    - \frac{\sin (\bar{\nu}\phi)}{\bar{\nu}} \frac{\partial \ln \nu}{\partial E^x} \right) \right) \bar{\lambda}^2 \frac{\sin (2 \bar{\lambda} K_\varphi)}{2 \bar{\lambda}}
    \nonumber\\
    &&\qquad\qquad
    + \frac{\nu^2}{\bar{\nu}^2} \frac{P_\phi{}^2}{(E^\varphi)^2 \cos^2 (\bar{\nu}\phi)} \bar{\lambda}^2 \frac{\alpha_3}{4 E^x} \cos^2 (\bar{\lambda} K_{\varphi}) \Bigg)
    \nonumber\\
    &&\quad
    - \frac{1}{E^\varphi} \frac{\bar{\nu}^2}{\nu^2} \left(  \left(\frac{\sin (\bar{\nu}\phi)}{\bar{\nu}}\right)'
    + (E^x)' \left( \frac{\nu}{\bar{\nu}} c_{h3}
    - \frac{\sin (\bar{\nu}\phi)}{\bar{\nu}} \frac{\partial \ln \nu}{\partial E^x} \right) \right)^2 \frac{E^x}{\alpha_3}
    - \frac{\lambda^2}{\bar{\lambda}^2} E^\varphi V
    \Bigg]
    \nonumber\\
    &&
    + \lambda_0^2 \cos^4 (\bar{\nu} \phi) \frac{E^x}{2} \sqrt{q_{xx}} V_q
    + \frac{\lambda^2}{\bar{\lambda}^2} \frac{(E^\varphi)^2}{2} \sqrt{q^{xx}} V^q
    \ ,
    \label{eq:Hamiltonian constraint - DF - scalar polymerization}
\end{eqnarray}
with structure function 
\begin{eqnarray}
    q^{x x}
    =
    \frac{\bar{\lambda}^2}{\lambda^2} \lambda_0^2 \left( \left( c_f
    + \left(\frac{\bar{\lambda} (E^x)'}{2 E^\varphi}\right)^2 \right) \cos^2\left(\bar{\lambda} K_{\varphi}\right)
    - 2 \bar{\lambda}^2 \frac{\lambda}{\bar{\lambda}} q \frac{\sin \left(2 \bar{\lambda} K_{\varphi}\right)}{2 \bar{\lambda}} \right) \cos^4 (\bar{\nu} \phi) \frac{E^x}{(E^\varphi)^2}
    \ .
    \label{eq:Structure function - DF - scalar polymerization}
\end{eqnarray}
The constraint (\ref{eq:Hamiltonian constraint - DF - scalar polymerization})
has been succesfully polymerized. It is periodic in both $K_\varphi$ and
$\phi$ and allows for non-constant holonomy parameters $\lambda$ and $\nu$.

The mass observable in vacuum associated with (\ref{eq:Hamiltonian constraint - DF
  - scalar polymerization}) is given by
\begin{eqnarray}
    \mathcal{M}
    &=&
    d_0
    + \frac{d_2}{2} \left(\exp \int {\rm d} E^x \ \left(\frac{\alpha_2}{2 E^x}
        - \frac{\partial \ln \lambda^2}{\partial E^x}\right)\right)\notag\\
  &&\qquad\quad \times
    \left(
    \frac{\sin^2\left(\bar{\lambda} K_{\varphi}\right)}{\bar{\lambda}^2}
    + 2 \frac{\lambda}{\bar{\lambda}} q \frac{\sin \left(2 \bar{\lambda}  K_{\varphi}\right)}{2 \bar{\lambda}}
    - \cos^2 (\bar{\lambda} K_\varphi) \left(\frac{(E^x)'}{2 E^\varphi}\right)^2
    \right)
    \notag\\
    &&
    + \frac{d_2}{4} \int {\rm d} E^x \ \left(\frac{\lambda^2}{\bar{\lambda}^2}\left( \Lambda_0
    + \frac{\alpha_0}{E^x} \right) \exp \int {\rm d} E^x \ \left(\frac{\alpha_2}{2 E^x} - \frac{\partial \ln \lambda^2}{\partial E^x}\right)\right)
    \ ,
    \label{eq:Gravitational weak observable - DF}
\end{eqnarray}
and, when $V=V_q=V^q=0$, the scalar-field observable is
\begin{eqnarray}
    G [\alpha] &=& \int {\rm d}^3 x\ \alpha \frac{\nu}{\bar{\nu}} \frac{P_\phi}{\cos(\bar{\nu}\phi)}
    \ ,
    \label{eq:Scalar field symmetry generator - DF}
\end{eqnarray}
where $\alpha$, $d_0$, and $d_2$ are constants.
The associated conserved matter current $J^\mu$ has the components
\begin{eqnarray}
    J^t &=& \frac{\nu}{\bar{\nu}} \frac{P_\phi}{\cos(\bar{\nu}\phi)}
    \ , \\
    J^x &=& \frac{\partial G}{\partial P_\phi} \frac{\partial H}{\partial \phi'}\nonumber\\
    &=& \frac{\bar{\nu}}{\nu} \frac{\bar{\lambda}}{\lambda} \lambda_0 \cos^2 (\bar{\nu} \phi) \frac{(E^x)^{3/2}}{\alpha_3 E^\varphi}  \left( \left(\frac{\sin (\bar{\nu}\phi)}{\bar{\nu}}\right)'
    + (E^x)' \left( \frac{\nu}{\bar{\nu}} c_{h3}
    - \frac{\sin (\bar{\nu}\phi)}{\bar{\nu}} \frac{\partial \ln \nu}{\partial E^x} \right) \right)
    \ .
    \label{eq:Conserved matter current - DF}
\end{eqnarray}

\subsubsection{Classical limits and conditions}

The constraint (\ref{eq:Hamiltonian constraint - DF - scalar polymerization})
cannot reproduce the classical-matter limit because it does not have the
necessary structure-function terms. It is not compatible with the limit of
reaching the classical constraint surface  because it does not have the $\bar{h}_4$-term.

The following limits can be realized:
\begin{itemize}
\item The classical-geometry limit is given by first performing the canonical
  transformations (\ref{eq:Inversion of scalar periodic CT}) and
  (\ref{eq:Inversion of gravitational periodic CT}), which
  eliminate $\bar{\nu}$ and $\bar{\lambda}$, respectively,
  followed by  the limit $\lambda_0 , c_f \to 1$ and $\lambda,\nu \to 0$.
In this limit, one can absorb $q$ into $\Lambda_0$ via a canonical transformation.
\item The classical-gravity limit is given by the classical-geometry
  limit together with $\Lambda_0 \to - \Lambda$, $\alpha_0 , \alpha_2 \to 1$
  and $\alpha_4$ becoming a constant, while $V(E^x,\phi)\to V(\phi)$,
  $V_q(E^x,\phi)\to V_q(\phi)$, $V^q(E^x,\phi)\to V^q(\phi)$ are turned into
  free functions of $\phi$ only.
\item The vacuum limit is given by $P_\phi , \phi , V_q \to 0$, recovering the
  vacuum constraint (\ref{eq:Hamiltonian constraint - vacuum}).
\end{itemize}
The constraint (\ref{eq:Hamiltonian constraint - DF - scalar polymerization})
can be Abelianized by imposing the condition (\ref{eq:Partial
  Abelianization condition on q}), which simply requires that we set $q=0$.

As will be shown in Section~\ref{sec:Homogeneous spacetime}, the constraint
(\ref{eq:Hamiltonian constraint - periodic - CCSL - scalar polymerized}) is
the only one of the three classes derived here that is non-singular at a
maximum-curvature surface of spatially homogeneous dynamical solutions.

\section{Dynamical solutions with  homogeneous spatial slices}
\label{sec:Homogeneous spacetime}

It is always difficult to find sufficiently many analytical solutions for
inhomogeneous scalar field theories on a curved background in order to display
characteristic physical effects.  In our case, the different versions of
consistent Hamiltonian constraints contain several new terms that distinguish
them from minimally coupled theories of scalar fields on a modified background
and which remain in the constraints even for spatially constant fields and
backgrounds. In this homogeneous setting, suitable for instance for
large-scale cosmological evolution or models of non-rotating interiors of
black holes in a specific slicing, the original partial differential equations
are reduced to ordinary differential equations that can often be solved
exactly. As we will show now, their implications help us to distinguish
between different versions of modified constraints.

\subsection{Hamiltonian constraint compatible with the classical-matter limit}

Our first class of modified theories is given by Hamiltonian constraints that
are compatible with the classical-matter limit. These theories are most closely
related to minimally coupled classical matter on a modified background.

\subsubsection{Equations of motion}

To be specific, we use the Hamiltonian constraint (\ref{eq:Hamiltonian
  constraint - CML - scalar polymerization}) compatible with the existence of
matter observables and a polymerized scalar field.  We only consider the
simple case $q = V_q = \Lambda_0 = 0$, $c_f = \alpha_i = 1$ and constant
$\lambda_0$, $\lambda=\bar{\lambda}$ and $\nu=\bar{\nu}$,  looking for 
homogeneous solutions where
$(E^x)'=(E^\varphi)'=P_\phi'=K_x'=K_\varphi'=\phi'=0$.  We call the time
coordinate $t_{\rm h}$ and the spatial coordinate $x_{\rm h}$.  A dot refers
to a derivative with respect to $t_{\rm h}$, and a prime to  a derivative with
respect to $x_{\rm h}$.  It can be shown that the partial gauge
fixing
\begin{equation}
    N^x = 0
    \ , \hspace{1cm}
    N' = 0
    \label{eq:Partial gauge fixing - Homogeneous solution}
\end{equation}
allows initial homogeneous data to remain homogeneous during evolution.  The
conserved scalar charge is $G=P_\phi/\cos(\bar{\nu}\phi)$, such that
$\Dot{G}=0$ even locally thanks to spatial homogeneity.

We now restrict ourselves to on-shell solutions.  The diffeomorphism
constraint is automatically satisfied by the homogeneity condition, and we
solve the Hamiltonian constraint \eqref{eq:Hamiltonian constraint - CML} for
\begin{eqnarray}
    K_x =
    - \frac{E^\varphi}{4 E^x} \frac{2 \bar{\lambda}}{\sin \left(2 \bar{\lambda} K_{\varphi}\right)}
    \left( 1 + \frac{\sin^2 \left(\bar{\lambda} K_{\varphi}\right)}{\bar{\lambda}^2} - \frac{G^2}{(E^\varphi)^2} |\cos\left(\bar{\lambda} K_{\varphi}\right)| \right)
    \ .
\end{eqnarray}
We write equations of motion with respect to  $K_\varphi$, 
using $\dot{A} / \dot{K}_\varphi = {\rm d} A / {\rm d} K_\varphi$ for
any phase-space function $A$: We have
\begin{eqnarray}
  &&  \frac{{\rm d} \ln \left((E^\varphi)^2/\cos^2\left(\bar{\lambda} K_{\varphi}\right)\right)}{{\rm d} \left( \sin \left(\bar{\lambda} K_{\varphi}\right) / \bar{\lambda} \right)}\nonumber\\
    &=&
    2 \frac{\bar{\lambda}}{\sin \left(\bar{\lambda} K_{\varphi}\right)} \left(
    \left(1-\frac{\sin^2(\bar{\lambda} K_\varphi)}{\bar{\lambda}^2}\right) \cos^2\left(\bar{\lambda} K_{\varphi}\right) \frac{(E^\varphi)^2}{\cos^2\left(\bar{\lambda} K_{\varphi}\right)}
    - G^2 \frac{\cos\left(2\bar{\lambda} K_{\varphi}\right)}{|\cos\left(\bar{\lambda} K_{\varphi}\right)|} \right)
    \nonumber\\
    && \times
    \left( \left(1+\frac{\sin^2(\bar{\lambda} K_\varphi)}{\bar{\lambda}^2}\right) \cos^2\left(\bar{\lambda} K_{\varphi}\right) \frac{(E^\varphi)^2}{\cos^2\left(\bar{\lambda} K_{\varphi}\right)} + G^2 |\cos\left(\bar{\lambda} K_{\varphi}\right)|\right)^{-1}
    \label{eq:E^phi EoM - Homogeneous - CML}
\end{eqnarray}
for the combination $(E^\varphi)^2/\cos^2(\bar{\lambda} K_{\varphi})$ that
  appears in the emergent space-time metric, 
\begin{eqnarray}
    \frac{{\rm d} \ln E^x}{{\rm d} K_{\varphi}}
    &=&
    - 4 \frac{\sin \left(2 \bar{\lambda} K_{\varphi}\right)}{2 \bar{\lambda}}
    \left(
    1
    + \frac{\sin^2\left(\bar{\lambda} K_{\varphi}\right)}{\bar{\lambda}^2}
    + \frac{\cos^2(\bar{\lambda} K_{\varphi})}{(E^{\varphi})^2} G^2 |\cos\left(\bar{\lambda} K_{\varphi}|\right)
    \right)^{-1}
    \label{eq:E^x EoM - Homogeneous - CML}
\end{eqnarray}
for the second configuration variable, and
\begin{eqnarray}
    &&\frac{{\rm d}}{{\rm d} K_{\varphi}} \left(\frac{\sin
       (\bar{\nu}\phi)}{\bar{\nu}}\right)\nonumber\\ 
    &=&
    - 2 G \frac{\cos^3\left(\bar{\lambda} K_{\varphi}\right)}{E^{\varphi}}
    \Bigg(
    |\cos\left(\bar{\lambda} K_{\varphi}\right)| \left(1 + \frac{\sin^2\left(\bar{\lambda} K_{\varphi}\right)}{\bar{\lambda}^2}\right)
    + G^2 \frac{\cos^2\left(\bar{\lambda} K_{\varphi}\right)}{(E^{\varphi})^2} 
    \Bigg)^{-1}
    \label{eq:phi EoM - Homogeneous - CML}
\end{eqnarray}
for the scalar field.
The dependence on $t_{\rm h}$ is then given by using the solution of
\begin{equation}
    \dot{K}_\varphi
    =
    - \lambda_0 N \frac{\cos^2(\bar{\nu} \phi)}{2 \sqrt{E^x}} \left(
    1 + \frac{\sin^2\left(\bar{\lambda} K_{\varphi}\right)}{\bar{\lambda}^2}
    + \frac{G^2}{(E^\varphi)^2} |\cos\left(\bar{\lambda} K_{\varphi}\right)|
    \right)
    \ .
    \label{eq:K_phi time evolution - homogeneous - CML}
\end{equation}
Multiplying equations (\ref{eq:phi EoM - Homogeneous - CML}) and
(\ref{eq:K_phi time evolution - homogeneous - CML}), we see that the time
derivative of $\sin(\bar{\nu}\phi)/\bar{\nu}$ vanishes at the extrema of the
sine function. Evolution therefore respects the bounds of this function.

Equation~\eqref{eq:E^phi EoM - Homogeneous - CML} can be solved for
$E^\varphi$ such that
\begin{eqnarray}
  &&  \frac{(E^\varphi)^2}{\cos^2(\bar{\lambda} K_\varphi)}\label{eq:E^phi - Homogeneous - simple case}\\
  &=&
    \frac{c_\varphi^2}{4} \frac{\sin^2(\bar{\lambda} K_\varphi)}{\bar{\lambda}^2} \left(1 + \frac{\sin^2(\bar{\lambda} K_\varphi)}{\bar{\lambda}^2}\right)^{-2}
    \left( 1 + \sqrt{ 1 + \frac{4}{c_\varphi^2} \frac{G^2}{|\cos(\bar{\lambda} K_\varphi)|} \left(1 + \frac{\bar{\lambda}^2}{\sin^2(\bar{\lambda} K_\varphi)}\right)} \right)^2
    \ ,\nonumber
\end{eqnarray}
where $c_\varphi$ is the integration constant. We chose the sign of the square
root so as to obtain a non-vanishing vacuum limit at $G \to 0$. The classical
limit sets $c_\varphi^2 = 2 M$.  The right-hand side of (\ref{eq:E^phi -
  Homogeneous - simple case}) diverges as $\sec (\bar{\lambda} K_\varphi)$ at
$K_\varphi = - \pi / (2 \bar{\lambda})$. Multiplying with
$\cos^2(\bar{\lambda} K_{\varphi})$ shows that $E^{\varphi}$ approaches zero at
$K_\varphi = - \pi / (2 \bar{\lambda})$ and does not diverge, but the relevant
combination of $E^{\varphi}$ and $K_{\varphi}$ in the spatially homogeneous
emergent space-time metric is $(E^{\varphi})^2/\cos^2(\bar{\lambda} K_{\varphi})$,
given by the left-hand side of (\ref{eq:E^phi - Homogeneous - simple case})
without multiplication by $\cos^2(\bar{\lambda} K_{\varphi})$. This combination
diverges at $K_\varphi = - \pi / (2 \bar{\lambda})$.

We can then use this result in (\ref{eq:E^x EoM - Homogeneous - CML}) and
(\ref{eq:phi EoM - Homogeneous - CML}) and directly integrate to get $E^x$ and
$\phi$.  The exact integrations are too complicated.  However, it suffices
to note that the right-hand sides of (\ref{eq:E^x EoM - Homogeneous - CML})
and (\ref{eq:phi EoM - Homogeneous - CML}) remain finite even at
$K_\varphi = - \pi / (2 \bar{\lambda})$. The right-hand side of the $\phi$-equation
vanishes at this value, such that $\phi$ remains finite,
independently of the bounded range of the sine function, and reaches a local
maximum at $K_\varphi = - \pi / (2 \bar{\lambda})$ if it has initially been
increasing.  The $K_{\varphi}$-derivative of $\ln(E^x)$ reaches a negative
value at $K_\varphi = - \pi / (2 \bar{\lambda})$, such that $E^x$ continues to
decrease from its initial value in a collapse model, staying finite. The
crucial factor in the radial component $q_{xx}$ of the emergent space-time
metric is therefore (\ref{eq:E^phi - Homogeneous - simple case}) which
diverges at $K_\varphi = - \pi / (2 \bar{\lambda})$. 

\subsubsection{Internal time gauge}

Instead of integrating (\ref{eq:K_phi time evolution - homogeneous - CML}), we
can complete the gauge by choosing a new homogeneous time coordinate
$t_\varphi = - K_\varphi$.  The resulting consistency equation
$\dot{K}_\varphi=-1$ then determines the lapse function
\begin{equation}
    N
    =
    \frac{1}{\lambda_0} \frac{2 \sqrt{E^x}}{\cos^2(\bar{\nu} \phi)} \left(
    1 + \frac{\sin^2\left(\bar{\lambda} K_{\varphi}\right)}{\bar{\lambda}^2}
    + \frac{G^2}{(E^\varphi)^2} |\cos\left(\bar{\lambda} K_{\varphi}\right)|
    \right)^{-1}
    \ .
    \label{eq:Lapse - homogeneous - internal time gauge - CML}
\end{equation}
Since $(E^\varphi)^2 \propto \cos (\bar{\lambda} K_\varphi)$ as
$K_\varphi = - \pi / (2 \bar{\lambda})$, the lapse function remains finite at
$K_\varphi = - \pi / (2 \bar{\lambda})$ provided $\phi \neq \pi / (2\bar{\nu})$.

The Ricci scalar of a spatially homogeneous metric of the form
${\rm d} s^2 = - N^2 {\rm d} t_\varphi^2 + q_{xx} {\rm d} x^2 + E^x {\rm d}
\Omega^2$ is given by
\begin{eqnarray}
    R &=&
    - \frac{1}{2 N^2}
    \Bigg(
    (\partial_{t_\varphi} \ln E^x)^2
    + \left(\partial_{t_\varphi} \ln \frac{N^2}{(E^x)^2}\right) (\partial_{t_\varphi} \ln q_{xx})
    + (\partial_{t_\varphi} \ln q_{xx}))^2 - 2 \frac{\Ddot{q}_{xx}}{q_{xx}{}^2}
    \nonumber\\
    &&
    - 4 \left( \frac{N^2}{E^x} - (\partial_{t_\varphi} \ln N) (\partial_{t_\varphi} \ln E^x) + \frac{\Ddot{E}^x}{(E^x)^2}\right)
    \Bigg)
    \ .
    \label{eq:Ricci scalar of homogeneous metric}
\end{eqnarray}
At the maximum-curvature surface, $K_\varphi = - \pi / (2 \bar{\lambda})$, the
Ricci scalar diverges as 
\begin{eqnarray}
    R |_{K_\varphi \approx - \pi / (2 \bar{\lambda})} &\propto&
    \left(\partial_{t_\varphi} \ln \frac{N^2}{(E^x)^2}\right) (\partial_{t_\varphi} \ln q_{xx})
    + (\left(\partial_{t_\varphi} \ln q_{xx})\right)^2 - 2 \frac{\Ddot{q}_{xx}}{q_{xx}{}^2}
    \nonumber\\
    &\sim&
    \tan^2 (\bar{\lambda} K_\varphi)
    \ .
\end{eqnarray}
Thus, there is a physical singularity even though $K_{\varphi}$ remains finite.

\subsection{Hamiltonian constraints compatible with the limit of reaching the
  classical constraint surface}

Constraints compatible with the limit of reaching the classical constraint
surface are closest to modifications obtained from linear combinations of the
classical constraints with phase-space dependent coefficients. They may
therefore be considered matter versions of the non-singular black-hole models
analyzed in \cite{SphSymmEff,SphSymmEff2}. However, here we will find that
matter implies the existence of a physical singularity.

\subsubsection{Equations of motion}

We use the minimally coupled, polymerized version of the Hamiltonian
constraint (\ref{eq:Hamiltonian constraint - periodic - CCSL - scalar
  polymerized}) compatible with the existence of a gravitational observable,
considering only the case of $V_q = \lambda_0 = 0$,
$\alpha_i^x = 1$ and constant $\lambda_0$.  As in the previous
example, we look for a homogeneous solution where
$(E^x)'=(E^\varphi)'=P_\phi'=K_x'=K_\varphi'=\phi'=0$. The time coordinate is
again $t_{\rm h}$, and the spatial coordinate $x_{\rm h}$, a dot referring to
derivative with respect to $t_{\rm h}$, a prime to a derivative with respect
to $x_{\rm h}$, and the partial gauge fixing $N^x = 0$ and $N' = 0$ allows
initial homogeneous data to remain homogeneous during evolution.  The
conserved charge is $G=P_\phi/\cos(\bar{\nu}\phi)$, such that $\Dot{G}=0$.

We now turn to on-shell solutions.  The diffeomorphism
constraint is automatically satisfied by the homogeneity condition, and we
solve the Hamiltonian constraint \eqref{eq:Hamiltonian constraint - periodic - CCSL - scalar
  polymerized} for
\begin{eqnarray}
    K_x =
    \frac{E^\varphi}{4 E^x} \frac{2 \bar{\lambda}}{\sin \left(2 \bar{\lambda} K_{\varphi}\right)}
    \left( 1 + \frac{\sin^2 \left(\bar{\lambda}
  K_{\varphi}\right)}{\bar{\lambda}^2} - \frac{G^2}{(E^\varphi)^2} \right) 
    \ .
    \label{eq:K_x - Homogeneous - CCSL}
\end{eqnarray}
We write the equations of motion using 
$K_\varphi$ as the evolution parameter using $\dot{A} / \dot{K}_\varphi = {\rm
  d} A / {\rm d} K_\varphi$ for any phase space function $A$:
\begin{eqnarray}
&&    \frac{{\rm d} \ln \left((E^\varphi)^2/\cos^2\left(\bar{\lambda} K_{\varphi}\right)\right)}{{\rm d} K_{\varphi}}\nonumber\\
    &=&
    \left(
    2 \left(1-\frac{\sin^2(\bar{\lambda} K_\varphi)}{\bar{\lambda}^2}\right) \frac{\bar{\lambda} \cos^2\left(\bar{\lambda} K_{\varphi}\right)}{\tan \left(\bar{\lambda} K_{\varphi}\right)}
    - G^2 \frac{\cos^2\left(\bar{\lambda} K_{\varphi}\right)}{(E^\varphi)^2} \frac{2 \bar{\lambda}\left(1-3\cos (2\bar{\lambda} K_\varphi)\right)}{\sin (2 \bar{\lambda} K_\varphi)} \right)
    \nonumber\\
    &&\times
    \left( \left(1+\frac{\sin^2(\bar{\lambda} K_\varphi)}{\bar{\lambda}^2}\right) \cos^2 \left(\bar{\lambda} K_{\varphi}\right)
    + G^2 \frac{\cos^2\left(\bar{\lambda} K_{\varphi}\right)}{(E^\varphi)^2} \right)^{-1} \,,
    \label{eq:E^phi EoM - Homogeneous - CCSL}\\
&&    \frac{{\rm d} \ln E^x}{{\rm d} K_{\varphi}}
    =
    - 4 \frac{\sin \left(2 \bar{\lambda} K_{\varphi}\right)}{2 \bar{\lambda}}
    \left(
    1 + \frac{\sin^2\left(\bar{\lambda} K_{\varphi}\right)}{\bar{\lambda}^2} +
   \frac{G^2}{(E^\varphi)^2} \right)^{-1} \,,
    \label{eq:E^x EoM - Homogeneous - CCSL}\\
  &&  \frac{{\rm d}}{{\rm d} K_{\varphi}} \left(\frac{\sin (\bar{\nu}\phi)}{\bar{\nu}}\right)
    =
    - 2 \frac{G}{E^\varphi} \left(
    1 + \frac{\sin^2\left(\bar{\lambda} K_{\varphi}\right)}{\bar{\lambda}^2} + \frac{G^2}{(E^\varphi)^2}\right)^{-1}
    \label{eq:phi EoM - Homogeneous - CCSL}
\end{eqnarray}
where $K_{\varphi}$ depends on $t_{\rm h}$ according to the solution of
\begin{equation}
    \dot{K}_\varphi
    =
    - \lambda_0 N \frac{\cos^2(\bar{\nu} \phi)}{2 \sqrt{E^x}} \left(
    1 + \frac{\sin^2\left(\bar{\lambda} K_{\varphi}\right)}{\bar{\lambda}^2} + \frac{G^2}{(E^\varphi)^2}
    \right) \,.
    \label{eq:K_phi time evolution - homogeneous - CCSL}
\end{equation}

Equation~\eqref{eq:E^phi EoM - Homogeneous - CCSL} can be solved for
\begin{eqnarray}
    \frac{(E^\varphi)^2}{\cos^2 (\bar{\lambda} K_\varphi)} &=&
    \frac{c_\varphi^2}{4} \frac{\sin^2(\bar{\lambda} K_\varphi)}{\bar{\lambda}^2} \left(1 + \frac{\sin^2(\bar{\lambda} K_\varphi)}{\bar{\lambda}^2}\right)^{-2}
    \nonumber\\
    &&\times
    \left( 1 + \sqrt{1
    + G^2 \frac{4}{c_\varphi^2} \left(\frac{2 \bar{\lambda}}{\sin(2 \bar{\lambda} K_\varphi)}\right)^2 \left(1 + \frac{\sin^2(\bar{\lambda} K_\varphi)}{\bar{\lambda}^2}\right)} \right)^2
    \ ,
    \label{eq:E^phi - Homogeneous - CCSL}
\end{eqnarray}
where $c_\varphi$ is an integration constant, and we chose the sign of the
square root to obtain a non-vanishing vacuum limit if $G \to 0$. The classical
limit determines $c_\varphi^2 = 2 M$.  The expression (\ref{eq:E^phi -
  Homogeneous - CCSL}) diverges as $\sec^2 (\bar{\lambda} K_\varphi)$, such that
$E^\varphi$ now remains finite at $K_\varphi = - \pi / (2 \bar{\lambda})$. There is
a clear distinction between this behavior for $G\not=0$ and the vacuum limit of
$G=0$, where $E^{\varphi}$ approaches zero at $K_\varphi = - \pi / (2
\bar{\lambda})$. For $G=0$, this model is equivalent to the minimal-coupling
extension from \cite{SphSymmMinCoup} of the models analyzed in
\cite{SphSymmEff,SphSymmEff2}. Sincgularity-freedom observed in the latter
papers is therefore shown to be unstable under the inclusion of minimally
coupled matter.

Equation (\ref{eq:K_x - Homogeneous - CCSL}) then shows that $K_x$ diverges as
$\sec (\bar{\lambda} K_\varphi)$.  Equations (\ref{eq:E^x EoM - Homogeneous - CCSL})
and (\ref{eq:phi EoM - Homogeneous - CCSL}) imply that $E^x$ and $\phi$ remain
finite with $E^x$ achieving its minimum value (in a collapse model) at
$K_\varphi = - \pi / (2 \bar{\lambda})$, while the value of $\phi$ depends on
initial conditions.

\subsubsection{Internal-time gauge}

We complete the gauge by choosing a new time coordinate as
$t_\varphi = - K_\varphi$.  The consistency equation $\dot{K}_\varphi=-1$
determines the lapse function
\begin{equation}
    N
    =
    \frac{1}{\lambda_0} \frac{2 \sqrt{E^x}}{\cos^2(\bar{\nu} \phi)} \left(
    1 + \frac{\sin^2\left(\bar{\lambda} K_{\varphi}\right)}{\bar{\lambda}^2} + \frac{G^2}{(E^\varphi)^2}
    \right)^{-1}
    \label{eq:Lapse - homogeneous - internal time gauge - CCSL}
\end{equation}
which remains finite at $K_\varphi = - \pi / (2 \bar{\lambda})$ provided $\phi \neq \pi / (2\bar{\nu})$.

Using (\ref{eq:Ricci scalar of homogeneous metric}) for the expression of the
Ricci scalar of a spatially homogeneous metric, we find that at the maximum-curvature
surface, $K_\varphi = - \pi / (2 \bar{\lambda})$, the Ricci scalar diverges as
\begin{eqnarray}
    R |_{K_\varphi \approx - \pi / (2 \bar{\lambda})} &\propto&
    \left(\partial_{t_\varphi} \ln \frac{N^2}{(E^x)^2}\right) (\partial_{t_\varphi} \ln q_{xx})
    + (\left(\partial_{t_\varphi} \ln q_{xx})\right)^2 - 2 \frac{\Ddot{q}_{xx}}{q_{xx}{}^2}
    \nonumber\\
    &\sim&
    \tan^2 (\bar{\lambda} K_\varphi)
    \ .
\end{eqnarray}
Thus, this hypersurface is a physical singularity for $G\not=0$, as in the previous
example. We emphasize again that the behavior of phase-space functions at the
maximum-curvature hypersurface depends significantly on whether $G$ is zero or
non-zero. The non-singular example of $G=0$ is therefore unstable under
perturbation by matter terms.

\subsection{Singularity-free Hamiltonian}

Our last example of a class of consistent Hamiltonians was not motivated by
the existence of specific limits or observables, but we now show that it
improves the singularity behavior of the previous two examples.

\subsubsection{Equations of motion}

We again consider a special case, given by $c_f=\alpha_i=1$, and
$\lambda_0=V=V_q=V^q=q=0$, and look for a homogeneous solution where
$(E^x)'=(E^\varphi)'=P_\phi'=K_x'=K_\varphi'=\phi'=0$ in terms of a time
coordinate $t_{\rm h}$ and a spatial coordinate $x_{\rm h}$, where a dot
refers to a derivative with respect to the former and a prime to a derivative
with respect to the latter. The partial gauge fixing $N^x=0$ and $N'=0$ allows
initial homogeneous data to remain homogeneous during evolution.  The
conserved charge is $G=P_\phi/\cos(\bar{\nu}\phi)$, such that $\Dot{G}=0$.

For on-shell solutions, the diffeomorphism constraint is automatically
satisfied by the homogeneous condition, and we solve the Hamiltonian
constraint, \eqref{eq:Hamiltonian constraint - DF - scalar polymerization}, for 
\begin{eqnarray}
    K_x &=&
    - \frac{E^\varphi}{4} \frac{2 \bar{\lambda}}{\sin \left(2 \bar{\lambda} K_{\varphi}\right)} \left(
    \frac{1}{2 E^x}
    + \frac{1}{E^x} \frac{\sin^2 \left(\bar{\lambda} K_{\varphi}\right)}{\bar{\lambda}^2}
    \right)
    + \frac{G^2}{E^\varphi} \frac{1}{4 E^x} \frac{\bar{\lambda}}{\tan \left(\bar{\lambda} K_{\varphi}\right)}
    \ .
    \label{eq:K_x - Homogeneous - DF}
\end{eqnarray}
We now obtain the equations of motion with respect to $K_\varphi$, as the
evolution parameter using
$\dot{A} / \dot{K}_\varphi = {\rm d} A / {\rm d} K_\varphi$ for any phase
space function $A$.  After some simplifications, the equations are
\begin{eqnarray}
    \frac{{\rm d} \ln \left((E^\varphi)^2\right)}{{\rm d} K_\varphi}
    &=&
    2 \frac{2 \bar{\lambda}}{\sin\left(2 \bar{\lambda} K_{\varphi}\right)}
    \frac{\cos \left(2 \bar{\lambda} K_\varphi\right)
    - \frac{\sin ^2\left(\bar{\lambda}  K_\varphi\right)}{\bar{\lambda}^2}
    - G^2 \frac{\cos^2\left(\bar{\lambda} K_{\varphi}\right)}{(E^\varphi)^2}}{1
    + \frac{\sin^2\left(\bar{\lambda}  K_\varphi\right)}{\bar{\lambda}^2}
    + G^2 \frac{\cos^2\left(\bar{\lambda} K_{\varphi}\right)}{(E^\varphi)^2}}
    \ ,
    \label{eq:E^phi EoM - Homogeneous - DF}
\end{eqnarray}
\begin{eqnarray}
    \frac{{\rm d} \ln E^x}{{\rm d} \left( \sin \left(\bar{\lambda} K_{\varphi}\right)/\bar{\lambda}\right)}
    &=&
    - 4 \frac{\sin \left(\bar{\lambda} K_{\varphi}\right)}{\bar{\lambda}}
    \Bigg(
    1
    + \frac{\sin^2\left(\bar{\lambda} K_{\varphi}\right)}{\bar{\lambda}^2}
    + \frac{G^2}{E_{\varphi}^2} \left( 1 - \sin^2\left(\bar{\lambda} K_{\varphi}\right) \right)
    \Bigg)^{-1}
    \ .
    \label{eq:E^x EoM - Homogeneous - DF}
\end{eqnarray}
and
\begin{eqnarray}
    \frac{{\rm d} (\sin (\bar{\nu}\phi)/\bar{\nu})}{{\rm d} \left( \sin \left(\bar{\lambda} K_{\varphi}\right)/\bar{\lambda}\right)}
    &=&
    - 2 \frac{G}{E^{\varphi}} \cos \left(\bar{\lambda} K_{\varphi}\right)
    \Bigg(
    1
    + \frac{\sin^2\left(\bar{\lambda} K_{\varphi}\right)}{\bar{\lambda}^2}
    + G^2 \frac{\cos^2\left(\bar{\lambda} K_{\varphi}\right)}{(E^{\varphi})^2}
    \Bigg)^{-1}
    \ ,
    \ ,
    \label{eq:phi EoM - Homogeneous - DF}
\end{eqnarray}
where dependence on $t_{\rm h}$ is given by the solution of
\begin{equation}
    \dot{K}_\varphi
    =
    - \lambda_0 N \frac{\cos^2(\bar{\nu} \phi)}{2 \sqrt{E^x}} \left(
    1 + \frac{\sin^2\left(\bar{\lambda} K_{\varphi}\right)}{\bar{\lambda}^2}
    + G^2 \frac{\cos^2\left(\bar{\lambda} K_{\varphi}\right)}{(E^\varphi)^2}
    \right)
    \ .
    \label{eq:K_phi time evolution - homogeneous - DF}
\end{equation}

Equation~(\ref{eq:E^phi EoM - Homogeneous - DF}) can be solved for
\begin{eqnarray}
  &&    \frac{(E^\varphi)^2}{\cos^2(\bar{\lambda} K_\varphi)}\label{eq:E^phi - Homogeneous - DF}\\
  &=&
    \frac{c_\varphi{}^2}{4} \left(1+\frac{\sin^2 (\bar{\lambda} K_\varphi)}{\bar{\lambda}^2}\right)^{-2}
    \left(
    \frac{\sin(\bar{\lambda} K_\varphi)}{\bar{\lambda}}
    + \sqrt{\frac{\sin^2(\bar{\lambda} K_\varphi)}{\bar{\lambda}^2}
    + \frac{4 G^2}{c_\varphi^2} \left(1+\frac{\sin^2 (\bar{\lambda} K_\varphi)}{\bar{\lambda}^2}\right)}
    \right)^2
    \ ,\nonumber
\end{eqnarray}
where $c_\varphi$ is the integration constant, and we chose the sign of the
square root to obtain the correct vacuum limit at $G \to 0$. Unlike in the
previous two examples, this function does not diverge at
$\bar{\lambda} K_{\varphi}=\pi/2$.

We can then use this result in \eqref{eq:E^x EoM - Homogeneous - DF} to solve
for $E^x$ by the complicated function
\begin{eqnarray}
    E^x
    &=&
    \frac{- c_x c_\varphi{}^{2 / \sigma}}{2 c_\varphi{}^2\sigma^2}
    \left(- \sigma \frac{\sin(\bar{\lambda} K_\varphi)}{\bar{\lambda}}
    + \sqrt{\frac{\sin^2(\bar{\lambda} K_\varphi)}{\bar{\lambda}^2}
    + \frac{4 G^2}{c_\varphi{}^2} \left(1+\frac{\sin^2 (\bar{\lambda}
        K_\varphi)}{\bar{\lambda}^2}\right)} \right)^{2 + 2/\sigma}
    \notag\\
    && \times
    \bigg( 1
    + \sigma \left(1+\frac{\sin^2 (\bar{\lambda} K_\varphi)}{\bar{\lambda}^2}\right)
    - \frac{\sin (\bar{\lambda} K_\varphi)}{\bar{\lambda}} \sqrt{ \frac{\sin^2(\bar{\lambda} K_\varphi)}{\bar{\lambda}^2}
    + \frac{4 G^2}{c_\varphi{}^2} \left(1+\frac{\sin^2 (\bar{\lambda} K_\varphi)}{\bar{\lambda}^2}\right)} \bigg)
    \notag\\
    && \times
    \bigg( 1
    - \sigma \left(1+\frac{\sin^2 (\bar{\lambda} K_\varphi)}{\bar{\lambda}^2}\right)
    + \frac{\sin (\bar{\lambda} K_\varphi)}{\bar{\lambda}} \sqrt{ \frac{\sin^2(\bar{\lambda} K_\varphi)}{\bar{\lambda}^2}
    + \frac{4 G^2}{c_\varphi{}^2} \left(1+\frac{\sin^2 (\bar{\lambda} K_\varphi)}{\bar{\lambda}^2}\right)} \bigg)^{-1}
    \notag\\
    && \times
    \left(1+\frac{\sin^2 (\bar{\lambda} K_\varphi)}{\bar{\lambda}^2}\right)^{-1}
    \Bigg(
    \frac{\sin^2(\bar{\lambda} K_\varphi)}{\bar{\lambda}^2}
    + \frac{2 G^2}{c_\varphi{}^2} \left( 1
    + 2 \frac{\sin^2(\bar{\lambda} K_\varphi)}{\bar{\lambda}^2} \right)
    \notag\\
    &&
    - \frac{\sin(\bar{\lambda} K_\varphi)}{\bar{\lambda}} \sigma
    \sqrt{\frac{\sin^2(\bar{\lambda} K_\varphi)}{\bar{\lambda}^2}
    + \frac{4 G^2}{c_\varphi{}^2} \left(1+\frac{\sin^2 (\bar{\lambda} K_\varphi)}{\bar{\lambda}^2}\right)}
    \Bigg)^{-1}
    \ ,
    \label{eq:E^x - Homogeneous - DF}
\end{eqnarray}
where $c_x$ is a constant of integration and we have introduced $\sigma=\sqrt{1+4G^2/c_{\varphi}^2}$.
Equation~(\ref{eq:phi EoM - Homogeneous - DF}) can be solved for $\phi$ by
\begin{eqnarray}
    \frac{\sin (\bar{\nu}\phi)}{\bar{\nu}} &=&
    \frac{\sin (\bar{\nu}\phi_{\rm H})}{\bar{\nu}}
    + \frac{2 G}{c_\varphi\sigma} \ln \left( \frac{\sqrt{4
                                               G^2/c_\varphi^2 + \sigma^2
                                               \sin^2(\bar{\lambda}
                                               K_\varphi)/\bar{\lambda}^2} 
    - \sigma |\sin(\bar{\lambda} K_\varphi)|/\bar{\lambda}}{2 G/c_\varphi}\right)
    \ ,
    \label{eq:phi - Homogeneous - DF}
\end{eqnarray}
where $\phi_{\rm H}$ is the integration constant, representing the value of the
scalar field at $K_\varphi=0$.  Because this expression is
bounded by $|\phi| = \pi / (2 \bar{\nu})$, it implies a bound on the conserved
quantity $G$ and the initial condition $\phi_{\rm H}$.

\subsubsection{Vacuum limit and homogeneous Schwarzschild gauge}

In order to understand the role of the constants of integration, we look at the vacuum
limit $\phi , G \to 0$ where the expressions reduce to
\begin{subequations}
\begin{eqnarray}
    E^\varphi &\to&
    c_\varphi \left(1+\frac{\sin^2 (\bar{\lambda} K_\varphi)}{\bar{\lambda}^2}\right)^{-1}
    \frac{\sin(2 \bar{\lambda} K_\varphi)}{2 \bar{\lambda}}
    \ ,
    \label{eq:E^phi - Homogeneous - simple case - vacuum limit}
\end{eqnarray}
and
\begin{eqnarray}
    E^x
    &=&
    \frac{- c_x c_\varphi^{ 2
    + O(G^2)}}{2 c_\varphi^2 \left(1
    + O (G^2) \right)}
    \left(
    \frac{2 G^2}{c_\varphi^2} \frac{\bar{\lambda}}{\sin(\bar{\lambda} K_\varphi)}
    + O(G^4)
    \right)^{ 4
    + O(G^2)}
    \bigg( 2
    + O (G^4) \bigg)
    \notag\\
    && \times
    \bigg( - \frac{2 G^4}{c_\varphi^4} \frac{\bar{\lambda}^2}{\sin^2 (\bar{\lambda} K_\varphi)} \left(1 + \frac{\sin^2 (\bar{\lambda} K_\varphi)}{\bar{\lambda}^2} \right) + O (G^6) \bigg)^{-1}
    \notag\\
    && \times
    \left(1+\frac{\sin^2 (\bar{\lambda} K_\varphi)}{\bar{\lambda}^2}\right)^{-1}
    \Bigg(\frac{2 G^4}{c_\varphi^4} \frac{\bar{\lambda}^2}{\sin^2 (\bar{\lambda} K_\varphi)}
    + O (G^6)
    \Bigg)^{-1}
    \notag\\
    &\to&
    c_x
    \left(1 + \frac{\sin^2 (\bar{\lambda} K_\varphi)}{\bar{\lambda}^2} \right)^{-2}
    \ ,
    \label{eq:E^x - Homogeneous - simple case - vacuum limit}
\end{eqnarray}
where $K_{\varphi}$ depends on $t_{\rm h}$ according to
\begin{equation}
    \dot{K}_\varphi
    \to
    - \lambda_0 \frac{N}{2 \sqrt{c_x}} \left(
    1 + \frac{\sin^2\left(\bar{\lambda} K_{\varphi}\right)}{\bar{\lambda}^2}
    \right)^2
    \label{eq:K_phi time evolution - homogeneous - simple case - vacuum limit}
\end{equation}
    \label{EoMs - homogeneous - simple case - vacuum limit}
\end{subequations}
for a given $N$.

Alternatively, we can complete the gauge by assuming the (homogeneous)
Schwarzschild condition $E^x = t_{\rm h}^2$ suitable for a black-hole
interior. We then obtain $K_\varphi$ as a function of $t_{\rm h}$ by imposing
the consistency equation $\dot{E}^x=2t_{\rm h}$, using (\ref{eq:E^x -
  Homogeneous - DF}), and then inverting for $K_\varphi$.
We simplify the required inversion by working with the vacuum equations
(\ref{eq:E^x - Homogeneous - simple case - vacuum limit}). 
We obtain
\begin{align}
    \frac{\sin^2 (\bar{\lambda} K_\varphi)}{\bar{\lambda}^2}
    = \frac{\sqrt{c_x}}{t_{\rm h}} - 1
    \,.
    \label{eq:K_phi - Homogeneous - simple case - vacuum - Schwarzschild gauge}
\end{align}
Inverting this expression for $K_{\varphi}(t_{\rm h})$, we have
\begin{eqnarray}
    \frac{E^\varphi}{\sqrt{E^x}}
    &=&
    \left(1 - \bar{\lambda}^2 \left( \frac{\sqrt{c_x}}{t_{\rm h}} - 1\right)\right)^{-1/2} \left(\frac{\sqrt{c_x}}{t_{\rm h}} - 1\right)^{-1/2} \frac{c_\varphi}{\sqrt{c_x}}
    \ ,
    \label{eq:E^phi - Homogeneous - simple case - vacuum - Schwarzschild gauge}
\end{eqnarray}
and
\begin{eqnarray}
    N
    &=&
    \lambda_0^{-1} \left(1 - \bar{\lambda}^2 \left(\frac{\sqrt{c_x}}{t_{\rm h}} - 1\right)\right)^{- 1/2}
    \left(\frac{\sqrt{c_x}}{t_{\rm h}} - 1\right)^{-1/2}
    \ ,
    \label{eq:K_phi - Homogeneous - simple case - vacuum - Schwarzschild gauge2}
\end{eqnarray}

The structure function is
\begin{eqnarray}
    q^{x x}
    &=&
    \left(\frac{\sqrt{c_x}}{t_{\rm h}} - 1\right) \frac{\lambda_0^2}{c_\varphi^2}
    \ ,
\end{eqnarray}
and the emergent space-time metric reads
\begin{eqnarray}\label{eq:Spacetime metric - homogeneous - Schwarzschild - DF}
&&    {\rm d} s^2\\
    &=&
    - \left(1 - \bar{\lambda}^2 \left(\frac{\sqrt{c_x}}{t_{\rm h}} - 1\right)\right)^{- 1}
    \left(\frac{\sqrt{c_x}}{t_{\rm h}} - 1\right)^{-1} \frac{{\rm d} t_{\rm h}^2}{\lambda_0^2}
    + \left(\frac{\sqrt{c_x}}{t_{\rm h}} - 1\right)^{-1} \frac{c_\varphi{}^2 {\rm d} x_{\rm h}^2}{c_x \lambda_0^2}
    + t_{\rm h}^2 {\rm d} \Omega^2
    \ .\nonumber
\end{eqnarray}
A comparison with the exterior metric (or an application of the mass
observable) then sets $c_x = 4 M^2$ and $c_\varphi = 2 M \lambda_0 / \mu$ with
$\mu$ a constant that scales the metric.

\subsubsection{Internal time gauge}

We can instead complete the gauge by using $K_{\varphi}$ as our time
coordinate, $t_\varphi = - K_\varphi$.  The lapse function can then be
obtained from the consistency equation $\dot{K}_\varphi = -1$:
\begin{align}
    N =&
    \frac{2 \sqrt{E^x} \sec^2(\bar{\nu} \phi) / \lambda_0}{1 + \sin^2\left(\bar{\lambda} t_{\varphi}\right)/\bar{\lambda}^2
    + (G^2/(E^{\varphi})^2) \cos^2\left(\bar{\lambda} t_{\varphi}\right)}
    \ .
\end{align}
The emergent space-time metric is determined by using this lapse function in
the time component and the structure function \eqref{eq:Covariant structure
  function} in the radial component,  replacing the solutions \eqref{eq:E^phi
  - Homogeneous - DF}, \eqref{eq:E^x - Homogeneous - DF}, and \eqref{eq:phi -
  Homogeneous - DF} with $K_\varphi = - t_\varphi$.  The expression is quite
lengthy, but we can obtain meaningful results as follows.

Since the internal time takes the same values as $K_{\varphi}$, up to a sign
difference, it suffices to restrict ourselves to the range
$t_\varphi \in (0 , \pi / \bar{\lambda})$. Comparing with the classical
situation, $t_\varphi = 0$ represents the hypersurface matching the horizon of
the black hole, while the mid-point $t_\varphi = \pi / (2 \bar{\lambda})$ is a
new hypersurface with maximum-curvature effects (the would-be classical
singularity). Continuing through the allowed range,
$t_\varphi = \pi / \bar{\lambda}$ would be the hypersurface matching the
horizon of the black (or white) hole on the other side of the classical
singularity in the spirit of
\cite{SphSymmEff,SphSymmEff2,EmergentFluid}. These characteristic
hypersurfaces allow us to take specific limits in their proximity, resulting
in tractable equations.  For an infalling matter field, we assume $P_\phi < 0$
initially, such that $\phi$ starts growing as a function of $K_{\varphi}$ but
remains bounded.

The geometry of this process is described by the emergent space-time metric,
in which the inverse radial component, given by the structure 
\begin{eqnarray}
    q^{x x}
    &=&
    \lambda_0^2 \cos^4(\bar{\nu} \phi) E^x
    \cos^2 \left(\bar{\lambda} K_{\varphi}\right)
    \frac{1}{(E^\varphi)^2}
    \notag\\
    &=&
    \lambda_0^2 \cos^4(\bar{\nu} \phi)
    \frac{4}{c_\varphi^2} E^x \left(1+\frac{\sin^2 (\bar{\lambda} K_\varphi)}{\bar{\lambda}^2} \right)^{2}
    \notag\\
    && \times
    \left(
    \frac{\sin(\bar{\lambda} K_\varphi)}{\bar{\lambda}}
    + \sqrt{\frac{\sin^2(\bar{\lambda} K_\varphi)}{\bar{\lambda}^2}
    + \frac{4 P_\phi^2}{c_\varphi^2} \left(1+\frac{\sin^2 (\bar{\lambda} K_\varphi)}{\bar{\lambda}^2}\right)}
    \right)^{-2}
    \notag\\
    &\to&
    \frac{4 \lambda_0^2}{c_\varphi^2} \cos^4(\bar{\nu} \phi_0)
    \left(1+\bar{\lambda}^2 \right)^{2} \frac{E^x_0}{\bar{\lambda}^2}
    \left(
    1
    + \sqrt{1
    + \frac{4 P_\phi^2}{c_\varphi^2} \left(1+\bar{\lambda}^2\right)}
    \right)^{-2}
    \label{eq:Structure function - homogeneous - minimum}
\end{eqnarray}
remains regular, provided $\phi_0 \neq \pi / (2\bar{\nu})$. This conclusion is
stable with respect to matter perturbations.

\subsubsection{Near the maximum-curvature hypersurface}

Using the equations of motion, the behavior near the maximum-curvature hypersurface is given by
\begin{eqnarray}
    \frac{(E^\varphi)^2}{\cos^2(\bar{\lambda} K_\varphi)} &\approx&
    \frac{c_\varphi{}^2}{4} \left(\frac{\bar{\lambda}}{1+\bar{\lambda}^2}\right)^2
    \left( 1 + \sqrt{1
    + \frac{4 G^2}{c_\varphi^2} \left(1+\bar{\lambda}^2\right)}
    \right)^2
    \ ,
    \label{eq:E^phi - Homogeneous - DFNearMax}
\end{eqnarray}
and
\begin{eqnarray}
    E^x &\approx&
    - \frac{\bar{\lambda}^2 c_x}{4 (1 + \bar{\lambda}^2) c_\varphi^2 \left(c_\varphi^2 + 4 G^2\right)} \left(\frac{\sqrt{c_\varphi^2+4 \left(1+\bar{\lambda}^2\right) G^2}-\sqrt{c_\varphi^2+4 G^2}}{\bar{\lambda}}\right)^{2 c_\varphi/\sqrt{c_\varphi^2+4 G^2}}
    \\
    &&\times
    \frac{\bar{\lambda}^2 c_\varphi \sqrt{c_\varphi^2+4 G^2}
    + \left(1+\bar{\lambda}^2\right)\left(c_\varphi^2+4 G^2\right)
    - \sqrt{c_\varphi^2+4 G^2} \sqrt{c_\varphi^2+ \left(1+\bar{\lambda}^2\right) 4 G^2}}{\bar{\lambda}^2 c_\varphi \sqrt{c_\varphi^2+4 G^2}
    - \left(1+\bar{\lambda}^2\right)\left(c_\varphi^2+4 G^2\right)
    + \sqrt{c_\varphi^2+4 G^2} \sqrt{c_\varphi^2+ \left(1+\bar{\lambda}^2\right) 4 G^2}}\notag
\end{eqnarray}
for the gravitational fields, and
\begin{eqnarray}
    \frac{\sin(\bar{\nu}\phi)}{\bar{\nu}} \approx
  \frac{\sin(\bar{\nu}\phi_{\rm H})}{\bar{\nu}} + \frac{2 G /
  c_\varphi}{\sqrt{1 + 4 G^2/c_\varphi^2}} \ln \left( \frac{\sqrt{1 +
  (4 G^2/c_\varphi^2) \left(1 + \bar{\lambda}^2\right)} - \sqrt{1 +
  4 G^2/c_\varphi^2}}{2 \bar{\lambda} |G| / c_\varphi} \right) 
    \label{eq:Maximum scalar field - Homogeneous}
\end{eqnarray}
for the scalar field. The lapse function, appearing in the time component of
the emergent space-time metric, has the limit
\begin{eqnarray}
    N \approx
    \frac{2}{\lambda_0} \frac{\bar{\lambda}^2}{1+\bar{\lambda}^2} \sqrt{E^x} \sec^2(\bar{\nu} \phi)
\end{eqnarray}
\begin{subequations}
and we have the time derivatives
\begin{eqnarray}
    \frac{{\rm d} \left((E^\varphi)^2/ \cos^2\left(\bar{\lambda} t_{\varphi}\right)\right)}{{\rm d} t_\varphi}
    &\approx&
    0
    \ ,
    \label{eq:E^phi EoM - Homogeneous - NMCS}
\end{eqnarray}
\begin{eqnarray}
    \frac{{\rm d} E^x}{{\rm d} t_\varphi}
    &\approx& 0
    \ .
    \label{eq:E^x EoM - Homogeneous - NMCS}
\end{eqnarray}
\begin{eqnarray}
    \frac{{\rm d}}{{\rm d} t_\varphi} \left(\frac{\sin(\bar{\nu}\phi)}{\bar{\nu}}\right)
    &\approx& 0
    \label{eq:phi EoM - Homogeneous - NMCS}
\end{eqnarray}
at first order, as well as
\begin{eqnarray}
&&    \frac{{\rm d}^2 \left((E^\varphi)^2/\cos^2(\bar{\lambda} t_\varphi)\right)}{{\rm d} t_\varphi{}^2}\\
    &=&
    - \frac{(E^\varphi)^2}{\cos^2\left(\bar{\lambda} t_{\varphi}\right)} \Bigg( 2 \bar{\lambda}^2 \sec^2(\bar{\lambda} t_\varphi) - 8 \bar{\lambda}^2 \frac{\cos \left(2 \bar{\lambda} t_{\varphi}\right)}{\sin^2\left(2 \bar{\lambda} t_{\varphi}\right)} \frac{\cos \left(2 \bar{\lambda} t_\varphi\right) - \sin^2\left(\bar{\lambda}  t_\varphi\right)/\bar{\lambda}^2 - G^2 \cos^2\left(\bar{\lambda} t_{\varphi}\right)/(E^\varphi)^2}{1 + \sin^2\left(\bar{\lambda}  t_\varphi\right)/\bar{\lambda}^2 + G^2 \cos^2\left(\bar{\lambda} t_{\varphi}\right)/(E^\varphi)^2}
    \nonumber\\
    && + 4 \frac{  2 (1+\bar{\lambda}^2) - G^2 
       \bar{\lambda}\cos^2\left(\bar{\lambda} t_{\varphi}\right)/(\sin\left(2
       \bar{\lambda} t_{\varphi}\right) (E^\varphi)^2)
       {\rm d} \ln \left((E^\varphi)^2/\cos^2(\bar{\lambda}
       t_\varphi)\right)/{\rm d} t_\varphi}{1 +
       \sin^2\left(\bar{\lambda}  t_\varphi\right)/\bar{\lambda}^2 +
       G^2 \cos^2\left(\bar{\lambda} t_{\varphi}\right)/(E^\varphi)^2} 
    \nonumber\\
    &&
    + 2 \frac{\cos \left(2 \bar{\lambda} t_\varphi\right) -
       \sin^2\left(\bar{\lambda}  t_\varphi\right)/\bar{\lambda}^2 -
       G^2 \cos^2\left(\bar{\lambda}
       t_{\varphi}\right)/(E^\varphi)^2}{\left(1 +
       \sin^2\left(\bar{\lambda}  t_\varphi\right)/\bar{\lambda}^2 +
       G^2 \cos^2\left(\bar{\lambda}
       t_{\varphi}\right)/(E^\varphi)^2\right)^2} \left( 4 - \frac{2 G^2
       \bar{\lambda}}{\sin\left(2 \bar{\lambda} t_{\varphi}\right)} \frac{{\rm
       d} \ln \left((E^\varphi)^2/\cos^2(\bar{\lambda} t_\varphi)\right)}{{\rm
       d} t_\varphi}\right) \Bigg) 
    \nonumber\\
    &\approx&
    - c_\varphi{}^2 \frac{\bar{\lambda}^6}{(1+\bar{\lambda}^2)^3}
    \left( 1 + \sqrt{1
    + \frac{4 G^2}{c_\varphi^2} \left(1+\bar{\lambda}^2\right)}
    \right)^4
    \nonumber\\
    &&\times
    \left( \left( 1 + \sqrt{1
    + \frac{4 G^2}{c_\varphi^2} \left(1+\bar{\lambda}^2\right)}
    \right)^2
    + \frac{4}{c_\varphi{}^2} G^2 (1+\bar{\lambda}^2) \right)^{-1}
    \nonumber\\
    &=&
    - \bar{\lambda}^2 \frac{(E^\varphi)^2}{\cos^2(\bar{\lambda} K_\varphi)} \frac{{\rm d}^2 E^x}{{\rm d} t_\varphi{}^2}
    \ ,
    \label{eq:E^phi EoM2 - Homogeneous - DF}
\end{eqnarray}
\begin{eqnarray}
&&    \frac{{\rm d}^2 E^x}{{\rm d} t_\varphi{}^2} \label{eq:E^x EoM2 - Homogeneous - NMCS}\\
    &\approx&
    4 E^x \frac{\bar{\lambda}^2}{1+\bar{\lambda}^2} \left(
    1 + \sqrt{1 + \frac{4 G^2\left(1+\bar{\lambda}^2\right)}{c_\varphi^2}}
    \right)^2 \left( \left(
    1 + \sqrt{1 + \frac{4 G^2\left(1+\bar{\lambda}^2\right)}{c_\varphi^2}}
    \right)^2 + \frac{4 G^2 (1+\bar{\lambda}^2)}{c_\varphi{}^2} \right)^{-1}
    \ .
   \nonumber
\end{eqnarray}
\begin{eqnarray}
  &&  \frac{{\rm d}^2}{{\rm d} t_\varphi{}^2}
     \left(\frac{\sin(\bar{\nu}\phi)}{\bar{\nu}}\right)\nonumber \\
    &\approx&
    - 4 G \bar{\lambda} \sin \left(\bar{\lambda} t_{\varphi}\right) \frac{E^{\varphi}}{\cos \left(\bar{\lambda} t_{\varphi}\right)} \Bigg( \left(1 + \frac{\sin^2\left(\bar{\lambda} t_{\varphi}\right)}{\bar{\lambda}^2}\right) \frac{(E^{\varphi})^2}{\cos^2\left(\bar{\lambda} t_{\varphi}\right)} + G^2 \Bigg)^{-1}
    \nonumber\\
    &\approx&
    - 4 \bar{\lambda}^2 \frac{2 G}{c_\varphi}
    \left(
    1 + \sqrt{ 1 + \frac{4 G^2 \left(1+\bar{\lambda}^2\right)}{c_\varphi^2}}
    \right)
    \Bigg( \left(
    1 + \sqrt{1 + \frac{4 G^2 \left(1+\bar{\lambda}^2\right)}{c_\varphi^2}}
    \right)^2 + \frac{4 G^2 \left(1+\bar{\lambda}^2\right)}{c_\varphi{}^2} \Bigg)^{-1}
    \nonumber\\
    &=&
    - \frac{1+\bar{\lambda}^2}{E^x} \frac{2 G}{c_\varphi}
    \left( 1 + \sqrt{ 1 + \frac{4 G^2 \left(1+\bar{\lambda}^2\right)}{c_\varphi^2}} \right)^{-1}
    \frac{{\rm d}^2 E^x}{{\rm d} t_\varphi{}^2}
    \label{eq:phi EoM2 - Homogeneous - NMCS}
\end{eqnarray}
\label{eq:Equations of motion - Homogeneous - NMCS}
\end{subequations}
at second order.
It will also be useful to compute
\begin{eqnarray}
    \dot{q}^{xx} &=&
    q^{xx} \left( \frac{\dot{E}^x}{E^x}
    - \frac{\cos^2(\bar{\lambda} t_\varphi)}{(E^\varphi)^2} \frac{{\rm d} \left((E^\varphi)^2/\cos^2(\bar{\lambda} t_\varphi)\right)}{{\rm d} t_\varphi}
    - 4 \bar{\nu}^2 \frac{\tan(\bar{\nu}\phi)}{\bar{\nu} \cos (\bar{\nu}\phi)} \frac{{\rm d}}{{\rm d} t_\varphi} \left(\frac{\sin(\bar{\nu}\phi)}{\bar{\nu}}\right)\right)
    \nonumber\\
    &\approx& 0
\end{eqnarray}
and
\begin{eqnarray}
    \Ddot{q}^{xx} &\approx&
    q^{xx} \left( \frac{\Ddot{E}^x}{E^x}
    - \frac{\cos^2(\bar{\lambda} t_\varphi)}{(E^\varphi)^2} \frac{{\rm d}^2 \left((E^\varphi)^2/\cos^2(\bar{\lambda} t_\varphi)\right)}{{\rm d} t_\varphi{}^2}
    - 4 \bar{\nu}^2 \frac{\tan(\bar{\nu}\phi)}{\bar{\nu} \cos (\bar{\nu}\phi)} \frac{{\rm d}^2}{{\rm d} t_\varphi{}^2} \left(\frac{\sin(\bar{\nu}\phi)}{\bar{\nu}}\right)\right)
    \nonumber\\
    &\approx&
    (1+\bar{\lambda}^2) q^{xx} \frac{\Ddot{E}^x}{E^x} \Bigg[ 1
    + 4 \bar{\nu}^2 \frac{\tan(\bar{\nu}\phi)}{\bar{\nu} \cos (\bar{\nu}\phi)} \frac{2 G}{c_\varphi}
    \left( 1 + \sqrt{ 1 + \frac{4 G^2 \left(1+\bar{\lambda}^2\right)}{c_\varphi^2}} \right)^{-1}
    \Bigg]
    \ ,
\end{eqnarray}
which is finite.

Using (\ref{eq:Ricci scalar of homogeneous metric}) for the expression of the
Ricci scalar of a spatially homogeneous metric we find that it is finite at
the maximum-curvature hypersurface:
\begin{eqnarray}
    R &\approx&
    \left(
    \frac{2}{E^x}
    + 2 N^{-2} \frac{\Ddot{E}^x}{(E^x)^2}
    - N^{-2} \Ddot{q}^{xx}\right)
    \ .
    \label{eq:Ricci scalar of homogeneous metric - DF}
\end{eqnarray}
Thus,  the Ricci scalar is finite even in the presence of matter
and when $\phi = \pi / (2\bar{\nu})$.
In this limiting case, we obtain
\begin{eqnarray}
    R |_{t_\varphi=\pi/(2 \bar{\lambda}) ,\ \phi = \pi/(2 \bar{\nu})}
    &=& \frac{2}{E^x} \bigg|_{t_\varphi=\pi/(2 \bar{\lambda})}
    \ .
\end{eqnarray}
We conclude that the coordinate singularity at $\phi = \pi / (2\bar{\nu})$ is
due to the scalar field reaching its maximal value when or before $K_\varphi$
reaches its own maximum. In this case, $\phi$ would be a better indicator of
the transition if it were used as an internal time instead of $K_\varphi$.
The equations of motion with $\phi$ as the internal time are more complicated,
but the solution would be qualitatively similar, just replacing $K_\varphi$
with $\phi$ as the time coordinate.

\subsubsection{Bounded-curvature and bounded-scalar effects}

The solution (\ref{eq:phi - Homogeneous - DF}) can be inverted,
\begin{eqnarray}
    \frac{\sin(\bar{\lambda} K_\varphi)}{\bar{\lambda}} &=&
    - \frac{\sinh \left(\left(\sin (\bar{\nu}\phi)/\bar{\nu} -
    \sin (\bar{\nu}\phi_{\rm H})/\bar{\nu}\right) \sqrt{1 + c_\varphi^2/(4
                                                            G^2)}
                                                            \right)}{\sqrt{1+c_\varphi^2/4
                                                            G^2}} 
    \nonumber\\
    &=&
    - \frac{\sinh \left(\left(\sin (\bar{\nu}\phi)/\bar{\nu} -
    \sin (\bar{\nu}\phi_{\rm H})/\bar{\nu}\right) \sqrt{1 + M^2
        \lambda_0^2/(\mu^2 G^2)} \right)}{\sqrt{1+M^2 \lambda_0^2/(\mu^2
        G^2)}} 
    \ .
\end{eqnarray}
If the mass $M$ is considered to be supplied primarily by the scalar field, then
we must have $\mu |G| \sim M$.  Furthermore, the left-hand side of this
expression is bounded, implying the inequality
\begin{eqnarray}
    \frac{1}{\bar{\lambda}} &\gtrsim&
    \frac{1}{\sqrt{2}} \sinh \left(\frac{1}{\sqrt{2}}\frac{\sin (\bar{\nu}\phi)}{\bar{\nu}} -
    \frac{1}{\sqrt{2}}\frac{\sin (\bar{\nu}\phi_{\rm H})}{\bar{\nu}}\right)
    \ .
\end{eqnarray}
(The right-hand side is also bounded as a function of $\phi$, but the
gravitational bound is more universal as it may apply to multiple matter
fields, and it is more instructive regarding space-time singularities.)  The
maximum effect of the scalar field is achieved at
$\phi = \phi / (2 \bar{\nu}) \gg \phi_H$.  In this extreme case we have
\begin{eqnarray}
    \frac{1}{\bar{\lambda}} &\gtrsim&
    \frac{1}{\sqrt{2}} \sinh \left(\frac{1}{\sqrt{2} \bar{\nu}} -
    \frac{1}{\sqrt{2}} \frac{\sin (\bar{\nu}\phi_{\rm H})}{\bar{\nu}}\right)
    \ .
\end{eqnarray}
Since we expect $\bar{\lambda}, \bar{\nu} \ll 1$, we can approximate this
expression by
\begin{eqnarray}
    \frac{1}{\bar{\lambda}} &\gtrsim&
    \frac{1}{2\sqrt{2}} \exp \left( \frac{1}{\sqrt{2}\bar{\nu}}\right)
    \ .
\end{eqnarray}
This result imposes a theoretical limit on the value of $\bar{\nu}$ in terms
of $\bar{\lambda}$, as given by the specific dynamical solution to this model.
Since $\bar{\lambda}$ is then exponentially smaller than $\bar{\nu}$, we can
expect its effects to be in general much weaker. Non-classical matter
properties are therefore more pronounced in the extreme case of $\phi$
reaching its maximal value,  compared with gravitational effects,
in parallel with standard quantum effects that are usually more relevant for
matter than for gravity, as seen for instance in various applications of
quantum matter fields on a curved background in early-universe
cosmology. Scalar collapse into a black hole should therefore be a promising
line of research in emergent modified gravity.

\section{A new outlook on scalar-tensor theories}
\label{sec:Outlook}

We have demonstrated that there are many interesting and previously
unrecognized theories of spherically symmetric emergent modified gravity
coupled to a scalar field. This outcome suggests several new options for
scalar-tensor theories that may be useful for phenomenological studies in
astrophysics and cosmology. Our new theories do not go beyond the second-order
nature of field equations and do not encounter the Ostrogradski problem
\cite{OstrogradskiProblem}. In some cases, they have intriguing new features
such as the absence of physical singularities and, as shown in
\cite{HigherMOND}, make it possible to implement intermediate-scale
modifications of general relativity such as MOND.

A new challenge that so far has not been explored much, but could be the
origin of new and useful physical effects, is a possible dependence of the
emergent space-time metric on the scalar field. Such a dependence is not
always necessary but may be implied indirectly by additional physical
requirements, as demonstrated in our specific classes of modified theories. In
some of these cases, the same conditions also imply deviations of consistent
scalar-field couplings from minimal coupling to the emergent space-time
metric.

We found that physical conditions on the combined gravity-matter theory
sometimes rule out minimal coupling of a scalar field, as seen for instance in
\eqref{eq:Hamiltonian constraint - CML} for constraints compatible with the
classical-matter limit, where the matter terms
\begin{equation}
  \sqrt{q^{xx}} \left( \alpha_3P_\phi{}^2/E^x +
    \alpha_3^{-1}(\phi')^2E^x \right)+ \lambda_0^2 E^x \sqrt{q_{xx}} V_q
\end{equation}
must separate different dependencies on $E^x$ and $\phi$, such that $\alpha_3$
may depend only on $E^x$ and $\lambda_0$ on both $E^x$ and $\phi$. In terms of
the spatial part $q_{ab}$ of the emergent space-time metric, the factors of
$\sqrt{q^{xx}}/E^x=1/\sqrt{\det q}$ in the kinetic term, of
$\sqrt{q^{xx}}E^x=q^{xx}\sqrt{\det q}$ in the spatial-derivative term and of
$\sqrt{q_{xx}}E^x=\sqrt{\det q}$ are as expected for minimal coupling, even in
cases in which the structure function $q^{xx}$ depends on $K_{\varphi}$ and
$\phi$. However, all terms considered, we do not have minimal coupling unless
$\alpha_3=1$ and $\lambda_0=1$. Polymerization of the scalar field,
\eqref{eq:Hamiltonian constraint - CML - scalar polymerization - periodic},
then generates completely new terms in the modified Hamiltonian constraint,
such as those linear in $P_{\phi}$.

In other classes of modified constraints, minimal coupling is completely ruled
out, for instance in the constraints \eqref{eq:Hamiltonian constraint -
  periodic - CCSL} which requires a term of the form $P_{\phi}\phi'$ for any
modification with $\bar{\lambda}\not=0$, or in the singularity-free
constraints \eqref{eq:Hamiltonian constraint - DF} which have a simple
$1/\alpha_3$-modification of the $(\phi')^2$-term
\begin{equation}
  \frac{\bar{\lambda}\lambda_0}{\lambda} \frac{(E^x)^{3/2}}{E^{\varphi}} \frac{(\phi')^2}{\alpha_3}
\end{equation}
with the
classical-type metric factor $q_{\rm
  class}^{xx}\sqrt{\det q_{\rm class}}=(E^x)^{3/2}/E^{\varphi}$ {\em not} using the
emergent spatial metric,
but a more
complicated $P_{\phi}^2$-term
\begin{eqnarray}
&& \frac{\alpha_3\bar{\lambda}\lambda_0}{\lambda}\frac{P_{\phi}^2}{E^{\varphi}\sqrt{E^x}} \left( \left(c_f +
    \left(\frac{\bar{\lambda}(E^x)'}{2 E^{\varphi}}\right)^2\right) \cos^2 (\bar{\lambda} K_\varphi)
    - 2 \frac{\lambda}{\bar{\lambda}} q \bar{\lambda}^2 \frac{\sin (2
   \bar{\lambda} K_\varphi)}{2 \bar{\lambda}}  \right) \nonumber\\
  &=& \frac{\alpha_3\lambda}{\bar{\lambda}\lambda_0}  q^{xx}
      \frac{E^{\varphi}}{(E^x)^{3/2}} P_{\phi}^2=
      \frac{\alpha_3\lambda}{\bar{\lambda}\lambda_0} \frac{\sqrt{\det q_{\rm
      class}}}{\det q}
\end{eqnarray}
that makes use of a combination of the emergent and the classical spatial
metric.  The classical-type potential term
$(\lambda\lambda_0/\bar{\lambda})\sqrt{E^x}E^{\varphi}V$ in this case just
uses the classical volume element
$\sqrt{\det q_{\rm class}}=\sqrt{E^x}E^{\varphi}$ rather than the emergent
spatial metric, as in the $(\phi')^2$-term, but it has an extra factor of
$\lambda^2\bar{\lambda}^2$ compared with the latter, potentially
changing its $E^x$-dependence through $\lambda$. Moreover, there is a
possibility of two new scalar potentials in
\begin{equation}
  \lambda_0^2 \frac{E^x}{2} \sqrt{q_{xx}} V_q
    + \frac{\lambda^2}{\bar{\lambda}^2} \frac{(E^\varphi)^2}{2} \sqrt{q^{xx}}
    V^q
\end{equation}
that do make use of the emergent spatial metric $q_{xx}$, one with the
expected emergent
spatial volume element $E^x\sqrt{q_{xx}}$ and one with the combination
$(E^{\varphi})^2\sqrt{q^{xx}}= \det q_{\rm class}/\sqrt{\det q}$ of a
geometric mean of the two determinants. Some of these equations resemble
bimetric theories, but only for spatial metric tensors in non-standard
couplings in the constraints. These theories are not bimetric in the usual
meaning because only the emergent metric $q_{xx}$ has a consistent space-time
extension in our theories, but not the classical metric $q_{xx}^{\rm class}$.

So far, it remains unclear how emergent modified gravity could be constructed
explicitly without restrictions such as symmetry reduction. However, in cases
in which the emergent space-time metric does not depend on the scalar field, it is
possible to use a spherically symmetric modified solution as a
background for a non-spherical scalar field provided back-reaction can be
ignored. For a scalar-independent emergent space-time line element, the scalar
coupling can be minimal and derived from a standard action
\begin{equation}
    S [\phi] =  \frac{1}{2} \int {\rm d}^4 x \sqrt{-\det g}\left( g^{\mu \nu}
  (\nabla_\mu \phi) (\nabla_\nu \phi) + V(\phi) \right) 
\end{equation}
with a spherically symmetric emergent space-time metric $g_{\mu\nu}$ and a
non-spherical scalar field $\phi$. 
More generally, it is possible to use a scalar-dependent emergent space-time
metric as a spherically symmetric background for additional minimally coupled
scalar fields that do not back-react on the background and may be
non-spherical. The background can then be considered a scalar-tensor
description of space-time geometry, on which other matter scalar fields
evolve. If the emergent metric depends on the first scalar field, minimal
coupling of matter scalar fields then implies characteristic coupling terms
between all the scalar fields. There are therefore many new possibilities for
scalar-tensor theories and their phenomenology.

\section{Discussion}
\label{sec:Discussion}

We have extended emergent modified gravity in spherically symmetric
space-times by including a scalar matter field, suggesting several consistency
conditions for physically meaningful modifications of general relativity
coupled to a Klein--Gordon field.  Most importantly, we derived the condition
that the Hamiltonian constraint must satisfy for both gravity and the scalar
field to be covariant, given by equation~\eqref{eq:Spatial covariance
  condition - second reduced form}.  We studied implications of the
hypersurface-deformation brackets \eqref{eq:Hypersurface deformation algebra -
  spherical - Scalar field} and the specific covariance conditions
\eqref{eq:Covariance condition on K_x - modified - spherical2},
\eqref{eq:Covariance condition - modified - spherical2} and
\eqref{eq:Covariance condition on phi - modified - spherical} in the general
second-order Hamiltonian constraint \eqref{eq:Hamiltonian constraint ansatz}
for spherically symmetric models with a scalar field.  These conditions,
together with factoring out diffeomorphism-preserving canonical
transformations \eqref{eq:Diffeomorphism-constraint-preserving canonical
  transformations - Spherical - general}, completely determine the general
Hamiltonian constraint \eqref{eq:Hamiltonian constraint ansatz - Matter
  observable} and its structure function \eqref{eq:Covariant structure
  function} up to several free functions of the radial configuration variable
$E^x$ and the scalar field $\phi$. The structure function, together with a
lapse function according to gauge conditions or solutions of the equations of
motion, determines the emergent space-time metric \eqref{eq:ADM line element -
  spherical - modified2}.

As a new observation, the emergent space-time metric in general depends not
only on the gravitational phase-space degrees of freedom but also on the
scalar field through some of the free functions of a modified theory. This
unexpected feature is realized even at the kinematical level before any field
equations are solved, using only covariance conditions for the space-time line
element. While it is possible to assume that all free functions of a modified
Hamiltonian constraint that also appear in the emergent space-time metric are
independent of the matter field, this property is not generic and therefore
not representative of an effective theory of gravity coupled to scalar matter.
Moreover, we have shown in specific classes of modified theories that this
choice violates physically desirable conditions, mainly the existence of
certain limits and observables.  Therefore, if we view emergent modified
gravity as a collection of possible effective theories that can describe
covariant implications of quantum gravity, our result implies that a
quantum-gravity theory coupled to scalar matter cannot have a space-time
geometry derived solely from the fundamental gravitational degrees of freedom,
assumed to set up the canonical theory by a phase-space formulation. The
off-shell constraint system, rather than the kinematical phase space alone,
determines the meaning of gravity, geometry and matter.

This outcome presents a new viewpoint on possible implications of modified or
quantum gravity.  One of the most important features we have come to
understand from general relativity is that gravity is the geometry of
space-time, which may be dynamically affected by matter but does not directly
depend on the matter fields. Implicitly, higher-curvature or other traditional
effective actions use this observation as an assumption because they are built
on the basic statement that there is a space-time metric that directly appears
as a fundamental degree of freedom for a gravitational action, coupled in
different ways to matter fields.  Our result shows that this assumption is not
necessary, so far at least in spherically symmetric models, and rules out a
large class of emergent modified theories. The kinematical equivalence of
gravity and space-time geometry need no longer hold in quantum gravity,
depending on the quantization procedure: According to the examples of
\eqref{eq:Structure function - periodic - general - Scalar field
  polymerization}, \eqref{eq:Structure function - periodic - CCSL - scalar
  polymerized} and \eqref{eq:Structure function - DF - scalar polymerization}
space-time geometry is gravity {\em and} matter in particular in covariant
models with characteristic modifications suggested by loop quantum gravity.

For instance, if one computes the volume of a certain space-time region in
emergent modified gravity, one must know the gravitational field \emph{and}
the scalar field in that region. In practice, we would have two independent
measurements, one of the volume in terms of distances and one of the energy or
density of matter. In general relativity, volume measurements allow us to draw
conclusions about the metric in a given coordinate system, with a direct
connection with the gravitational field in this case. The same field appears
in energy or density expressions for matter, which allow us to compute the
values of matter fields from volume and density measurements. In emergent
modified gravity, however, the metric and density depend non-trivially on both
the gravitational and matter fields. Extracting the field values from
measurements is therefore a more involved procedure. The new property also
implies that field equations for matter are more challenging even for a free
field without self-interactions and if dynamical back-reaction on the
gravitational degrees of freedom is ignored. If metric coefficients in the
field equations depend not only on a background gravitational fields but also
on the matter field, even free-field equations on a background are non-linear.

Conceptually, the result is a step towards unification of gravity and matter,
given by a relational theory in which space-time is an emergent concept
derived from the fundamental fields on phase space. Space-time becomes
identical with gravity only in the vacuum limit, but even in this case the
emergent metric depends non-trivially on both the configuration and momentum
degrees of freedom of gravity. The dependence simplifies to the well-known
configuration dependence of general relativity only in the complete classical
limit of gravity and matter.

In some cases, we were able to obtain complete solutions of the field
equations, but given the complexity of gravity-matter coupling in this
framework, this was possible only in the simple (yet instructive) case of a
space-time slicing that allows spatially homogeneous fields. In this setting,
we found that different classes of scalar couplings in emergent modified
gravity imply different conclusions about the fate of classical
singularities. In specific examples, we demonstrated instability of vacuum results
about singularity avoidance under matter perturbations, while one new class was
able to maintain a singularity-free homogeneous behavior even in the presence
of unrestricted matter.

Given the vast set of new covariant theories in spherical symmetry, many
physical implications can now be explored. It remains to be seen how covariant
modified gravity, for instance with terms such as point holonomies or partial
Abelianizations motivated by loop quantum gravity, describes cosmological
inhomogeneity in an expanding universe, the collapse of matter into a black
hole, a modified form of Hawking radiation in models of black-hole
evaporation, or critical properties of gravitational collapse studied numerically in
\cite{LoopCollapse} using a model now known to violate covariance \cite{NonCovPol}. We
expect that the kinematical dependence of space-time on the scalar field will
imply new and previously unforeseen challenges to these questions, such as a
suitable treatment of Hawking radiation.

\section*{Acknowledgements}

We thank Asier Alonso-Bardaji and David Brizuela for discussions. This work
was supported in part by NSF grant PHY-2206591.


\begin{thebibliography}{10}

\bibitem{MultiMess1}
LIGO Scientific, Virgo, Fermi GBM, INTEGRAL, IceCube, IPN, Insight-Hxmt,
  ANTARES, Swift, Dark Energy~Camera GW-EM, DES, DLT40, GRAWITA, Fermi-LAT,
  ATCA, ASKAP, OzGrav, DWF (Deeper Wider~Faster Program), AST3, CAASTRO,
  VINROUGE, MASTER, J-GEM, GROWTH, JAGWAR, CaltechNRAO, TTU-NRAO, NuSTAR,
  Pan-STARRS, KU, Nordic~Optical Telescope, ePESSTO, GROND, Texas~Tech
  University, TOROS, BOOTES, MWA, CALET, IKI-GW Follow-up, H.E.S.S., LOFAR,
  LWA, HAWC, Pierre Auger, ALMA, Pi~of~Sky, DFN, ATLAS Telescopes, High Time
  Resolution~Universe Survey, RIMAS, RATIR, SKA South~Africa/MeerKAT
  Collaborations, AstroSat Cadmium Zinc Telluride~Imager Team, AGILE Team, 1M2H
  Team, Las Cumbres~Observatory Group, MAXI Team, TZAC Consortium, SALT Group,
  Euro~VLBI Team, and Chandra~Team at~McGill University (Abbott B.~P.\~et al.),
\newblock Multi-messenger Observations of a Binary Neutron Star Merger,
\newblock {\em Astrophys.\ J.} 848 (2017) L12

\bibitem{MultiMess2}
LIGO Scientific, Virgo, Fermi-GBM, and INTEGRAL Collaborations (Abbott B.~P.\
  et~al.),
\newblock Gravitational Waves and Gamma-Rays from a Binary Neutron Star Merger:
  GW170817 and GRB 170817A,
\newblock {\em Astrophys.\ J.} 848 (2017) L13

\bibitem{MultiMess3}
D.~A.\ Coulter, R.~J.\ Foley, C.~D.\ Kilpatrick, M.~A.\ Drout, A.~L.\ Piro,
  B.~J.\ Shappee, M.~R.\ Siebert, J.~D.\ Simon, N.\ Ulloa, D.\ Kasen, B.~F.\
  Madore, A.\ Murguia-Berthier, Y.-C.\ Pan, J.~X.\ Prochaska, E.\ Ramirez-Ruiz,
  A.\ Rest, and C.\ Rojas-Bravo,
\newblock Swope Supernova Survey 2017a (SSS17a), the optical counterpart to a
  gravitational wave source,
\newblock {\em Science} 358 (2017) 1556--1558

\bibitem{MultiMess4}
A.\ Murguia-Berthier, E.\ Ramirez-Ruiz, C.~D.\ Kilpatrick, R.~J.\ Foley, D.\
  Kasen, W.~H.\ Lee, A.~L.\ Piro, D.~A.\ Coulter, M.~R.\ Drout, B.~F.\ Madore,
  B.~J.\ Shappee, Y.-C.\ Pan, J.~X.\ Prochaska, A.\ Rest, C.\ Rojas-Bravo,
  M.~R.\ Siebert, and J.~D.\ Simon,
\newblock A Neutron Star Binary Merger Model for GW170817/GRB 170817A/SSS17a,
\newblock {\em Astrophys.\ J.} 848 (2017) L34

\bibitem{HigherCov}
M.\ Bojowald and E.~I.\ Duque,
\newblock Emergent modified gravity: Covariance regained,
\newblock {\em Phys.\ Rev.\ D} 108 (2023) 084066, [arXiv:2310.06798]

\bibitem{SphSymmMinCoup}
A.\ Alonso-Bardaj\'{\i} and D.\ Brizuela,
\newblock Spacetime geometry from canonical spherical gravity,
  [arXiv:2310.12951]

\bibitem{ReviewEff}
M.\ Bojowald,
\newblock Quantum Cosmology: Effective Theory,
\newblock {\em Class.\ Quantum Grav.} 29 (2012) 213001, [arXiv:1209.3403]

\bibitem{LoopRep}
C.\ Rovelli and L.\ Smolin,
\newblock Loop Space Representation of Quantum General Relativity,
\newblock {\em Nucl.\ Phys.\ B} 331 (1990) 80--152

\bibitem{SphSymmCov}
M.\ Bojowald, S.\ Brahma, and J.~D.\ Reyes,
\newblock Covariance in models of loop quantum gravity: Spherical symmetry,
\newblock {\em Phys.\ Rev.\ D} 92 (2015) 045043, [arXiv:1507.00329]

\bibitem{GowdyCov}
M.\ Bojowald and S.\ Brahma,
\newblock Covariance in models of loop quantum gravity: Gowdy systems,
\newblock {\em Phys.\ Rev.\ D} 92 (2015) 065002, [arXiv:1507.00679]

\bibitem{SphSymmMatter}
A.\ Alonso-Bardaj\'{\i} and D.\ Brizuela,
\newblock Holonomy and inverse-triad corrections in spherical models coupled to
  matter,
\newblock {\em Eur.\ Phys.\ J.\ C} 81 (2021) 283, [arXiv:2010.14437]

\bibitem{SphSymmMatter2}
A.\ Alonso-Bardaj\'{\i} and D.\ Brizuela,
\newblock Anomaly-free deformations of spherical general relativity coupled to
  matter,
\newblock {\em Phys.\ Rev.\ D} 104 (2021) 084064, [arXiv:2106.07595]

\bibitem{EffLine}
M.\ Bojowald, S.\ Brahma, and D.-H.\ Yeom,
\newblock Effective line elements and black-hole models in canonical (loop)
  quantum gravity,
\newblock {\em Phys.\ Rev.\ D} 98 (2018) 046015, [arXiv:1803.01119]

\bibitem{LapseGauge}
J.~M.\ Pons, D.~C.\ Salisbury, and L.~C.\ Shepley,
\newblock Gauge transformations in the Lagrangian and Hamiltonian formalisms of
  generally covariant theories,
\newblock {\em Phys.\ Rev.\ D} 55 (1997) 658--668, [gr-qc/9612037]

\bibitem{CUP}
M.\ Bojowald,
\newblock {\em Canonical Gravity and Applications: Cosmology, Black Holes, and
  Quantum Gravity},
\newblock Cambridge University Press, Cambridge, 2010

\bibitem{SymmRed}
M.\ Bojowald and H.~A.\ Kastrup,
\newblock Symmetry Reduction for Quantized Diffeomorphism Invariant Theories of
  Connections,
\newblock {\em Class.\ Quantum Grav.} 17 (2000) 3009--3043, [hep-th/9907042]

\bibitem{SphSymm}
M.\ Bojowald,
\newblock Spherically Symmetric Quantum Geometry: States and Basic Operators,
\newblock {\em Class.\ Quantum Grav.} 21 (2004) 3733--3753, [gr-qc/0407017]

\bibitem{SphSymmHam}
M.\ Bojowald and R.\ Swiderski,
\newblock Spherically Symmetric Quantum Geometry: Hamiltonian Constraint,
\newblock {\em Class.\ Quantum Grav.} 23 (2006) 2129--2154, [gr-qc/0511108]

\bibitem{ReggeTeitelboim}
T.\ Regge and C.\ Teitelboim,
\newblock Role of surface integrals in the Hamiltonian formulation of general
  relativity,
\newblock {\em Ann.\ Phys.} 88 (1974) 286--318

\bibitem{SphSymmEff}
A.\ Alonso-Bardaj\'{\i}, D.\ Brizuela, and R.\ Vera,
\newblock An effective model for the quantum Schwarzschild black hole,
\newblock {\em Phys.\ Lett.\ B} 829 (2022) 137075, [arXiv:2112.12110]

\bibitem{SphSymmEff2}
A.\ Alonso-Bardaj\'{\i}, D.\ Brizuela, and R.\ Vera,
\newblock Nonsingular spherically symmetric black-hole model with holonomy
  corrections,
\newblock {\em Phys.\ Rev.\ D} 106 (2022) 024035, [arXiv:2205.02098]

\bibitem{EmergentFluid}
E.~I.\ Duque,
\newblock Emergent modified gravity: The perfect fluid and gravitational
  collapse, [arXiv:2311.08616]

\bibitem{LoopSchwarz}
R.\ Gambini and J.\ Pullin,
\newblock Loop quantization of the Schwarzschild black hole,
\newblock {\em Phys.\ Rev.\ Lett.} 110 (2013) 211301, [arXiv:1302.5265]

\bibitem{LoopSchwarz2}
R.\ Gambini and J.\ Pullin,
\newblock Hawking radiation from a spherical loop quantum gravity black hole,
\newblock {\em Class.\ Quant.\ Grav.} 31 (2014) 115003, [arXiv:1312.3595]

\bibitem{HigherMOND}
M.\ Bojowald and E.~I.\ Duque,
\newblock MONDified gravity,
\newblock {\em Phys.\ Lett.\ B} 847 (2023) 138279, [arXiv:2310.19894]

\bibitem{MOND1}
M.\ Milgrom,
\newblock A modification of the Newtonian dynamics-Implications for galaxies,
\newblock {\em Ap.\ J.} 270 (1983) 371--383

\bibitem{MONDRotation}
M.\ Milgrom and E.\ Braun,
\newblock The rotation curve of DDO 154-A particularly acute test of the
  modified dynamics,
\newblock {\em Ap.\ J.} 334 (1988) 130--133

\bibitem{MOND2}
S.~S.\ McGaugh and W.\ De~Blok,
\newblock Testing the hypothesis of modified dynamics with low surface
  brightness galaxies and other evidence,
\newblock {\em Ap.\ J.} 499 (1998) 66

\bibitem{FermionHiggs}
T.\ Thiemann,
\newblock Kinematical Hilbert Spaces for Fermionic and Higgs Quantum Field
  Theories,
\newblock {\em Class.\ Quantum Grav.} 15 (1998) 1487--1512, [gr-qc/9705021]

\bibitem{AshVar}
A.\ Ashtekar,
\newblock New Hamiltonian Formulation of General Relativity,
\newblock {\em Phys.\ Rev.\ D} 36 (1987) 1587--1602

\bibitem{AshVarReell}
J.~F.\ Barbero~G.,
\newblock Real Ashtekar Variables for Lorentzian Signature Space-Times,
\newblock {\em Phys.\ Rev.\ D} 51 (1995) 5507--5510, [gr-qc/9410014]

\bibitem{Regained}
S.~A.\ Hojman, K.\ Kucha\v{r}, and C.\ Teitelboim,
\newblock Geometrodynamics Regained,
\newblock {\em Ann.\ Phys.\ (New York)} 96 (1976) 88--135

\bibitem{LagrangianRegained}
K.~V.\ Kucha\v{r},
\newblock Geometrodynamics regained: A Lagrangian approach,
\newblock {\em J.\ Math.\ Phys.} 15 (1974) 708--715

\bibitem{KucharHypI}
K.~V.\ Kucha\v{r},
\newblock Geometry of hypersurfaces. I,
\newblock {\em J.\ Math.\ Phys.} 17 (1976) 777--791

\bibitem{KucharHypII}
K.~V.\ Kucha\v{r},
\newblock Kinematics of tensor fields in hyperspace. II,
\newblock {\em J.\ Math.\ Phys.} 17 (1976) 792--800

\bibitem{KucharHypIII}
K.~V.\ Kucha\v{r},
\newblock Dynamics of tensor fields in hyperspace. III,
\newblock {\em J.\ Math.\ Phys.} 17 (1976) 801--820

\bibitem{OstrogradskiProblem}
R.~P.\ Woodard,
\newblock Avoiding Dark Energy with $1/R$ Modifications of Gravity,
\newblock {\em Lect.\ Notes Phys.} 720 (2007) 403--433, [astro-ph/0601672]

\bibitem{LoopCollapse}
F.\ Ben\'{\i}tez, R.\ Gambini, L.\ Lehner, S.\ Liebling, and J.\ Pullin,
\newblock Critical collapse of a scalar field in semiclassical loop quantum
  gravity,
\newblock {\em Phys.\ Rev.\ Lett.} 124 (2020) 071301, [arXiv:2002.04044]

\bibitem{NonCovPol}
M.\ Bojowald,
\newblock Non-covariance of ``covariant polymerization'' in models of loop
  quantum gravity,
\newblock {\em Phys.\ Rev.\ D} 103 (2021) 126025, [arXiv:2102.11130]

\end{thebibliography}

\end{document}